\documentclass[12pt]{iopart} 
\usepackage{epsfig} 
\def\bit{\begin{itemize}} 
\def\fit{\end{itemize}} 
\def\beq{\begin{equation}} 
\def\feq{\end{equation}} 
\def\beqar{\begin{eqnarray}}
\def\feqar{\end{eqnarray}}
\def\benu{\begin{enumerate}} 
\def\fenu{\end{enumerate}} 

\begin{document} 
\sloppy   
\newpage  

\title[Author guidelines for IOPP journals]
{Solar Neutrino Physics: historical evolution, present status and perspectives}  

\author{Lino Miramonti\footnote[1]{e-mail:miramonti@mi.infn.it}} 
\address{Dipartimento di Fisica dell'Universita' di Milano and\\ 
Istituto Nazionale di Fisica Nucleare di Milano}  

\author{Franco Reseghetti\footnote[2]{e-mail: reseghetti@santateresa.enea.it}  
                         \footnote[3]{now at ENEA-CRAM, Pozzuolo di Lerici}  } 
\address{Istituto Nazionale di Fisica Nucleare di Milano}  

\begin{abstract} 
Solar neutrino physics is an exciting and difficult field of research for physicists,  
where astrophysics, elementary particle and nuclear physics meet. \\
The Sun produces the energy that life has been using on Earth for many years,  
about $10^9$ y, emits a lot of particles: protons, electrons, ions, electromagnetic quanta... 
among them neutrinos play an important role allowing to us to check our knowledge on solar 
characteristics. \\ 
The main aim of this paper is to offer a practical overview of various aspects concerning the solar 
neutrino physics: after a short historical excursus, the different detection techniques and 
present experimental results and problems are analysed. Moreover, the status of art of solar modeling, 
possible solutions to the so called solar neutrino problem (SNP) and planned detectors are reviewed.  
\end{abstract}  

\newpage  
\tableofcontents  
\newpage  
\maketitle 

\section {Introduction.}
\label{sect:intro} 
The Sun, "our own" star, has been shining for many years (last data, based on  meteoritic age, are suggesting 
$\sim$ 4.6$\cdot 10^9$ y) so it is a fundamental question to find where the source of this energy is. \\ 
At the beginning of XIX$^{th}$ century it was supposed the  gravitational energy was the true source of solar energy. 
In this case the Sun would  have a very short life because the total amount of gravitational energy is $\sim 10^{41}$ J: 
if the presently measured solar luminosity ($L_\odot\sim 10^{26}$ W) has been constant during the Sun's life 
we obtain as a solar living time  
$\sim 10^7$ y, a value too short to allow any biological evolution. Different proposed models including chemical 
or other reactions gave  very non-physical results (the Sun would shine for $\sim 10^4$ y).\\ 
To obtain right figures we must consider the relativistic energy of the star: in this case one finds for the 
Sun a life of $\approx 10^{9}-10^{10}$ y. The nuclear forces allow a 
right analysis of this problem but   scientists knew their existence only starting from 1920.\\
Immediately before 1940 a  realistic 
mechanism of energy production was suggested after the discovery of  the quantum mechanical tunnel effect, the 
formulation of 
$\beta$-decay theory  and the "creation" of a particle, the neutrino, invented to preserve the  conservation laws. 
H.A.Bethe, \cite{BET38a,BET38b,BET39}, realised a model in which the  solar energy is produced by thermonuclear 
reactions 
(specifically by the fusion  of 4 atoms of H in He) {\it {via}} the so called p-p chain or the CNO cycle. \\ 
The confirmation of the robustness of this theory and of all its following developments was acquired only after 1968 by 
direct detection of solar  neutrinos, $\nu_\odot$'s hereafter. At the same time an interesting problem remained: 
$\nu_\odot$  flux showed a deficit with respect to the predictions of the solar models (SMs).
This is known as solar neutrino problem, SNP.\\ 
Up to now a robust and right answer seems to be out of our knowledge even if experimental results are suggesting a 
possible explanation based on new neutrino properties. \\
The number of articles and books concerning the Sun and $\nu_\odot$'s is huge therefore it  is hard to list or 
select them: we underline \cite{BAH88,BAH95,SCI96,GUE97,TUR98a,BRU99, FIO00,BAH01a,COU02} with respect to the Sun 
and helioseismological features while we  remember \cite{BAH89,BAH92,TUR93,BAH95,RIC96,CAS97,BAH98a,BAH01a,BAH02f} in 
$\nu_\odot$ sector. For a complete analyses see also \cite{BET39,BUR57,PON67,CLA68,FOW84,ROL88,RAF96} while 
\cite{CRE93} presents both theoretical and experimental aspects concerning $\nu_\odot$ physics at the beginning of 90s'.  
For a nice and detailed review concerning experimental aspects and perspectives in neutrino physics 
see also \cite{BET01}.\\   
We do not include images and sketches concerning detectors: we refer the interested reader to the available WEB-pages
of running and planned experiments that we have listed in Appendix.\\ 
In short, the organisation of this article is the following: 
sect. 2 shows a journey through the evolution of knowledge on how the Sun works, on neutrinos and on $\nu_\odot$'s; 
the main features of SMs and of $\nu_\odot$'s production are shown in sect. 3; 
in sect. 4 methods to detect $\nu_\odot$'s are exposed; in sect. 5  present experimental results are reviewed; 
in sect. 6 solar neutrino problems are detailed while in sect. 7 proposed solutions are analysed; 
in sect. 8 the aim  and the features of the next starting detectors are presented; in sect. 9 the  
characteristics of the proposed next generation of detectors are listed; 
sect. 10 sums  up the perspectives in $\nu_\odot$ physics at the beginning of this century.\\
This paper was prepared just before the communication of the first experimental result from KAMLAND detector
which seems to exclude all solutions to the SNP but the so called Large Mixing Angle (LMA) solution. We mention
this conclusion only in section 2. We refer to the "Neutrino Unbound" WEB page, see. Appendix, for a complete list
of published papers on the subjet.
\section{Historical development.} 
\label{sect:history}
A list of some steps of the knowledge on solar characteristics and on neutrino properties
follows: its main aim is to underline how difficult was the evolution toward a model well reproducing 
phenomena related to the Sun, which is the nearest star. 
\bit 
\item 
In the 1840's, Mayer and Waterson suggested the conversion of 
gravitational energy into heat as a probable source of solar radiation. 
\item 
In 1854, von Helmholtz, \cite{HEL56}, 
officially proposed the gravitational  contraction of a mass as the origin of the energy emitted by the Sun. 
\item 
In 1859, C.Darwin ,\cite{DAR59}, made an estimation of the age of the Earth by  erosion of some valley and the 
evolution of biological species: 3$\cdot 10^8$ y. 
\item 
Kelvin supposed that the impact of meteors on solar surface was the true mechanism of energy production; 
in 1862 he estimated the Sun is not older than 2$\cdot 10^7$ y and ruled out a chemical solution to this problem 
($\approx 10^4$ y is the allowed age), \cite{KEL62}. Kelvin also calculated the lifetime of an object as massive as 
our Sun and with the same radius which radiates the same amount of energy produced by gravitational contraction: he 
found $\approx 10^7$ y. 
\item 
In this period, the theoretical physicists were not able to find processes  other than the previous ones so 
that there was a big challenge to reconcile  the physically estimated value ($\approx 10^7$ y) and the 
biologists  and geologists which proposed $\approx 10^8$ y. 
\item 
In 1899, E.Rutherford experimentally discovered the $\beta$-decay. 
\item 
In 1903, P.Curie and Laborde discovered the radium salt to be an emitter of heat constant in time without its 
cooling down to the surrounding medium temperature. G.Darwin and Wilson immediately proposed the  radioactivity as 
a probable solar energy source. 
\item 
In 1904, E.Rutherford, not much time after the discovery of the emitted  $\alpha$ particle energy, 
strengthened that hypothesis but no astronomical observational data indicated an abundance of radioactive elements in 
the Sun as high  as required to reproduce the present $L_\odot$. Furthermore the rate of radioactive  
emission is independent from stellar temperature but the experimental data  showed the opposite behaviour. 
\item 
In 1905, A.Einstein, \cite{EIN05}, presented his theory of special relativity  and the famous equation 
$E\,=\,mc^2$ which allows a conversion of mass in  energy and $\it{viceversa}$. 
\item 
In 1911-12, O.Von Bayer, O.Hahn and L.Meitner, \cite{BAY11a,BAY11b,BAY12}, measured the  spectrum of 
electron emitted in $\beta$ radioactivity and found it to be discrete. 
\item 
In 1914, J.Chadwick, \cite{CHA14}, from experimental measurements on Radium E, deduced the  electron 
energy spectrum to be continuous. This was an unexpected result because $\alpha$ and $\gamma$ radioactivities 
showed mono-energetic particles. It was even
proposed an emission of more than one electron in each decay. 
\item 
In 1919, H.N.Russell, \cite{RUS19}, supposed that the physical process which allows  the Sun to shine is 
strictly related to T$_c$, the temperature of the core of  the star, and it yields a substantial stability 
over long periods of time. 
\item 
In 1920, F.W.Aston, \cite{AST20}, measured the mass of a lot of elements; in addition   
he discovered that "{\it{..an $\alpha$ particle (a He nucleus) is lighter
than 4 protons}}". Then, A.S.Eddington, \cite{EDD20a,EDD20b}, suggested that the Sun could shine  by transforming 
protons in helium; the mass difference should be converted in  energy. At this point he estimated Sun lifetime: 
$\approx 10^{11}$ y. 
\item 
In 1927, C.D.Ellis and W.A.Wooster, \cite{ELL27}, by a calorimetric measurement, deduced the mean value of 
energy liberated in $\beta$-decay of radium-E: 350 $\pm$ 40 keV; this result was confirmed in 1930 by L.Meitner: 
337 $\pm$ 20  keV, \cite{MEI30}. Moreover, no $\gamma$ emission was detected. A problem arose: the maximum energy 
for that spectrum was 1.05 MeV and the electron showed a continuum  energy spectrum. To solve this problem N.Bohr 
suggested that in microphysics the energy 
is conserved only at statistical level; he also supposed this process as a possible source of energy emitted in stars.  
\item 
In 1928, G.Gamow, \cite{GAM28}, discovered that two nuclei interact with a non-zero  probability; he 
introduced the so called "Gamow factor" for a technical  description of this process which is allowed only 
by quantum wave mechanics. 
\item  
In 1929, Atkinson and Houtermans, \cite{ATK29}, gave a first estimation of the  nuclear reactions rate 
that the temperature in the stellar core allows. 
\item 
In 1930, W.Pauli, \cite{PAU30}, as a "{\it{desperate expedient for saving the  WECHSELSATZ of statistics 
and energy conservation}}" in $\beta$-decays in which electrons are emitted, invented a neutral particles, 
having spin 1/2, obeying the exclusion principle, with a mass much lower than the proton mass. Then, He  
supposed such a particle to be always confined inside a nucleus.  
\item 
In 1932, J.Chadwick discovered the neutron, a neutral nuclear particle  having a proton-like mass. 
\item 
In 1933, F.Perrin, \cite{PER33}, supposed both the momenta of the electron and  the Pauli's particle 
to be equal and deduced a mass much lower than the electron one, the null value also included; he thought 
this particle as a photon with half-integer spin. In 1933-34 E.Fermi, \cite{FER33,FER34a,FER34b},  included this 
particle, which he re-called "neutrino",  in his $\beta$-decay theory. The electron energy spectrum is strongly  
neutrino-mass dependent and from the radium E spectrum he suggested for $\nu$ mass a value much lower than the 
electron mass. The sensitivity to $\nu$ mass in the  $\beta$-decay arises clearly because the larger $m_\nu$ 
the less available kinetic  energy remains for the decay products, and hence the maximum electron energy is 
reduced. Pauli changed his opinion about the localisation of the neutrino. 
\item 
In 1934, H.A.Bethe and R.Peierls, \cite{BET34}, using the Fermi weak-interaction Hamiltonian and the 
Fermi coupling $G_F$, as estimated from radioactive  elements, computed the cross-section for the 
inverse $\beta$-decay: $\sigma\,\sim\,2.3\cdot 10^{-44}\,cm^2\,(\frac {p_eE_e}{m_e^2})$. This result 
astonished both the authors which concluded "{\it{..This meant that one obviously would never be able 
to see a neutrino.}}" \\ 
As a comment if one calculates the mean free path in water for a 1-2 MeV neutrino,  a typical energy when 
a neutrino is emitted in a $\beta$-decay process, the result is $\lambda = \frac {1}{n \sigma} 
\sim 2.5\cdot 10^{18}$ m, a "swimming-pool" as long as  the thickness of the Galactic disc. 
\item 
In 1936, G.Gamow and E.Teller, \cite{GAM36}, modified the Fermi Hamiltonian, where only vector currents 
are present, to a more general operator involving scalar, vector, axial vector, pseudoscalar and tensor currents but 
five different  coupling constants were needed to fit the experimental data. N.Bohr changed his  
opinion about the energy non-conservation in $\beta$-decay process. H.Bethe wrote  "{\it{..It seems therefore 
probable that the neutrino does not have any magnetic moment at all...The recoil of the nucleus, which can 
be observed in principle, will decide  definitely between the hypothesis of non-conservation of energy and 
the neutrino hypothesis}}", \cite{BET36}. 
\item 
In 1937, E.Majorana, \cite{MAJ37}, found that for neutral particles there was  "{\it{..no more any reason
to presume the existence of antiparticles}}" and "{\it{..it was possible to modify the theory of $\beta$ 
emission, both positive and negative, so that it came always associated with the emission of a neutrino}}". 
\item 
In 1938, von Weizs\"{a}cker, \cite{WEI37,WEI38}, discovered the CNO cycle, in  which protons are "burned", 
the carbon nuclei acting as a catalyst. In 1938-39 H.A.Bethe, \cite{BET38a,BET38b,BET39}, realized a set of papers 
reviewing the knowledge on nuclear reactions in the stars. He also derived the basic  nuclear process in stars not 
greater than the 
Sun: the proton-proton  reaction chain, p-p. The deduced $L_\odot$ and other parameters were in agreement  with the 
observational data available at that time. L.Alvarez firstly detected electronic captures, 
\cite{ALV38}: he showed that $\beta$-emissions and electrons are the  same particles. 
\item 
In 1939, H.R.Crane made the first radiochemical experiment by measuring $^{35}$Cl  $\rightarrow ^{35}$S: 
he detected no signal and deduced an upper limit on cross-section, \cite{CRA39a}. H.R.Crane and J.Halpern, \cite{CRA39b}, 
found an experimental qualitative evidence  of the emission of a third particle in $\beta$-decay. 
\item 
In 1942-46, S.Sakata and K.Inoue, \cite{SAK42,SAK46}, proposed the so called  "two meson theory" which 
claims the existence of another pair of leptons.  
\item 
In 1946, B.Pontecorvo, \cite{PON46}, suggested Cl or Br as useful chemical elements to detect neutrino; he
also wrote "{\it{.. direct proof of the existence of the neutrino...must be based on experiments, 
the interpretation of which does not require the law of the conservation of energy, i.e. an experiment in 
which some characteristic process produced by free neutrinos...is observed}}".   
\item 
In 1948, H.R.Crane proposed different physical processes to experimentally detect $\nu$'s,  \cite{CRA48}: 
\benu 
\item electromagnetic interaction by magnetic moment   
\item inverse reaction of electron capture 
\item (in present terms) coherent diffusion by neutral current interaction 
\item nuclear excitation followed by $\gamma$ emission or fission 
\item mesons production at energies as high as cosmic rays have 
\fenu 
\item 
In 1949, L.Alvarez, \cite{ALV49}, presented a list of background reactions, the main  experimental challenge 
in neutrino detection. 
\item 
After the second world war, many physicists and astrophysicists developed  theoretical aspects: 
Epstein, \cite{EPS50}, Salpeter, \cite{SAL52,SAL54,SAL57b}, Frieman and  Motz, \cite{FRI53}, 
upgraded and completed the 
Bethe's work.  Moreover, T.D.Lee and C.N.Yang, \cite{LEE56}, proposed the parity violation in weak interactions; 
as a consequence all weak interacting $\nu$'s are left-handed and  $\overline{\nu}$'s are right-handed. 
\item  
In 1952-53, Langer and Moffat, \cite{LAN52}, and Hamilton {\it{et al.}}, \cite{HAM53}, gave an upper limit 
to $\nu$ mass by measuring the Tritium $\beta$-decay: $m_\nu \leq$ 250 eV. C.L.Cowan and F.Reines, 
\cite{COW53,REI53,COW56}, discovered $\overline{\nu_e}$'s  through the inverse $\beta$-decay using a 
scintillator near a nuclear reactor.  In 1955 R.Davis, \cite{DAV55}, studied the feasibility of Pontecorvo's 
proposal by setting a tank of cleaning fluid outside a nuclear reactor. 
\item   
In 1957-58, B.Pontecorvo, \cite{PON57,PON58}, following the description given by M.Gell-Mann  
and A.Pais to $K^0$-$\overline {K^0}$ system, \cite{GEL55}, suggested  that a $\overline {\nu}$ produced 
in a reactor could oscillate into a $\nu$  and be detected by a detector such as the Davis's one. He 
supposed "{\it{..If 2-component $\nu$ theory should turn out to be incorrect, which at present seems to 
be rather  improbable, and if conservation law of $\nu$ charge would not apply then 
$\nu \rightarrow \overline {\nu}$ transitions could take place in vacuo}}". He also proposed to define mixed 
particles as $\nu = 2^{-1/2}(\nu_1 + \nu_2)$ and $\overline{\nu} =  2^{-1/2}(\nu_1 - \nu_2)$, 
where $\nu_1$ and $\nu_2$ (which are truly neutral Majorana particles) are mass eigenstates. C.S.Wu {\it{et al.}}, 
\cite{WUC57},
experimentally  discovered the parity violation in the $\beta$-decay. L.D.Landau, T.D.Lee and C.N.Yang, A.Salam  
independently proposed a theory  of two-component $\nu$, \cite{LAN57,LEE57,SAL57a}. As a toy comment let 
us remember that if $\nu$'s were to  look into a mirror they would be unable to see their reflected images. 
\item 
In 1958, M.Goldhaber, \cite{GOL58}, measured the $\nu$ helicity: he observed the K-electron capture in $^{152}$Eu 
which produces $^{152}$Sm$^*$ and a $\nu$. \item In 1959 R.Davis communicated a possible detection 
of the reaction $\overline {\nu_e} + ^{37}Cl \rightarrow e^- + ^{37}Ar$ which is allowed only  if lepton number is 
not conserved. 
\item 
In 1962, R.Leighton, \cite{LEI62}, analysed "solar vibrations" starting helioseismological studies. G.Danby et 
al., \cite{DAN62}, experimentally detected  $\nu_\mu$'s, the particles foreseen in the old Sakata's theory. 
\item In 1962-63 S.Sakata and his collaborators proposed a model including two  kinds of $\nu$'s, 
\cite{MAK62,NAK63a,NAK63b,OKO63}. Their main assumptions were: 
\benu 
\item 
$\nu$'s should be  4-component spinors in order to be the seeds of the massive baryons so that $\nu_1$ and 
$\nu_2$ should have their own masses. These are the TRUE $\nu$'s. 
\item 
$\nu_e$ and $\nu_\mu$ (the WEAK $\nu$'s), which are coupled to $e$ and  $\mu$ in the weak currents, 
should be mixing states of $\nu_1$ and $\nu_2$, as many people usually think today. 
\fenu 
\item 
In 1963-64, a Californian astrophysicist group (Bahcall, Fowler, Iben and Sears) realised a first 
solar model in realistic agreement with observations, \cite{BAH63,SEA64}. 
\item 
In 1964, J.N.Bahcall and R.Davis, \cite{BAH64,DAV64}, proposed to build an  underground Cl detector 
searching for $\nu_\odot$. The fascinating motivation they gave was "{\it{..to see into the interior of a star 
and thus verify directly the hypothesis of nuclear energy generation in stars}}".  
V.A.Kuzmin, \cite{KUZ64,KUZ65}, suggested to use the Gallium (Ga) as a target for $\nu_\odot$. 
\item 
In 1967, B.Pontecorvo, \cite{PON67}, described processes violating leptonic and muonic charge: 
$\nu \leftrightarrow \overline {\nu}$ and $\nu_e \leftrightarrow \overline {\nu_\mu}$.   
\item 
In 1968, the first experimental results from Homestake detector showed a deficit 
in $\nu_\odot$ flux: the value was at a level of 1/3 of the expected one, \cite{DAV68}. This was the first SNP. 
\item 
In 1969, V.Gribov and B.Pontecorvo gave a good formulation of  $\nu$ flavour oscillation in vacuum \cite{GRI69}. 
\item 
In the 70s', a lot of theoretical and experimental discoveries highly increased  the knowledge in particle 
physics sector (J/$\psi$ and B, quarks called charm and beauty,  the supersymmetry and the grand unification theories... 
but nothing new on $\nu_\odot$ from the experimental point of view). 
\item 
In 1985-86, S.P.Mikheyev and A.Smirnov, \cite{MIK85,MIK86}, developed the Wolfenstein's  proposal, 
\cite{WOL78,WOL79}, and presented a model of $\nu$ oscillation in matter. 
\item 
At the end of  80s', the Japanese detector KAMIOKANDE, originally a proton-decay  dedicated experiment, 
realised the first "neutrinography" and showed that $\nu$'s come from the Sun, \cite{HIR89}. The comparison between 
the results from Homestake  and KAMIOKANDE implied a second SNP, \cite{BAH90a}. 
\item 
In the 90s', satellites and experiments on Earth surface strongly enhanced the knowledge on solar 
inner features $\it{via}$ seismological measurements, \cite{SCI96}.  
\item 
In the middle of 90s', Gallium calibrated experiments detected low energy  $\nu_\odot$'s and (finally) confirmed the 
$\nu$ production $\it{via}$ p-p reactions and the solar machinery, see \cite{GAV97,HAM99,GAV01a} and references therein. 
These results seemed to be  not compatible with other experimental data so that a third SNP arose.  
\item 
At the end of the 2${nd}$ millennium, the SuperKAMIOKANDE's (SK) solar  and atmospheric $\nu$ data 
suggested the "neutrino flavour oscillation" as a possible solution of SNP, \cite{FUK01a,FUK01b}. 
If this is true, $\nu$ should have a non-zero mass and constraints in particle physics and cosmology are allowed. 
\item 
In 2001, a Canadian experiment (SNO) detected different $\nu_\odot$ flux  depending on interaction
process, \cite{AHM01}: the flavour oscillation solution was enhanced. At the end of the year a terrible accident 
occurs at SK detector. 
\item 
In 2002, new results from SNO concerning neutral current interactions strengthen the oscillation hypothesis as "THE" 
right way out to solve the SNP, \cite{AHM02a,AHM02b}. In December 2002, KAMLAND reactor experiment first result confirms
LMA solution as "the" solution, see sect. 7 for details. In an exposure of about 150 days, the
ratio of observed inverse $\beta$-decay events to the expected one without disappearance is
0.611$\pm$0.085(stat)$\pm$0.041(syst) for $\bar\nu_e$ energy greater than 3.4 MeV.
This deficit is incompatible with standard predictions at the 99.95$\%$ confidence level. In the context
of two-flavour neutrino oscillations with CPT invariance, all oscillation solutions to the SNP but the LMA
solution are excluded, \cite{EGU02}. The best fit is obtained with $\Delta m^2$= 6.9$\cdot 10^{-5}$ (eV)$^2$ and 
maximal mixing, see later for details. BOREXINO detector should finish its building step.
\fit    

\section{How the sun shines.} 
\label{sect:sun}
We shortly review properties and characteristics of the Sun and aspects of the proposed mechanism it uses to produce 
energy and neutrinos. We refer for instance to \cite{BAH89} for details and a complete 
description of these arguments. 

\subsection{Introduction.} 
\label{sect:sunintro}
The Sun is a G 2 type main-sequence star; it seems to be a "normal" star but X-ray  and UV observations show 
interesting features. It has complex rotation, an unexplained magnetic activity, anomalies in surface chemical  
composition, an unknown mechanism acting in its corona and accelerating solar wind. \\ 
Detailed observations were carried out of the solar surface rotation by  tracking the motion of surface 
characteristics such as sunspots and by Doppler-velocity  measurements. Sun is not rotating as a solid body: 
at the equator the rotation period 
is $\sim$ 25 days, but it increases gradually towards the poles, where the period is estimated to be $\sim$ 36 days. 
The rotational velocity at the surface of the Sun is $\sim$ 2 km/s, dropping off rather smoothly towards higher 
latitudes. However, it has been  found that bands of faster and slower rotation, a few metres per second  
higher or lower than the mean 
flow, are superimposed, see \cite{KOS97,SCH99}.  The origin of this behaviour is unknown. \\ 
The differential rotation seems to be linked to  the dynamic nature of the solar external regions: in the outer 
29 \% of the Sun's radius ($\sim 2\cdot 10^8$ m), energy is  transported by convection, in rising elements of 
warm gas and sinking elements of  colder gas, \cite{CHR91,KOS91}. These motions also transport angular momentum, and 
hence provide a link  between rotation in different parts of convective zone. Convection is even affected  by rotation, 
which may introduce anisotropy in the angular momentum transport. \\
At the base of the convection zone, a transition occurs: the variation of rotation  rate with latitude disappears, 
so that the inner region rotates  essentially as a rigid body, at a rate corresponding to 
the surface rate at mid-latitudes.  The region over which the transition occurs is very narrow, no more than a 
few per cent of the total solar radius: it is called tachocline, see \cite{KOS96,CHA99}. It was proposed 
that a large-scale weak magnetic field ($B_\odot$) permeates inner regions and enforces nearly rigid rotation by dragging 
the gas along at a common rate, \cite{GOU98}.\\  
Astrophysical data concerning young solar-type stars show faster rotations, by a factor up to 50: this means 
that the solar angular momentum probably decreased its initial value (for a complete review on rotation of 
solar interior see \cite{CHR01} and references therein). The contribution due to mass loss and electromagnetic 
emission is hard to compute from the stellar evolution code; moreover, $B_\odot$ could have modified 
such processes. \\
The Sun has a magnetic activity probably  due to a "dynamo" mechanism acting within the convective surface 
layers and connected with its non-uniform rotation. A strictly  related effect is the sunspots 
appearing on the solar surface: they are modulated by a period of $\sim$ 11 y (the first detected cycle 
started in 1755 A.D.) and are produced by magnetic flux tubes crossing the solar surface. Their origin  
could be due to a subsurface toroidal $B_\odot$ having different directions  in northern and southern solar 
hemisphere. This small component ($\sim$ 5 $\%$  of the total intensity) sums up to a much stronger dipole field. 
The polarity is  reversed every $\sim$ 11 y so that the complete cycle is as long as 22 y.\\ 
The solar interior is a plasma: it is essentially neutral close to the photosphere
while it is in practice fully ionised down to the centre. \\  
The structure of the Sun is the result of an equilibrium between the energy loss at the surface and its generation 
in the core (with a stationary energy transport between core and surface) while, if the forces acting in the system 
are analysed, the hydrostatic equilibrium provides a relation between pressure gradient and gravitational acceleration
(which is connected with density distribution inside the Sun). Properties of solar matter are expressed by equations  
of state relating pressure P, temperature T, density $\rho$ and chemical composition, often characterised in terms 
of fractional mass of hydrogen (X), helium (Y), and other elements, Z. The value of T inside the Sun is fixed by 
the energy balance. \\
In a great part of the Sun  the energy transport is radiative and depends on the matter opacity, \cite{EDD26}; a steep 
temperature gradient ($\nabla$) is required leading to convective instability, \cite{SCH06}: then, it is 
slightly steeper than adiabatic, where P and $\rho$ are related by:  
\beq
P \simeq K \rho ^{\Gamma_1} 
\label{eq_P}
\feq 
where K is constant depending on convective efficiency and the exponent $\Gamma_1$ is: 
\beq 
\Gamma_1 = \left( \frac {\partial ln P}  {\partial ln \rho} \right)_{ad} 
\label{eq_Gamma1}
\feq 
Generally speaking, $\nabla$ overcomes the adiabatic value near the solar  surface where $\rho$ is very low 
and conductive velocities are therefore much higher to sustain the flux of energy. \\ 
A detailed knowledge of physical phenomena inside the Sun is needed to achieve a realistic solar model.  
The energy production mechanism is supposed to be a series of nuclear exothermic  
reactions induced by thermal motion: light nuclei fuse among them to form heavier ones. The average binding 
energy per nucleon is a useful parameter to study the possible  behaviour; it gets to a maximum in coincidence 
with $^{56}$Fe: heavier nuclei take part  to fission reactions, for light nuclei the fusion is allowed. Both 
these processes reach  the end-point when they approach the maximum value of the binding energy per nucleon and, 
at the same time, this condition is the natural end-point  of the nuclear burning processes inside the stars. \\ 
In any case, the difference between the masses of $in$  and $out$ nuclei is converted 
into energy, following the Einstein's relation: $\sim$ 27 MeV are produced for the basic nuclear reaction  
in the Sun, the conversion of 4 protons in a He nucleus.\\ 
The fusions in the core of a star supply the required radiated luminosity and the thermal  pressure, due to 
the motion of electrons and ions that is needed to support the star  against the gravity force. Moreover, the 
nuclear reactions imply chemical composition and temperature distribution variations inside the star.  

\subsection{Basic nuclear reactions.} 
\label{sect:sunbasicreactions}
Following the description given by solar models, the solar energy production is 
done by the conversion of H into He: the so called p-p chain produces the 
main contribution, the remaining one being due to the CNO cycle. It is hard 
to give a right evaluation of all solar nuclear reaction rates because of our partial  
knowledge of interaction cross-sections at energy as low as in 
the solar core but  only if T$_c\geq 6\cdot 10^6$ K the p-p reaction becomes "efficient". \\ 
Then, we detail both the cycles: in table \ref{tabpp} the main features of p-p chain nuclear  
reactions are shown while table \ref{tabcno} summarises the CNO cycle, which shows a "knot" 
due  to different reactions the $^{15}$N nucleus has.\\ 
The p-p chain starts when two protons interact to form deuterium, D, (the process  is 
dominated by weak interactions, its cross-section is very low, $\sigma\,\sim  10^{-47}cm^2$). 
Alternatively a 3-bodies process can occur: 2 protons and an electron interact producing 
D (p-e-p reaction), but this reaction is highly disfavoured. Some seconds later, D captures 
a proton forming $^3$He; for a detailed description of D importance in p-p chain see 
\cite{BAH97c}. The D production is quite a crucial point: in fact D does not accumulate 
in  the solar interior, on the contrary only a great abundance of $^3$He 
nuclei  allows further nuclear reactions (even if higher T$_c$ values are required).  
D and $^3$He are burned quickly in the Sun, their lifetime being respectively  
$\sim$ 10$^{-8}$y and $\sim$ 10$^5$y: both these values are very short compared to lifetime 
of a proton, which is destroyed by the p-p reaction ($\sim$10$^{10}$y), \cite{BAH89}. Hence, 
it is assumed that both D and $^3$He are in local kinetic equilibrium: this means that the 
production of D $via$ p-p and p-e-p reactions is balanced by its destruction $via$ p+D reaction, 
see for instance \cite{GAU97} for a discussion on D/H ratio in astrophysics.\\ 
\begin{table}[ht]
\caption{\it{ Reactions of the proton-proton chain: the probability of each 
step and the maximum kinetic energy are reported.}}
\begin{center} 
\begin{tabular}{|l|c|c|c|}  
\hline 
REACTION& Probability&Max. Kinetic \\         
&   ($\%$)& Energy (MeV)\\ \hline 
p + p $\rightarrow$ D + e$^+$ + $\nu_e$ & 99.76&0.42341\\ \hline 
p + e$^-$ + p $\rightarrow$ D + $\nu_e$& 0.24&1.445\\ \hline\hline\hline 
p + D $\rightarrow $ $^3$He + $\gamma$& 100&5.49 \\ \hline\hline\hline 
$^3$He + $^3$He $\rightarrow$ $^4$He + 2p & 81.03&12.86 \\P-P I & & \\ \hline 
$^3$He + p $\rightarrow $e$^+$ + $\nu_e$ + $^4$He&0.00002 & 18.778 \\ \hline  
$^3$He + $^4$He $\rightarrow$ $^7$Be + $\gamma$& 18.97&1.59 \\ \hline\hline\hline 
$^7$Be + e$^-$ $\rightarrow$ $^7$Li + $\nu_e$& 18.95&0.8631(89.7$\%$)\\   
&  & 0.3855(10.3$\%$)\\ $^7$Li + p $\rightarrow$ $^4$He + $^4$He & &17.35 \\
P-P II & & \\ \hline 
$^7$Be + p $\rightarrow$ $^8$B + $\gamma$&0.02&0.137 \\ 
$^8$B $\rightarrow$ 2 $^4$He + e$^+$ + $\nu_e$ & &14.06\\ 
P-P III & & \\ \hline 
\end{tabular} 
\end{center} 
\label{tabpp}
\end{table}
\\
When T$_c\geq 8\cdot 10^6$ K other fusion reactions are allowed:  
\bit 
\item 
Two $^3$He nuclei could interact originating a $^4$He nucleus and 2 residual protons (this is the P-P I branch). 
\item 
If $^4$He is present and T$_c\geq 1.5\cdot 10^7$ K, $^3$He could interact  with $^4$He producing $^7$Be 
(a radioactive isotope) which transforms into $^7$Li by  free electron capture. This reaction does not depend on 
nuclear Coulombian barrier. Some time later, this nucleus interacts with a free proton  originating two $^4$He 
nuclei (P-P II branch). Another option is possible: $^7$Be  can capture a proton originating $^8$B which then 
breaks into two $^4$He nuclei  (P-P III). This reaction is dominant at 
T$_c\geq 2\cdot 10^7$ K. Even $^7$Li has a very short lifetime so that the
equilibrium abundance in temperature and density range in the solar core 
is quickly reached.
\item 
$^3$He  captures a proton forming $^4$He, a positron and a neutrino, the most energetic one in $\nu_\odot$ 
spectrum, or produces $^4$Li, plus a photon. Then, $^4$Li transforms into $^4$He, a positron and a $\nu$. 
Both these processes have a very low cross-section (weak interaction). The
$^3$He abundance profile predicted by SMs shows a peak at d $\sim$ 0.28 
R$_\odot$ from the centre, with a bell-like curve due to the competition 
between its creation and its destruction. Its abundance is expected to have
an influence on the structure of the solar core at a level where 
helioseismology can detect it.
\fit 
The second way producing energy in the Sun is the CNO (or  Bethe-Von Weizs\"{a}cker) cycle 
occurring at T$_c$ higher than in the p-p chain: it starts when a $^{12}$C 
nucleus captures a proton producing 
$^{13}$N and a photon, \cite{WEI37,WEI38}.  Its contribution to solar energy production was estimated 
as the  main one in Bethe's seminal papers, see {\it{e.g.}} \cite{BET39}, because of incorrect T$_c$ values.  
When more recent and precise T$_c$ estimates were adopted, the CNO 
contribution diminished at a level of some percent; moreover, the Sun is a 
low-mass star, see \cite{TUR01c} for a discussion of the influence of CNO
cycle reactions on the solar structure. \\
\begin{table}[ht]
\caption{\it{ Reactions of the CNO chain.}}
\begin{center}
\begin{tabular}{|l|c|c|}  
\hline 
REACTION& Max. Kinetic \\     
& Energy (MeV)\\\hline 
$^{12}$C + p $\rightarrow \gamma$ + $^{13}$N& 1.94 \\\hline 
$^{13}$N $\rightarrow$ e$^+$ + $^{13}$C + $\nu$& 1.1982\\\hline 
$^{13}$C + p $\rightarrow \gamma$ + $^{14}$N & 7.55  \\\hline 
$^{14}$N + p $\rightarrow \gamma$ + $^{15}$O& 7.30 \\\hline 
$^{15}$O $\rightarrow$ e$^+$ + $\nu$ + $^{15}$N& 1.7317\\\hline 
$^{15}$N + p $\rightarrow \gamma$ + $^{16}$O(1$\%$)&12.13 \\ 
\hspace{1.0cm} $\rightarrow \alpha$ + $^{12}$C(99$\%$)& \\\hline 
$^{16}$O + p $\rightarrow \gamma$ + $^{17}$F&0.60 \\\hline 
$^{17}$F $\rightarrow$ e$^+$ + $\nu$ + $^{17}$O&1.7364 \\\hline 
$^{17}$O + p $\rightarrow \alpha$ + $^{14}$N& 5.61 \\ 
\hspace{1.0cm} $\rightarrow \gamma$ + $^{18}$F& \\\hline 
$^{18}$F $\rightarrow$ e$^+$ + $\nu$ + $^{18}$O& 0.63\\\hline 
\end{tabular} 
\end{center} 
\label{tabcno}
\end{table}
\\
In table \ref{tabenergy} the energy produced in solar reactions is shared following the production mechanism; 
the fraction carried out by $\nu_\odot$'s is also shown. \\ 
\begin{table}[ht]
\caption{\it{ Thermal energy produced by different nuclear reactions and fraction of $L_\odot$ carried 
out by $\nu_\odot$'s.}} 
\begin{center} 
\begin{tabular}{|l|c|c|c|}  
\hline 
REACTION& Q$_{eff}$ & Energy $\nu$&L$_{\odot}$\\     
& (MeV)& $\%$&$\%$\\\hline\hline\hline 
PP I&26.19&1.95&84.87\\\hline 
PP II&25.65&4.00&13.52\\\hline 
PP III&19.12&28.46&0.01\\\hline 
CNO&25.03&15.66&1.60\\\hline 
\end{tabular} 
\end{center}
\label{tabenergy}
\end{table} 

\subsection{Some technical aspects.} 
\label{sect:suntechnical}
\begin{figure} [ht]
\begin{center} 
\mbox{\epsfig{file=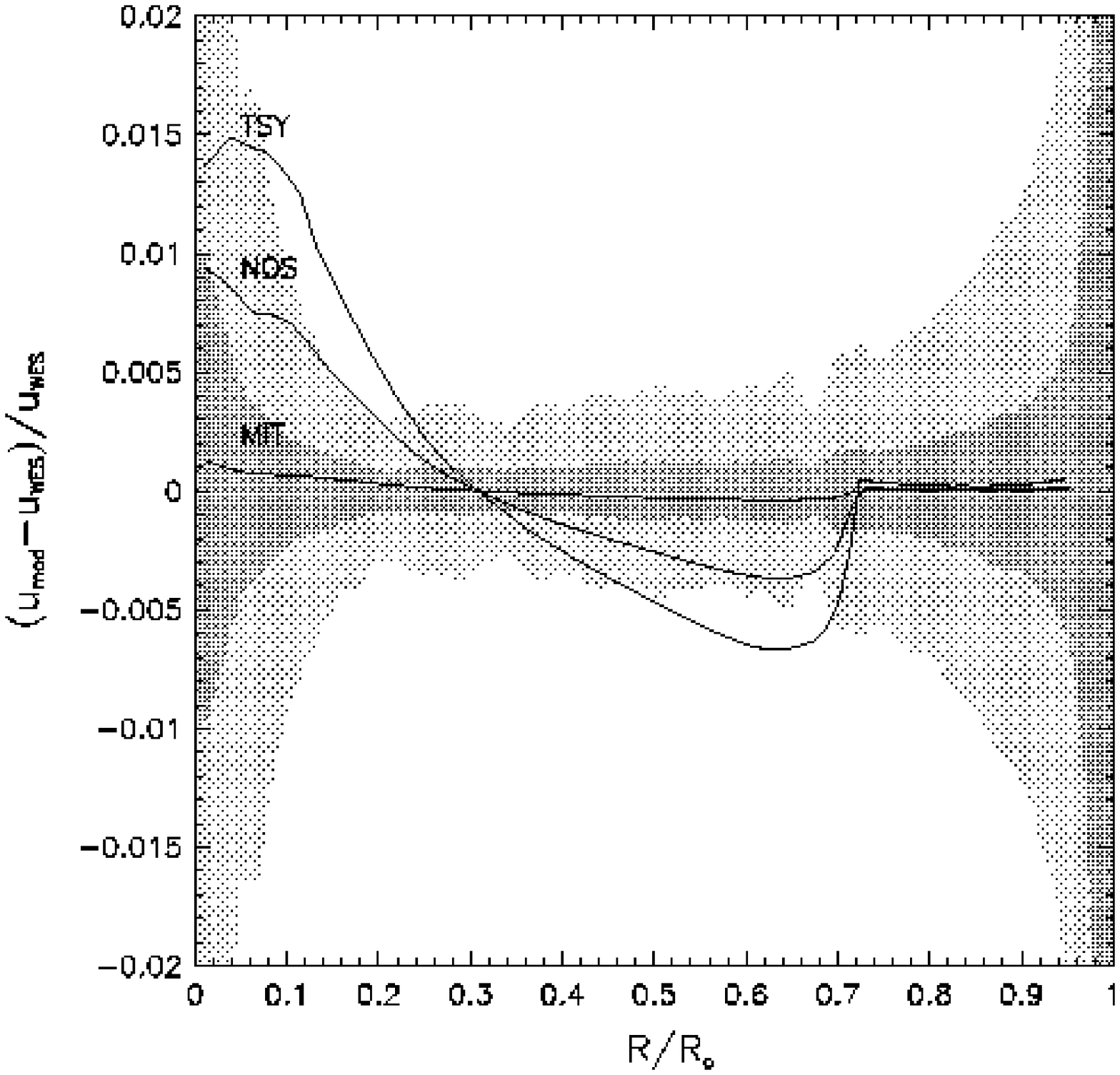,width=0.5\textwidth}} 
\end{center}
\caption{ Fractional isothermal sound speed difference: the shaded region represents  
experimental and a conservative 3$\sigma$ uncertainties. Three different 
screening factors are drawn: TSY - Tsytovich; MIT - Mitler; NOS - without
screening term, from \cite{FIO01a}. }
\label{fig:FIOscreen}
\end{figure}

Let us point out the main features of physical processes inside the Sun. 
Charged nuclear particles interact only if they are so close each to other 
that strong nuclear forces become dominant with respect to Coulomb 
repulsive forces: typical distance is $d\approx 10^{-15}$m (the nuclear radius is 
$r_N\simeq 1.3\,A^{\frac {1}{3}}\cdot 10^{-15}$m, where A is the  atomic mass number).\\ 
An interaction between two nuclei is usually described by a potential well with average 
depth of $\sim$ 30 MeV followed by a Coulomb barrier: 
\beq 
E_{Coul}\,=\,\frac{Z_1Z_2e^2}{r_N}\sim Z_1Z_2 \;MeV \approx 10 MeV
\label{eq_ECoul}
\feq 
where Z$_1$ and Z$_2$ are the electric charges of interacting nuclei (the thermal energy of a 
particle in solar plasma at T$_c\sim 1.5\cdot 10^7$ K is 
$E_k\approx$ 10 keV).\\  
A Maxwell-Boltzmann distribution function describes the energy distribution of 
particles in solar plasma while the probability of the interaction process is: 
\beq 
P(E)\,=\,\frac{2}{\sqrt {\pi}} (kT)^{-3/2}\,\sqrt{E_k}exp \left(-\frac{E_{Coul}}{kT} \right) 
\sim exp \left(-\frac{E_{Coul}}{kT}\right)\,\ll 10^{- 400} 
\label{eq_Prob}
\feq 
This value for fusion reactions is so small that, even taking  into account an extreme 
tail of the distribution, all solar nucleons could never  interact in a "reasonable" time 
(nor all the baryonic matter in the Universe).\\ 
Solar nuclear fusion reactions occur $via$ a strong interaction at energies  ranging from 
$\sim$ 5 keV to 30 keV; its energy dependence in non-resonant reactions is expressed as: 
\beqar  
\sigma(E) = {S(E) \over E} {exp} \left\{ -2 \pi \eta(E) \right\}  \protect \\
\eta(E) = \frac{Z_1 Z_2 e^2}{\hbar v}  ~~~~~
\mu = \frac{A_1\,A_2}{A_1+A_2} ~~~~~ v = \sqrt{\frac{2E}{\mu}}
\nonumber
\label{eq_SigmaE}
\end{eqnarray} 
$\eta$ is the Sommerfeld parameter, E is the centre-of-mass energy, $\mu$ is the reduced mass 
where A$_1$ and A$_2$ are the atomic mass number, v the relative velocity in the entrance channel. 
The exponential Gamow penetration factor, which dominates the energy dependence, describes 
quantum-mechanical tunneling through the Coulomb barrier. 
The WKB approximation, which is usually adopted to evaluate the Gamow factor, is  valid if 
$2\pi \eta \geq 1$ and becomes even more accurate at low energy. S(E), which is slowly 
varying (except for resonances), is then usually written as: 
\beq      
S(E) \simeq S(0) + S^\prime(0)E + {1\over2} S^{\prime\prime}(0)E^2. \protect 
\label{eq_Sastro}
\feq 
The coefficients can often be deduced by a fitting procedure to laboratory 
measurements or by theoretical calculations of the cross-section made at 
higher  energies and then extrapolated to the solar energies.\\ 
In this picture the probability P(E) depends also on the Gamow factor G:  
\beq 
P(E)\sim exp(-G)\,=\ exp \left(-\frac{4\pi}{h}\sqrt{2\mu}\int_{R_-}^{R_+} 
\sqrt{\frac{Z_1Z_2e^2}{r}\,-\,E}\,dr \right) 
\label{eq_PGamow}
\feq 
where E is the relative kinetic energy between two nuclei in their centre of mass.
The previous relation is usually written as:
\beq
P(E)\sim exp \left[\left(-31.29 Z_1 Z_2)\sqrt{\frac {\mu}{E(keV)}}
\right)\right]
\feq
When T$\sim 10^7$ K and Z$_1$Z$_2 \leq 2$, as in the solar core, P(E) is $\sim 10^{-20}$;  
higher temperatures are required for reactions including heavier nuclei.\\ 
The rate of a non-resonant fusion reaction can be written, \cite{BAH89}: 
\beq  
\left<\sigma v\right> = 1.3005 \cdot 10^{-15} \left[Z_1 Z_2 \over A \mu T_6^2
\right ]^{1/3} f S_{eff} exp\left(-\tau \right) cm^{3}s^{-1} 
\label{eq_Reacrate}
\feq 
where T$_6$ is the temperature in units of $10^{6}$K. 
$S_{eff}$ (in keV~b) is the effective cross-section factor for the  fusion reaction and 
is evaluated at the most probable interaction energy (E$_0$); to first order in $\tau^{-1}$: 
\beq 
S_{eff} = S(E_0) \left\{ 1 + \tau ^{-1} \left[{5 \over 12} + {5 S^\prime E_0 \over 2 S } + 
{S^{\prime \prime }E_0^2 \over S} \right]_{E = E_0} \right\}
\label{eq_Sastro1}
\feq 
where $S^\prime = \frac{{d}S}{{d}E}$.\\ 
The quantity $f$, the screening factor, was first calculated by \cite{SAL54}. 
The exponent $\tau$, which is varying in the range 15 - 40, dominates the dependence of the 
reaction rate from T; it is given by:  
\beq
\tau = 3 \frac{E_0}{kT} = 42.487
\left(\frac {Z_1^2 Z_2^2 \mu}{T_6} \right)^{1/3} 
\label{eq_tau}
\feq
The most probable energy (the Gamow energy, $E_0$) at which the reaction occurs is: 
\beqar
E_0 & = &\left[(\pi \alpha Z_1 Z_2 kT)^2 (\frac{m \mu c^2} {2}) \right]^{1/3}\nonumber
\\ & = &1.2204(Z_1^2  Z_2^2 A \mu T_6^2)^{1/3} ~{keV} 
\label{eq_GamowEn}
\feqar 
For example $^3He(^3He,2p)^4He$ E$_0 \sim$ 21 keV, $^{14}N(p,\gamma)^{15}O$ E$_0 \sim$ 26 keV.\\ 
In most analyses the values of $S$ and derivatives are quoted  at E = 0 not at $E_0$: in order 
to relate the equations, one has to express the quantities in terms of their values at E = 0: 
\beqar 
S_{eff} (E_0) \simeq
S(0) \left[1 + {5 \over 12 \tau} + {S^\prime (E_0 +{35\over 36} kT) \over S } + 
{S^{\prime \prime} E_0 \over S} \left({E_0 \over 2} + {89 \over 72}kT \right) \right]_{E=0} 
\label{eq_Sastroeff}
\feqar 
$S_{eff}(E_0)$ is often referred to as simply "the low-energy $S$-factor"; in literature S is
usually labelled by introducing the electric charge of reacting particles: S$_{11}$, S$_{34}$ 
and so on.\\  
In practice, the electrostatic repulsion between interacting nuclei is 
diminished by this screening: the effective energy in centre of mass is 
increased by a constant term $-u_0$. Consequently, the value of 
interaction rate is enhanced, from $\sim$ 4$\%$ for the first reactions 
of p-p chain up to $\sim$ 40$\%$ for the  $^{14}$N + p interaction. \\
The weak screening description is usually introduced, see \cite{GRU98,BAH00b}. This approach 
is valid if $kTR_D\gg Z_1Z_2e^2$: in the solar case $Z_1Z_2\leq$ 10.  
The enhancement factor is given by: 
\beqar
f = exp \left({Z_1Z_2e^2\over kTR_D}\right), \protect \\
R_D=\left(\frac{4\pi ne^2\zeta^2}{kT}\right)^{-1/2}  \nonumber \\
\zeta = \left\{\Sigma_i X_i {Z^2_i\over A_i} + \left({f^\prime\over f}\right)  
\Sigma_i X_i {Z_i\over A_i}\right\}^{1/2}  \nonumber
\label{eq_Ffactor}
\feqar
where $R_D$ is the Debye radius, $n$ is the baryon number density 
$(\rho/m_{amu})$, $X_i$, $Z_i$, and $A_i$ are, respectively, 
the mass fraction, the nuclear charge  and the atomic weight 
of nucleus of type $i$.  The quantity ${f^\prime\over f} \simeq 0.92$ describes 
electron degeneracy, \cite{SAL54}. \\ 
Thus, plasma screening corrections are known with uncertainties of a few percent. 
Corrections of the order of a few percent to the Salpeter formula  come from the 
non-linearity of the Debye screening and from the electron  degeneracy,
see \cite{JOH92,DZI95b,BRO97,GRU97} for further discussions.\\ 
Moreover, the Salpeter relation also describes screening effects on the $^7$Be  
electron capture rate with an accuracy better than 1\%, \cite{GRU97}.\\
The screening treatment is based on the mean field approximation which is the field 
that a  particle in the plasma sees "averaged" on thermodynamically long times. 
Following the  ergodic hypothesis, this value is equal to a mean over all particles 
in the system at  any chosen time. In solar core $R_D\sim$ 0.87 $d$, where $d$ is the 
mean interparticle  distance, so that the mean number of particles in a Debye sphere ranges 
from 2 up to 5,  depending on ions electric charge. The mean field approximation 
supposes this number  is constant, then it deduces the mean field potential. \\ 
Many authors estimated this is a poor approximation  and proposed different solutions 
trying to overcome the Salpeter's picture, see 
\cite{MIT77,CAR88,SHA96,GRU98,BAH00a,SHA00,WEI01,BAH02d}.
In \cite{BRO97,GRU97,BAH00a} the validity of "classic" treatment were confirmed. 
Physical interpretation of these  results is currently unknown; an "antiscreening" 
effect, as proposed in \cite{TSY00}, is strongly disfavoured, \cite{FIO01a}.  
In table \ref{tabfactorFIO} the influence of different screening factors
on each branch of the p-p chain nuclear reactions is shown,
while in fig. \ref{fig:FIOscreen} the variations induced by screening factors on the
sound speed profile inside the Sun are drawn (this is a test usually adopted to check the 
goodness of a SM). \\
\begin{table}[ht]
\caption{\it{ Screening factors in  solar core, for weak screening (W-S), 
Mitler model (MIT), no screening (NO-S) and Tsytovitch model (TSY), adapted 
from \cite{FIO01a}.}}
\begin{center}
\begin{tabular}{|l|c|c|c|c|}  
\hline 
             &W-S   &MIT   &NO-S&TSY\\ \hline
p+p          & 1.049& 1.045& 1.0& 0.949\\
$^3$He+$^3$He& 1.213& 1.176& 1.0& 0.814\\
$^3$He+$^4$He& 1.213& 1.176& 1.0& 0.810\\
$^7$Be+p     & 1.213& 1.171& 1.0& 0.542\\\hline
\end{tabular}
\end{center}
\label{tabfactorFIO}
\end{table}
\\
At present, helioseismological data restrict the variability of corrective screening factor: 
0.95 $f_{Salp}\leq f \leq 1.1\,f_{Salp}$, where $f_{Salp}$ is the value in Salpeter's 
description. In any case, no direct measurements of such a parameter will be possible.
\\
If resonances are present in the reaction cross-section $\sigma$(E), equations \ref{eq_SigmaE}
have to be substituted by Breit-Wigner relation: 
\beqar
\sigma(E)\,=\,\frac{\omega \pi}{k^2} \frac{\Gamma_{\alpha} \Gamma_{\beta}} {(E-E_{res})^2\,+\, 
\frac{\Gamma^2}{4}} \\
\omega\,=(1+\delta_{1,2})\frac{2I_c+1}{(2i+1)(2I+1)} \nonumber 
\label{eq_sigmareson}
\feqar
where $k$ is the wave vector, $I_c$, $i$, $I$ are, respectively, the compound nucleus, the 
incoming nucleus and the target nucleus spins, $E_{res}$ is the energy at which the resonance 
occurs, $\Gamma_\alpha, \Gamma_\beta, \Gamma$ are the widths of the initial $\alpha$ channel, 
of the compound nucleus in the $\beta$ channel and of the decay of the compound nucleus. After 
substitutions and taking also  into account that the Maxwell-Boltzmann  distribution function 
varies its value  when the energy is near the resonance value, the interaction rate equation 
becomes: 
\beq 
r\,=\,\frac{n_1 n_2}{1+\delta_{1,2}} \left(\frac {2\pi}{\mu kT}\right)^{3/2}\,h^2  
\omega \frac{\Gamma_{\alpha} 
\Gamma_{\beta}}{\Gamma} exp \left(-\frac{E_{res}}{kT}\right)  
\label{eq_rateresonance}
\feq 
When T is fixed, the reactions with energy near $E_{res}$ give the main contribution in the  
previous equation so that a good knowledge of their characteristics at low energy is  needed.

\subsection{Solar Models.(SMs)} 
\label{sect:sunsolarmodels}
Details about the energy production and other aspects of the solar physics are provided by  SMs; among them we 
mention the so called "standard solar models", SSMs. Their definition is often  changing in time but in general 
they could be thought as models offering a description of  the solar properties under a set of hypotheses picking 
out the values of physical and  chemical inputs within the range of uncertainty of experimental data. SMs are 
required to  fit the measured $L_\odot$, $R_\odot$ and Z  at surface. No constraints deduced from helioseismology,
see later in section \ref{sect:sunhelio}, are imposed in SSMs.\\  
Many non-standard SMs were proposed, see  
\cite{BAH89,BAH97d,GUE97,CAS97,BRU99,BAH01a,BAS00,TUR01a,COU02} for detailed analyses. It was demonstrated 
the inconsistency of a part of these models  with the deduced solar sound speed c$_s$ 
in \cite{BAH97d} or with p-mode frequencies in 
\cite{GUE97}. Non-standard SMs have been used even to constrain the cross-section for the p-p interaction, to 
estimate the mass loss from the Sun, \cite{GUZ95}, or the anomalous  energy transport by WIMPS, \cite{CHR92}.\\ 
We underline that the so called "seismic" SMs, that presently better  agree with helioseismological measurements, 
are strictly speaking non-standard SMs.\\ 
The fundamental assumptions of SMs are: 
\bit 
\item 
At the first stage 
of its evolution the Sun is a homogeneous spherical protostar formed by the contraction of interstellar gas. The  
released gravitational  energy heats the solar matter. After $\sim 5\cdot 10^7$y, the Sun goes into  the 
"main sequence" and the reactions of the p-p chain lead to a hydrostatic  equilibrium (gravitational force is 
balanced by the pressure due to the thermal motion). 
\item 
The time evolution of the Sun in the "main sequence" 
is described by a set of  differential equations where the distance from the centre of the Sun and the time  
from the nuclear reactions beginning are independent variables. Asymmetric distributions,  angular momentum 
and $B_\odot$ usually are not introduced in computations. Nuclear reactions do not change the Z value but 
only X and Y. Moreover, a smooth variation of the remaining physical parameters  is produced. The physical quantities 
under analysis are $m(r,t)$, the mass inside a sphere  whose radius is r at time t; $P(r,t)$, $L(r,t)$, 
$T(r,t)$ and $X_i(r,t)$, the chemical mass fraction of i-element, at distance r from the  centre at time t.
\fit
Boundary conditions are imposed: the abundance of heavy elements at the entrance in the 
"main sequence" is supposed to be equal to the present meteoritic one; the remaining functions 
at the present time are  required to be coincident with data measured at solar surface.\\
The solar evolution is then considered in the context of "standard" particle physics model.\\ 
Three different time scales are present: 
\bit 
\item 
{\it{Hydrostatic time scale}} ($t_{hyd}\approx R^{1.5}G^{-0.5}M^{-0.5}$) gives the typical time after 
which a star can reach a hydrostatic equilibrium   after a small perturbation. For the Sun $t_{hyd}\sim$ 30
 minutes so that  it is very near to the hydrostatic equilibrium. 
\item 
{\it{Kelvin-Helmholtz time scale}} ($t_{KH}\equiv \frac {E_G}{L}  \approx 0.5 \,GM^2R^{-1}L^{-1}$) 
describes the evolution time of a star for which gravitational energy $E_G$ is the only  available one (the 
absolute value of $E_G$ has to be inserted).  For the Sun $t_{KH}\sim 1.7\cdot 10^7$ y. 
\item 
{\it{Nuclear time scale}} ($t_N\equiv \frac {E_N}{L}$) refers to a star balancing its luminosity by 
release of nuclear energy $E_N$  (in the previous relation the absolute value of this energy has to be used).
In the case of the Sun, if one supposes H is the sole component, an its complete conversion in He and a constant 
luminosity, a $t_N\sim 10^{11}$ y is required. 
\fit 
Let us show some characteristics of the differential equations adopted in SMs.\\ 
The first equation imposes the conservation of a mass inside a sphere: 
\beq  
\frac {\partial m}{\partial r}\,=\,-\,{4\pi r^2\rho} 
\label{eq_ssmequat1}
\feq 
The second equation, which states the energy conservation, is:  
\beq 
\frac{\partial L}{\partial m}\,=\,\epsilon\,-\,C_P\frac{\partial T}{\partial t}\,-\,\frac{1}{\rho} 
\left(\frac{\partial ln \rho}{\partial ln T}\right)_P \frac  {\partial P}{\partial t} 
\label{eq_ssmequat2}
\feq 
where $C_P$ is the specific heat at constant pressure and $\epsilon$ (the net energy produced 
per time and mass unity) is the dominant term: consequently the energy produced by nuclear 
reactions is balanced by its flux emerging  from the production side. Moreover, the energy 
production per mass and time is correlated  to all the specific nuclear reaction rates and to 
their specific Q-value, which is the maximum energy "injected" in the solar matter by each reaction. \\ 
The third equation gives the relations between gravitation and pressure forces: 
\beq  
\frac{\partial^2 r}{\partial t^2}\,=\,-\frac {mG}{r^2}\,-\,4\pi r^2\frac{\partial P}{\partial m} 
\label{eq_ssmequat3}
\feq 
When hydrostatic equilibrium is present, the term at left side vanishes. \\ 
The energy flux is described by the following equation: 
\beq 
\Phi_c\,=\,\frac{16 \sigma T^3} {3 k_c \rho} \frac {dT}{dr} 
\label{eq_energyflux}
\feq 
where $k_c$ is the conductive opacity.\\ 
The "transport" equation shows the outward energy flux in terms of temperature gradient 
($\nabla$) which depends on typical physical processes originating those temperature 
variations: 
\beq 
\frac{\partial T}{\partial m}\,=\,-\frac {T}{P}\frac{Gm}{4\pi r^2}\nabla\,= 
\,-\frac {T}{P}\frac{Gm}{4\pi r^2}(\nabla_{R}\,+\,\nabla_{C}\,+\,\nabla_{co}) 
\label{eq_transport}
\feq 
where $\nabla_{R}$, $\nabla_{C}$ and $\nabla_{co}$ are respectively the radiative, the 
conductive and the convective component of temperature gradient. \\
The Sun is supposed to be "stratified", depending on the dominant physical process  which 
allows the energy propagation in matter. Energy is produced in the solar core where the 
transport is radiative: in this region the radiative gradient is the dominant term  
therefore $\nabla_{R}\gg \nabla_C$, the conductive component which is due 
to  the electron plasma thermal motion. The last term ($\nabla_{co}$) is 
acting only in  the convective region, at the distance from the centre 
$r\geq 0.7 R_\odot$.\\ 
The energy transport is based on diffusion and $\nabla_R$ is  calculated under the assumption 
of local thermodynamic equilibrium: 
\beq 
\nabla_{R}\,=\, - {3\over 64} \frac {kL} {\pi \sigma G m T^3} 
\label{eq_transradiative}
\feq 
where $k$ is the Rosselland coefficient describing the radiative opacity. 
The inner solar region is opaque for photons: their mean free path has been  estimated 
at a level lesser than 10$^{-3}$m, see \cite{FIO00}, the mean absorption coefficient being at a level 
of 0.4 $cm^2 g^{-1}$.\\
When convective motions are present, the stability of the system depends on adiabatic 
temperature  gradient ($\nabla_A$) (Schwartzschild criterion). The possible solution requires: 
\beq 
\nabla_{R} \leq \nabla_{A}\,=\,\left(\frac{\partial ln P} {\partial ln T}\right)_S 
\label{eq_schwartz}
\feq 
The analysis of convective motion is not trivial: in the usual description viscosity terms 
are neglected and the convection is assumed to be due to the motion of convective elements. 
At the start of the solar evolution, particles are in equilibrium with the medium, then 
they  move adiabatically and release the thermal energy 
surplus into the new medium in which they are.\\ 
Up to now, only approximate descriptions of convective motions have been possible: a 
phenomenological approach introduces a "mixing length", see \cite{BOE58}, the distance over 
which a unit of gas can be identified before it mixes. This value strictly depends on P: 
\beq 
l = \alpha \frac {d\,r} {d\,ln P} 
\label{eq_mixingl}
\feq 
where $\alpha$ is a free parameter: when it increases the convection becomes more efficient, 
$\nabla$ lowers and T at surface raises. As a further effect, being fixed the $L_\odot$ value, 
$R_\odot$ has to decrease. In any case, this is an over-simplified approach which does not allow any knowledge  
on the stellar radius and on the convective region features. \\
The initial chemical composition of Sun, which is unknown, has a great influence 
on solar  evolution, therefore the present values of $\rho$, T and chemical composition in the  
solar core are strictly correlated to the radiative opacity which controls $\nabla$.
Heavy elements are not completely ionised and their abundance on solar surface  decreases with 
time due to the gravitational settling and to diffusion processes.\\ 
When compared with other time scales, motions of gas particles in convective zone are so  rapid that this region 
may be taken chemically homogeneous. However, if macroscopic motions are absent, heavy elements tend to sink 
relative to lighter ones while composition gradients are smoothed by diffusion. The time scale of these processes 
in radiative zone is much longer. \\
SMs including settling and diffusion of elements were firstly detailed in 
\cite{COX89} while the influence of these processes on quantities 
measured by helioseismological techniques was pointed 
out in \cite{CHR93}.
Its inclusion in SMs gave results in better agreement with helioseismological  data, but it has increased the 
discrepancy between $\nu_\odot$ flux predictions and  experimental measurements on Earth. \\ 
The mixing-length description of the convective flux is a parameterisation with weak physical basis and the 
dynamical effect of convective motions on solar structure is generally ignored. Inclusion of convection in more 
realistic solar calculations was also attempted, \cite{CAN91}. \\
The last "solar" equation describes the evolution of the chemical composition but it gives a good representation 
only in the inner radiative region, where no matter exchanges between neighbouring layers are present and nuclear  
reactions mainly modify the hydrogen abundance. Diffusive processes do not give significant
contribution because of their slow temporal evolution; the equation is then: 
\beq 
\frac{\partial x_i}{\partial t}\,=\,\frac {m_i}{\rho}(\sum_{j} r_{ji} -\sum_{k} r_{ik} ) 
\label{eq_ssmequat4}
\feq 
where $r_{ik}$ is the rate of reaction in which nucleus "i" is transformed in nucleus "k".\\ 
In the convective zone turbulent motions change the chemical composition which is usually  assumed as a constant: 
this approximation is allowed by the local low temperature which  forbids nuclear fusion reactions. It is 
useful to remember that the mass of the convective region is less than 2$\%$ of the total solar mass 
($M_\odot \sim 2\cdot 10^{30}$ kg). \\ 
The introduction of the physics of the solar matter into the model is thus 
necessary to complete the description of the Sun. Auxiliary equations are 
usually equations of state and the equations for opacity and for nuclear 
reaction rates. The evolution in a SM is determined by microscopic 
properties of solar matter, at first thermodynamics relations, such as 
the dependence of P on $\rho$, T, chemical composition and nuclear 
parameters.\\ 
Taking into account the thermal pressure dominance over other pressures 
(mainly radiative), equations of state of a fully ionised perfect gas can 
describe the solar matter in regions where the energy is produced within 
an accuracy better than 1 \%,\cite{DEG97,FIO00}: 
\beq 
kT\,=\,u\mu \hspace{1.5cm} \mu\,=\,\frac{m_P}{1.5X\,+\,0.25Y\,+\,0.5} 
\label{eq_gas}
\feq 
where $\mu$ is the mean molecular weight and the isothermal sound speed squared $u$ is given by:
\beq
u\,=\,\frac{P}{\rho}\,=\,\frac{c_s^2}{\Gamma_1} 
\label{eq_isothermalspeed}
\feq
where $\Gamma_1$ is the adiabatic exponent. \\
In general, two approaches (chemical or physical) are possible. The former introduces atoms  
and ions while ionisation is described as a chemical reaction, see EFF formulation \cite{EGG73}. 
An evolution of this description is done in MHD description, \cite{DAP88,HUM88,MIH88}: 
modifications of atomic states are "heuristic" but the probability that a state is a function 
of the parameters of the surrounding plasma. \\
The latter provides a method to include non-ideal 
effects. OPAL project, \cite{IGL96}, presently the most adopted in computations, is based on a 
general description of the basic constituents (electrons and nuclei). Configurations 
corresponding to bound combinations, such as ions, atoms and molecules, are considered as a 
clustering process while plasma effects are introduced in terms of statistical mechanics.
Among non-ideal processes, the screening effect of positive charges by surrounding electrons 
and the interaction between bound particles (pressure ionisation) are noteworthy. 
In 2001 an upgrading to OPAL96 equations has been proposed where a relativistic treatment of 
electrons is combined with an improved activity expansion method for repulsive interactions, 
\cite{ROG01}.\\ 
Photons are responsible of the energy transport  while convective motions, energy  generation 
and chemical composition variations are produced only by nuclear reactions.  The opacity of 
solar material controls the energy flow through the Sun and in practice its luminosity. 
Different ways to construct opacity values were followed by OPAL group, 
\cite{ROG92,IGL92,IGL95,IGL96}, and by Opacity Project, \cite{SEA94}, but their results agree 
at a level better than 10 \%. In early SMs, opacities computed in \cite{HUE77} were used. \\
To develop SMs calculations, experimental values have to be introduced as input parameters; 
usually $L_\odot$, $R_\odot$ and Z values are chosen:  
\bit 
\item 
The present luminosity $L_\odot=\,3.842\,(1\,\pm\,0.004)\,\cdot 10^{26}W $, \cite{CRO96,FRO98}. 
\item 
The present radius $R_\odot\simeq\,6.9599\,(1\,\pm\,0.0007)\,\cdot 10^8\,m$. 
\item 
The detailed element composition for nuclei heavier than He as deduced from photospheric determinations 
and meteoritic analysis. 
\fit  
After a numerical integration of "solar" differential equations, T,
$\rho$, P and X are then computed as a function of their distance from the solar centre. 

\subsection{Nuclear cross-sections.}
\label{sect:suncrosssections}

The estimation of interaction cross-sections inside the Sun is a persistent and not easy 
to overcome problem: in most cases laboratory measurements are possible only at high energy 
and "solar" cross-sections are estimated by an extrapolation. \\ 
It was stressed the overwhelming importance of astrophysical factor S$_{11}$: if 
its value is higher than the usually adopted in SMs then T$_c$ is lower than  presently 
estimated, \cite{DEG98b}.\\   
Let us consider the $^3$He - $^3$He and $^3$He - $^4$He fusion rates. If 
the cross-section of the first reaction were higher than the extrapolated 
one presently used   or if the electron capture rate 
were lower or the $^7$Be + p  cross-section were higher, a significant and measurable 
modification in the  $\nu_\odot$ flux would be expected, see later 
in section \ref{sect:sunsolarnu}. \\
Experimentally speaking, the analysis of the region near 
the Gamow peak (the product of the Maxwell-Boltzmann distribution by the  cross-section, 
E$\sim$ 20 keV), is a very difficult task because of the smallness of the  cross-sections 
and the cosmic rays background.\\ 
A few years ago, a program (LUNA) of  deep underground nuclear measurements 
has started at Laboratori Nazionali del Gran Sasso, Italy, (LNGS). 
A resonance at thermal solar energy was proposed to explain the 
first solar neutrino problem, \cite{FET72,FOW72}: it would deplete the 
contribution of the reaction $^3$He($\alpha , \gamma$)$^7$Be, which 
starts the $^7$Be and $^8$B production. 
The latest LUNA results are based on measurements at a projectile energy 
of 16.5 keV, a value as low as at the solar Gamow peak, \cite{BON99}: the obtained
cross-section is 0.02 $\pm$  0.02 pb  and this means an interaction rate 
of 2 events per month! Resonances that could enhance the  $^3$He-$^3$He fusion 
rate were not found. It is impossible to lower the projectile energy because of the 
background reaction $^3$He(D,p)$^4$He which has a cross-section much 
larger than $^3$He($^3$He,2p)$^4$He reactions. \\
In any case, further studies and  measurements are needed for a better 
knowledge of the electron screening effect on fusion cross-sections at 
stellar energies: the experimental uncertainties on low-energy 
$^3$He($^4$He,$\gamma$)$^7$Be cross-section give in practice the dominant 
contribution to the uncertainty in 
$^7$Be and $^8$B $\nu_\odot$ flux calculation. 
Very recently experimental measurements concerning 
$^7$Be(p,$\gamma$)$^8$B reaction have been published, \cite{JUN02}: a higher value of 
astrophysical factor, even if within the  previous uncertainty range, has been computed 
and then inserted  in SMs computations, see \cite{BAH02c}. 
Significant variations on physical quantities have been excluded  but the $^8$B 
$\nu_\odot$ flux increases, \cite{BAH02c}. See also \cite{HAS99,GIA00,DAV01,HAM01,STR01a} 
for a complete review on $^7$Be + p cross-section measurements and relative astrophysical 
factor values.  It has to be stressed that newest results do not agree well each other 
so that, as a conservative prescription, previously quoted values, as done in 
\cite{ADE98,ANG99}, have to be presently preferred.

\subsection{Comments and problems.} 
\label{sect:suncomments} 
\begin{figure}[ht]
\begin{center} 
\mbox{\epsfig{file=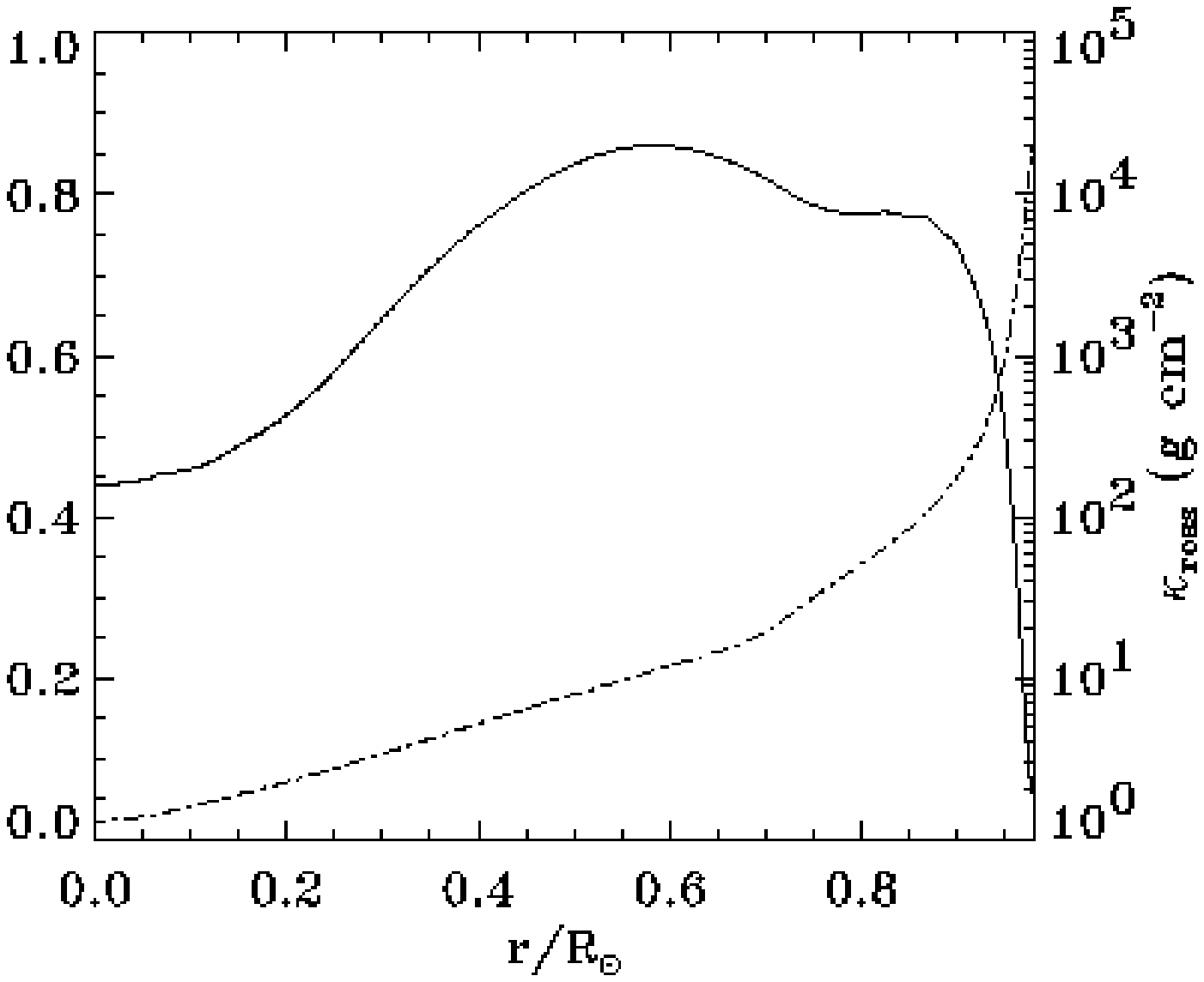,width=0.5\textwidth}} 
\end{center}
\caption{ The contribution of the heavy elements to the opacity (full line). Superimposed is 
the profile of the Rosseland opacities $\kappa_{ross}$ (dot-dashed line). Solar composition 
at $4.6$ Gyr, from \cite{COU02}. }
\label{fig:COUopacity}
\end{figure}

As a first comment, it has to be stressed that even if SMs are an "easy" way to 
describe the solar evolution, neglecting rotation, $B_\odot$ ... 
and despite many solar quantities have no direct measurement, the 
agreement between experimental and predicted values is very well, maybe surprising. 
Differences enhance the rough knowledge  of the physics acting inside the solar core.\\ 
The transport equation states both a temperature gradient, which induces an energy flux,  and an energy surplus, 
producing a gradient, enable to slow down it. In other words if  a radiative flux emerging from the solar core is 
present, then T must monotonically  increase toward the centre as long as the energy source will rise T in the core. 
On the contrary, that zone must be in thermal equilibrium. When $\nabla_A \leq \nabla_R$,  T decreases faster in the 
surrounding matter: an element adiabatically moving becomes  hotter than the medium so that it receives a further 
contribution to its motion. 
If the criterion on radiative equilibrium is violated, all the regions 
become  unstable with respect to any kind of motion of material elements. 
Consequently, this  is a criterion 
to establish when a zone is convectively stable and a "mixing zone"  between stable and unstable zones is present 
near the boundaries. This will be important in helioseismological analysis and in SMs skill to reproduce the 
behaviour of physical quantities inside the Sun.\\ 
Concerning the opacity, the main problem is a right estimation of the solar photon absorption 
cross-section: many different interactions and terms have to be considered, 
even if the basic contribution is due to the classic energy-independent Thompson cross-section. 
In radiative region the term related to heavy elements is dominant: 
at d = 0.6 $R_\odot$ its contribution is $\sim$ 85 $\%$ of the total value; 
on the contrary, in the solar core, where all the elements are fully 
ionized, except iron, its value decreases down to $\sim$ 45 $\%$ of the total. 
The opacities in the core are  mainly due to electron scattering and inverse brehmsstrahlung 
while at d = 0.6 $R_\odot$  the bound-bound processes dominate, see fig. \ref{fig:COUopacity}. 
The main sources of uncertainty for opacity are the metal content of the solar central region, 
which is without experimental data, and the actual calculation when one assumes a fixed chemical composition. \\ 
We list for instance processes which should contribute to the opacity value, for 
detailed description see \cite{BOE87,GOU90,IGL91,BAH95,IGL95,TSY95}: 
\bit 
\item 
Photon scattering on free electrons, taking also into account Doppler and collisional shifting 
of Raman resonance and relativistic corrections in the non-linear response due to the 
polarization.
\item 
Inverse bremsstrahlung, with or without collective and relativistic effects, collective 
effects from ion-ion and electron-electron interactions.
\item 
Electron degeneracy, quantum effects in scattering, stimulated scattering, frequency  
diffusion, plasmons contributions, refractive index corrections, variable density ... 
\fit 
These terms produce an uncertainty up to $\sim$ 10 - 15$\%$ on the Rosselland's  coefficient 
value and this is a crucial point because opacity affects T$_c$ and  therefore the nuclear 
interaction rates: the higher the opacity the slower is the  photon diffusion through the 
solar matter, so that T$_c$ rises.  In fact T$_c$ depends on the radiative opacity, on Z, on 
the astrophysical factor S$_{11}$ (or S$_{pp}$) and on the solar age.  
We underline that the present error on the last two quantities is at a level of 1$\%$ while 
the uncertainty on opacity and chemical composition is $\sim$ 10$\%$.\\ 
Uncertainties affect even the radiative zone description: convective motions penetrates  beyond 
the region of instability and can change $\nabla$. Induced gravity waves might lead to a 
mixing over time scale of solar evolution while  additional contributions to mixing could 
be produced by instabilities due to the solar rotations, \cite{ZAH92,MON94,CHA95}. 

\subsection{Helioseismology.} 
\label{sect:sunhelio}

\subsubsection{Introduction.} 
\label{sect:sunhelio_intro}
Helioseismology has deeply increased our knolwedge of the Sun and of other sun-like stars: it  
analyses the mechanical seismic waves, produced  by turbulences in the outer region of the 
convective zone, which travel through the solar matter with reflection on surfaces or 
refraction during the propagation toward the centre. Thanks to the movements of solar granules 
at the surface, the Sun  forms a spherical acoustic resonator producing a broad spectrum of 
random acoustical  noise; its period is of about 5 minutes. This phenomenon was  
first analysed 40 years ago, \cite{LEI62}.\\ 
The possibility of measuring the wave travel time inside the Sun was  discovered in 
the early 90's and was applied to probing subsurface flows, \cite{DUV93,DUV96,KOS96}. 
The accurate measurements of vibrations penetrating inside the Sun down to different depths 
allow to deduce information on thermodynamical quantities as a function of depth.\\ 
Solar oscillations are detected by measuring Doppler shift 
of a spectrum line or intensity of optical radiation but their
observations is a challenging task: the velocity amplitude of  
an acoustic p-mode is $\sim$ 1 cm/s while the associated brightness variation  
is $\sim 10^{-7}$, (the expected amplitude of gravity modes is about 10 times smaller). 
The errors in helioseismological frequencies have been reduced at a level of $\sim 10^{-5}$ 
while the errors on deduced quantities are higher, $\sim 10^{-4}$. \\
Data concerning solar oscillations were inverted to compute 
c$_s$ and $\rho$ profiles, \cite{DZI90,DAP91,ANT94,KOS97}, and the adiabatic index $\Gamma_1$, 
\cite{ANT94,ELL96,ELL98}; c$_s$ and $\rho$ radial profiles were also used to test opacity 
computations, \cite{KOR89,TRI97}, equations of state, \cite{ULR82,ELL98,BAS98b}, and elemental 
abundance profiles inside the Sun, \cite{ANT98,TAK98}. \\ 
Before the helioseismological measurements, three parameters were completely free in solar  
models: the starting Y and Z values and the mixing length. As output three  observable 
quantities were deduced: the present $R_\odot$, $L_\odot$ and $Z_{surf}$ .  During the last 
fifteen years results based on inversion of  helioseismological data provided strong 
constraints: the solar convection zone depth has  been derived by c$_s$ inversion, 
\cite{CHR91,BAS97}, while $Y_{surf}$ was estimated by inversion of p-mode data, 
\cite{BAS98a}. \\ 
In any case, helioseismology can only check the robustness of SMs because of its 
impossibility to directly reconstruct the profiles of physical quantities used in SMs. 

\subsubsection{Technical features.}
\label{sect:sunhelio_technical}  
The equations describing adiabatic and linear oscillations are hermitian, \cite{CHA64}, and the oscillations are 
thought as combination of standing waves, the eigenmodes of vibration. As usual procedure in wave mechanics, spherical 
harmonics $Y^m_l (\theta,\phi)$ are introduced.  Scalar perturbations and displacement are written in terms of 
physical parameters  and eigenfrequencies $\omega_{n,l,m}$, where $n$ is the number of nodes along the  radius, 
$l$ is the horizontal waves number on the surface and $m$ is the number  of nodes along the equator. Solar surface 
oscillations have a superposition of modes  with $0\leq l\leq 1500$. If only adiabatic oscillations and small radial  
wavelengths, compared to $R_{\odot}$, are considered, a dispersion relation  concerning the squared length of wave 
vector is valid: 
\beq  
|k|^2\,=\,k^2_r\,+\,k^2_h =
\frac {1}{c^2_s} \left(F^2_l\,(\frac{N^2}{\omega^2_{n,l,m}}\,-\,1)\, 
+\,\omega^2_{n,l,m}\,-\,\omega^2_c \right) + \frac{F^2_l}{c^2_s}  
\label{eq_heliok2}
\feq  
where the Lamb frequency is: 
\beq 
F^2_l\,=\,\frac {l^2c^2_s}{r^2}
\label{eq_Lamb} 
\feq 
the Brunt-V\"{a}is\"{a}l\"{a} frequency is:  
\beq 
N^2\,=\,g \left(\frac {1}{\Gamma_1} \frac {d}{dr} ln\, P\,-\,\frac  {d}{dr} ln\,\rho \right) 
\label{eq_Brunt}
\feq 
the acoustic cut-off frequency ($\sim$ 5.8 mHz) is: 
\beq 
\omega^2_c\,=\,c^2\,\left(1\,-\,2\frac {d}{dr}H_\rho\right) 
\label{eq_cutoff}
\feq 
the density scale height is: 
\beq 
H^{-1}_\rho\,=\,-\frac {d}{dr} ln\,\rho 
\label{eq_height}
\feq 
the sound speed is: 
\beq  
c^2_s\,=\,\Gamma_1 \frac {P}{\rho} 
\label{eq_soundspeed}
\feq 
Different oscillatory solutions are possible: the acoustic waves, when  
$\omega_{n,l,m} \geq N,F_l$, the gravity ones otherwise. The major restoring forces 
responsible for the solar oscillations are pressure and 
buoyancy. Pressure fluctuations are important at high frequencies $via$ 
the production of acoustic waves; at lower frequencies, buoyancy is much  
stronger and produces internal gravity waves which can propagate only in  
convective stable regions (solar atmosphere and beneath the convective 
zone). Standing acoustic waves are known as p-modes; standing internal 
gravity waves are called  g-modes. Their magnitude depends on local 
$\rho$, on P and on chemical composition. 

\subsubsection{Detection techniques.}
\label{sect:sunheliodetection}  
Ground instruments, like GONG (Global Oscillation Network Group),  consisting of 6 Doppler 
imaging equal instruments, and satellite experiments,  like SOHO (SOlar and Heliospheric 
Observatory), the project of an international  collaboration (NASA and ESA) carrying 3 
different detectors (GOLF, MDI, SOI), measure  these 
waves (for a nice overview see \cite{SCI96}). 
The seismic waves are detected by different techniques: the simplest one 
is the  whole-disk integrated light measurement, as presently in VIRGO 
experiment and in future satellite detectors. 
A second technique uses velocity oscillations by detecting Doppler shift 
of spectral lines. Two kinds of observation are possible: the 
detection of the global Doppler velocity, measured by the displacement of 
specific absorption lines (Na or K), or local velocity, separating the 
Sun into pixels. In the former, low degree modes ($0\leq l\leq 3$), 
which are the most penetrating ones, are detected (BiSON, IRIS and GOLF use
this method).  The global Doppler velocity shift 
of sodium lines is under analysis. The latter,  isolating modes of higher 
degree, is used by MDI, which analyses Ni spectral lines.\\ 
When integrated light coming from the whole solar disk is detected, the 
sensitivity is limited to few modes having wavelength as large as the 
solar diameter. \\
At ground level, GONG network, which  is sensitive to low l-modes, 
is continuously observing the solar surface. Even SOHO satellite allows 
continuous data taking and a reduction of detectable amplitude at small 
values, where influences due to the turbulence at solar surface are not 
present; in particular GOLF and MDI analyse the 
time variability of the Doppler velocity. \\
The observations have to be made over a long time without any 
interruptions; solar oscillation frequencies are deduced from time series 
of the intensity or from radial velocity data (more frequently used).  
In this case modes with very small amplitudes and velocity not greater 
than some mm/s can be identified.

\subsubsection{Data analysis and results.}
\label{sect:sunhelioresults} 
\begin{figure}[ht]
\begin{center} 
\mbox{\epsfig{file=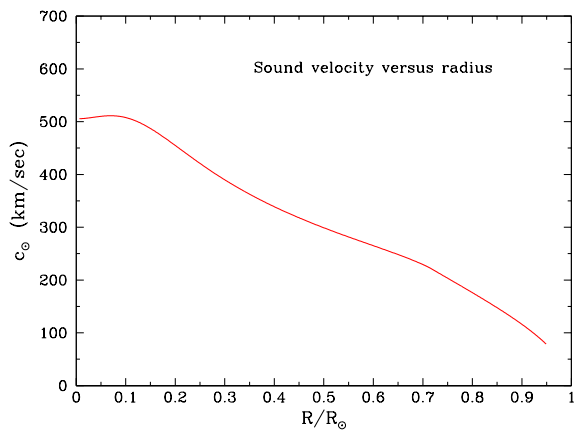,angle=-90,width=0.5\textwidth}} 
\end{center}
\caption{ Calculated solar sound speed versus radius for the reference SM, \cite{BAH01a}. 
The difference between computed and observed values is overall lesser than 
0.5 km/s, from \cite{BAH01a}.} 
\label{fig:BAH346sound}
\end{figure}

\begin{figure}[ht]
\begin{center} 
\mbox{\epsfig{file=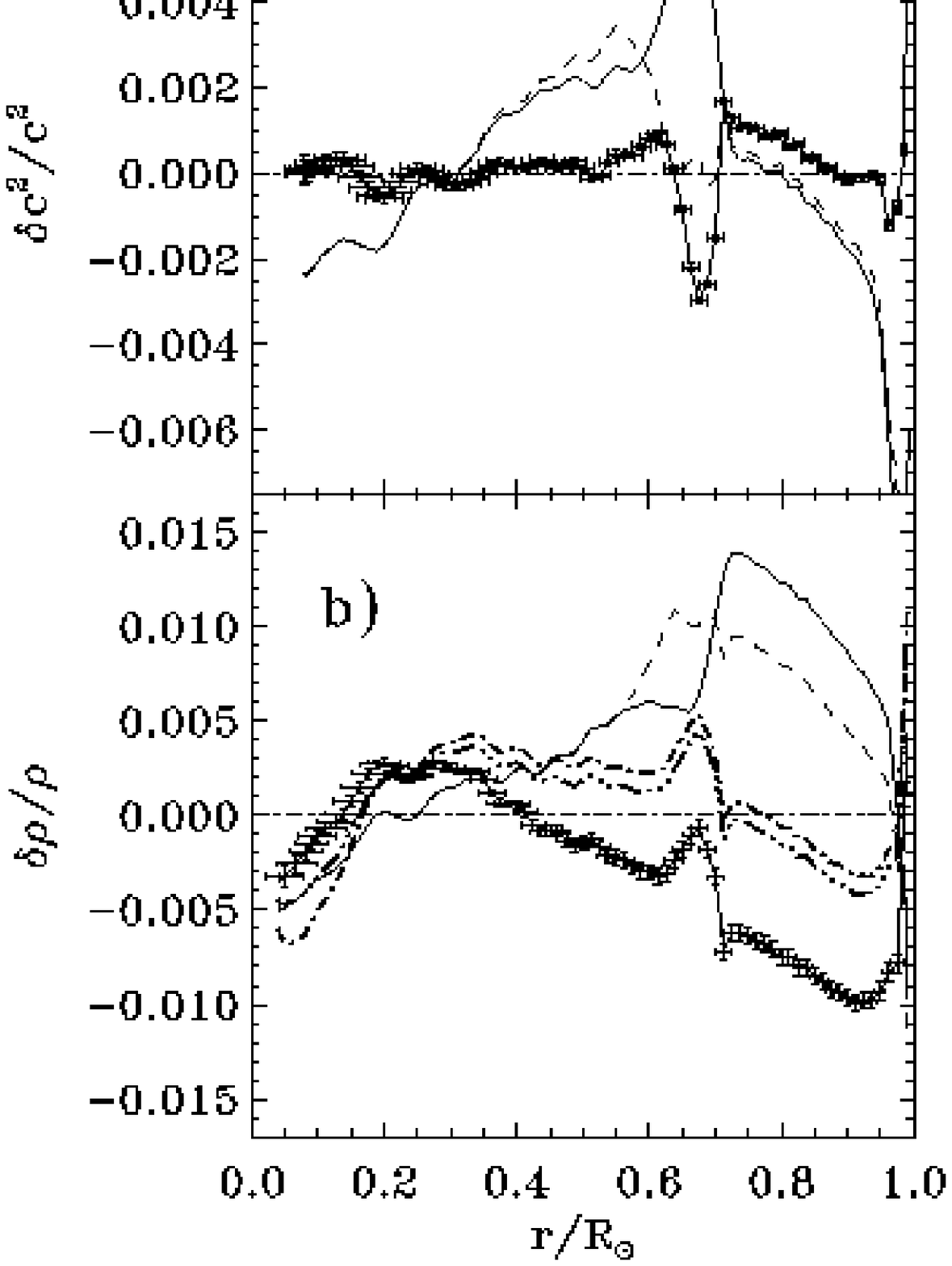,width=0.5\textwidth}} 
\end{center}
\caption{ In a) the difference in the square of the sound speed between 
the Sun and the model \cite{COU02}(full line with errors) while in  b) is shown the 
difference in the density.
The plain lines with no error bar are for the Saclay standard SM, \cite{BRU98}, 
while the dashed lines are for the Btz model, \cite{BRU99}. 
Two other SMs based on \cite{COU02} are considered on the density 
figure: a model with the $^{3}He(^{4}He,\gamma)^{7}Be$ reaction rate reduced 
by $10\%$ (dot-dot-dot-dashed line) and a model with the CNO poly-cycle 
reaction rates reduced by $70\%$ (dot-dashed line), from \cite{COU02}. }
\label{fig:COUseismic1}
\end{figure}

Presently about thousand of p-modes could be used in inversion relations 
(they are a  fraction of the total number of measured frequencies) but 
only modes with l$\leq$ 150 are useful to study solar interior because 
higher frequencies propagate exclusively on the shallow outermost layers. 
P-mode frequencies are measured with an accuracy better than 
0.01$\%$ but no more than 5$\%$ of their total amount is produced in the
solar core; on the contrary the most part of gravity waves should come 
from the solar core.\\ 
Two ways are possible to extract information from experimental data: 
\bit 
\item 
The direct method, when detected and expected acoustic frequencies are compared. 
\item 
The indirect method, when inversion relations are used to reconstruct the radial  profile 
of variables as c$^2_s$, $\rho$ or $\Gamma_1$ which are then compared to  theoretical ones. 
\fit 
As example in fig. \ref{fig:BAH346sound} the c$_s$ profile is shown: values vary from 
$\sim$ 510 km/s in solar core down to $\sim$ 80 km/s at 0.95 $R_\odot$, \cite{BAH01a}. 
\\
The sensitivity of c$_s$ to different "ingredients" of a SM is strong: a 2$\%$ 
variation  in the p-p reaction rate implies a 0.2$\%$ variation on c$_s$, as much as a 
similar change of the opacity coefficient. Moreover, a 10$\%$ variation of this parameter 
along the solar radius has a negligible effect on c$_s$ but T$_c$ is modified at a level 
of 1$\%$ and significant variations in $\nu_\odot$ flux are foreseen,
see section \ref{sect:sunsolarnu}.
\begin{figure}[ht]
\begin{center} 
\mbox{\epsfig{file=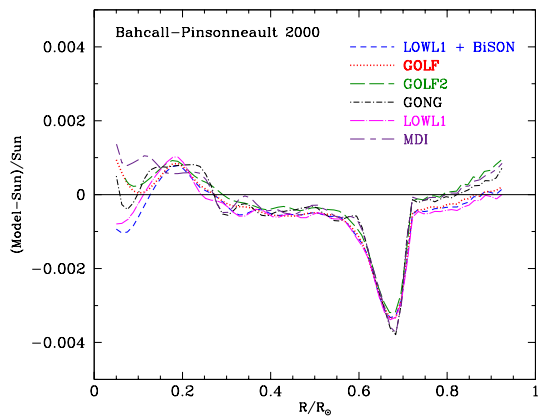,angle=-90.,width=0.5\textwidth}} 
\end{center}
\caption{ Fractional difference between the sound speed calculated in  
\cite{BAH01a} and the values measured in six helioseismological experiments. Systematic 
uncertainties due to the assumed reference SM and the width of the inversion kernel are 
each $\sim 0.0003$, from \cite{BAH01a}.}
\label{fig:BAH346csdiff}
\end{figure}
\\
The measured oscillation frequencies give information on solar interior  $via$ "inversion" 
procedure. If $f(r)$ is a function describing a solar inner property, $\nu^{exp}_{l,n}$ is 
an observed frequency and $\nu^{the}_{l,n}$ the corresponding frequency as computed in a 
SM then: 
\beq 
\frac {\delta\,\nu_{l,n}} {\nu_{l,n}}\,=\,\int^R_0 K^n(r) \frac  {\delta\,f(r)}{f(r)} dr   
\label{eq_inversion1}
\feq   
where $K^n(r)$ is a function of SM relating variation in frequency  
$\delta\nu_{l,n}$ with $\delta f(r)$. Inversion technique tries to form 
linear combinations of terms, as expressed in previous equation, by 
varying l and n with a weight $c_{l,n}(r_0)$ chosen to obtain a value 
$\frac{\delta f}{f}$ localised in $r_0$. 
The "averaging kernel" $H(r_0,r)$ indicates the localisation of the 
measure of  $\frac{\delta f}{f}$:
\beqar
\int^R_0 H(r_0,r) \frac {\delta\,f_1(r)}{f_1(r)} dr \,=\,\sum_{l,n}  a_{l,n}(r_0) 
\frac {\delta\,\nu_{l,n}(r)} {\nu_{l,n}(r)}  \\    
H(r_0,r)\,=\,\sum_{l,n} a_{l,n}(r_0) K^n(r) \nonumber
\label{eq_inversion2}
\feqar
Hence, it is possible to calculate 
the difference  between $\frac{\delta f}{f}$ as computed by using a SM 
and the observed  values. It has to be stressed that this procedure 
becomes difficult going toward the solar core.\\ 
Usually, the rms difference  between the measured (as deduced from 
helioseismological inversion formula) and the  SMs estimated one 
is plotted.
Uncertainties from measured frequencies are smaller than those allowed by 
present inversion  relations; in any case they are at a level lower than 
0.5$\%$ in the total analysed region.
A constant uncertainty of $\sim $ 3 $\cdot 10^{-4}$ comes from each source 
of the c$_s$  profile: the adopted SM, the width of the kernel of the 
inversion procedure and the  experimental errors. The finite resolution 
of the inversion kernel leads to those rms  systematic uncertainties 
although errors are much larger in the solar core 
and at the base of the convective zone. 
On the other hand, the uncertainties in $\rho$ profile are an  order of 
magnitude larger than the previous, see for instance figures \ref{fig:BAH346csdiff} and
\ref{fig:COUseismic1}. 
Among a lot of interesting results, a sharp transition between differential rotation in 
the external convective zone and a practically rigid rotation in the  radiative zone has 
been found. This transition appears as a peak in c$_s$ difference profile and allows an 
accurate evaluation of the base of the  convective zone where a change between temperature 
gradient is "physically" present. Up to now the origin of this behaviour is not clear. 
The bump close to the transition zone between  convective and radiative regions 
can be reduced by a local enhancement of  the opacity coefficient introduced in SMs, 
but this is a trick, and the true physical motivation is unknown.\\ 
The depth of the convective zone was analysed starting from 90s', \cite{DEM91}, 
firstly without element diffusion, \cite{BAH90b,PRO91,BAH92}. \\
The measurements of the g-modes is needed to further progress in the solar core analysis: 
these frequencies, which are proportional to Brunt-V\"{a}is\"{a}l\"{a} value and mainly 
produced in the region below 0.2 R$_{\odot}$, are about 10 times more sensitive to core 
perturbations. The main experimental difficulties come from the estimated amplitude of 
their speed, less than 1 mm/s, and  from broad fluctuation of the solar noise in that 
frequency range. A long time of  data taking (several years) is needed to improve the 
signal to noise ratio. \\ 
The sensitivity of g-modes to physical parameter variations is low: a slight opacity 
enhancement and modified nuclear reaction rates change the g-modes frequencies at a 
level  below the experimental resolution. In any case, g-modes must 
be identified at a lower  order to be able to give useful information. 
The dependence of the g-modes on physical processes is not easy to 
anticipate therefore the identification of such modes 
is difficult. Up to now measurements of c$_s$ are possible at distance from the solar 
core greater than 0.07 $\pm$ 3.6$\%$ $R_\odot$, \cite{TUR01c}, (only g-modes may improve 
this situation). Presently few "candidate" frequencies seem to be present in SOHO's data.\\ 
Temporal variations of rotation rate in solar interior and possible 
correlation between solar activity, namely the sunspots  number, and 
variation on $R_\odot$ value have been recently analysed. In the first 
case, a slightly different  rotation rate was detected into the outer 
part of the convective zone, from the solar surface down to 0.9 R$_\odot$
from the solar centre; nothing significant seems to be present at greater
depth. Moreover, no statistically important
variation in solar rotation rate near the base  of the convective zone was found, 
\cite{ANT00}. \\ 
Sometines, SMs are directly compared with the p-mode  oscillation 
frequencies, see \cite{GUE97}, while 
in the most part of the SMs the c$_s$ profile is used: in fact the inversion procedure 
allows to remove uncertainties arising from  external solar region where turbulence and 
non adiabatic effects are present.\\ 
The possibility of estimating the solar age $via$ helioseismological data 
was shown in \cite{DZI99} by 
using Small Frequencies Separation Analysis (SFSA). The analysed quantities 
have a strong sensitivity to the c$_s$ 
gradient near the solar centre and a weak  correlation with the description of outer layers. 
The so computed value is $t_\odot = 4.66 \pm 0.11 \cdot 10^9$ y (the lifetime on the pre-main 
sequence has to be added), see also \cite{MOR00}. This value, which is $\approx 2.9\%$ greater 
than the standard one, produces a rather important change in c$_s$ 
profile and reduces the differences with respect to the "real" Sun.\\
A good agreement between computed SMs  and solar properties as deduced from helioseismology 
has been obtained by including relativistic corrections in equations of state: in fact 
the adiabatic exponent and c$_s$  are better reproduced, see \cite{ELL98,BON01}. 
In \cite{BON02a} new results have been computed by using updated equations 
of states, including relativistic contributions in OPAL and MHD equations: 
the so obtained solar age is $t_\odot$ = 4.57$\pm$ 0.11 $10^9$y, in 
excellent agreement with the meteoritic age $t_\odot$ = 4.57 $\pm$ 0.02 
$\cdot 10^9$ y, \cite{BAH95}. As a further result, the most 
favoured value for the zero-energy astrophysical factor has been deduced 
$S_{11}$ = 4.00$\cdot 10^{-25}$ Mev b.\\ 
Many SMs reproduce p-mode frequencies within an uncertainty of 0.1$\%$  
without any special adjustment of the input parameters. Over the entire 
region of the Sun  for which the helioseismological values are determined 
($0.05\leq x \leq 0.95\, R_\odot$) the computed profiles of P and $\rho$ 
agree very well with helioseismological data, the difference being lesser 
than 1$\%$. In these regions T varies by a factor of 20 while the molecular 
weight changes at a level of few percent. The agreement is worse in the 
deepest solar core and in the most external layer. 
A direct reconstruction from helioseismology either of T$_c$ and of 
$\nu_\odot$ flux is impossible: data only constrain 
the range of the allowed values, checking the agreement of different SMs 
with measurements ({\it{e.g.}} the $^8$B and $^7$Be $\nu$ fluxes are 
differently constrained, at a level not greater than 25$\%$ for 
the latter, \cite{RIC99}). 

\subsection{Reference solar model.} 
\label{sect:sunBAH01a}
\begin{figure}[ht]
\begin{center} 
\mbox{\epsfig{file=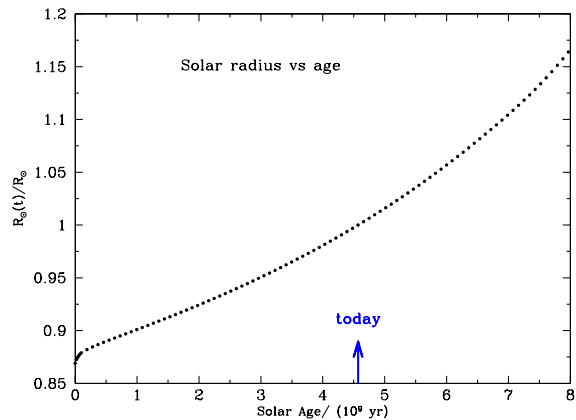,angle=-90.,width=0.5\textwidth}} 
\end{center}
\caption{ $R_\odot(t)$ as a function of the solar age with respect to the present value, 
from \cite{BAH01a}.}
\label{fig:BAH346radius}
\end{figure}

\begin{figure}[ht]
\begin{center} 
\mbox{\epsfig{file=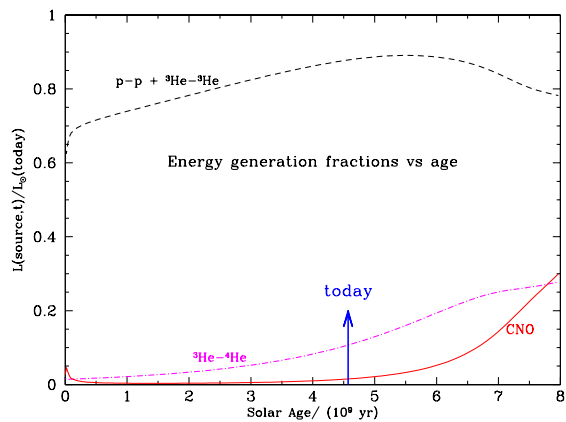,angle=-90.,width=0.5\textwidth}} 
\end{center}
\caption{ The fraction of $L_\odot$ with respect to the present value produced by nuclear 
fusion reactions versus solar age, from \cite{BAH01a}: the $p-p$ branch, terminated by the 
$^3$He-$^3$He reaction,  (dashed curve); the $^3$He-$^4$He branch (dot-dashed curve); the 
CNO cycle (solid line), from \cite{BAH01a}.}
\label{fig:BAH346enfrac}
\end{figure}

\begin{figure}[ht]
\begin{center} 
\mbox{\epsfig{file=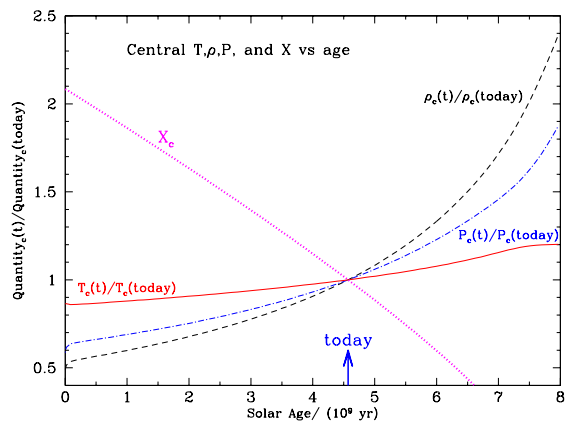,angle=-90.,width=0.5\textwidth}} 
\end{center}
\caption{ The temporal evolution of T$_c$ (solid line), $\rho_c$ (dash line), P$_c$ 
(dot-dash line) and X$_c$ (dotted line), from \cite{BAH01a}.}
\label{fig:BAH346central}
\end{figure}

For many years Bahcall's SMs have been seen as "reference" model or "the" SSM, 
\cite{BAH88,BAH92,BAH95,BAH98a}. The last one updates and refines the analyses within a 
"classic" treatment, \cite{BAH01a,BAH02c}.\\  
As input parameters the model takes cross-sections from \cite{ADE98,ANG99}, the OPAL96 
opacities from \cite{ROG96a}, the low temperature opacities from \cite{ALE94}, equations 
of state from \cite{IGL96}, the solar composition from \cite{GRE98} (even if the OPAL96 
equations of state are not available with these values), the electron and ion weak 
screening treatment from \cite{GRU98}; the He and heavy metal diffusion from \cite{THO94,BAH95}. 
Furthermore, a new electron density profile and time evolution are done and the mixing 
length treatment has been applied. The pre-main sequence evolution is not included because 
its effect on the internal solar  structure is estimated negligible on $\nu_\odot$ production, 
\cite{BAH94a,MOR00}. The physical  variables under analysis (M, R, T, $\rho$, 
P, L, source densities of 8 different $\nu$  fluxes, 
H-$^3$He-$^4$He-$^7$Be-$^{12}$C-$^{14}$N-$^{16}$O mass fraction,    
electron density) are given in 875 separate radial shells in the Sun.
$L_\odot=3.844 \cdot 10^{26}\, W$, (3.842 in $\nu_\odot$ flux calculations), 
$R_\odot= 6.9598\cdot 10^8\, m$, $t=4.57 \cdot 10^9\, y$ are the input values.\\ 
The structure of the Sun is estimated to be affected only very slightly by 
the solar radius: in fact the available values differ as much as less than 
1 part in 10$^3$, \cite{BAH97b}. The time evolution of many parameters, 
including the depth of the convective zone, has been calculated from the 
Sun entrance into the  main sequence up to an age of 8$\cdot 10^9$ y. 
The derived c$_s$ well agrees with helioseismological data: the fractional 
difference is less than $ 10^{-3}$  over the interval 0.05 - 0.95 
$R_\odot$ while in the region $x\leq \, 0.25\,R_\odot$, where $\nu_\odot$'s 
are produced, this discrepancy is lesser than 7 $\cdot 10^{-4}$, see fig. \ref{fig:BAH346csdiff}.  \\
We resume the most important conclusions: 
\bit 
\item 
$R_\odot$ increases monotonically with a rise of 13$\%$ up to now; it will further increase 
by a factor of 1.17, see fig. \ref{fig:BAH346radius}.
\item 
$L_\odot$ rises its value from the $\sim$ 67$\%$ up to $\sim$ 136$\%$ of the present one. 
\item 
The $^3$He-$^3$He termination reactions produce 87.8$\%$ of the present $L_\odot$, the 
$^3$He-$^4$He and CNO cycle contributing respectively at a level of the 10.7$\%$ and 1.5$\%$: 
at an age of 8$\cdot 10^9$ y these values will become 57.6$\%$, 20.4$\%$ and 22.0$\%$. 
The energy loss due to solar expansion varies from 0.03 up to 0.07$\%$, see fig. 
\ref{fig:BAH346enfrac}. 
\item 
T$_c$ increases from $\sim$ 1.4$\cdot 10^7$K up to $\sim$ 1.9$\cdot 10^7$K, see fig.
\ref{fig:BAH346central}; $R_\odot$ shows the same behaviour, in fact the ratio 
$R_\odot$/T$_c$ is almost constant, within $\sim$ 1.5$\%$. 
\item 
The hydrogen mass fraction in the solar core decreases from the starting value of  $\sim$ 
0.7 to the present one (0.34) to the final value of $\sim$ 0.15 of the total solar mass. 
\item $\nu_\odot$ flux values are given in table \ref{tabsmflux}; some uncertainties are 
increased  with respect to the previous Bahcall's SSM, \cite{BAH98a}. 
\fit   
Different astrophysical factors have been used in \cite{BAH02c} and a new $^8$B $\nu_\odot$ 
flux has been computed (without significant variations in the remaining solar quantities); 
in following papers, see \cite{BAH02f}, this upgrading has been not confirmed, due 
to a more conservative analysis of the presently available experimental 
measurements of related fusion cross-sections.

\subsection{Other solar models.}
\label{sect:sunotherSM} 
\begin{figure}[ht]
\begin{center} 
\mbox{\epsfig{file=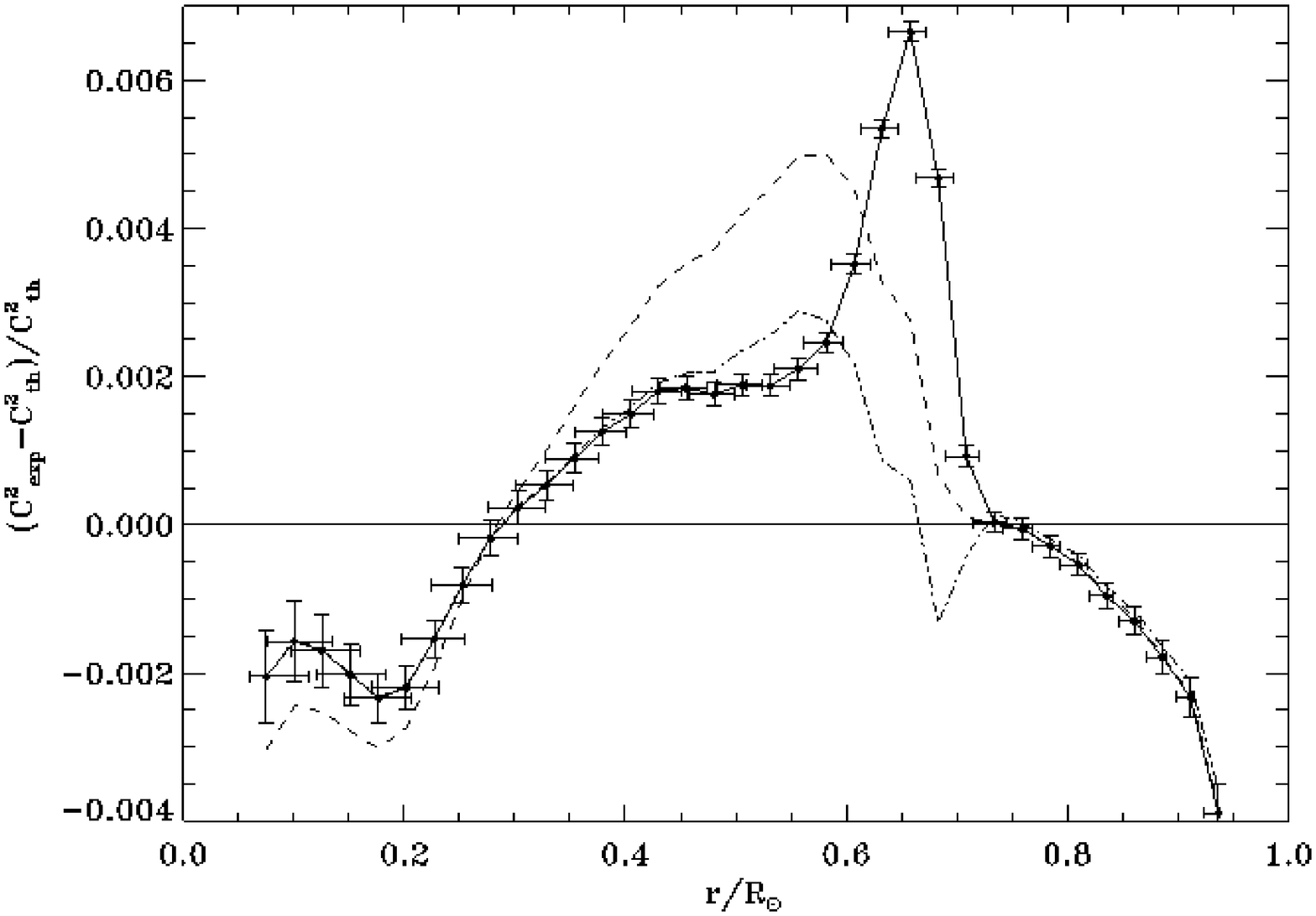,angle=90.,width=0.5\textwidth}} 
\end{center}
\caption{ The effect of calibration on the sound square speed difference between the SM 
\cite{BRU98} and two SMs including a macroscopic diffusion term: $B_t$ calibrated in Z/X (- - -) 
and $B_{tz}$ with a non calibrated Z$_0=$Z$_0^{std}=0.01959$ (dash dot line), from \cite{BRU99}.}
\label{fig:BRUfig6}
\end{figure}

Many SMs were presented and it is impossible to list them or their features; we mention  
\cite{PRO94,CHR96,CAS97,GUE97,DAR98,BRU99,DAR99,SCH99,TUR01c,WAT01,COU02} among models proposed in the 
last decade. We refer to \cite{BOO02}for a detailed analysis of the individual influence of parameters
like radius, time evolution, luminosity, heavy element abundance, equations of state ... on SM
calculations.\\
Even non-standard SMs were suggested searching for the explication of the 
experimental $\nu_\odot$ deficit compared with the theoretical expectation based on the standard 
evolutionary SMs, see
sections \ref{sect:sunsolarnu} and \ref{sect:astrosolution}.
We mention models having a core with low Z, 
\cite{JOS74,LEV94,JEF97}: in this case low opacity and lower T$_c$ have to be expected.\\ 
Presently, many researchers are working on the so  called "seismic" SMs, a dynamical evolution of 
standard SMs in  which strong constraints on solar parameters and on evolution are deduced from 
helioseismology. In general, "seismic" SMs are directly derived from seismic 
measurements and use primary inversion of data producing c$_s$ and 
$\rho$ profiles. After the addition of "normal"  physical conditions, P 
and T profiles and the Y value are computable, 
\cite{KOS91,DZI95a,BAS96a,BAS96b,SHI96,ANT98}.\\ 
In SMs the chemical abundance distribution is obtained by following the time evolution, where the nuclear reactions 
and the diffusion processes are taken into  account. The chemical composition is assumed to be uniform at zero-age; 
in many non-standard SMs other time evolution processes  are introduced and/or the initial conditions at t = 0 are 
different. Seismic SMs do not follow the time evolution: the helioseismologically determined c$_s$ profile
is imposed as a constraint.\\ 
If only the X, Y and Z value are introduced, c$_s$ is a function of P, T, X and Z  
\beq 	
c (P, T, X, Z) = c_{obs}(r)
\label{eq_soundrelation1}
\feq 
This inversely relates X with P, T, Z and c$_s$,  
\beq 	
X = X (P, T, Z, c_{obs})
\label{eq_soundrelation2}
\feq 
If Z is given, the basic equations can be solved with the proper boundary 
conditions. The helioseismically reconstructed $\rho_{obs}(r)$ profile  
can be used as further constraint in computing a seismic SM and in 
addition to c$_s$ it determines the Z profile as a part of the solution. 
Both the weak dependence of the equations of state upon 
Z and the error of $\rho_{obs}(r)$, which is much larger than that of 
$c_{obs}(r)$, even by a factor of 10, have to be pointed out.\\ 
Very recently, the contributions of solar rotations and $B_\odot$ have been added in solar codes 
to better estimate the evolution, \cite{COU02}; in any case the way the  convection acts is not clear. \\ 
In \cite{BRU98,BRU99} a new term is introduced in the "standard solar equations"  describing physical processes which 
better reproduce properties in the layer between convective and radiative zone. The aim is a change in rotation 
rate which is supposed to contribute to a turbulent mixing in a thin layer where $B_\odot$ could maximally act. 
A macroscopic diffusion, describing the mixing in a shear layer between the convective zone (where a differential 
rotation is present) and the radiative zone (which is rotating as a "solid")
is added $via$ a diffusivity term in equation describing chemical evolution: 
\beq 
\frac{\partial X_i}{\partial t}\,=\,-\frac{\partial (4\pi \rho r^2 X_i V_i)} {\partial m}\,+\,nuclear \,terms 
\label{eq_brundiff}
\feq 
where the velocity $V_i$ of species $i$ with respect to the centre of mass  depends 
both on microscopic and macroscopic diffusion and on concentration gradient.  It is possible to imagine an anisotropic 
turbulence with viscous transport much stronger in the horizontal than in vertical direction.\\ 
In order to solve the solar equations, simplifying assumptions concerning the strength of velocities, $\rho$ 
variation and tachocline dimension are introduced. The mixing in the tachocline depends on rotation and 
differential rotation rate; moreover, it is related to horizontal diffusivity and  Brunt-V\"{a}is\"{a}l\"{a} frequency. 
The tachocline width evolution in time and the  efficiency of the mixing are also analysed. \\ 
Opacities from \cite{IGL96}, the equations of state from \cite{ROG96a}, the solar chemical  composition from 
\cite{GRE96}, the reaction rates from \cite{ADE98,ANG99} (with the  exception of proton capture on Li from 
\cite{ENG92}), the nuclear screening from \cite{MIT77,DZI95b}, the microscopic diffusion from \cite{MIC93}; an estimated  
solar age of 4.55$\cdot 10^9$ y after a 5$\cdot 10^7$ y of pre-main sequence phase have been used as input parameters. 
In fig. \ref{fig:BRUfig6} the solar sound speed difference profile computed for either Z calibrated or Z free SMs are 
shown.
Among many interesting results, a good $^7$Li depletion has been obtained without a  corresponding $^9$Be deficit, 
according with newest photospheric observation, \cite{BAL98}.  These results seem to be dependent on the adopted 
Brunt-V\"{a}is\"{a}l\"{a} frequency.  This SM presents a mixing at the base of the convective zone, an increased value 
both for Z at the surface and for the opacity; the discrepancy between the seismic c$_s$ profile and the reconstructed 
one  is lower. In any case, Z at surface is within the experimental uncertainties, \cite{TUR00,TUR01a}. \\ 
A "seismic" SM has been proposed in \cite{TUR01c}. Thanks to the long duration of SOHO data taking, low order modes 
(n$\leq$ 9) of lowest degree ($l=0,1,2$), which are more insensitive to the turbulence of upper solar layers, have 
allowed an upgrade of the inversion method used to compute the c$_s$ profile. The new c$_s$ fractional error is 
$\sim 2\cdot 10^{-4}$, with a small increase up to $\sim 5\cdot 10^{-4}$  in the more external region. The spatial 
uncertainty remains at a level ranging from 1 up to 3.5  $\%$. These constraints are imposed to the \cite{BRU99} model; 
moreover, the p-p reaction rate  has been enhanced at a level of 1$\%$ while Z$_0$ is 3.5$\%$ greater than in \cite{BRU99}. 
A smaller radius ($R_\odot = 6.95865\cdot 10^8$  m) and a 10$^7$ y pre-main sequence time have been introduced in 
computations. A good agreement with helioseismological c$_s$ profile is obtained, even if a large difference is present 
in very external solar layers, $x\geq 0.95\,R_\odot$. The computed $\nu_\odot$ flux does not show significant deviations 
from the previous estimated one in \cite{BRU99}.\\ 
An upgraded "seismic" SM has been proposed in \cite{COU02}: it is computed 
with a 1D quasi-static stellar evolution code which includes pre-main 
sequence and  solves the stellar structure equations by a spline 
coefficient method as in \cite{BRU99,TUR01c}. The main characteristics are: the nuclear reaction rates from \cite{ADE98}, 
the screening treatment from \cite{MIT77}, the  astrophysical factor from \cite{ENG92,HAM98a}; the time step and the 
rotation law were modified to take into account the Li burning in pre-main sequence, \cite{PIA02}; 
the opacities from \cite{IGL96} for T$\geq$ 5600 K, for lower T values from \cite{ALE94}; the equation of state from 
\cite{ROG96b} while the microscopic diffusion is  from \cite{MIC93} and the turbulent mixing at the base of the 
convection zone is from \cite{BRU99}. Large screening and mixing in the 
central region are excluded following \cite{TUR01b}.\\ 
The inversion procedure for c$_s$ and $\rho$ profiles is based on the 
Optimally Localised Averaging method, \cite{KOS99}. Hydrodynamic prescriptions are 
introduced to define tachocline: they reduce the influence of the diffusion of heavy elements toward the solar centre 
and allow to burn $^{7}$Li on the main sequence. A correct $^7$Li abundance at the solar surface is computed, the value 
according to \cite{GRE93}. \\  
The p-p cross-section, which has a strong influence on c$_s$, is increased by 1$\%$ so that T$_c$ slows down; 
Z$_0$ is higher (3.9$\%$) therefore the Rosselland opacities and the mean molecular weight are different. In fig. 
\ref{fig:COUopacity} the profiles of heavy metal contribution and Rosselland opacities are plotted as a function
of the distance from the solar core. In such a model, c$_s$ decreases in 
the core because the raise of $\mu$ dominates the T raise due to a greater 
opacity value. On the contrary, at d$\geq$ 0.3 $R_\odot$ the raise in T overcomes the increase of $\mu$ 
so that  c$_s$ goes up. The higher Z initial does not modify the c$_s$ profile.\\ 
In order to reduce the differences at the base of the convective zone, the parameters defining the tachocline have been 
modified by introducing a lower width, a slightly greater rotation rate and a higher Brunt-V\"{a}is\"{a}l\"{a} 
frequency.\\ 
However, a better agreement with solar parameters has been searched for 
by calibrating this SM at a radius 
value different from the standard photometric one. Recent analyses based on f-mode frequencies have given respectively 
$R_\odot$ = 6.9578  $\cdot 10^8$m, \cite{SCH97}, and $R_\odot$ = 6.9568$\cdot 10^8$m, \cite{ANT98}, while optical  
determination indicates $R_\odot$ = 6.9551$\cdot 10^8$m, \cite{BRO98}. Presently there is neither an answer explaining 
these discrepancies, nor the effect of the solar cycle on the radius. 
The value used in computation is $R_\odot$ = 6.95936$\cdot 10^8$m. \\
The impact of solar age has been checked: if $t_\odot$ increases, the 
discrepancy on c$_s$ below 0.2 $R_\odot$ diminishes while the differences 
in $\rho$ are greater. $t_\odot=4.6$ Gyr seems to be a 
satisfactory value (the value needed to best reduce the discrepancy on $\rho$ below 0.6  $R_\odot$ is 4.55 Gyr). A 
higher $t_\odot$ value is strongly disfavoured by $\rho$ while c$_s$ does not rule it out (but does not strongly 
support it too). \\
In fig. \ref{fig:COUseismic1} the solar sound speed difference profile is shown for the basic model; 
for comparison also $\rho$ difference profiles concerning different SMs are drawn. \\ 
Different $B_\odot$ profiles have been introduced in the calculations but none of the so computed models 
greatly modify the $\nu_\odot$ flux predictions: it seems that $B_\odot$ 
has a quite slight impact.\\  
Italian contributions to solar modeling and their critical evolutions were done by  V.Castellani, G.Fiorentini, 
O.Straniero and coll., \cite{STR89,CAS97}. In \cite{CAS97} a "helioseismological" constrained SM was proposed, 
FRANEC being the code for computation. Opacity from \cite{ROG95}, equations of state from \cite{ROG96b}, Z value from 
\cite{GRE93}, molecular opacity at T$\leq 10^4$K from \cite{ALE94}, He and heavy metal diffusion from \cite{THO94}, a 
time evolution step of $\sim 3\cdot 10^7$y,  updated nuclear cross-sections are the main input parameters.\\ 
Papers concerning refinements and constraints both in solar and $\nu_\odot$ physics were published, 
\cite{RIC97,DEG97,DEG98a,FIO99a,FIO99b,CAS99,DZI99,RIC99,FIO00,RIC00,FIO01a,FIO01b,FIO01c,FIO02}: 
the influence of screening factor, plasma collective effects, solar age, solar radius and  opacity on $\nu_\odot$ flux, 
characteristics of $^7$Be $\nu_\odot$ and $^8$B $\nu_\odot$ component and 
relations between them, uncertainties in solar parameters have been analysed in detail.\\ 
The model done by A.Dar and G.Shaviv, \cite{DAR96,DAR98,DAR99}, uses quite 
different values for nuclear reaction rates, modified screening and diffusion description. Each element is separately 
diffused inside the Sun and diffusion  
coefficients are calculated for the local ionisation state; both these values are in agreement with \cite{RIC96}. 
Moreover, Z value from \cite{GRE96} was introduced, Y value being an adjustable parameter.\\ 
Watanabe and Shibahashi have realized "seismic" SMs both with homogeneous Z and with low-Z  core, see \cite{WAT01} 
and references therein. In this case, the low-Z solar core is larger  than in SSMs, therefore the opacity is decreased as 
well as T$_c$, $\mu$  is smaller to balance the lower T$_c$; the X value is increased, the core density is higher to  
compensate the lower nuclear reaction rate due to the lower T$_c$; P is higher. All the other parameters and equations 
are "standard". The effect of various uncertainties in microphysics upon the seismic SM with homogeneous  Z and the 
theoretically expected $\nu_\odot$ fluxes (which were investigated by a Monte Carlo simulation) are summarised 
in table \ref{tabuncwat}.  \\ 
In \cite{WIN01} SMs which upgrade models previously computed in \cite{GUE97} are analysed. Routines and equations are 
"standard" but an enhanced  contribution due to heavy elements, slightly different values of $L_\odot$ and shells have 
been introduced: 16 non-standard SMs and 4 SSMs have been computed. Among non-standard SMs, only two models are 
marginally compatible with  observations: in any case an upper limit to Z 
contribute to the accretion of the Sun during its early 
main sequence phase is provided. A "good" SSM, having predictions in 
agreement with experimental results within the errors, is present but 
Z$_{surf}$ is high and $^8$B $\nu_\odot$ flux is large. \\  
Table \ref{tabsm1} and table \ref{tabsm2} summarize characteristics of different SMs. \\
\begin{table}[ht]
\caption{\it{ Characteristics of solar models. The quantities T$_c$ (in units of  10$^6$ K), 
$\rho_c$ $(10^2~{gm~cm^{-3}})$, and P$_c$ $(10^{17}~{erg~cm^{-3}})$  are the present central temperature, density, and 
pressure; Y and Z are the He and heavy  element mass fractions, the subscript "$_0$" shows the zero-age main sequence 
value while "$_c$" indicates the values in the solar core. The symbol "-" means not indicated.}}
\begin{center} 
\begin{tabular}{|l|c|c|c|c|c|c|c|}  \hline 
Model&T$_c$&$\rho_c$&$P_c$&$Y_0$&$Z_0$&$Y_c$&$Z_c$\\ \hline\hline\hline 
BAH01a&15.696&152.7&2.342&0.2735&0.0188&0.6405&0.0198\\\hline 
NACRE&15.665&151.9&2.325&0.2739&0.0188&0.6341&0.0197\\\hline 
ASP00&15.619&152.2&2.340&0.2679&0.0187&0.6341&0.0197\\\hline 
GRE93& 15.729 & 152.9 & 2.342 & 0.2748 & 0.02004 & 0.6425 & 0.02110\\\hline 
Pre-MS& 15.725 & 152.7 & 2.339 & 0.2752 & 0.02003 & 0.6420 & 0.02109\\\hline 
Rot& 15.652 & 148.1 & 2.313 & 0.2723 & 0.01934 & 0.6199 & 0.02032\\\hline 
Rad$_{78}$& 15.729 & 152.9 & 2.342 & 0.2748 & 0.02004 & 0.6425 &0.02110\\\hline 
Rad$_{508}$& 15.728 & 152.9 & 2.341 & 0.2748 & 0.02004 & 0.6425 &0.02110\\\hline 
No-Diff.& 15.448 & 148.6 & 2.304 & 0.2656 & 0.01757 & 0.6172 &0.01757\\\hline 
Old & 15.787 & 154.8 & 2.378 & 0.2779 & 0.01996 & 0.6439 &0.02102\\\hline 
S$_{34} = 0$& 15.621 & 153.5 & 2.417 & 0.2722 & 0.02012 & 0.6097 &0.02116\\\hline 
Mixed& 15.189 & 90.68 & 1.728 & 0.2898 & 0.02012 & 0.3687 &0.02047\\\hline 
PRO94&15.66&154.7&-- &0.2707&0.01907&0.6370&0.02013\\\hline 
RIC96&15.67&154.53&2.350&0.2793&-- &0.6465&-- \\\hline 
CAS97&15.69&151.8&-- &0.2690&0.0198&0.630 &0.0198  \\\hline 
GUE97&15.74&153.11&2.355&-- &0.0200&-- &0.0211\\\hline 
BRU99&15.71&153.1&-- &0.2722&0.01959&0.6405&0.02094\\\hline 
DAR99&15.61&155.4&-- &0.2509 &0.01833 &0.6380 &0.01940 \\\hline 
MOR99&15.73&153.8&-- &0.2723&0.0197&0.6418&0.0210\\\hline 
SCH99&15.7&152.0&-- & 0.275&0.020&-- & 0.018\\\hline 
WAT01&15.61&156.0&2.378&-- &-- &0.6437 &0.0180 \\\hline 
WIN01&15.885 & 154.17&2.3605 &-- &0.0220 &-- &0.0232 \\\hline 
COU02&15.739  &153.02&2.3375&0.2759&0.02035&0.6445 &0.02168\\\hline    
\end{tabular} 
\end{center} 
\label{tabsm1}
\end{table}
\\
BAH01a is the reference Bahcall's SSM; values up to Rad$_{508}$ are concerning SSMs which use only different input 
parameters with respect to \cite{BAH01a}. Model NACRE uses the charged particle fusion cross-sections from \cite{ANG99}; 
model ASP00 uses a lower abundance $Z/X= 0.0226$ \cite{ASP00}. Models from 
GRE93 up to Rad$_{508}$ use the value of Z quoted in \cite{GRE93}. Model 
Pre-MS is evolved from the pre-main sequence  stage; model Rot 
incorporates mixing induced by rotation and is a "reasonable" upper bound  
to the degree of rotational mixing consistent with the observed 
Li solar depletion. Rad$_{78}$ is a model using a solar radius of 
6.9578$\cdot 10^8$ m, \cite{ANT98}, while in Rad$_{508}$ the adopted value 
is 6.95508$\cdot 10^8$ m, \cite{BRO98}. 
These models are consistent with the helioseismological data; their rms sound speed  differences with respect to 
\cite{BAH01a} are: $0.03$\% (Pre-MS), $0.08$\% (Rot), $0.15$\% (Rad$_{78}$), 
and $0.03$\% (Rad$_{508}$). \\
No-Diff model does not use  diffusion of elements; Old  model uses older 
equations of state and opacities;  $S_{34}=0$ model introduces 
a null $^7$Be production, so that no $\nu_\odot$'s from $^7$Be or $^8$B 
are present; Mixed model modifies the solar core composition following 
\cite{CUM96}. The remaining SMs are quoted in references.\\ 
\begin{table}[ht]
\caption{\it{ Characteristics of the convective zone and surface: Y$_{surf}$ and Z$_{surf}$ are the 
He and heavy element abundances at surface, $\alpha$ is the mixing length parameter, R$_{CZ}$  and T$_{CZ}$ 
are the radius and temperature at the base of the convective zone,  and M$_{CZ}$ is the mass included within the 
convective zone. The symbol "--" means not  indicated.}} 
\begin{center} 
\begin{tabular}{|l|c|c|c|c|c|c|} \hline 
Model&$Y_{surf}$&$Z_{surf}$&$\alpha$&$R_{CZ}$&$M_{CZ}$&$T_{CZ}$\\ 
&&&&$(R_\odot)$&$(M_\odot)$&$(10^6~K)$\\\hline 
BAH01a&0.2437&0.01694&2.04&0.7140&0.02415&2.18\\\hline 
NACRE&0.2443&0.01696&2.04&0.7133&0.02451&2.19\\\hline 
ASP00&0.2386&0.01684&2.05&0.7141&0.02394&2.18\\\hline 
GRE93&0.2450 & 0.01805 & 2.06 & 0.7124 & 0.02457 & 2.20\\\hline 
Pre-MS&0.2455 & 0.01805 & 2.05 & 0.7127 & 0.02443 & 2.20\\\hline 
Rot&0.2483 & 0.01797 & 2.03 & 0.7144 & 0.02388 & 2.15\\\hline 
Rad$_{78}$& 0.2450 & 0.01806 & 2.06 & 0.7123 & 0.02461 & 2.20\\\hline 
Rad$_{508}$& 0.2450 & 0.01806 & 2.06 & 0.7122 & 0.02467 & 2.20\\\hline 
No-Diff.&0.2655 & 0.01757 & 1.90 & 0.7261 & 0.02037 & 2.09\\\hline 
Old &0.2476 & 0.01796 & 2.04 & 0.7115 & 0.02455 & 2.21\\\hline 
S$_{34} = 0$&0.2422 & 0.01811 & 2.03 & 0.7151 & 0.02309 & 2.17\\\hline 
Mixed& 0.2535 &0.01782 & 1.85 & 0.7315 & 0.01757 & 2.02\\\hline 
PRO94&0.2422&0.01758&1.677&0.715&0.02340&-- \\\hline 
RIC96&0.2584&-- &1.768&0.716&-- &2.175\\\hline 
CAS97&0.238&0.0182&1.90&0.716&-- &2.17\\\hline 
GUE97&-- &0.018&-- &0.716&0.0234&-- \\\hline 
BRU99&0.2508&0.01858&1.755&0.7141&-- &2.194\\\hline 
DAR99&0.2308&0.0170&-- &0.7130&-- &2.105\\\hline 
MOR99&0.2436&0.0181&1.924&0.7138&-- &-- \\\hline 
SCH99&0.245&0.18&-- &0.713&0.0264&2.19\\\hline 
WAT01&0.2455&0.018&-- &-- &-- &-- \\\hline 
WIN01&0.2519&0.0198&-- &0.7128&0.02449&-- \\\hline 
COU02&0.2508&0.01918&1.934&0.7115&0.025&2.22 \\\hline  
\end{tabular} 
\end{center}
\label{tabsm2}
\end{table}

\subsection{Solar neutrinos.} 
\label{sect:sunsolarnu}
\begin{figure}[ht]
\begin{center} 
\mbox{\epsfig{file=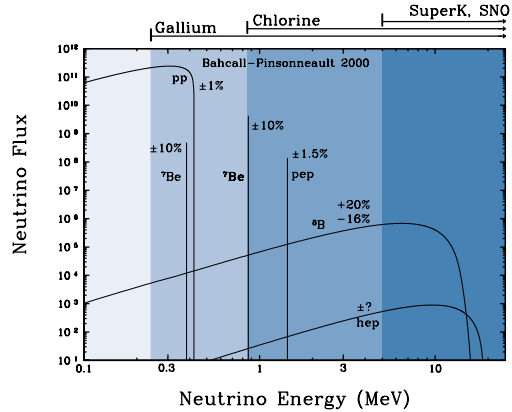,angle=-90.,width=0.5\textwidth}} 
\end{center}
\caption{ The p-p neutrino energy spectrum with related uncertainties as computed in 
reference solar model, \cite{BAH01a}; the detection energy range allowed for different 
experiments is also shown, from \cite{BAH01a}. }
\label{fig:BAHnuflux}
\end{figure}

\begin{figure}[ht]
\begin{center} 
\mbox{\epsfig{file=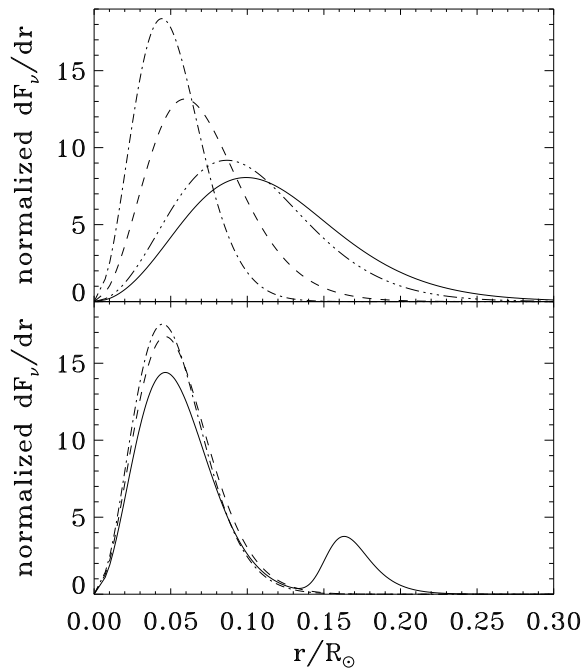,width=0.5\textwidth}} 
\end{center}
\caption{ Neutrino production as a function of the distance from the solar centre. 
In the upper figure: p-p (plain curve), $^8$B (dot-dashed curve), $^7$Be (dashed curve), pep 
(dot-dot-dot-dashed curve) neutrinos are shown.
In the lower figure: $^{13}$N (plain curve), $^{15}$O (dashed curve), $^{17}$F (dot-dashed curve) 
neutrinos are reported. $(1/F_{t}) \, (dF/dr)$ is drawn for each component where $F$ is 
the flux in s$^{-1}$, $r$ the fractional radius, $F_{t}$ the total flux for the selected component, from
\cite{COU02}. }
\label{fig:COUnuprodrad}
\end{figure}

\begin{figure}[ht]
\begin{center} 
\mbox{\epsfig{file=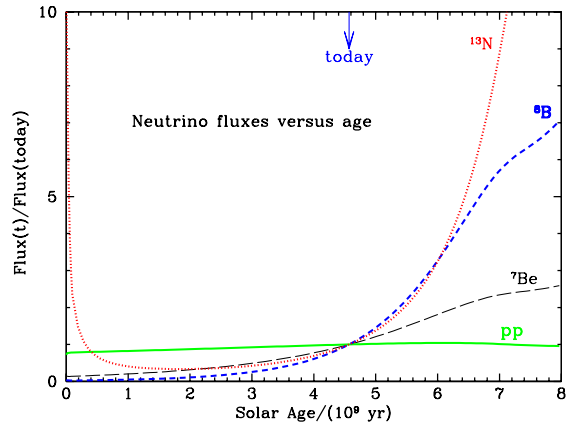,angle=-90.,width=0.5\textwidth}} 
\end{center}
\caption{ P-p (solid line), $^7$Be (long dashes), $^8$B (short dashes), and $^{13}$N (dotted line) 
neutrino fluxes as a function of solar age with respect to the present value are shown, from \cite{BAH01a}. }
\label{fig:BAH346fluxage}
\end{figure}

Following the SMs, even $\nu$'s are produced in p-p and CNO reaction chains, see table \ref{tabpp} and table \ref{tabcno}.
Energy conservation law constrains the total $\nu_\odot$  flux to be fixed by $L_\odot$. If the Sun is supposed to be in 
steady state with nuclear  energy production equal to $L_\odot$, the basic reaction 
\(4\,p^+\,+\,2\,e^-\,\rightarrow  ^4_2He\,+\,2\,\nu_e\) produces 26.732 MeV, the Q value of the reaction.\\ 
A roughly estimated $\nu_\odot$ flux on the Earth is thus:
\beqar 
\Phi\,\simeq\, \frac{2\,S_\odot} {Q\,-\,2\,E_\nu} \sim \, 6.5 \cdot 10^{10}\,
{\nu_e}\,cm^{-2}\,s^{-1} \\
S_\odot\,=\,\frac{L_\odot}{4\,\pi\,d^2}\sim\,1.367\,kWm^{-2}=\,8.533\cdot 10^{11} \,MeV\,cm^{-2}s^{-1} 
\nonumber
\label{eq_nuflux}
\feqar
where $d$ is the Earth-Sun distance and $S_\odot$, the present "solar constant"
is affected by an uncertainty of $\sim$ 0.4$\%$.\\ 
The $\nu_\odot$ energy spectrum can be thought as a "superposition" of $\beta$-decay spectra 
with an  addition due to the electron capture processes, see fig. \ref{fig:BAHnuflux}; it is 
computed in the standard way:
\beq
\frac{dN}{dp_e}\propto \left(E\,-\,E_{kin,e}\right)^2\,p^2_e F(Z,E_{kin,e})
\label{eq_nuspectrum}
\feq
where $p_e$ is the momentum of the electron.
\\
Eight $\nu$ sources are available within the p-p and CNO reaction chains: six of them give  rise to continuous 
energy spectra while the remaining two, p-e-p and $^7$Be, produce  monochromatic lines.
The global energy fractions carried out by  $\nu_\odot$'s are different for the various reaction 
chains, see table \ref{tabenergy}, but the most part of the energy released to solar matter is produced in the p-p I 
branch: since only one $\nu$ source is present, its amount is strictly related to $L_\odot$ and therefore less dependent  
on adopted SM.\\ 
The relative contribution of various  $\nu_\odot$ sources is depending on 
inner solar conditions, mainly T$_c$, $\rho$ and  chemical composition. \\ 
In \cite{BAH96a} a power law between nuclear fusion reactions, 
$\nu_\odot$ flux and T$_c$ was found: $\Phi(pp)\propto T_c^{-1.1\pm 0.1}$, $\Phi(pep)\propto 
T_c^{-2.4\pm 0.9}$, $\Phi(Be)\propto T_c^{10\pm 2}$, $\Phi(B)\propto T_c^{24\pm 5}$, $\Phi(N)\propto  T_c^{24.4\pm 0.2}$, 
$\Phi(O)\propto T_c^{27.1\pm 0.1}$ and $\Phi(F)\propto T_c^{27.8\pm 0.1}$. 
Hence, T$_c$ indicates the dominant process in  p-p chain: the p-p I when T$_c\leq 1.6\cdot 10^7$K; p-p II if 
$1.6\cdot 10^7K\leq  T_c\leq 2.3\cdot 10^7K$ and p-p III at higher temperature. All $\nu_\odot$ flux 
components, with the exclusion of p-p term, are strongly  T$_c$ dependent, so that $\nu_\odot$ flux and solar features 
are strictly related. \\
Produced $\nu_\odot$'s are of electronic type: they escape from the Sun without any  interaction with the solar matter 
and reach the Earth. Following the hypotheses done in SMs, this flux is not reduced and without changes in energy 
spectrum. Their experimental  detection and the determination of their energy spectrum enable us to check the solar  
interior and T$_c$; moreover, a test for solar and stellar astrophysics is allowed. In  practice $\nu_\odot$'s are 
the most efficient instruments to evaluate the nuclear aspects  into the central solar regions.\\ 
Fig. \ref{fig:COUnuprodrad} presents the $\nu_\odot$ production as a function of the distance from the solar centre and 
of the mass fraction inside this sphere. The right location of these maxima  is weakly SM dependent: it is possible to 
estimate more than 90$\%$ of the total p-p  $\nu_\odot$ production at distance x$\leq$ 0.2 $R_\odot$; $^7$Be $\nu_\odot$'s 
are produced at x$\leq$ 0.06 $R_\odot$, $^8$B $\nu_\odot$'s at x$\leq$ 0.05 $R_\odot$.
\\
The number of produced $\nu_\odot$'s depends on their reaction rate $r$ which is related  to the value of all physical 
parameters in each inner solar point; therefore it is the  selected SM that changes their flux.\\ 
If $N_i$ is the number of $\nu_\odot$'s produced in a specific nuclear reaction $i$, then: 
\beq 
N_i\,=\,4\pi \int^{R_\odot}_0 x^2 r_i(x,P,T,..) dx 
\label{eq_nufluxreaction}
\feq 
where the distance from the solar centre is $x$ while $r_i(x,P,T,..)$ is the rate of the i-reaction. The integral 
has to be computed at each distance the reaction occurs: the maximum distance from the solar centre depends on 
the selected nuclear process but it is in any case lesser than $R_\odot$ because of the threshold temperature, see 
fig. \ref{fig:COUnuprodrad}.\\
The p-p $\nu_\odot$'s are the most abundant: they are produced in the first step of the fusion reaction chain and have a 
continuous energy spectrum up to 420 keV. Their flux is practically SM independent, being related to $L_\odot$. A small 
fraction is due to the 3-bodies reaction, p-e-p, with the emission of a monoenergetic $\nu$ at E = 1.44 MeV. \\
$^7$Be $\nu_\odot$'s are monoenergetic; the prediction of their flux is quite  stable;
they are produced in a secondary branch of the cycle; $\sim$ 90$\%$ of them  are emitted in a 
line at 863 keV, the remaining at E = 386 keV. \\
$^8B$ $\nu_\odot$ flux is affected by a large  uncertainty. The suppression of this 
component  would have no impact on $L_\odot$; they are produced in a very small branch of the p-p chain, but 
their energy extends up to 14 MeV and therefore they are more  easily detected than the previous ones.\\
CNO $\nu_\odot$'s give a small contribution in the ongoing experiments because their flux and energy 
are low; $^{13}$N and $^{15}$O  $\beta$-decays correspond to the thermal energy derived from the reactions 
having $^{12}$C and $^{13}$C as starting nuclei. On the other hand, the $\nu_\odot$ flux from $^{17}$F $\beta$-decay is 
a potential measure of the primordial $^{16}$O abundance but it  gives a very small contribution to 
$L_\odot$, \cite{BAH82}. \\
The so called Hep component of the $\nu_\odot$ flux is the most energetic and mysterious. Its value is affect by a
great uncertainty ($\sim 100\%$);  it is the less abundant in $\nu_\odot$ flux by many orders of magnitude, 
\cite{BAH98b}. \\
In table \ref{tabsolarprod} the $\nu_\odot$ production in the Sun is
quoted,\cite{COU02}.\\
\begin{table}[ht]
\caption{\it{ The $\nu_\odot$ production in the Sun, adapted from \cite{COU02}. }}
\begin{center} 
\begin{tabular}{|l|c|} 
\hline 
Reaction&SUN\\ 
source    &$10^{33}\nu\,s^{-1}$ \\\hline
p-p       & 167300\\\hline 
p-e-p     &393.6\\\hline 
$^7$Be    &13720\\\hline 
$^8$B     &14.08\\\hline 
$^{13}$N  &1631 \\\hline 
$^{15}$O  &1406\\\hline 
$^{17}$F  &8.72\\\hline 
\end{tabular} 
\end{center} 
\label{tabsolarprod}
\end{table}
The reference model, \cite{BAH01a}, also proposes a time evolution of the $\nu_\odot$ flux up to an age of 
8$\cdot 10^9$ y, see fig. \ref{fig:BAH346fluxage}.  
The p-p $\nu_\odot$ flux remains almost constant in  time: at the beginning of the main sequence its value was 
the 75$\%$ of the present one  and its variation is $\sim$ 4$\%$ per $10^9$ y. It will reach its maximum  
(4$\%$ higher than the present value) at the age of 6.0$\cdot 10^9$ y and then it will decline very slowly.  
The $^7$Be $\nu_\odot$ flux started at a low level (14$\%$ of the present one) and it will increase 
monotonically by a factor of 2.6 . At the  beginning of main sequence, the $^8$B $\nu_\odot$ flux was very small 
($\sim$ 3$\%$ of the present one); it will increase up to a value greater by a factor of 8.1 with a fast rise of 
120$\%$ per $10^9$ y. The $^{13}$N $\nu_\odot$ flux was 11 times larger than the present one during the first 
$10^8$ y on the main sequence; then it decreased down to 1/3 to the present one at an age of 1.8$\cdot 10^9$ y 
and it started to go up steadily as T$_c$ increases. Its final value will be larger by 
a factor of 18 with respect to the present value. 

\subsection{Solar neutrino flux on Earth.}
\label{sect:sunnufluxearth} 
Usually all SMs assume standard model  for electroweak interactions 
without mass and magnetic moment for $\nu$'s; moreover, the  energy emission from the Sun is supposed to be isotropic. 
The solar matter is in practice  totally transparent for $\nu$'s (the mean free path for $\nu_\odot$'s is 
$\approx 10^{18}$ m,  a value much greater than $R_\odot \approx 10^9$ m).\\ 
The flux which should be measured on Earth surface is then: 
\beq 
\Phi_i\,=\,\frac{N_i} {4 \pi d^2} 
\label{eq_nuflux2}
\feq 
The distance $d$ between Earth and Sun is variable from 1.471$\cdot 10^{11}$ m 
in January (perihelion) up to 1.521$\cdot 10^{11}$ m in July (aphelion) 
because of the ellipticity of the terrestrial orbit, therefore: 
\beq
d\,=\,d_0(1+\epsilon\,\cos \pi t)
\label{eq_eccentricity}
\feq
where $\epsilon=$0.0165 is the eccentricity. Consequently, the $\nu_\odot$ 
flux varies at a level of some percent between January and July.\\
The total interaction rate of $\nu_\odot$'s inside a detector 
having a mass M is: 
\beq 
I\,=\,MN_{Av}\frac {X}{A}\sum_i \Phi_i \int^{E_{max}}_{E_{thr}}  \frac {d\lambda_{i}} {d E}\sigma(E)dE 
\label{eq_interactionrate}
\feq 
$\sigma$(E) is the $\nu$ capture cross-section, $\Phi_i$ is the integral flux on  Earth of $i$-type $\nu_\odot$'s, 
X is the isotopic abundance of nucleus with mass number A,  N$_{Av}$ is the Avogadro's number and 
$\frac {d\lambda _{i}}{dE}$ is the differential  normalized energy spectrum of $\nu_\odot$'s produced by $i$ reaction.\\  
The interaction rate per target atom is: 
\beq 
R\,= \,\sum_i \Phi_i \int^{E_{max}}_{E_{thr}} \frac {d\lambda_{i}}{d E} \sigma(E)dE \,= \,\sum_i \Phi _{i} 
\overline{\sigma _i} 
\label{eq_ratesnu}
\feq 
where $\overline{\sigma _i}$ is the capture cross-section averaged on the energy spectrum. \\ 
R is usually measured in SNU, the Solar Neutrino Unit, which means 1 capture per second 
per  10$^{36}$ target nuclei: this is the not a flux unit but an interaction rate unit. \\
In table \ref{tabsmflux} the values of $\nu_\odot$ flux on Earth from different SMs are listed.
There is a general good agreement among results computed under different working
hypotheses: if models with unusual assumptions are excluded, p-p $\nu_\odot$ values agree 
within 1\% while the differences on $^7$Be and $^8$B components are respectively 
at a level not greater than 10\% and 35\%.\\
\begin{table}[ht]
\caption{\it{ Solar neutrino fluxes on Earth from different solar models, the unit is
$\nu\,cm^{-2}s^{-1}$; the multiplying factor is also reported. Interaction rates 
on Cl and Ga are shown in the last two columns.  
The symbol "-" means not computed or not indicated.}}
\begin{center} 
\begin{tabular}{|l|c|c|c|c|c|c|c|c|c|c|}
\hline Model&p-p&p-e-p&Hep&$^7$Be&$^8$B&$^{13}$N&$^{15}$O&$^{17}$F&Cl&Ga\\  
 &10$^{10}$&10$^8$&10$^3$&10$^9$&10$^6$&10$^8$&10$^8$&10$^6$&SNU &SNU \\ \hline 
BAH01a&5.95&1.40&9.3&4.77&5.05&5.48&4.80&5.63&7.6&128\\ \hline 
NACRE&5.97&1.39&9.4&4.85&5.54&4.93&4.24&5.39&8.0&128\\ \hline 
ASP00&5.99&1.41&9.4&4.62&4.70&5.25&4.56&5.33&7.1&126\\ \hline 
GRE93& 5.94 & 1.39 & 9.2 & 4.88 & 5.31 & 6.18 & 5.45 & 6.50 & 8.0 &130 \\\hline 
Pre-MS& 5.95 & 1.39 & 9.2 & 4.87 & 5.29 & 6.16 & 5.43 & 6.47 & 7.9 &130 \\\hline 
Rot.& 5.98 & 1.40 & 9.2 & 4.68 & 4.91 & 5.57 & 4.87 & 5.79 & 7.4 &127 \\\hline 
Rad$_{78}$& 5.94 & 1.39 & 9.2 & 4.88 & 5.31 & 6.18 & 5.45 & 6.50 &8.0 & 130 \\\hline 
Rad$_{508}$& 5.94 & 1.39 & 9.2 & 4.88 & 5.31 & 6.18 & 5.45 & 6.50 &8.0 & 130 \\\hline 
No-Diff& 6.05 & 1.43 & 9.6 & 4.21 & 3.87 & 4.09 & 3.46 & 4.05 & 6.0 &120 \\\hline 
Old & 5.95 & 1.41 & 9.2 & 4.91 & 5.15 & 5.77 & 5.03 & 5.92 & 7.8 &130 \\\hline 
S$_{34} = 0$& 6.40 & 1.55 & 10.1 & 0.00 & 0.00 & 6.47 & 5.64 & 6.70 & 0.8 &89 \\\hline 
Mixed& 6.13 & 1.27 & 6.2 & 3.57 & 4.13 & 3.04 & 3.05 & 3.61 & 6.1 & 115\\\hline 
PRO94&5.98&1.42&1.23&4.79&5.46&5.02&4.27&5.16&7.71&130.1\\\hline 
RIC96&5.94&1.38& --   &4.80&6.33&5.59&4.81&6.18&8.49&132.8\\\hline 
CAS97&5.99&1.40&  --  &4.49&5.16&5.30 &4.50 & --   &7.4 &128  \\\hline 
GUE97&5.95&1.40&1.22&4.94&5.96&5.64&4.89&5.92&8.35&133  \\\hline  
BRU99&5.98 &1.41   & --   &4.70  &5.99& 4.66  & 3.97  & --   &7.04&127.1\\\hline 
DAR99&6.10&1.43&  -- &3.71&2.49&3.82&3.74&4.53&4.10&115.6\\\hline 
MOR99&5.91&1.40&  -- &4.90&5.68&5.73&4.96&6.41&8.31&130.1\\\hline 
SCH99&  -- & --   &  -- & --   &5.18& --   & --   & --   &7.79&128.7\\\hline 
WAT01&5.98&1.44&2.11&4.72&4.77&4.43&4.15&5.22&7.17&126  \\\hline 
WIN01&5.90&1.38&2.03&5.07&5.63&7.30&6.48&7.50&8.16&136  \\\hline  
BAH02c&5.95&1.40&9.3&4.77&5.93&5.48&4.80&5.63&8.6&129.9\\\hline 
COU02&5.92&1.39& --   &4.85&4.98&5.77&4.97&3.08&7.48&128.1\\\hline 
\end{tabular} 
\end{center} 
\label{tabsmflux}
\end{table}

\subsubsection{Time variations.} 
\label{sect:sunnufluxtime}
It is usually assumed that the Sun is in a steady state therefore SMs do not predict $\nu_\odot$ flux variation on 
short time-scale. In any case, different effects are suspected to produce modulations:
\bit 
\item 
{\bf{Day-Night effect}} = The line of sight between a detector 
and the Sun intersects  different part of Earth so that a little variation 
in $\nu_\odot$ flux could be present due  to different density the $\nu_\odot$'s cross.
\item 
{\bf{Winter-Summer effect}} =  The distance between Sun and Earth is not constant because of eccentricity of 
terrestrial orbit; a difference at a level of some \% is expected but it is below the present detectors sensitivity, 
with the exception of SuperKAMIOKANDE. 
\item 
{\bf{Other solar features}} = Many authors have looked for a correlation 
between the $\nu_\odot$ flux and solar  features: sunspots number, 
\cite{BAH87,BIE90,DOR91}, $B_\odot$ strength at surface, \cite{MAS93,OAK94}, 
green-line corona intensity, \cite{MAS95b,MAS95c,MAS96}, solar wind flux,  \cite{MCN95}. \\
If $\nu_\odot$'s have a non-standard interaction with $B_\odot$ a further effect is possible. In fact solar 
equatorial plane forms an angle of 7$^{\circ}$.25 with the Earth's orbital plane: therefore 
the solar core, where $\nu_\odot$'s are  produced, is viewed from the Earth by different position. The line of 
sight crosses the solar equator where $B_\odot$ weaker than at higher latitude should be present. 
\fit  

\subsection{Uncertainties in the neutrino flux.} 
\label{sect:sunnufluxuncertainties}
The procedure for estimating the uncertainties in $\nu_\odot$ flux calculation was described  in detail in 
\cite{BAH89} and then upgraded in \cite{BAH92,BAH95,BAH98a,HAX99,BAH01a}.\\ 
The equations describing the solar evolution depend on many quantities 
which are not exactly known. The $\nu_\odot$ flux is affected by 
many uncertainties; among them we underline: 
\bit 
\item 
Fusion cross-sections give a dominant contribution because of the  competition between different solar 
nuclear reactions. Typical solar energy per  nucleus, from 5 to 30 keV, are well below the 
laboratory limits so that extrapolations  $via$ nuclear calculations are needed. 
Only the interaction cross-section between two $^3$He  nuclei has been measured down 
to $\sim$ 15 keV (these values indicate centre of  mass energy). Uncertainties are present due to the extrapolation 
procedure, to the atomic electron screening and to the plasma effects. Terms concerning $^3$He and $^7$Be 
interactions are fundamental because they establish the relative importance among the different terms in p-p chain; 
on the other hand the proton capture by $^{14}$N constrains the CNO cycle.  
Nuclear matrix treatment also presents uncertainties mainly when transitions involving excited states have to be
included in computations, see \cite{GOO80,FUJ96,AKI97}.
\item 
$L_\odot$ changes its value at a level of $\sim$ 0.1$\%$ during the solar  cycle and affects T$_c$; 
the $\nu_\odot$ flux, which depends on T$_c$, is modified. The pre-main sequence solar phase  
($\sim 5\cdot 10^7$ y) is an estimated value; the contribution of $R_\odot$ and $B_\odot$ and their
variability are presently under analyses.
\item 
Solar chemical composition at the beginning is an important term because of the influence of heavy elements on 
opacity and on $\nabla$. Different results have been obtained within an uncertainty of about $6\%$. 
In table \ref{tabfactorscre} the dependence of each branch of p-p chain from different screening factors and the 
related variations in $\nu_\odot$ flux are reported.\\
\begin{table}[ht]
\caption{\it{ The fractional differences, (model - W-S )/W-S, in \%, of several solar
quantities computed in SMs with different screening factors:
Helium abundance at surface($Y_{surf}$), depth of the convective zone
($R_{CZ}$), central temperature (T$_c$), isothermal sound speed squared at the solar core ($u_c$), 
$\nu_\odot$ fluxes ($\Phi_i$).
Labels W-S, MIT, NO-S and TSY describe respectively weak screening, Mitler, no screening 
and Tsytovich treatment of screening processes,  
adapted from \cite{FIO01a}.}}
\begin{center}
\begin{tabular}{|l|c|c|c|}
\hline
 &MIT&NO-S&TSY \\ \hline
Y$_{surf}$  & - 0.076 & - 0.86 & - 1.4  \\
R$_{CZ}$    & + 0.037 & + 0.34 & + 0.59 \\
T$_c$       & + 0.45  & + 0.54 & + 1.4  \\
u$_c$       & + 0.10  & + 1.0  & + 1.4  \\ \hline\hline\hline
$\Phi$(pp)  & + 0.033 & + 0.45 & - 0.35 \\ \hline
$\Phi$($^7$Be)& - 0.19  & - 2.4  & - 5.9  \\ \hline   
$\Phi$($^8$B) & - 2.7   & -12.0  & +11.0  \\ \hline\hline\hline
\end{tabular}
\label{tabfactorscre}
\end{center}
\end{table}
\item 
Radiative opacity depends on the inverse of the photon diffusion length 
in the solar plasma: it constrains T$_c$ and the $\nu$ flux. Difficulties 
related to the opacity calculations been detailed in section 
\ref{sect:suncomments}. 
\fit
It has to be stressed that Hep $\nu_\odot$ flux is presently quoted without uncertainties due to
the difficulty in nuclear cross-section calculations; a conservative estimated error is at a 
level of $\sim$ 100\%, \cite{BAH01a}.\\ 
Frequently the computed errors are asymmetric because of the asymmetry in the uncertainties of the input parameters. \\ 
In table \ref{tabuncbah} and table \ref{tabuncwat}  contributions to the $\nu_\odot$ flux uncertainty are  
reported from two different SMs, \cite{WAT01,BAH02c}. \\
\begin{table}[ht]
\caption{\it{ Average uncertainties in $\nu_\odot$ flux as computed in the reference solar model. 
The $^7$Be electron capture rate increases by 2 $\%$ the  uncertainty in $^7$Be $\nu_\odot$ flux, 
adapted from \cite{BAH01a,BAH02c}.}}
\begin{center} 
\begin{tabular}{|l|c|c|c|c|c|c|c|c|c|}  
\hline 
Frac.&p-p&$^3$He$^3$He&$^3$He$^4$He&$^7$Be+p&Z/X&Opac.&$L_\odot$&Age&Diffus.\\\hline 
Uncert.(\%)&1.7&6.0&9.4&4.0&6.1&-&0.4&0.4&15.0\\\hline 
$\Phi$(p-p)\%&0.2&0.2&0.5&0.0&0.4&0.3&0.3&0.0&0.3\\\hline 
$\Phi$($^7$Be)\%&1.6&2.3&8.0&0.0&3.4&2.8&1.4&0.3&1.8\\\hline 
$\Phi$($^8$B)\%&4.0&2.1&7.5&4.0&7.9&5.2&2.8&0.6&4.0\\\hline\hline\hline 
Cl (SNU)&0.3&0.2&0.6&0.3&0.6&0.4&0.2&0.4&0.3\\\hline 
Ga (SNU)&1.3&1.0&3.3&0.6&3.1&1.8&1.3&0.2&1.5\\\hline 
\end{tabular} 
\end{center} 
\label{tabuncbah}
\end{table}
\\
\begin{table}[ht]
\caption{\it{ Sensitivity of neutrino fluxes,  central density
and the surface helium abundance to the  uncertainties in the input physics in a seismic SM. 
The first ten entries  are due to the nuclear cross-section factors, adapted from \cite{WAT01}.} }
\begin{center} 
\begin{tabular}{|l|c|c|c|c|c|} 
\hline       & Cl    & Ga    & $^8$B & $\rho_{c}$ & $Y_{conv}$ \\       
& 7.17  & 126   & 4.77$\cdot 10^6$  & 156            & 0.246  \\       
& SNU & SNU & cm$^{-2}$s$^{-1}$ & g cm$^{-3}$  &  \\\hline 
p-p: 4.00(1$^{+0.021} _{-0.013}$)$\cdot$10$^{-22}$  keV b & $_{+0.372} ^{ }$ &   
$_{+1.8} ^{-3.1}$ & $_{+0.282} ^{-0.477}$ & $_{+1.6} ^{-3.0}$ & $^{+0.0008}_{-0.0003}$\\ \hline
p-e-p: $\pm$1\%   & $\pm$0.002 & 0.0 & 0.000  & 0.0 & 0.0000  \\  \hline
$^3$He$^3$He: 5.4$\pm$0.32 MeV b & $\mp$0.131 & $\mp$0.8 & $\mp$0.094 & $\pm$0.3 &$\mp$0.0001\\  \hline
$^3$He$^4$He: 0.53$\pm$0.05 keV b & $\pm$0.408 & $\pm$2.6 & $\pm$0.293 & $\mp$1.0 &$\pm$0.0003\\  \hline
$^7$Be+e: $\pm$2\% (1$\sigma$) & $\mp$0.106 & $\mp$0.2 & $\mp$0.093 & 0.0 & 0.0000 \\  \hline
$^7$Be+p: 19$^{+4} _{-2} (1\sigma ) ^{+8}_{-4} (3 \sigma )$ eV b & $^{+1.057} _{-0.456}$ & 
  $^{+2.2} _{-1.0}$ & $^{+0.927} _{-0.400}$ & 0.0 & 0.0000  \\  \hline
$^{12}$C+p: 1.34$\pm$0.21 keV b & 0.000 & 0.0 & 0.000 & 0.0 & 0.0000      \\  \hline
$^{13}$C+p: 7.6$\pm$1  keV b & 0.000 & 0.0 & 0.000 & 0.0 & 0.0000      \\  \hline
$^{14}$N+p: 3.5$^{+1.0}_{-2.0}(3 \sigma)$keV b&$^{+0.023}_{-0.069}$&    
$^{+0.5} _{-1.6}$ & $_{+0.021} ^{-0.007}$ & $_{+0.6} ^{-0.2}$ & $^{+0.0001}_{-0.0002}$\\   \hline
$^{16}$O+p: 9.4$\pm$1.7 keV b & $\pm$0.001 & 0.0 & 0.000 & 0.0 & 0.0000 \\  \hline \hline
$(Z/X)_{s}$: 0.0245$\pm$0.0006 & $\pm$0.120 & $\pm$0.5 &$\pm$0.094&$\mp$0.2&$\pm$0.0026\\ 
$L_\odot$: 3.844(1$\pm$0.004)$\cdot$10$^{26}$ W & $\pm$0.203 & $\pm$1.3 &    $\pm$0.153 & $\pm$0.5 & $\pm$0.0005 \\   
Sound-speed profile&  $\pm$0.030 & $\pm$0.1 & $\pm$0.022 & $\pm$0.3 & $\pm$0.0001 \\ 
$r_{CZ}/R_\odot$: 0.713$\pm$0.001 & $\mp$0.031 & $\mp$0.2 & $\mp$0.023 & 0.0 & $\pm$0.0031\\ 
$\nu$ cross-section& $\pm$0.173 & $^{+5.1} _{-2.4}$ &  -- & -- & --   \\  \hline
Opacity: $\pm 5\%$& $\pm$0.083 &$\pm$0.4 & $\pm$0.062 & $\pm$0.2  & $\pm$0.0075 \\ 
EOS: ideal ($\Gamma_1 = \frac{5}{3}$)& $-$0.007 & $-$0.3 & $+$0.006 & $-$0.6 & $+$0.012 \\ \hline  \hline
Total: $\sqrt{\sigma ^2 + \sigma ^2 + \cdots }$ & $^{+1.24} _{-0.95}$ & $^{+6.6} _{-5.3}$ &    
$^{+1.04} _{-0.72}$  & $^{+2.0} _{-3.2}$  & $\pm$0.004  \\\hline 
\end{tabular}
\end{center}
\label{tabuncwat}
\end{table}

\subsection{Conclusions.} 
\label{sect:sunfinal}
Solar models assume the Sun at thermal equilibrium: they relate the present  $\nu_\odot$ flux to the present 
$L_\odot$ and to nuclear energy production rate; in any case they have to predict a $\nu_\odot$ flux as great as 
measured by a Gallium detector if it is supposed  that all $\nu_\odot$'s are produced by p-p interaction. 
In fact such an experiment can measure low energy $\nu_\odot$'s produced in the first step of p-p chain.\\ 
Temperature values computed by different SMs agree within 
1\% and helioseismological data provide a strong support to this finding.
There is a general good agreement between experimental c$_s$ and the values computed in SMs.
Z$_{surf}$ agrees very well with the meteoritic composition but  the present accuracy does not exclude a small 
effect of diffusion  between the initial composition and the photospheric observations.\\  
A microscopic diffusion was introduced in SMs: the so obtained
$\nu_\odot$ flux predictions at higher energy are increased of $\sim$ 
20$\%$. Consequently, "turbulent" terms, which partially reduce the flux, 
were added. Even if different treatments of diffusion processes are 
possible, results are very similar,  \cite{THO94,BAH95,TUR98b}. 
SMs without element diffusion are not consistent with helioseismology. 
Despite these improvements, the observed surface abundance of elements 
such as $^7$Li and $^9$Be shows in general a depletion even strong. \\
The density $\rho$ is a little bit more sensitive than c$_s$ to the p-p reaction 
rate and depends on many physical parameters whose uncertainties can be 
quite large: S$_{11}$, opacities, microscopic diffusion processes, Z, $t_\odot$ and $R_\odot$.\\
The structure of radiative zone is very sensitive to these quantities: they 
are closely related since a change in Z modifies the opacity and the microscopic diffusion, which in turn 
re-modifies Z, as a consequence of an iterative calibration process.\\
Along the solar evolution $Z_c$ varies mainly because  of the CNO cycle and the microscopic diffusion 
therefore the opacity has to be modified. 
There are SMs using higher Z$_0$ value: this could indicate that some 
forgotten hydrodynamic phenomenon could act: the use of $\rho$, rotation 
profiles in the core and gravity modes may be helpful to clarify this point. \\ 
A crucial progress has been achieved in the He abundance evaluation: Y$_{surf}$ = 0.249 $\pm$ 0.003; this 
value is not far from the cosmological value  but it is smaller than Y$_0$ as computed by SMs.
Other physical processes have to be introduced to enable us to good
reproduce all the observed elemental abundances.\\
The small differences in solar radial profiles of physical quantities seem to be constrained by $ad$ $hoc$ opacity 
modifications, always in the present uncertainty range.\\  
If one uses non-standard SMs other possibilities are open. It could be possible a mild mixing in nuclear region, 
for example close to the $^3$He peak, in order to improve the solar structure but helioseismological 
results reject this solution: a strong disagreement (at a level of some \%) is present in c$_s$ profile.\\
In the low-Z core models, the p-mode frequencies require a central core with mass 
$M<\,0.06\,M_{\odot}$. All mixed core models with $M>\,0.02\,M_{\odot}$ are 
excluded by p-mode frequencies; if  $M<\,0.02\,M_{\odot}$ 
the calculated $\nu_\odot$ flux is much higher, see \cite{GUE97}.\\  
The influence of p-mode frequencies on mixed-shell SMs depends on mixing features but the $\nu_\odot$ flux is at 
a level well above than in SSMs. Helioseismology does not  completely rule out SMs in which $^3$He and other trace 
elements are slowly mixed in a  region where the $^4$He abundance is practically uniform. In this case the ratio  
$^7$Be/$^8$B might be changed without substantial modifications of c$_s$ in that region,  \cite{GUE97}.\\  
There is a good agreement between solar c$_s$ and estimated profiles below 0.6 $R_\odot$, but tachocline and 
upper layers are poorly described: in the first zone the turbulent pressure is as important as
the gas pressure; moreover, tachocline and external
solar region  are shear layers in which rotation rates change rapidly. \\ 
At present, only a static description of the radiative region is possible, contrary to the convective one, and 
the history of the angular momentum is not introduced in the stellar equations. 
A 1-D stellar evolution code cannot provide an efficient treatment of these dynamic  regions. Neither the 
rotation of the Sun nor $B_\odot$ are taken into  account, whereas it is widely thought that the 
tachocline is the base of the  magnetic dynamo process. Of course, the $\nu_\odot$ emission and the solar core  
physics are rather insensitive to the tachocline and beyond, but the $\nu_\odot$  behaviour may depend on these 
layers ({\it{e.g.}} if $\nu$'s have a magnetic moment). \\ 
If $B_\odot$ acts, a magnetic pressure term $P_{mag}$ is added in the stellar structure equations
($P_{mag}=\frac {B^{2}}{8\pi}$, in cgs units) and the basic hydrostatic equilibrium equation is modified:
\begin{equation}
\frac{dP_{gas}}{dm} = -\frac{G M}{4\pi r^{4}} -\frac{dP_{mag}}{dm}
\end{equation}
therefore the wave velocity changes in two ways, \cite{COU02}. In fact the gas pressure is modified because of 
the hydrostatic equilibrium and a part of the Alfv\'en speed $v_{a} = \frac {B^2}{4 \pi\rho}$ (in cgs units) 
is added. $B_\odot$ has a 3-D structure and the angle between the field lines and the seismic waves determines 
the way these latter are accelerated: the wave velocity is no more an isotropic  
quantity but a 1-D stellar code cannot reproduce these events. 
The main problem is to select a reliable $B_\odot$(r) for the solar interior: in fact large scale field should 
be important either in radiative and tachocline regions and below the solar surface. Despite the present accuracy, 
c$_s$ is not suited to the determination of the large scale magnetic features of the Sun. In any case, the magnetic 
contribution is small when compared to large uncertainties the remaining parameters have. \\
As a final remark, the global consistency of the solar description has to be stressed as well as the stability of 
the values of $\nu_\odot$ flux done by different SMs along the last 30 years, mainly 
concerning the p-p contribution. In fact, even if "seismic" constraints are imposed on 
the "classical" treatment for the solar properties, numerical results do not vary at 
large level. Hence, SMs enable us to reproduce the measured physical solar parameters within a small 
uncertainty range even if simplified assumptions are done.

\section{Solar neutrino detection.}
\label{sect:nudetection}
Before the analysis of the experimental aspects concerning $\nu_\odot$'s physics, let us remember the conclusion 
given by H.Bethe on the detectability of $\nu_\odot$'s, \cite{BET34}:\\
{\it{"..This meant that one obviously would never be able to see a neutrino."}} \\
Experiments aiming the detection of $\nu_\odot$'s were, and still are, a real challenge for many reasons:
\bit
\item 
The reaction rate is very low, much less than one event per day and per ton of target  material; the mean 
energy is low ($\sim$ MeV) and in an energy region where many sources  of background are presents 
(natural radioactivity, secondary cosmic rays...).
\item 
It is impossible to have "beam-off" measurements.
\item 
There are no artificial $\nu$ sources with an energy spectrum as $\nu_\odot$'s have. 
\fit
No viable suggestions concerning the  $\nu$ detection were available until 
the end of the second world war: in 1946 B.Pontecorvo suggested a  
radiochemical method to capture $\nu$'s emitted by a nuclear reactor using 
Cl or Br as a target, \cite{PON46}. Three years later L.Alvarez wrote a 
paper detailing technical procedures to realize a Cl detector near a pile 
and he stressed out the background problem, \cite{ALV49}.
Both these works supposed $\nu$'s and $\overline{\nu}$'s to be equivalent. \\ 
Because of the big amount of technical difficulties, $\nu_\odot$'s were experimentally  detected only in the 70s', by 
Homestake Cl experiment, the sole $\nu_\odot$ experiment up to 1985, \cite{DAV68,DAV70,DAV71}. \\ 
In the middle of 80s', KAMIOKANDE, in 
Japan, started to operate and  revealed that ... $\nu_\odot$'s come from the Sun, \cite{HIR89,KAJ94}:   
it realized a "neutrinography" of the Sun by using an exposure time of some years. In the 90s', the Ga 
experiments, SAGE and GALLEX, which detected  low-energy $\nu_\odot$'s coming from the p-p chain, \cite{ANS92,ANS95}, 
showed that the Sun  produces energy through the conversion of H into \(^4\)He and confirmed the main hypothesis  
on the stellar energy production and the deficit in $\nu_\odot$ flux. The latest SK and SNO measurements, 
\cite{FUK01a,FUK01b,AHM01,AHM02a,AHM02b},  strongly suggest $\nu$ properties beyond the particle standard model. 

\subsection{Interaction processes.} 
\label{sect:interactionproc}
We will now summarize the main features of the $\nu_\odot$'s detection. The 
basic physical processes are the $\nu$ capture on nucleus and the $\nu$ - 
electron scattering; it is  possible a more refined classification:
\benu
\item $\nu_e$ + (A,Z) $\rightarrow\,e^-$ + (A,Z+1)
\item $\nu_x$ + (A,Z) $\rightarrow\,\nu_x$ + (A,Z)
\item $\nu_e\,+\,e^-\,\rightarrow \nu_e\,+\,e^-$
\item $\nu_x\,+\,e^-\,\rightarrow \nu_x\,+\,e^-$
\fenu
The odd cases are charged current (CC) interactions, the even ones describe neutral current (NC)
processes. Concerning NC and CC interaction on
nucleus, we detail reactions on D, a target material which presently a canadian experiment (SNO) uses.\\
We resume some characteristics of different interactions:
\benu
\item 
Electron elastic scattering (ES):$\qquad \nu_x + e^- \rightarrow \nu_x + e^-$ ; this reaction is 
predominantly sensitive to $\nu_e$'s and strongly  directional: the electron is emitted 
preferentially in the forward direction respect to the propagation of incoming $\nu_\odot$. 
The angle between the directions of incoming $\nu_\odot$ and the electron depends on energy of $\nu_\odot$ and on
detector threshold (if the E$_{thr}$ = 5 MeV, the angle fast increases from 0 up to $\sim$ 20$^\circ$ when the energy
of $\nu_\odot$ goes up to $\sim$ 20 MeV then it gets to its asymptotic value of $\sim$ 24 $^\circ$).
\item 
Charged current interaction (CC):$\qquad \nu_e + d \rightarrow p + p + e^- \; (Q = 1.44 \,MeV)$ ; 
this reaction has a relatively large cross-section.  
\item 
Neutral current interaction (NC):$\qquad \nu_x + d \rightarrow p + n + \nu_x 
\; (Q = 2.2 \,MeV)$ ; this reaction is sensitive to all $\nu$ flavours.
The signature of the reaction is given by the detection of a neutron but 
it is still a difficult process to be revealed and depends on the final 
radiopurity of the whole apparatus. 
\fenu 
The ratios between interaction rates are: 
\beqar
\frac {CC}{ES}\,=\,\frac {\nu_e} {\nu_e+0.154(\nu_\mu + \nu_\tau)}\\
\frac {CC}{NC}\,=\,\frac {\nu_e} {\nu_e+\nu_\mu+\nu_\tau}
\label{eq_ccncesratio}
\feqar
If the incoming particle is a $\nu_e$, both NC and CC interactions are allowed while only NC 
interactions are possible when $\nu$ flavour is other than electronic one. It is  impossible 
to distinguish between $\nu_\mu$ and $\nu_\tau$ interaction. Under standard assumptions both 
these ratios are equal to the unity.\\
The ratio between NC and CC is presently used as a test for $\nu_\odot$ 
flavour oscillations, see later in section \ref{sect:partsolutions}; 
its value depends on energy. \\
Table \ref{tabrateclga} and table \ref{tabrates}  show the composition of 
the $\nu_\odot$ flux and the expected interaction rate as 
measurable on terrestrial detectors for interaction on nuclei.\\
\begin{table}[ht]
\caption{\it{ Neutrino fluxes and interation rates on Cl and Ga computed 
with different nuclear reaction cross-sections: \cite{ADE98} (left side), 
\cite{ANG99} (right side). The cross-sections for neutrino absorption on 
Cl are from \cite{BAH96c}, the cross-sections for Ga are from \cite{BAH97a}.
Errors on flux are quoted in first column; the Hep component is without 
errors due to uncertainties in calculations, adapted from \cite{BAH01a}.}}
\begin{center}
\begin{tabular}{|l|c|c|c||c|c|c|}
\hline
Source&Flux&Cl&Ga&Flux&Cl&Ga\\
\& error(\%)&$\nu\, cm^{-2}s^{-1}$&(SNU)&(SNU) &
$\nu\, cm^{-2}s^{-1}$&(SNU)&(SNU) \\\hline
p-p($\pm$ 1.0)&5.95$\cdot 10^{10}$ &--&69.73 &5.96$\cdot 10^{10}$&--&69.85\\\hline
p-e-p($\pm$ 1.5)&1.40 $\cdot 10^{8}$&0.22&2.86&1.39 $\cdot 10^{8}$&0.22&2.84 \\\hline
Hep &9.3 $\cdot 10^{3}$&0.04&0.07&9.4 $\cdot 10^{3}$&0.04&0.07 \\\hline
$^7$Be($\pm$ 10)&4.77 $\cdot 10^{9}$&1.15&34.20&4.81 $\cdot 10^{9}$&1.15&34.49 \\\hline
$^8$B($^{+20}_{-16}$)&5.05 $\cdot 10^{6}$&5.76&12.12&5.44 $\cdot 10^{6}$&6.20&13.06 \\\hline  
$^{13}$N($^{+21}_{-17}$)&5.48 $\cdot 10^{8}$&0.09&3.31&4.87 $\cdot 10^{8}$&0.08&2.94 \\\hline
$^{15}$O($^{+25}_{-19}$)&4.80 $\cdot 10^{8}$&0.33&5.46&4.18 $\cdot 10^{8}$&0.28&4.75 \\\hline
$^{17}$F($\pm$ 25)&5.63 $\cdot 10^{6}$&0.0&0.06&5.30 $\cdot 10^{6}$&0.0&0.06\\\hline\hline\hline
Total&6.545$\cdot 10^{10}$ &$7.6^{+1.3}_{-1.1}$&$127.8^{+9}_{-7}$&
      6.546$\cdot 10^{10}$ &$8.0^{+1.4}_{-1.1}$&$128^{+9}_{-7}$ \\\hline
\end{tabular}
\end{center}
\label{tabrateclga}
\end{table}
\\
\begin{table}[ht]
\caption{\it{ Expected neutrino capture rates on nucleus (in SNU). The null value 
means not calculated or negligible, "-" means not measurable by the detector, 
"*" means not shared, adapted from \cite{BAH01a,EJI00b}.}}
\begin{center}
\begin{tabular}{|l|c|c|c|c|c|c|c|c|} 
\hline
Type&$^{37}$Cl&$^{71}$Ga&$^{100}$Mo&$^{127}$I&$^{2}$H&$^{40}$Ar&$^{7}$Li&
$^{115}$In \\\hline
p-p&--&69.73&639&--&--&--&--&468\\\hline
p-e-p&0.22&2.86&13&0.0&--&--&9.2&8.1\\\hline
Hep&0.04&0.07&0.0&0.0&0.02&0.02&0.1&0.05\\\hline
$^7$Be&1.15&34.20&206&9.4&--&--&9.1&116\\\hline
$^8$B&5.76&12.12&27&13.0&6.0&7.2&19.7&14.4\\\hline
$^{13}$N&0.09&3.31&22&0.0&--&--&2.3&32.1*\\\hline
$^{15}$O&0.33&5.46&32&0.0&--&--&11.8&32.1*\\\hline
$^{17}$F&0.0&0.06&0.0&0.0&--&--&0.1&0.2\\\hline\hline\hline
Total&7.59&127.81&939&22.4&6.02&7.22&52.3&638.85\\\hline
\end{tabular}
\end{center}
\label{tabrates}
\end{table}

\subsection{Detection techniques.}
\label{sect:detectechniques}
Different techniques were tried to detect $\nu_\odot$'s by interaction with target mass, see 
\cite{CRE93} for a detailed analysis.
In $\nu_\odot$ - nucleus interaction the energy threshold at which reaction occurs is 
defined by the Q-value of the detection reaction itself. If the detector is not able to record the time of 
each event and "integrates" each $\nu_\odot$ with $E\geq E_{thr}$,  
it is also impossible to distinguish the origin of different contribution.\\
\begin{description}
\item 
{\bf{Geochemical experiments}} = People should study a very long meanlife isotope produced
in a deep underground ore. In this case, it could even be possible to estimate the constancy of the $\nu_\odot$ flux 
over long time. $^{98}$Mo or $^{205}$Tl should be good candidates but many problems in the evaluation of 
background forbid the feasibility of experiments using such a technique.
\item 
{\bf{Radiochemical experiments}} = 
The isotopes produced by $\nu_\odot$'s are extracted using the different chemical behaviour of these atoms 
compared to the target ones. Presently, only experiments having Ga as target mass use this technique.
Due to the low interaction rate, the amount of analysed isotopes is very small so that a 
target mass of $\sim$ 100 tons is needed to have a production of 1 atom per day. The detectors are sensitive only to 
$\nu_e$'s by inverse $\beta$-decay. Radiochemical experiment indicates only the total rate at which $\nu_\odot$'s 
with $E\geq E_{thr}$ are captured; the energy spectrum cannot be measured. The nuclei produced  by $\nu_\odot$'s 
interactions are usually unstable for electron capture and their amount can be established by looking for their 
decay or by resonance ionisation  spectroscopy. In any case, a chemical separation from target nuclei is needed. 
Several isotopes enable us to build a detector: $^{37}$Cl, $^{71}$Ga, $^{127}$I, $^7$Li, $^{81}$Br; up to 
now $^{37}$Cl and $^{71}$Ga have been used. Ga experiments are sensitive to the 
low-energy component of the $\nu_\odot$ flux but cannot do any $\nu_\odot$ spectroscopy.  
A problem for this technique is the very small number of interesting nuclei produced
during each data taking session and the extremely efficient extraction yield needed 
for the separation.  
\item 
{\bf{Direct counting experiments}} = The main aim is to detect in real-time 
particles emitted after  $\nu_\odot$ interaction, measuring the $\nu_\odot$ energy  
spectrum and temporal variation on $\nu_\odot$ flux, if present.  Usually, 
scintillation or \v Cerenkov light after the $\nu$ interaction are 
detected; among possible solutions (not presently operating) time 
projection chambers, low-temperature detectors and scintillators are in 
R\&D. The most intriguing aspects that these options allow are the $\nu_\odot$ 
spectroscopy and the different flavour recognition.\\ 
An interesting idea to measure low energy $\nu_\odot$'s in real time is 
the coincidence technique for $\nu$ capture on nuclei. The target material 
should be either a large amount of $\beta\beta$ isotopes ($^{176}$Yb,
$^{100}$Mo, $^{160}$Gd) or highly forbidden $\beta$-decay emitters like 
$^{115}$In (4-fold forbidden).
The possibility to apply the same technique for CdTe(CdZnTe) 
semiconductor detectors, which have already a wide field of application 
in $\gamma$-ray astronomy and medical physics, has been recently suggested.
CdTe semiconductor detectors. Ge and GaAs semiconductor detectors for 
$\nu_\odot$ experiments were also considered relying on the detection of 
electrons from $\nu$-electron scattering.\\ 
The detection principle for $\nu_\odot$'s using coincidences relies on 
the reactions:
\beqar
\nu_e + (A,Z) \rightarrow (A,Z+1)_{g.s.} + e^- \rightarrow (A,Z+2) + e^- 
+ \overline\nu_e \\
\nu_e + (A,Z) \rightarrow (A,Z+1)^* + e^-  \rightarrow (A,Z+1)_{g.s.} + 
\gamma 
\label{eq_nuinteract}
\feqar
Therefore, either coincidence between two electrons for the ground state 
transitions or the coincidence of an electron with the corresponding 
de-excitation photon(s) is required.
The energy of electrons produced in the first part of the coincidence is: 
\beq
E_e = E_\nu - (E_f - E_i)
\label{eq_electronenergy}
\feq
with E$_\nu$ as $\nu$ energy, E$_f$ and E$_i$ as the energy of the 
final and initial nuclear state involved in the transition.\\
D, $^{40}$Ar, $^{81}$Br, $^{115}$In ... are good candidate isotopes: 
up to now only SNO, a Canadian detector which is using D, is running. 
\end{description}
The structure of the interaction cross-section is the following:
\beq
\sigma \propto G^2_F |M|^2 p_e E_e F \approx 10^{-46} \left(\frac
{E}{m_e}\right)^2 cm^2
\label{eq_crosssectinteract}
\feq
where $G_F$ is the Fermi constant, M is the nuclear matrix element, p$_e$ and E$_e$ are 
respectively the momentum and the energy of the electron, F includes Coulomb corrections.\\ 
In order to calculate the interaction rate R per target atom the uncertainties due  to the 
selected SM, to the absorption cross-sections  and to transitions toward excited nuclear 
levels (because of the great complexity in nuclear matrix sector) have to be taken into 
account, see tables \ref{tabuncbah}, \ref{tabuncwat} and \ref{tabgarates}.\\
On the other hand, if the experiment detects $\nu_\odot$ elastic scattering 
off electron, at present only \v Cerenkov effect in water is used. 
A track or a flash produced by a charged particle gives evidence of 
$\nu_\odot$ interaction. The emitted light is collected by an array of 
PMT's and a fiducial volume is defined. This technique allows the 
reconstruction of the direction of incoming $\nu_\odot$'s and the 
estimation of the energy spectrum and its shape.
The energy threshold is fixed by electronic setup.\\ 
A big challenge in these  detector is the analysis of the low-energy 
spectrum because background raises  exponentially at lower energy: 
up to now $E_{thr}\geq 5.0$ MeV so that only high energy components of
$\nu_\odot$ flux are detectable ($^8$B and Hep $\nu_\odot$'s).\\ 
KAMIOKANDE, SK and SNO, which use this technique, have confirmed that 
$\nu$'s come from the Sun; moreover, the energies of detected $\nu_\odot$'s 
are fully compatible with predictions given by SMs.

\subsection{Background.}
\label{sect:background}
The background is probably the main problem in $\nu_\odot$ experiments: it can  either hide 
the true signal or produce a false one. To avoid this serious trouble it is  needed to 
determine the features of all possible known sources of noise and to emphasize the expected 
signal as much as possible. The most important contributions to the background come 
from cosmic radiation and from radioactivity in structures, materials and environment.\\ 
The charged component of cosmic radiation and high energy atmospheric $\nu$'s can be  
identified with sufficient confidence but they can interact either in detector or in its  
structure and/or in surrounding medium originating "side reactions" which produce fake signals. 
Hence, detectors are homed deep underground, in mines or tunnels under the 
mountains, independently of the detecting technique: in this way, the 
cosmic ray flux ($\sim$ 180 events m$^{-2}$ s$^{-1}$ at sea level) 
reduces to a value less problematic in comparison with $\nu_\odot$  
interaction rate.\\ 
Further problems come from long-lived cosmogenic isotopes, produced {\it{en plein air}} and  from the decay of 
isotopes in the detector structure: in most cases critical isotopes  are $^{222}$Rn, $^{40}$K and all intermediate 
products of $^{232}$Th and $^{238}$U chains.\\ 
A contamination level not greater than $\sim 10^{-16}$ g/g is
often required, \cite{BOR98}: this is an example of the difficulties 
that the researchers have to overcome. \\
In order to summarize the energy dependence of background, the main term up to 3 MeV is 
coming from structure, wall and ore radioactivity, $\alpha$ and $\beta$ contributions being  complementary. 
At higher energy neutron capture processes are dominant, due to reactions  where $\alpha$'s from U and Th are 
involved. As a standard procedure, low  activity concrete, "clean" materials, water and lead are 
searched for.  

\section{Experimental Results.} 
\label{sect:experimental}

We now shortly expose the main features and results of experiments which 
have measured  $\nu_\odot$ flux. Up to 2002 seven detectors claimed 
evidence of $\nu_\odot$'s:  Homestake, KAMIOKANDE, GALLEX, SuperKAMIOKANDE, 
SAGE, GNO and SNO; the last three experiments are on run in spring 2002.  

\subsection{The Chlorine experiment.} 
\label{sect:homestake}
The first experiment searching for the $\nu_\odot$'s detection was proposed in 1964, 
\cite{DAV64}, immediately after the publication of the first SM. The building operations 
for Cl  detector began in 1965 at Homestake gold mine (USA), at a depth of 4200 mwe. \\ 
The detected reaction is a $\nu$ absorption on nucleus, 
$^{37}Cl\,(\,\nu_e\,,\,e^-\,)\,^{37}Ar$ at E$\geq$ 814 keV, \cite{DAV94}, 
and can be induced by $^7$Be, CNO and p-e-p 
$\nu_\odot$'s but the dominant contribution is from $^8$B component. \\
Among technical and economical motivations, the $^{37}$Cl  isotope has a 
good abundance in nature and it is cheap; moreover, $^{37}$Ar can be  
separated without problems from target mass and the inverse reaction 
has a right period  for radiochemical detection. \\
The target was a tank filled with 615 tons of perchloroethylene (C$_2$Cl$_4$). 
Helium stream was used to "extract" $^{37}$Ar and carrier Ar:
after an exposure time ranging up to 3 months, the produced $^{37}$Ar
nuclei, which are unstable and transform back into $^{37}$Cl ${\it{via}}$ electron 
capture (half-life = 35 days), were analysed in proportional counters
searching for Auger electrons emitted when $^{37}$Ar decays. Tritium, Kr
and Ar isotopes different from $^{37}$Ar were searched for to be removed, in fact they 
constitute the most serious background in the counting procedure. 
The event selections were based on the pulse amplitude and the rise time: only fast rising 
pulses were estimated as good signature of an event because they are electronic captures.
Periodical calibration were done both on preamplifiers and on counters. In 1970 an electronic system to analyse 
the pulse rise time enhanced the experimental sensitivity. Usually, only results after 1970 were quoted.\\
The detector took data continuously from 1967 to 1994 with the exception 
of a period from May 1985 to October 1986.\\
$^{37}$Ar atoms produced by cosmic rays ($\sim 0.28\pm$ 0.08 SNU) and  neutron contribution  (0.13 $\pm$ 0.13 SNU) 
were the main components of background; smaller terms
were due to $\alpha$ particles produced from U and Th and to cosmic ray $\nu$'s. 
When the counting rate in a single run was found to be formally negative, a zero value  was adopted. 
The average  value was computed with a maximum likelihood analysis. The total systematic error (1.5$\%$  
for extraction, 5$\%$ for neutron, 3$\%$ for cosmic ray background) was at a level of 6 $\%$.\\ 
In the middle of 50s', $\nu_\odot$ flux was  estimated at a level of $\sim 10^{14}  \nu\,cm^{-2}\,s^{-1}$; 
this value was based on CNO dominant contribution (it is equivalent  to $\sim$ 40000 SNU).
A more complete theoretical knowledge on solar parameters lowered  the value 
down to 3900 SNU, \cite{CAM58a,CAM58b,FOW58}, but only the formulation of a first  SM in 1963-64, \cite{BAH63,SEA64} 
gave a more realistic estimated interaction rate: 40 $\pm$ 20 SNU.\\ 
In 1968 the $\nu_\odot$ flux in the Cl experiment was estimated in the range 
8 - 49 SNU, \cite{BAH68}, but
first experimental results gave an upper limit of 3 SNU, \cite{DAV68}. This  disagreement was 
the origin of the solar neutrino problem or the SNP of first kind. \\   
The final interaction rate, based on 1970-1994 runs, was 2.56 $\pm$ 0.16 
$\pm$ 0.15 SNU, \cite{CLE98,LAN98}, the 1 $\sigma$ statistical error was 
$\sim$ 6 $\%$ of the measured rate. Let us stress the value predicted in 
\cite{BAH01a}: 7.59$^{+1.3}_{-1.1}$ SNU.\\ 
It was suggested a possible anticorrelation of   $\nu_\odot$ flux with the activity of the Sun, 
namely its sunspots number, \cite{DAV96} and reference therein, see also 
\cite{MAS95a,MAS95b,MAS95c}. It was pointed out that this conclusion was 
extrapolated from the data detected in 1979-80 without further confirmations in the following 
years. For a review on this problem see \cite{BOG00}: the main conclusion is that the claim of 
anticorrelation with sunspots number seems to be due to technical procedures in analyses.\\
Let us remark the importance of this experiment, even historically speaking, because it 
firstly detected $\nu_\odot$'s and confirmed the main features of energy production
mechanism in stars. Moreover, it showed the feasibility of radiochemical technique, it 
overcame a lot of troubles in background evaluation and suppression, it was 
continuously running over about 3 decades and the only experiment until 1987. We 
underline that anyone has never used this technique and no calibration was made.

\subsection{The KAMIOKANDE experiment.}
\label{sect:kamiokande}
KAMIOKANDE (Kamioka Nucleon Decay Experiment), originally designed to 
search  for proton decay, showed that the 
interacting $\nu$ are coming from the Sun and  that their energies are 
compatible with the predictions of SSMs.\\ 
This experiment, homed in a mine in Japan, was a real-time water detector 
measuring the \v Cerenkov light emitted by 
electron recoil produced in ES interactions from electrons at $E\geq$ 9.0 
MeV  (some time later the threshold was lowered down to 7.0 MeV), therefore 
only $^8$B and Hep  $\nu_\odot$'s were detectable, \cite{FUK96}.\\ 
This detector first showed the main challenge for the \v Cerenkov 
technique:  a good cut criterion for track recognition and timing.  
In fact the total trigger rate for all the detector was $\sim$ 150000 
events  per day while the expected signal was less than 1! The angular 
resolution in track reconstruction was $\sim$ 27$^{\circ}$ and a 
global systematic error ($\sim 10\%$) came from uncertainties in the  
angular resolution, in the energy scale and in the fiducial volume cut.\\  
>From 1987 to 1995, with 2079 days of running time $597^{+41}_{-40}$ 
events were achieved (the expected number was $\sim$ 1200). \\ 
The final measured flux was $ \Phi\,= \,2.80\pm 0.19\pm 0.33 
\cdot 10^{6}\,\nu\,cm^{-2}\,s^{-1}$, \cite{FUK96}.\\ 
Searching for a possible explanation of SNP in term of particle solution, 
see later in section \ref{sect:partsolutions}, the $\nu_\odot$ flux as 
measured when $\nu_\odot$'s do not cross the  Earth before the detection 
(Day time) and otherwise (Night-time) were computed. More precisely the D 
flux requires $\cos \theta_z <$ 0 while for N flux  $\cos \theta_z >$ 0, 
$\theta_z$ being the angle between the detector  vertical axis and the 
vector from the Sun to the Earth.\\  
No significant differences were found:  
$\Phi_D\,=\,2.70\,\pm\,0.27 \cdot 10^{6}\,\nu\,cm^{-2}\,s^{-1}$, 
$\Phi_N\,=\,2.87^{+0.27}_{-0.26} \cdot 10^{6}\,\nu\,cm^{-2}\,s^{-1}$. \\ 
In order to study the time variations correlated with solar activity, as Homestake experiment 
suggested, the events were divided into short time periods, 200 days, but this further analysis 
did not indicate any anticorrelation with sunspots number, \cite{HIR89,KAJ94}. \\ 
We underline the quality of the detector and the technical skill of the 
people operating on that detector; they were also able  
to observe the burst of $\nu$'s  emitted in SN1987a explosion. 

\subsection{The SuperKAMIOKANDE experiment.} 
\label{sect:superkam}
\begin{figure}[ht]
\begin{center} 
\mbox{\epsfig{file=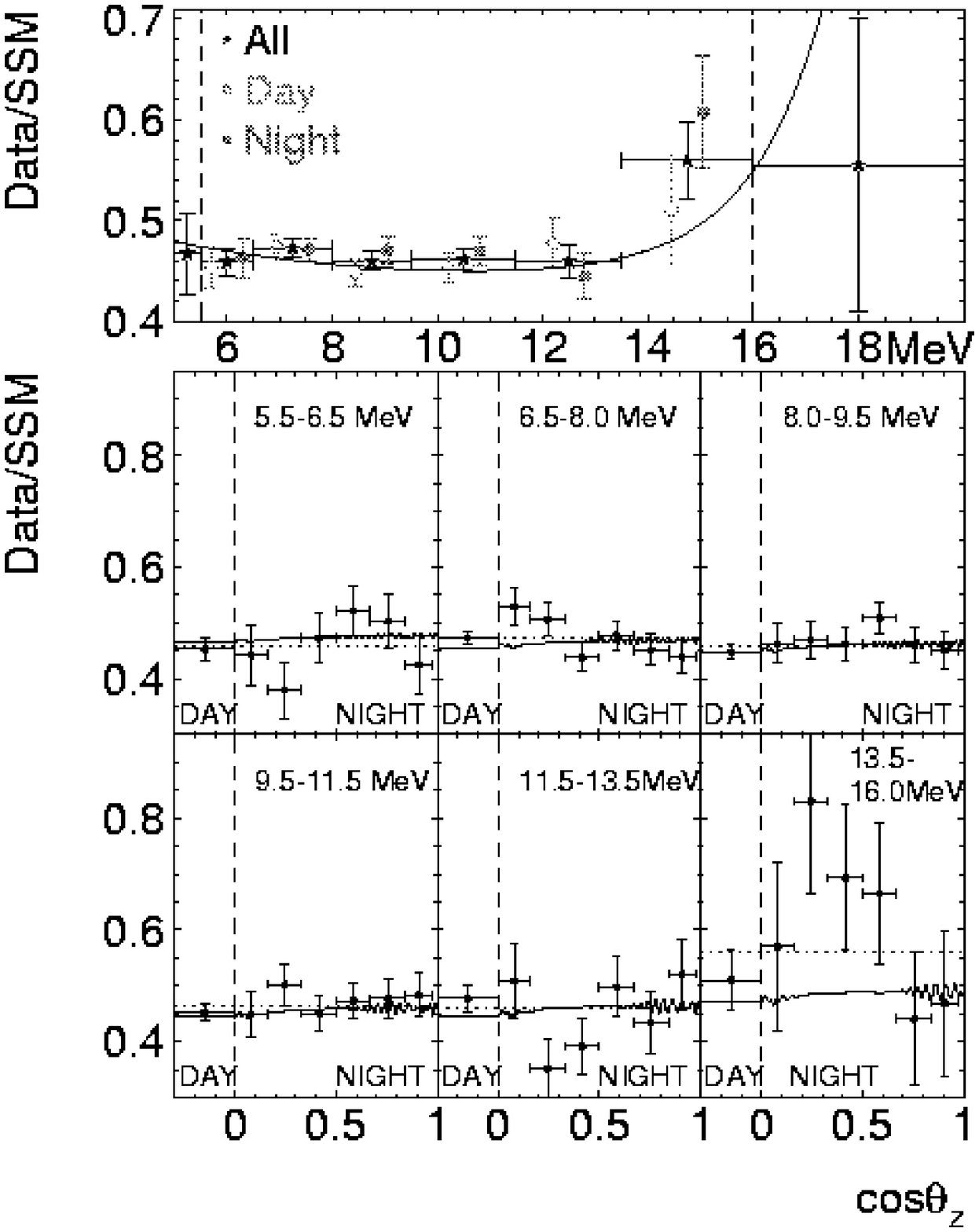,width=0.5\textwidth}} 
\end{center}
\caption{ Upper panel: SK energy spectrum; day and night values are also drawn. 
Lower panel: SK zenith-angle energy spectrum between 5.5 and 16 MeV. 
The dotted line is the combined rate in that bin. Error bars reflect the
statistical uncertainty only. Superimposed is an oscillation prediction 
($\tan^2\theta$=0.34 and $\Delta m^2$=6.0$\cdot10^{-5}$eV$^2$) near the best-fit point, 
from \cite{SMY02a}.}
\label{fig:SMYbw}
\end{figure}

SuperKAMIOKANDE, an enlarged version of KAMIOKANDE, began its data taking in spring 1996.  
Many physical items are under investigation with this real-time cylindrical detector: 
proton-decay, atmospheric and supernova $\nu$'s and, of course, $\nu_\odot$'s \cite{SUZ95,SUZ97}. 
Owing to the huge mass ($\sim$ 50000 tons even if the fiducial mass for $\nu_\odot$'s is $\sim$ 
22500 tons) and the low energy threshold, more accurate analyses are possible. The inner part of 
the cylinder ($\sim$ 32000 tons) is viewed by 11146 inward-facing 20"-PMT's while the external 
one has 1885 outward-facing 8"-PMT's which are employed as anti-counter.
The total coverage given by PMT's is at a level of $\sim$ 40 $\%$.\\ 
In any case, SK is sensitive only to $^8$B and Hep $\nu_\odot$'s but it enables us  to search 
for possible time modulations (D-N effect, seasonal variations and so on)  and to study the 
energy spectrum of recoil electrons and Hep $\nu_\odot$ flux, 
\cite{FUK98a,FUK98c,FUK99,SMY00,FUK01a,FUK01b}.\\ 
The analyses concern a sample of 1496 day of running-time, from  May 1996 to July 2001, with 
different energy threshold: in first step $E_{thr}$=  6.5 MeV, in latest analysis $E_{thr}$= 5.0 
MeV. This value is due to the background and to the trigger.\\
Each event has been classified depending on arrival direction and energy. The sample includes 
22400$\pm$ 800 events, corresponding to $\sim$ 15 events per day, \cite{FUK02,SMY02a,SMY02b,
SMY02c,SMY02d}; the calculated flux is $\Phi\,=\,2.35\,\pm\,0.02\,\pm 0.08 \cdot 10^6 
\,\nu\,cm^{-2}\,s^{-1}$ while the ratio with respect to the flux estimated in \cite{BAH01a} is 
\(R\,=\,0.465\,\pm 0.005\,_{-0.015}^{+0.016}\), see also table \ref{tabsk8bin} where the energy 
spectrum is shown. 
The total systematic error is $^{+3.5\%}_{-3.0\%}$: the main contribution comes from reduction 
cut efficiency, energy scale and resolution, systematic shifts in event vertex and angular 
resolution of the recoil electron momentum. The available SK zenith angle-recoil energy spectrum 
consists of six night + one day bins for six energy bins between 5.5 and 16 MeV electron recoil 
energy, plus two daily averaged points for the lowest ($5.0<E<5.5$ MeV) and the highest ($E>16$ 
MeV) energy bins. On the other hand, the D-N energy spectra are based on 19 energy bins each for 
D and N periods; even the running time has been  shared in D and N time, see fig. \ref{fig:SMYbw} 
and table \ref{tabsktheta}. The fractional difference is -0.021$\pm$ 0.020 $_{-0.012}^{+0.013}$, 
\cite{FUK02,SMY02b,SMY02c,SMY02d}. It seems that this difference assumes larger value at higher 
energy. \\
The data have been also divided either in 10 days of measurements or in 45 days of data taking.
The seasonal variation of $\nu_\odot$ flux due to the Earth orbit eccentricity is in good 
agreement with the hypothesis of $d^{-2}$ dependence, where $d$ is the Sun-Earth distance:
the flux averaged over 45 days of data taking has been computed and analysed:
$\chi^2$-test yields 4.7 to be compared with 10.7 under the assumption of constant flux. 
Even a 10 day binning has been checked but no statistically significant deviations from 
expected distribution have been found.
Other long-term variations, like anticorrelation with 
number of sunspots, seem to be not present. Let us remember that the sunspots number  
increased from 1998 to 2000 in coincidence with a solar activity maximum.\\  
The high energy component of $\nu_\odot$ flux has been searched for: 4.9 $\pm$ 
2.7 events with $E\geq$ 14.0 MeV were detected (1 event was expected) and an 
upper limit to this flux has been established ($\Phi_{Hep}\leq 7.9 
\Phi_{BAH01a}$). As a further result an upper limit to the solar $\overline 
\nu_e$ flux has been deduced ($\phi < 2\%$ of the total flux). 
\\
It has to be stressed that a severe accident occurred at the end of 2001 during 
technical operations: $\sim$ 6800 inner PMT's and $\sim$ 1100 outer PMT's were 
destroyed. The data taking will restart probably late 2002 after a partial
replacement of PMT's but with a higher energy threshold and a lower light 
coverage (SUPERKAMIOKANDE II).\\
The results SK has obtained can be summarized in this way:
\bit
\item 
Confirmation of a reduced $\nu_\odot$ flux but with the larger presently 
available statistics.
\item 
No distortion in the energy spectrum.
\item 
No relevant D-N asymmetry.
\item 
Time variation of the flux related to the terrestrial orbit.
\fit
\begin{table}[ht]
\caption{\it{ SK rates and uncertainties for eight energy bins. The quoted rates, 
statistical and systematic uncertainties (of the spectrum shape) in the third 
column are in units of flux predicted by \cite{BAH01a} . These systematic 
uncertainties are assumed to be uncorrelated in energy while uncertainties of 
the $^8$B $\nu_\odot$ spectrum, of the energy scale and of the  energy 
resolution are fully correlated in energy (but uncorrelated with each other), 
adapted from \cite{SMY02a}.}}
\begin{center} 
\begin{tabular}{|l|c|c|}
\hline Bin & Range [MeV] & Rate$\pm$stat$\pm$syst \cite{BAH01a} \\\hline 
1 &  5.0-5.5  & 0.4671$\pm$0.0404$^{+0.0165}_{-0.0138}$ \\\hline 
2 &  5.5-6.5  & 0.4580$\pm$0.0141$^{+0.0066}_{-0.0065}$ \\\hline 
3 &  6.5-8.0  & 0.4729$\pm$0.0084$\pm{0.0065}$ \\\hline  
4 &  8.0-9.5  & 0.4599$\pm$0.0093$\pm{0.0063}$ \\\hline  
5 &  9.5-11.5 & 0.4627$\pm$0.0103$\pm{0.0063}$ \\\hline 
6 & 11.5-13.5 & 0.4621$\pm$0.0168$\pm{0.0063}$ \\\hline 
7 & 13.5-16.0 & 0.5666$\pm$0.0390$\pm{0.0078}$ \\\hline 
8 & 16.0-20.0 & 0.5554$\pm$0.1458$\pm{0.0076}$ \\\hline 
Total  & 5.0-20.0  & 0.4653$\pm{0.0047}^{+0.0138}_{-0.0122}$ \\\hline 
\end{tabular} 
\end{center} 
\label{tabsk8bin}
\end{table}
\begin{table}[ht]
\caption{\it{ Subdivision of bins 2--7 flux ratio with respect to the \cite{BAH01a} 
model value, according to the solar zenith angle $\theta_z$. The range of $\cos \theta_z$ 
is  given for each bin: $\cos \theta_z<0$ is 'Day' and $\cos \theta_z>0$ is 'Night'.  
The rates are given in units of 0.001*\cite{BAH01a} SM. Only statistical uncertainties 
are quoted. All systematic uncertainties are assumed to be 
fully correlated in zenith angle, adapted from \cite{SMY02a}.}}
\begin{center} 
\begin{tabular}{|l|c|ccccc|c|} \hline      
Bin  & Day   & \multicolumn{5}{|c|}{Mantle} & Core \cr\hline  
     &-0.97--0.0 & 0.0--0.16 & 0.16--0.33 & 0.33--0.50 &
0.50--0.67 & 0.67--0.84 & 0.84--0.97 \cr \hline 
2 &453$\pm$20  & 442$\pm$53 & 379$\pm$49 &    472$\pm$45  & 522$\pm$45 & 503$\pm$49 & 426$\pm$52  \cr 
3 &474$\pm$12  & 530$\pm$34 & 506$\pm$30 &    438$\pm$26  & 478$\pm$26 & 451$\pm$28 & 439$\pm$31  \cr 
4 &448$\pm$13  & 463$\pm$36 & 470$\pm$33 &    462$\pm$29  & 509$\pm$29 & 461$\pm$32 & 451$\pm$35  \cr 
5 &453$\pm$15  & 449$\pm$40 & 502$\pm$38 &    451$\pm$32  & 473$\pm$32 & 477$\pm$35 & 483$\pm$40  \cr 
6 &477$\pm$25  & 509$\pm$67 & 351$\pm$55 &    391$\pm$49  & 498$\pm$53 & 434$\pm$56 & 521$\pm$64  \cr 
7 &511$\pm$54  & 570$\pm$150 & 831$\pm$167 &    694$\pm$131  & 665$\pm$127 &
441$\pm$118 & 469$\pm$131 \cr\hline 
1-8&459.9$\pm$6.7&483$\pm$18& 476$\pm$17&451$\pm$15&496$\pm$15&467$\pm$16&456$\pm$17 \cr\hline
\end{tabular} 
\end{center} 
\label{tabsktheta}
\end{table}

\subsection{The Gallium experiments.} 
\label{sect:gallium}
Three experiments, SAGE, GALLEX and GNO, have detected $\nu_\odot$'s interactions on Ga target, through a reaction 
suggested by Kuzmin, \cite{KUZ64,KUZ65}, which is only sensitive to CC interactions: 
\beq
^{71}Ga\,(\,\nu_e\,,\,e^-\,)\,^{71}Ge ~~~~~~~  E\geq 232.69 \pm 0.15 \,keV
\label{eq_gallium}
\feq 
The target mass of these experiments should require about one year world-production of Ga (the cost of Ga is 
$\sim 10^3$ US \$;  
let us remember that the isotopic abundance of $^{71}$Ga is 40 $\%$, the remaining 60 $\%$ is due to $^{69}$Ga.  
The produced $^{71}$Ge atoms, which turn back into $^{71}$Ga ${\it{via}}$ electron  capture 
with a half-life of 11.4 days, are extracted from Ga every 3 or 4 weeks and  its electron 
capture is detected by the observation of Auger electrons and X-rays within shielded proportional
counters having a volume of about 1 cm$^3$.\\ 
The known background sources are side reactions, $^{69}$Ge, Tritium (which is present  in 
germane gas) and $^{222}$Rn.\\ 
The number of Ge atoms extracted in a single run is very small, less than 10 atoms. Thanks to the low-energy threshold, 
interactions are mainly produced by p-p  $\nu_\odot$'s, see table \ref{tabrateclga} and table \ref{tabgarates}, and 
are from ground to ground nuclear state, so that the nuclear part of computations is "easier" to be done.\\ 
The measurement of the p-p $\nu_\odot$ flux is an important goal for these detectors  due to
the independence of such a component from SMs: therefore it is possible to check the 
consistency of the theories on nuclear reactions.  \\
\begin{table}[ht]
\caption{\it{ Characteristics of $\nu_\odot$ flux and interaction rates on Ga target, from \cite{BAH97a,BAH01a,COU02}. 
Flux 1 and Flux 2 are in$\nu\, cm^{-2}s^{-1}$ while rate 1 and rate 2 are expressed in SNU. 
Flux 1 and rates 1 are from \cite{BAH01a}, flux 2 and rates 2 are from \cite{COU02}. }}
\begin{center} 
\begin{tabular}{|l|c|c|c|c|c|c|} 
\hline 
Reac.& Uncer.&$\sigma_{Ga}$&Flux 1& Flux 2&Rate 1&Rate 2\\ 
source&$\sigma\,(\%)$&$10^{-45}cm^{-2}$&     &       & & \\\hline
p-p& $\pm$2.3&1.172&5.95$\cdot 10^{10}$&5.916 $\cdot 10^{10}$&69.73&69.34\\\hline 
p-e-p&$^{+17}_{-7}$&20.4&1.4$\cdot 10^{8}$&1.392$\cdot 10^{8}$ &2.86&2.84\\\hline 
Hep&$^{+32}_{-16}$&7140&9.3$\cdot 10^{3}$& - &0.07& -\\\hline 
$^7$Be&$^{+7}_{-3}$&7.17&4.77$\cdot 10^{9}$&4.853$\cdot 10^{9}$ &34.20&34.79\\\hline 
$^8$B&$^{+32}_{-15}$&2400&5.05$\cdot 10^{6}$&4.979$\cdot 10^{6}$ &12.12&11.95\\\hline 
$^{13}$N&$^{+6}_{-3}$&6.04&5.48$\cdot 10^{8}$&5.767$\cdot 10^{8}$ &3.31&3.483\\\hline 
$^{15}$O&$^{+12}_{-5}$&11.37&4.80$\cdot 10^{8}$&4.967 $\cdot 10^{8}$&5.46&5.648\\\hline 
$^{17}$F&$^{+12}_{-5}$&11.39&5.63$\cdot 10^{6}$&3.083$\cdot 10^{6}$ &0.06&0.0351\\\hline 
\end{tabular} 
\end{center} 
\label{tabgarates}
\end{table}
\subsection{SAGE.} 
\label{sect:sage}
Soviet American Gallium Experiment (SAGE) started its operations in 1990: its detector,  
consisting in 7 so called "reactors", is homed at Baksan (Russia), at a depth of  $\sim$ 
4700 mwe, \cite{ABD94,GAV97,ABD99b,SAG99}. It uses Ga in metallic  form as a target (Ga is 
liquid at T$\geq$ 300 K). Ge is extracted $via$ immersion in HCl and H$_2$O$_2$ solutions in 
teflon-lined reaction vessels (this  procedure is very difficult). The metallic form allows a 
compact detector with  a reduced background and an easier calibration. The Ga target mass 
has been variable,  up to 57 tons; the present value is 49 tons.\\ 
The extraction procedure is based on the separation of Ge into an aqueous phase when metallic 
Ga is mixed with an acid solution and an oxidizing agent. The mixture is  stirred, then Ga comes 
back into emulsion while Ge goes to the surface of the  emulsion droplets. When H$_2$O$_2$ is 
finished, the emulsion breaks down and the phases separate: then, Ge is concentrated by 
distillation and HCl is added; at this point the Ar purification starts. After a three times 
repeated Ge extraction in CCl$_4$, the obtained germane is inserted in proportional counters 
and the counting procedure begins. For each extraction the best estimate of the $^{71}$Ge 
production rate is done by measuring likelihood function.\\ 
It is interesting to point out that the first published interaction rate was very close to zero, 
but with a very large uncertainty; on the contrary the latest result, based  on January 1990 - 
December 2001 runs, is \(70.8\,^{+5.3}_{-5.2}\,^{+3.7}_{-3.2}\) SNU,  \cite{ABD02}.\\
It has to be stressed that problems in data acquisition were present during 1996-1999 so that
SAGE coll. has applied an {\it{a posteriori}} correction to the detection efficiency. In any case 
an analysis on systematic effects and uncertainties has been done: the total  systematic effects
has been computed at a level of $^{+3.7}_{-3.2}$ SNU, see also table \ref{tabsageunc}. \\ 
\begin{table}[ht]
\caption{\it{ Systematic effects in SAGE and their uncertainties (in SNU).
The values for extraction and counting efficiencies are based on a
rate of 70.8 SNU, adapted from \cite{ABD02}.}}
\begin{center} 
\begin{tabular}{|l|c|c|} 
\hline 
Extraction            & Ge carrier mass         & $\pm$1.5   \\
efficiency            & Extracted Ge mass       & $\pm$1.8   \\
                      & Residual carrier Ge     & $\pm$0.6   \\
                      & Ga mass                 & $\pm$0.2   \\\hline\hline\hline
Counting              & Counter effects         & $\pm$1.3   \\
efficiency            & Gain shifts             & +2.2       \\
                      & Resolution              & -0.4,+0.5  \\
                      & Rise time limits        & $\pm$0.7   \\
                      & Lead and exposure times & $\pm$0.6   \\\hline\hline\hline
Backgrounds           & Neutrons                & $<$-0.02   \\
                      & U and Th                & $<$-0.7    \\
                      & muons                   & $<$-0.7    \\
                      & Internal radon          & $<$-0.2    \\
                      & External radon          &  0.0       \\
                      & Other Ge isotopes       & $<$-0.7    \\\hline\hline\hline
Total                 &                         & -3.2,+3.7  \\ \hline
\end{tabular} 
\end{center} 
\label{tabsageunc}
\end{table}
\\
A calibration of the apparatus with a $^{51}$Cr source of 19.1 $\pm$ 0.2 PBq 
strengthened the detection of $\nu_\odot$'s and the efficiency of the 
analysis (the ratio observed/expected $^{71}$Ge production rate 
is \(R\,=\,0.97\,\pm\,0.12\), \cite{ABD96,ABD99a} . \\ 
Looking for time correlation, all the data have been analysed yearly, 
monthly and bimonthly, the same distance from the Sun being the selecting 
parameter, but no statistically significant effect has been found, 
\cite{ABD02}; in table \ref{tabsagetime} yearly and monthly average 
interaction rates are shown. The difference between winter and summer interaction rate
has been computed; its value is - 6.7$^{+10.7}_{-10.3}$ SNU.\\ 
This experiment should continue up to 2006, \cite{GAV01b}. \\
\begin{table}[ht]
\caption{\it{ Combined analysis of all runs during yearly
and monthly intervals. The quoted value in SNU is the best fit result of the likelihood 
procedure, adapted from \cite{ABD02}.}}
\begin{center} 
\begin{tabular}{|l|c|c||l|c|c|} 
\hline
Year   & Nr. of data    & SNU  & Month    & Nr. of data  & SNU \\\hline
1990               &  5 &   43 & January            & 11 &   58 \\  
1991               &  6 &  112 & February           & 12 &   60 \\
1992               & 13 &   76 & March              &  9 &  102 \\
1993               & 15 &   84 & April              &  9 &   54   \\
1994               & 10 &   73 & May                & 12 &   75   \\
1995               & 13 &  102 & June               & 11 &   79   \\
1996               & 10 &   55 & July               & 15 &   52   \\
1997               & 16 &   62 & August             & 15 &   78   \\
1998               & 12 &   56 & September          & 20 &   68   \\
1999               & 14 &   87 & October            & 17 &   73   \\
2000               & 22 &   67 & November           & 15 &   59   \\
2001               & 22 &   65 & December           & 12 &  105   \\\hline\hline
\end{tabular} 
\end{center} 
\label{tabsagetime}
\end{table}

\subsection{GALLEX.} 
\label{sect:gallex}
The GALLEX (GALLium EXperiment) detector was operating from 1991 to 1997 at LNGS, Italy.  The target was a 101 tons 
solution of GaCl\(_3\) in water and HCl containing 30.3 tons of natural Ga corresponding to \(\sim 10^{29}\) nuclei of 
$^{71}$Ga. \\ 
The tank containing 8-molar Ga chloride solution was equipped with provisions for a  N$_2$ purge and with a central 
tube for inserption of either a man-made $\nu$ source  or of a neutron monitor. The surrounding rock gave a 3600 mwe 
shield to cosmic rays.\\ In short the experimental procedure was the following \cite{ANS92}: 
\bit 
\item 
The solution was exposed to $\nu_\odot$'s for 3 or 4 weeks; at  the end $\sim$ 16 nuclei of $^{71}$Ge should be 
present as a volatile  GeCl\(_4\) (if SMs are correct and all $\nu$'s produced in the Sun  reach the Earth without 
flavour change). 
\item 
The \(^{71}\)Ge was chemically extracted and converted into  GeH\(_4\) (Germane gas) and introduced into 
miniaturized proportional counters, \cite{WIN93}, mixed with Xe as counting gas. At the end \( 95 - 98 \%\) 
of the $^{71}$Ge  present in the solution at the time of the extraction was in the counter.  
\item 
The \(^{71}\)Ge electron capture (meanlife 16.5 days)  \(^{71}Ge\,(\,e^-,\nu_e\,)^{71}Ga\) was observed for a 
period of 6 months. A good determination of the time constant counter background was also permitted. 
\item 
Data taken during the counting time were analysed with a maximum  likelihood technique to obtain the most 
probable number of \(^{71}\)Ge detected.  The mean background was less than 0.1 counts per day.  
\item 
A correction was made to account for contributions to the observed signal  from sources other than $\nu_\odot$'s 
(mainly interactions generated by high  energy muons from cosmic rays and by natural radioactivity). 7 SNU have to be 
subtracted from the measured \(^{71}\)Ge production rate.  
\fit 
GALLEX solution was also exposed to an artificial $\nu$ source $^{51}$Cr, a nucleus having a lifetime of 40.0 days and 
decaying by electron capture: \(^{51}Cr\,(e^-,{\nu}_e)\,^{51}V\). 
$\nu_e$'s are produced with a discrete energy spectrum having values similar to the solar spectrum. Two expositions
were done, 63.4 $\pm$ 0.5 PBq and 69.1 $\pm$ 0.6 PBq being the activities of the used sources: 
the first published ratio observed/expected interaction rate was  \(R\,=\,0.93 \pm 0.08\) but a new 
analysis of the counter efficiency and of source intensity has lowered the 
previous result. The latest upgraded value is \(R\,=\,0.89 \pm 0.07\), \cite{CAT02}. 
This means that in GALLEX unknown systematic errors were of about 10$\%$.\\ 
A further test was the calibration with $^{71}$As  to check the correctness 
of "chemistry" in the detector. It was found a complete agreement with 
the expected value: the experimental 
result was 99.9$\pm$ 0.8$\%$ of the estimated one, \cite{HAM98b}.\\
The final result of GALLEX experiment was 77.5 $\pm$ 6.2$^{+4.3}_{-4.7}$ 
SNU, \cite{HAM99}, see also \cite{ANS95,HAM96} 
for intermediate results. The statistical error was at a level of $\sim$ 
8$\%$ of the measured rate; the systematic term adds a contribution of 
$\sim$ 6$\%$.
	
\subsection{GNO.}
\label{sect:gno}
\begin{figure}[ht]
\begin{center} 
\mbox{\epsfig{file=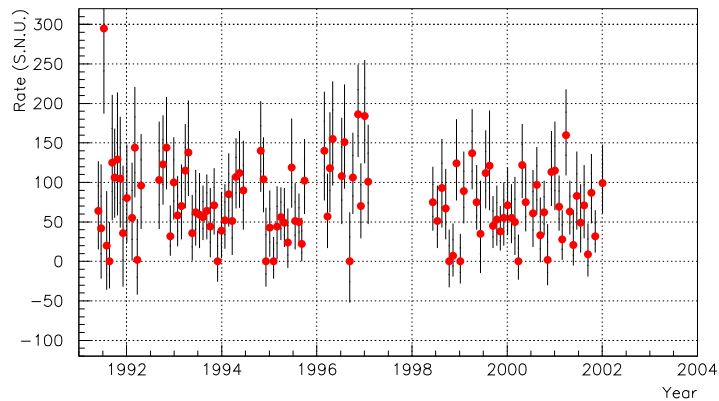,width=0.6\textwidth}} 
\end{center}
\caption{ Single extraction interaction rate of GALLEX+GNO experiments from 1991 up to May 2001, 
from \cite{KIR02}.}
\label{fig:GNOplot}
\end{figure}

Gallium Neutrino Observatory (GNO) experiment \cite{GNO96} is homed at LNGS 
and started in  1998; it upgrades of GALLEX experiment, in fact it uses 
the same 30 tons  target, \cite{ALT00,BEL01,CAT02}, but with a new data 
acquisition system and largely improved electronics. The Ge extraction from 
the Chloride solution procedure is not changed with respect to GALLEX.\\  
An improvement on the global error is expected to come from GNO: a new Rn 
cut procedure is still operative, then a reduction of systematic errors at 
a level of 3-5 \% by a direct measurement of the volume efficiency of all 
the counters and a neural network analysis technique are foreseen.\\ 
At present, the contributions to the systematic errors due to energy cuts, 
pulse shape cuts, event selection, Rn cut inefficiency and $^{68}$Ge are 
reduced with respect to GALLEX experiment so that the total systematic 
error in GNO is lowered from 4.5 SNU down to 3.0 SNU.\\
The GNO interaction rate, based on data performed from May 1998 to January 2002 (43 runs), 
is $65.2\pm  6.4(stat.) \pm 3.0(syst.)$ SNU, \cite{KIR02}; this value differs at a level of 1.7 
$\sigma$ from GALLEX result and could suggest a constant $\nu_\odot$ flux over 10 years. \\
The combined GALLEX+GNO (1991-2002) results is $70.8 \pm 4.5(stat.) \pm 3.8(syst.)$ SNU. 
The difference between winter and summer interaction rate over the whole period is -11 $\pm$ 9 
SNU, \cite{KIR02}.
In fig. \ref{fig:GNOplot} single extraction values with error bars are shown. 
\\
To further reduce the counting errors other aspects are under analysis: 
machined counters made by a plastic material, more uniform in shape, and a low temperature 
calorimeter to "read" the energy thermally deposited from a decay with a superconductor phase 
transition thermometer. The use of a new cryogenic detector which should allow an efficiency at a 
level of $\sim$ 100 \% (the present value is $\sim$ 70 \%), an energy resolution better by 
a factor of 7, no X-ray escape effects and no pulse-height degraded events is under evaluation,
\cite{FER01,CAT02}.\\
In 2003, a new calibration by inserting a $^{51}$Cr source will be carried 
out: a precise determination of absolute $\nu$ cross-section on Ga, within 
an uncertainty at a level of 5 \%, is expected.
In the original GNO proposal an increase of the target mass up to 100 tons with an intermediate 
step at a level of 60 tons, was foreseen by adding Ga in liquid form but these steps up to now 
are not funded. A joint experiment with SAGE collaboration has been even suggested but 
difficulties seem to rule out at present any development in this direction.

\subsection{SNO.}
\label{sect:sno}
\begin{figure}[ht]
\begin{center} 
\mbox{\epsfig{file=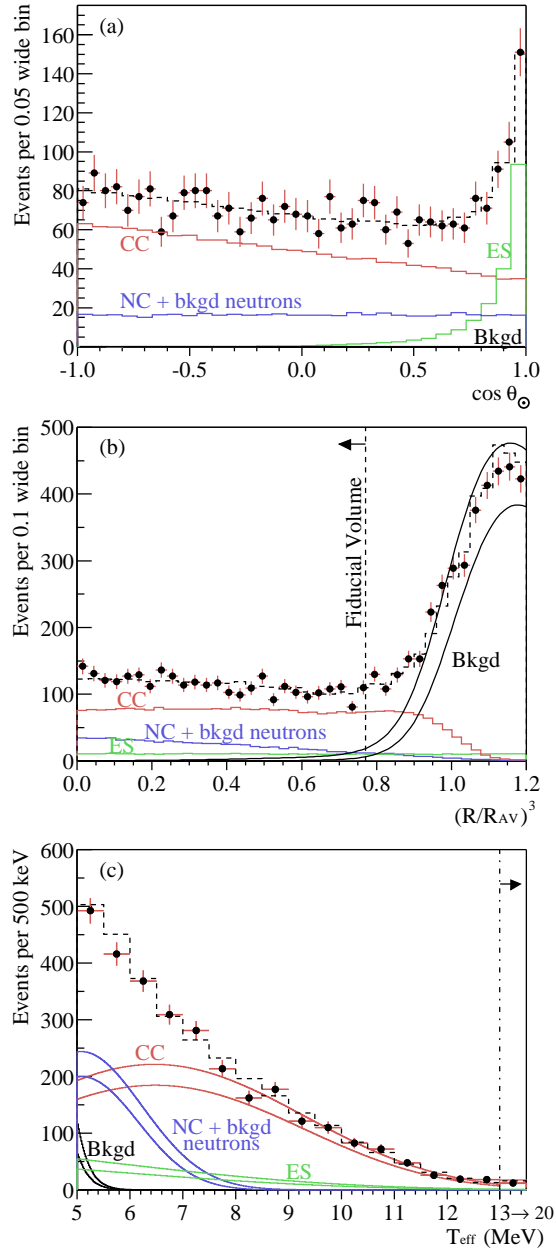,width=0.5\textwidth}} 
\end{center}
\caption{ SNO data:(a) Zenithal distribution of events at a distance R $\le$ 5.50 m. 
(b) Distribution of the volume weighted radial variable $(R/R_{AV})^{3}$.  
(c) Kinetic energy spectrum for R $\le$ 5.50 m.  Also shown are the Monte Carlo predictions 
for CC, ES and NC + bkgd neutron events scaled to the fit results, and the calculated 
spectrum of \v Cerenkov background (Bkgd) events.  The dashed lines represent the summed 
components, and the bands show $\pm 1\sigma$ uncertainties.  
All distributions are for events with $T_{\rm eff}$$\geq$5 MeV, from \cite{AHM02b}. }
\label{fig:SNOspectra}
\end{figure}

\begin{figure}[ht]
\begin{center} 
\mbox{\epsfig{file=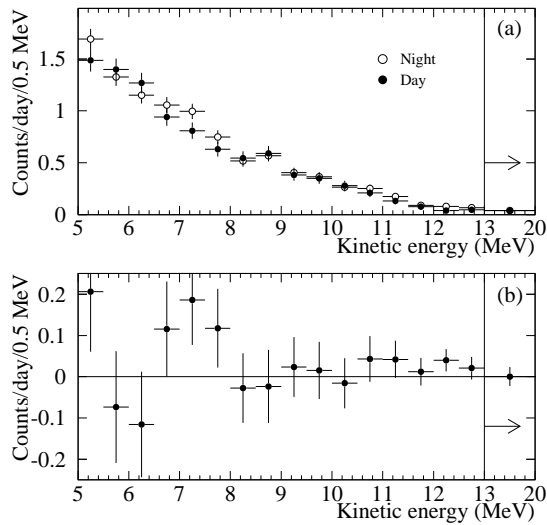,width=0.5\textwidth}} 
\end{center}
\caption{ SNO data: (a) Energy spectra for D and N events including signals and backgrounds. 
The final bin extends from 13.0 to 20.0 MeV.  (b) N - D difference  between the
spectra.  D rate = 9.23 $\pm$ 0.27 ev/day; N rate = 9.79 $\pm$ 0.24 ev/day, from \cite{AHM02a}.  }
\label{fig:SNOdnspec}
\end{figure}

The Sudbury Neutrino Observatory (SNO) experiment was proposed in the 80s', \cite{SNO87}: it measures 
at the same time and in the same detector CC and NC interaction events using \v Cerenkov technique. 
It is worth to underlining that physicists looked for results coming from 
this experiment from many years, \cite{YIN92,BAH96c,MAR00,BAH00a,BAH00b,BAH01d,BAR01a}.\\
The detector is homed at a depth of 6010 mwe in the "INCO Creighton Mine", near Sudbury (Canada): 
the inner part is an acrylic transparent spherical shell 12 m in diameter filled with 1000 tons of heavy water 
(D$_2$O) surrounded by 1700 tons of ultrapure $H_2O$. A further outer volume of 5300 tons of ultrapure $H_2O$ 
is used as a shield from neutrons and $\gamma$'s from the rock. The background due to U and Th contamination is
respectively at a level lesser than 4.5$\cdot 10^{-14}$ g/g and 3.5$\cdot 10^{--15}$ g/g.\\
The emitted \v Cerenkov light is collected by 9456 PMT's, 20 cm in diameter. The time resolution of 1.7 ns 
allows an event reconstruction within 30 cm of precision. About half of the light produced within 7 m of 
the detector centre strikes a PMT. The total coverage is at a level of $\sim$ 31 \%.
A fiducial volume was defined to accept events with vertices within 5.50 m from the detector centre.\\   
The energy threshold of 5.0 MeV allows contributions from CC interaction events in $D_2O$, ES events both in $D_2O$ 
and $H_2O$, capture of neutrons from NC interactions and backgrounds and low energy \v Cerenkov events. After a 
calibration with $^{252}$Cf source the neutron detection efficiency  was estimated at a level of 14.4\%. 
Only $^8$B and Hep $\nu_\odot$'s can produce detectable interactions; the expected count rate is $\sim$ 20 events 
per day, \cite{EWA92,MCD99,VIR00,SNO00,AHM01,SNO01}.\\
It is worthwhile to note that SNO experiment represents a milestone in the $\nu_\odot$ physics  thanks to 
many different parameters it could measure, \cite{BAH00a}: 
\bit 
\item CC and NC interactions. 
\item First and second moment of the recoil energy spectrum. 
\item D-N rates for both NC and CC interactions. 
\item Winter/summer CC interactions. 
\item NC to CC and ES to CC double ratio. 
\fit  
The value of $^8$B  $\nu_\odot$ flux or its flavour content, if any, can be extracted from CC and  ES reaction, 
independently from SM calculations. The comparison of CC and ES interactions, as measured in SNO itself 
and in SK, allows an estimate of the flux of $\nu_x$ (x$\neq$ e) and $\overline\nu$.\\ 
Moreover, it is possible to measure the spectral shape of $^8$B $\nu_\odot$'s with  good 
statistics and the energy of the recoil electrons for CC and ES interactions.\\ 
SNO provides a "{\it{verification}}" of the SK data on ES interactions with a detector  installed at 
a deeper depth and crossed by a much lower flux of  atmospheric muons.\\ 
During 1999 there was an intense program of electronics, optical and energy calibration and it 
was possible to lower the trigger threshold and the Rn below the  target levels. Phase 1 has 
provided an accurate measurement of $\nu_e$ flux {\it{via}} CC interaction, \cite{MCD01}.\\
The flashes recorded from November 1999 to May 2001 (306.4 days of running time)
have been analysed: 1967.7$^{+61.9}_{-60.9}$ CC events, 263.6$^{+26.4}_{-25.6}$ ES events and 
576.5$^{+49.5}_{-48.9}$ NC events were detected (only statistical errors are given), \cite{AHM02a}.
The estimated systematic uncertainties are: for CC events $\pm$ 5.2\% 
(exp.) and $\pm$ 1.8 \% (theor.); for NC events $^{+9.1}_{-8.9}$\% (exp.) 
and $\pm$ 1.3 \% (theor.); for ES events $^{+5.0}_{-4.8}$\% (exp.). \\
Recent improvements in theoretical calculation of the $\nu_e$-D cross-section for CC 
interactions, \cite{ORT00,BEA01,BUT01,KUR02,NAK01,NAK02}, have been introduced in  computations.\\ 
The deduced $^8$B $\nu_\odot$ fluxes are, \cite{AHM02a}, see also fig. \ref{fig:SNOspectra}: 
\beqar 
\Phi_{CC}\,=\,1.76^{+0.06}_{-0.05}(stat.)\,\pm 0.09(sys.)\cdot  10^6 \,cm^{-2}s^{-1} \nonumber\\
\Phi_{ES}\,=\,2.39^{+0.24}_{-0.23}(stat.)\pm 0.12(sys.)\cdot 10^6 \,cm^{-2}s^{-1} \\
\Phi_{NC}\,=\,5.09^{+0.44}_{-0.43}(stat.)^{+0.46}_{-0.43}(sys.)\cdot 10^6 \,cm^{-2}s^{-1} \nonumber
\label{eq_snofluxes}
\feqar 
The ES flux is fully compatible with SK result. The excess of the NC events with respect to CC and ES interactions 
has been explained as a strong signal of $\nu$ flavour transformation, see later in section \ref{sect:partsolutions}.
Under the assumption of undistorted $^8$B spectrum the non-electronic $\nu_\odot$ flux has been computed:
\beq
\Phi_{\mu+\tau}\,=\,3.41^{+0.45}_{-0.43}(stat.)^{+0.48}_{-0.45}(sys.)\cdot 10^6 \,cm^{-2}s^{-1}
\label{eq_snomutau}
\feq
If errors are added in quadrature this flux is 5.3 $\sigma$ above the null value; if the previous constraint on 
energy spectrum is removed, the $\nu_\odot$ flux, as deduced from the NC interaction data, is in agreement with 
\cite{BAH01a} but its value is higher. Moreover, a much larger statistical error is present:
\beq
\Phi_{NC}\,=\,6.42\pm 1.57 (stat.)^{+0.55}_{-0.58}(syst.)\cdot 10^6 \,cm^{-2}s^{-1} 
\label{eq_snoncfree}
\feq
Even D and N spectra and rates have been analysed, \cite{AHM02b}: the total livetimes were 128.5 days for D-time 
and 177.9 days for N-time. If a solution to SNP is given in terms of oscillations among known 
$\nu$ flavours, any asymmetry is expected for NC events unless a "sterile" component is present. In 
the same description CC interactions should have non-zero asymmetry. A calibration 
with $^{16}$N source puts in evidence a 1.3\% per year drift in the energy scale, 
which has been  removed in calculations. 
The results are reported in table \ref{tabsnodn}, see also fig. \ref{fig:SNOdnspec}.\\
\begin{table}[ht]
\caption{\it{ D-time and N-time fluxes in SNO data and A$_{D-N}$ asymmetry under the hypothesis of undisturbed $^8$B 
$\nu_\odot$ energy spectra are shown, adapted from \cite{AHM02b}.} }
\begin{center}
\begin{tabular}{|l|c|c|c|} 
\hline
signal & $\Phi_D$ & $\Phi_N$ &A(\%) \\  
       &$10^6\,cm^{-2}s^{-1}$& $10^6\,cm^{-2}s^{-1}$&        \\ \hline
CC & $1.62\pm0.08\pm 0.08$  & $1.87\pm 0.07\pm 0.10$ & $+14.0 \pm~6.3^{+1.5}_{-1.4}$ \\ 
ES & $2.64\pm0.37\pm 0.12$ & $2.22 \pm0.30\pm 0.12$ &  $-17.4 \pm 19.5^{+2.4}_{-2.2}$ \\ 
NC & $5.69\pm0.66\pm 0.44$  & $4.63\pm0.57\pm 0.44$  &  $-20.4 \pm
16.9^{+2.4}_{-2.5}$ \\ \hline
\end{tabular}
\end{center}
\label{tabsnodn}
\end{table}
In CC events a D-N asymmetry at a level of + 2.2$\sigma$ is 
present (- 0.9$\sigma$ for ES events and - 1.2 $\sigma$ for NC interactions). 
The signal extracted from $\nu_e$ component is $A_e$ = 12.8 $\pm$ 6.2$^{+1.5}_{-1.4}$ \% while
$A_{TOT}$ = -24.2 $\pm$ 16.1$^{+2.4}_{-2.5}$ \%. If the assumption of no-asymmetry is added, 
then $A_e$ = 7.0 $\pm$ 4.9$^{+1.3}_{-1.2}$ \%; this value has to be compared with the SK result
$A_e^{SK}$ = 5.3 $\pm$ 3.7$^{+2.0}_{-1.7}$ \%.\\
The deployment of NaCl to increase the NC detection capability started in May 2001. 
After 8 months of data taking, SNO will reduce its statistical error and should be able to give 
a more precise answer concerning the $\nu$ flavour 
oscillation solution.

\section{The solar neutrino problem. (SNP)} 
\label{sect:snp}

>From the first experimental result published in early 1968 it was possible 
to confirm the existence of solar neutrinos: all the experiments gave and 
currently confirm a positive signal but there was evidence of an 
interaction rate lower than the theoretical predictions; moreover, different 
data appeared even inconsistent. A puzzling situation was present until  
2002. SNO experiment, which is able to distinguish among ES, CC and NC 
interactions, has detected three different $\nu_\odot$ fluxes: this result
strongly supports an explanation of SNP in terms of new particle physics, 
as early recommended in \cite{BAH90a}.\\
Table \ref{tabexpflux} summarizes the presently available experimental results and theorethical 
fluxes or interaction rates.\\
\begin{table}[ht]
\caption{\it{ Solar neutrino experiments: energy threshold, experimental results and 
predictions from \cite{BAH01a}.}}
\begin{center}
\begin{tabular}{|l|c|c|c|c|} 
\hline
Experiment   & $E_{th}$  & Measurements& Predictions&Data  \\\hline
HOMESTAKE    & 0.814 & 2.56 $\pm$ 0.23 &7.6$^{+1.3}_{-1.1}$  &1970\\
615 t \(C_2Cl_4\)&  MeV  &SNU &SNU &1994\\ \hline
KAMIOKA     &  7 & 2.80 $\pm$ 0.38 & 5.05$^{+1.01}_{-0.81}$ &1987\\
3000 t \(H_2O\)& MeV & $10^6 \nu\,cm^{-2}s^{-1}$ & $10^6 \nu\,cm^{-2}s^{-1}$ &1995\\ \hline
GALLEX      &  0.233 & 77.5 $\pm$ 7.7 &127.8$^{+9}_{-7}$ &1991\\
101 t \(GaCl_3\)& MeV &SNU &SNU &1997\\ \hline
SAGE        &  0.233 & 70.8 $\pm$ 6.5  & 127.8$^{+9}_{-7}$ &1990\\
50 t $Ga$ Metal& MeV  &SNU &SNU &2001\\ \hline
GNO &  0.233 & 65.4 $\pm$ 7.1 & 127.8$^{+9}_{-7}$ &1998\\
101 t \(GaCl_3\)& MeV &SNU &SNU &2002\\ \hline
GALLEX+GNO &  0.233 & 70.8 $\pm$ 5.9 & 127.8$^{+9}_{-7}$ &1991\\
101 t \(GaCl_3\)& MeV &SNU &SNU &2002 \\ \hline
SK &  5.0 & 2.35 $\pm$ 0.08 & 5.05$^{+1.01}_{-0.81}$  &1996\\
22500 t \(H_2O\)& MeV &$10^6 \nu\,cm^{-2}s^{-1} $&$10^6 \nu\,cm^{-2}s^{-1}$ &2001\\ \hline
SNO& 5.0 &ES 2.39$\pm$0.268 &5.05$^{+1.01}_{-0.81}$  & 1999\\
   &  &CC 1.76$\pm$0.11& & \\
   &  &NC 5.09$\pm$0.64& & \\
1000 t $D_2O$& MeV&$10^6 \nu\,cm^{-2}s^{-1} $&$10^6 \nu\,cm^{-2}s^{-1}$ &2001\\ \hline
\end{tabular}
\end{center}
\label{tabexpflux}
\end{table}

\subsection{Present experimental situation.} 
\label{sect:experimentalsituation}
Homestake experiment first detected $\nu_\odot$'s and showed the first SNP: 
the flux was lower of  the predicted one by a factor of 3. Many years later 
KAMIOKANDE and SK measured a $^8$B $\nu_\odot$ reduced by a factor of 2 with
respect to the predictions of SMs. In Ga experiments the interaction rate
was $\sim$ 60 $\%$ of the expected one.
SNO, which has confirmed SK result concerning ES interactions, has identified CC and NC 
interaction events: it has measured three different values giving a robust indication that a  
component other than electronic one is present in $\nu_\odot$ flux reaching terrestrial detectors. \\
The presently available experimental results can be summarized in this way:
\benu 
\item 
The measured rates vary  from $\sim$  30$\%$ up to 100$\%$ of the values predicted by SMs but a more intriguing 
situation comes from the  analysis of different contribution to the $\nu_\odot$ flux. \\ 
Highest experimental errors are at level $\sim$ 15$\%$ (SNO) while the theoretical uncertainties 
on $\nu_\odot$ flux vary from $\sim$ 10$\%$ up to $\sim$ 25$\%$. Hence, it is difficult to 
eliminate the measured discrepancy in terms  of errors in experimental and theoretical procedures. 
\item 
SK "confirms" KAMIOKANDE; GALLEX, GNO and SAGE are each other compatible while SNO ES interaction rate 
strengthens SK by a different technique; Cl experiment is lonely and not reproduced.    
\item 
It seems unlikely that different experimental procedures and systematics could have 
unknown inefficiencies: the Ga experiments verified their total efficiency with strong  man-made 
\(^{51}\)Cr sources, yielding the expected results and reducing the probability of  any inefficiency 
at a level of $\sim$ 10$\%$.  Moreover, they are based on different operational techniques and their 
results, which are each other  compatible, confirm  the foundations of stellar structure and evolution theory 
through the experimental observation of p-p fusion in the solar interior. As a further remark, in 
the  limit of their modest statistics, Ga data do not yet exhibit any statistically significant  
$\nu_\odot$ flux variation in time which should imply a production rate constant in time.  
\item 
Using the computed $\nu$ interaction cross-sections 
on Cl and the $^8$B $\nu_e$ flux, as  observed in SK and SNO ES events, a rate of $\sim$ 
3.7 SNU would be foreseen  in Cl detector, a value well above the observed rate of 2.56 $\pm$ 0.23 SNU 
with a difference  as great as 6 $\sigma$. The net flux coming from p-e-p, $^7$Be and CNO $\nu_\odot$'s 
should be  completely disappeared: in fact its rate is negative (- 1.1 $\pm$ 0.5 SNU). It should be possible 
to conclude that something peculiar occurs to $\nu_\odot$'s having intermediate energy. 
\item 
L$_\odot$ (or the energy conservation law) fixes the $\nu_\odot$ production rate:  under the hypothesis 
that nothing happens to p-p and p-e-p $\nu_\odot$'s produced inside the Sun, the minimum interaction rate 
that Ga detectors should measure is 79.5 $\pm$ 2.0 SNU, \cite{BAH90a}. It is indeed necessary to add to the previous 
value a $^8$B  $\nu_\odot$ contribution as great as measured by SK and SNO ES flux: 5.6 $\pm$ 1.5 SNU, where the main 
component of the error is due to capture cross-section uncertainty. $^7$Be $\nu_\odot$'s, which are thought 
to be needed in reaction chain producing $^8$B  $\nu_\odot$, should add 5.9 $\pm$ 1.4 SNU therefore a minimum signal 
of 91.0 $\pm$ 3.0 SNU  is foreseen for Ga detectors, \cite{RIC99}. 
Moreover, if one supposes that SNO NC interaction result confirms 
SMs, or that part concerning $^8$B $\nu_\odot$ flux predictions, then it seems to be allowed to estrapolate the same
conclusion to the $^7$Be $\nu_\odot$ flux, which is the "parent" of $^8$B component. 
Consequently, the terms to be added in previous analysis are 12.1 $\pm$ 2.5 SNU ($^8$B 
$\nu_\odot$'s) and 34.2 $\pm$ 3.4 SNU ($^7$Be $\nu_\odot$'s) so that the deficit in Ga 
experiments becomes dramatic: the measured interaction rate by combining  
SAGE and GALLEX+GNO data is 70.8$\pm$ 4.5 SNU. 
\item 
Ga experiments should suggest an interaction rate 
essentially coincident with the  predictions concerning p-p and p-e-p components as given by SMs, 
therefore either the remaining nuclear reactions do not contribute to the $\nu_\odot$ production 
or a deficit in p-p and p-e-p $\nu_\odot$ flux is present. This is hard to accept, due to the 
strong correlation between p-p component and the present $L_\odot$. 
\item 
The NC interactions in SNO indicate a $^8$B $\nu_\odot$ flux in a good agreement with
values estimated in many SMs; on the other hand, ES and CC interaction results, which 
are depending on flavour of interacting $\nu_\odot$'s, show a strong  deficit. As a (natural) 
conclusion, a component other than electronic one seems  to be present in the $\nu_\odot$ flux 
interacting in detectors on Earth. 
\item 
The presently available results concerning temporal variations of $\nu_\odot$ flux are: 
\bit
\item 
{\bf{Day-Night effect}} = Radiochemical detectors, which integrate the  
$\nu_\odot$ flux over long exposure times, cannot detect D-N difference; 
only real-time  detectors enable us to analyse this effect.  
SK results do not show any significant difference between D and 
N fluxes while SNO data concerning CC interactions put
in evidence an asymmetry at a level of + 2.2$\sigma$.   
\item 
{\bf{Winter-Summer effect}} =  
There are no experimental indications of this effect. 
\item 
{\bf{Other solar features}} = 
Recently, data from GALLEX and GNO experiments have been analysed searching 
for correlation with solar modulations, \cite{STU99,STU01,STU02}. A bimodal 
distribution of measured rates per run has been obtained: its significance 
is not clear and this result is under further analysis. 
It has been suggested that the number of Ge atoms extracted in a single run 
has to be introduced in such statistical analysis not the interaction rate. 
A time-power spectrum analysis has enhanced periodicities either in Cl 
either in GALLEX+GNO data: they range from 24.5 up to 28.4 days, 28.4 and 
26.9 days being the most significant values (very similar to solar X-ray 
modulations).\\ 
No indications about non-standard interactions with B$_\odot$.
\fit
\fenu  

\section{SNP: Proposed solutions.} 
\label{sect:SNPsolutions}
Many different ideas were suggested to explain the SNP: we review some features 
of the most important solutions.  

\subsection{Astrophysical solution.} 
\label{sect:astrosolution}
The first proposed way-out to the SNP was a change 
in SMs $via$ the modifications  both in solar physical parameters needed to lower the 
$\nu_\odot$ flux and in technical  description (velocity distribution, screening 
treatment, initial  solar conditions, solar elemental abundances, solar age, 
cross-sections for stellar reactions,  opacities ...). 
Let us remember that many quantities are  theoretically estimated 
(usually by an extrapolation from experimental available results) 
therefore uncertainties are propagated in calculations.\\ 
Immediately after the publication of Homestake results,
Ezer and Cameron proposed that a mixing process in the solar core could have reduced 
the $\nu_\odot$ flux, \cite{EZE68}; in 1981 a similar approach was suggested by 
\cite{SCH81a}. In 90s'  helioseismological data put in evidence that SNP cannot be  
ascribed to the parameters entering SMs, see {\it{e.g.}} \cite{BAH90a,BAH02a},  
in fact T and $\rho$ theoretical profiles are in good agreement with measured data.  
In order to lower the predicted $\nu_\odot$ flux, T$_c$  should differ from the SMs 
expected values by more than 5$\%$ but theoretical profiles agree with helioseismological 
data at a level better than 0.1$\%$.\\ 
Other proposals were presented: a reduction of $^7$Be abundance in the  central 
solar region; changes in nuclear reaction rates at the extreme boundaries of the  
uncertainty range done by \cite{BAH01a}; deviation from the Maxwell-Boltzmann 
distribution  by introducing a corrective term 
$\sim exp \left[ - \frac {E}{kT} - \delta (\frac {E}{kT})^2\right]$; 
quantum treatments, different screening factors and opacities.\\ 
We underline that the effects of the surrounding plasma on nuclear reaction rates, 
namely  reactions concerning $^7$Be, are not known with accuracy, not only in solar 
plasma but  even in laboratory experiments. Many studies have suggested to modify the 
"weak screening"  approach. As a matter of fact interacting nuclei do have a kinetic 
energy   much larger than mean value and  the interaction rate is very sensitive to the 
high  energy tail of velocity distribution inside the Sun. Different factors could 
modify the Maxwell-Boltzmann distribution, which is assumed to be true: diffusion, 
radiative  flows, internal fluctuating electric and magnetic fields. SMs using these 
parameters were realized and their predictions
seem to be compatible with the helioseismological constraints: a consistent depletion on  
$\nu_\odot$ flux except for Ga experiments is produced. \\
We mention the works done by Quarati and coll. which 
developed a totally different  approach, \cite{KAN97,KAN98,COR99}, by using both a non 
extensive statistics in solar core and quantum uncertainty effects in the solar plasma 
analysis, \cite{LAV01}.  In their opinion Tsallis statistics, \cite{TSA88}, should give 
a better description of  particles behaviour in solar plasma, where strong interactions at 
small distances  among many particles could occur and the reaction collision time is 
comparable  with the inverse plasma frequency. Further effects are produced by random 
electric  microfields: a slow varying component due to the plasma oscillations, a fast 
random  component due to the diffusive cross-section and a further component due to the  
two-body Coulomb interaction. Equations relating these effects with  parameters entering 
the Tsallis distribution were deduced; electron screening contribution is neglected 
because of its smallness. Even quantum corrections were introduced, following  
\cite{GAL67,STA00}, and the equilibrium distribution function uses momentum rather energy:  
the computed c$_s$ profiles well agree with predictions given by usual SMs while the 
estimated $\nu_\odot$ flux is compatible with experimental data.\\
In  \cite{TUR00,TUR01a} the influence of these parameters on physical solar measurable 
quantities and on $\nu_\odot$ flux was analysed. The obtained interaction 
rates range from 2.8 to 6.7 SNU for Cl detector and from 102 to 125 SNU for Ga experiment; 
the $^8$B $\nu$ flux varies from 1.7 to 4.7 $10^6 \nu\,cm^{-2}s^{-1}$.\\ 
 
\subsection{Particle solution.} 
\label{sect:partsolutions}

\begin{figure}[ht]
\begin{center} 
\mbox{\epsfig{file=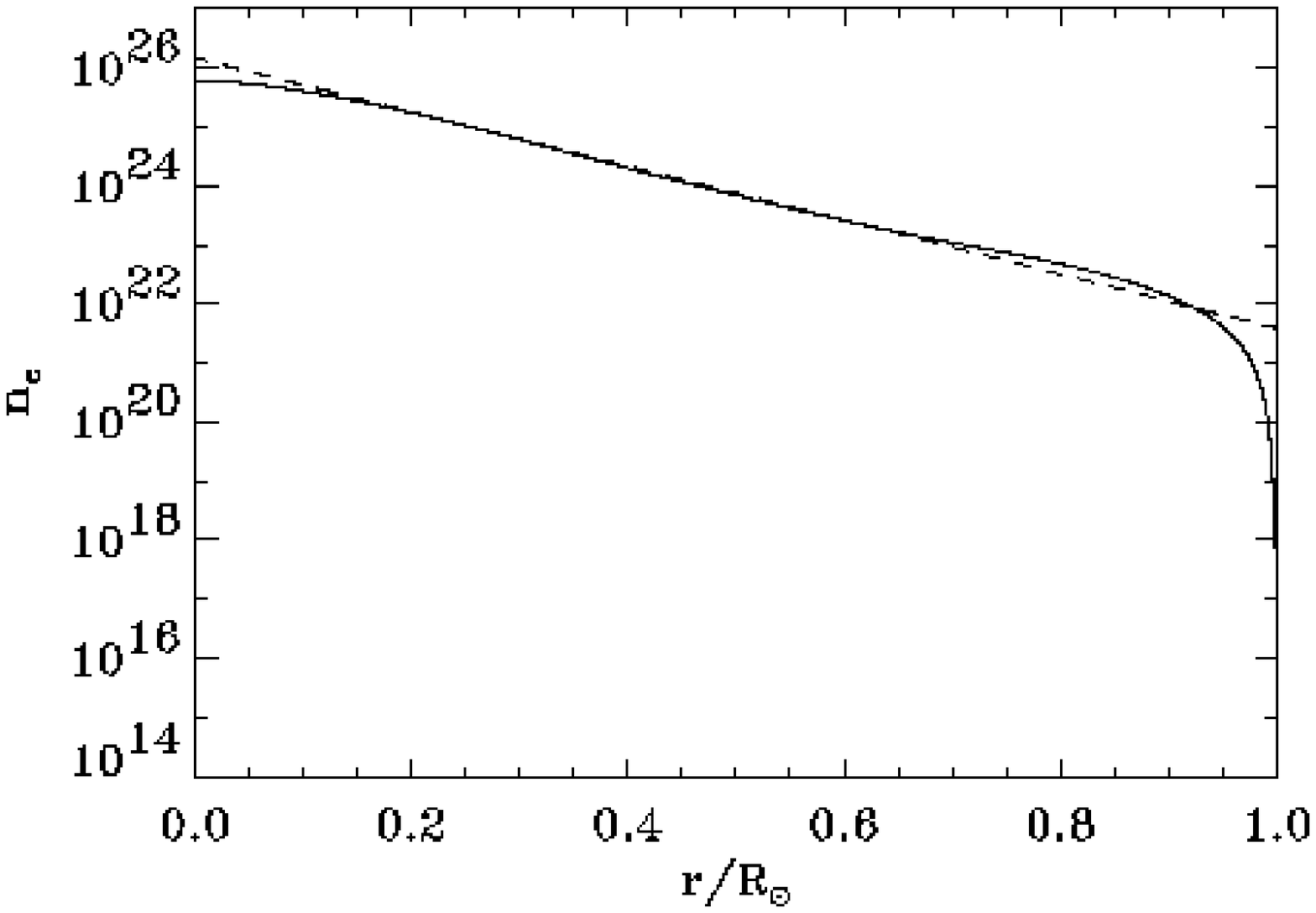,width=0.5\textwidth}} 
\end{center}
\caption{ The electron number density computed in \cite{COU02} as a function of distance from the solar
centre. The approximation proposed in \cite{BAH01a} (dashed line) is also drawn, from \cite{COU02}.}
\label{fig:COUeldens}
\end{figure}

\begin{figure}[ht]
\begin{center} 
\mbox{\epsfig{file=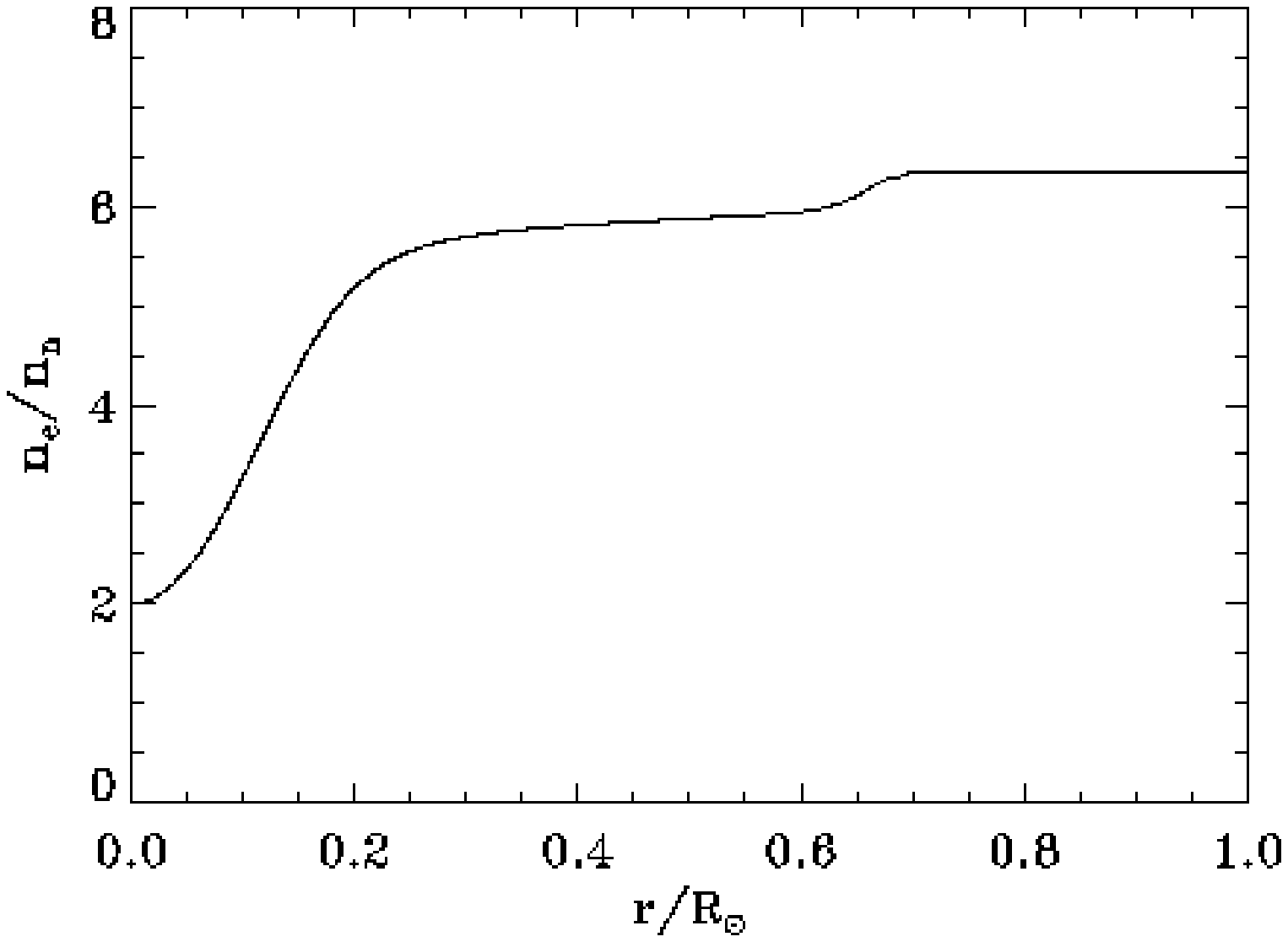,width=0.5\textwidth}} 
\end{center}
\caption{ The computed $n_e$/$n_n$ ratio as a function of distance from solar centre, from \cite{COU02}.}
\label{fig:COUprofneutron}
\end{figure}

A "popular" solution to SNP is within particle properties sector; among other  possibilities the most favoured one 
proposes $\nu$ flavour oscillations.\\ 
We mention a seminal Pontecorvo's paper on the subjet, \cite{PON67}:\\ 
"{\it{From an observational point of view the ideal object is the Sun. If the oscillation length is smaller than the 
radius of the Sun region effectively producing neutrinos...direct observations will be smeared  out and unobservable. 
The only effect on Earth's surface would be that the flux of observable sun neutrinos must be two times smaller 
than the total (active and sterile) neutrino flux.}}"\\ 
Neutrino flavour oscillations are quantum processes requiring both mass and  mixing of the $\nu$ flavours or, otherwise, 
a step beyond the standard particle model (we remember that in non-standard particle theories the flavour mixing  
can be defined even without massive $\nu$'s). If the weak interaction states are  not the mass eigenstates, the first 
ones are superposition of definite mass states which can be either Dirac or Majorana particles, either 
active or sterile $\nu$'s.\\ 
The mathematical description of the oscillation process is given as in quark sector by 
introducing a mixing matrix (similar to the CKM matrix); a CP violating phase term  is also possible but 
usually it is not included because of its smallness. For a good and exhaustive presentation see
\cite{KUO89}.\\
When solar data are under analysis, the simplest and "easier" explanation of SNP requires 2 flavour oscillation; 
in this picture 2 parameters, a mixing angle $\theta$ and a difference mass term $\Delta\,m^2$, are needed and sufficient 
to describe  this solution. Following \cite{FOG96,GOU00}, it is better to analyse the experimental data in 
term of  $\tan^2 \theta$ rather than $\sin^2 2\theta$ in order to study solutions with $\theta  \geq \pi /4$, the so 
called "dark side"; these analyses are given in "exclusion" plots. \\ 
The mixing matrix is real and orthogonal (a rotation by an angle $\theta$ is present).\\ 
The $\nu_\odot$ motion from the Sun to the Earth is  "in vacuum" and is described by a "Klein-Gordon" 
equation, see \cite{RAF96,HAX00}  for a complete review of the  technical treatment.
In the more general case: 
\beq
 |\nu(t=0)\rangle = a_e(t=0) |\nu_e \rangle + a_\mu(t=0) 
|\nu_\mu \rangle  
\label{eq_nuequatvacuum}
\feq
from which one calculates:
\beq
i {d \over dx} \left( \matrix { a_{\textstyle e} \cr
a_{\textstyle \mu} \cr} \right) = {1 \over 4E} \left ( \matrix{
- \Delta m^2 \cos 2 \theta_{V}~~~~~~~\Delta m^2\sin 2\theta_{V} \cr 
\Delta m^2\sin 2 \theta_{V} ~~~~~~~  \Delta m^2
\cos 2\theta_{V} \cr} \right) \left( \matrix {
a_{\textstyle e} \cr
a_{\textstyle \mu} \cr} \right)  
\label{eq_nuvacuum}
\feq
where $\theta_V$ is the mixing angle in vacuum and x = t, that is, set c = 1.\\  
Under usual approximations, a "Schr\"{o}dinger-like" time-dependent 
solution can be found, the Hamiltonian of the system being the kinetic 
term. \\
The probability to detect a different flavour is computable: the transition 
depends on distance, on momentum and on the square mass differences of the 
physical neutrinos. The appearance probability at a distance L is: 
\beq 
P(\nu_{1,2}) = \sin^2 2\theta_V \sin^2 \left (1.27 \Delta m^2(eV)^2
\frac{L(m)}{E(MeV)} \right)  
\label{eq_probappear2vac}
\feq 
where $\Delta m^2\,= |m^2_2 - m^2_1| >$0. The survival probability is complementary to the 
unity value. \\
The first oscillation maximum occurs when L/E $\sim \Delta m^{-2}$ while the term 
$\sin^2 2\theta$ (the oscillation probability amplitude) has a maximum when $\theta_V$ = $\pi$/4.
The time a $\nu$ has to transform its flavour (state) is proportional to $L/E$, where L is the 
distance between the $\nu$ source and the observer. The ratio E/L varies from $\sim 10^{-11}$  
for $\nu_\odot$'s up to $\sim$ 1 for high energy $\nu$'s from accelerators.  
The Sun-Earth distance (d=1.496 $\cdot 10^{11}$ m = 
7.58$\cdot 10^{23}$ MeV)  gives a lower bound on mass difference. \\
The evolution equation changes in the presence of matter to:
\beqar
i {d \over dx} \left( \matrix { a_{\textstyle e} \cr
a_{\textstyle \mu} \cr} \right) = \nonumber \\
{1 \over 4E} \left ( \matrix{
2E \sqrt2 G_F n_e(x) - \Delta m^2 \cos 2 \theta_{V}~~~~~~
\Delta m^2\sin 2\theta_{V} \cr 
\Delta m^2\sin 2 \theta_{V} ~~~~~~ -2E \sqrt2 G_F n_e(x) +
\Delta m^2 \cos 2\theta_{V} \cr} \right) 
\left( \matrix {a_{\textstyle e} \cr
		    a_{\textstyle \mu} \cr} \right) 
\label{eq_numsw}
\feqar
where G$_F$ is the weak coupling constant and $n_e$(x) the electron density (either solar or terrestrial).
The term $2 E \sqrt2 G_F n_e(x)$ represents the effective 
contribution to $M^2_\nu$  that arises from neutrino-electron scattering. The indices of refraction
of electron and muon neutrinos differ because the former scatter by CC and NC interactions, while the latter 
have only NC interactions.  The difference in the forward scattering amplitudes determines the density-dependent
splitting of the diagonal elements of the new matter equation. \\
The new Hamiltonian is not diagonal in the mass basis: a complexe 
treatment to diagonalize the operator and to compute the survival probability and the time evolution has to be applied, 
\cite{BAR80,OHL00,BAR01a}, but matter steady eigenstates do not coincide with the vacuum ones.
Sometimes, analytical solutions are possible; as example when $n_e$ is constant, the mixing angle is: 
\beq 
\tan 2 \theta_{M}\,=\,\tan 2 \theta_V \left(1\,-\,\frac{L_V}{L_e}  \frac{1} {\cos 2 \theta_V} \right)^{-1} 
\label{eq_thetavac}
\feq 
where the oscillation length in vacuum $L_V$ is: 
\beq 
L_V\,=\,\frac {4\pi E}{\Delta\,m^2} 
\label{eq_loscillvac}
\feq 
and the $\nu_e$ interaction leght is: 
\beq 
L_e\,=\,\frac {\sqrt {2} \,\pi} {G_F\,n_e} 
\label{eq_lnueinter}
\feq 
Independently of the smallness of $\theta_V$, $\theta_M$ gives a maximal mixing  
when a resonance condition is present: 
\beq 
\cos 2\theta_{V}\,=\, \frac {L_{V}}{L_e} 
\label{eq_thetavacmaxmix}
\feq 
The mass difference has to be positive; moreover, either $\nu$'s or $\overline \nu$'s can show this resonance, but not 
at the same time. Maximum mixing corresponds to the maximum mixing angle only in vacuum within 
a two flavour oscillation description: this conclusion is not valid when matter is present or 
within a more than two flavour analysis.\\ 
The appearance probability in matter becomes: 
\beq 
P(\nu_1\,\rightarrow \nu_2)\,=\,\sin^2\,2\theta_M\,\sin^2 \frac {\pi\,L_{V}}{L_{M(E)}} 
\label{eq_probappear2msw}
\feq  
where the energy dependent matter oscillation length  $L_{M(E)}$ is: 
\beq 
L_{M(E)}\,=\,L_{V} \cdot \left( 1\,-\,2\frac {L_V}{L_e}\,\cos 2\theta_{V} \,+ \, 
\left(\frac {L_V}{L_e} \right)^2 \right)^{-1/2} 
\label{eq_loscillmsw}
\feq 
The probability of flavour mixing  shows a typical resonance behaviour  and assumes the maximum 
value at the resonance energy.\\ 
If the density of the medium is variable, as in solar case, the evolution equation cannot be 
solved analitically with the exception of few peculiar cases. An adiabatic low varying density 
is an interesting situation: when an outgoing $\nu_e$ crosses regions with slow decreasing 
density $\theta_M$ can decrease and reach the resonance condition. A further  propagation  
toward  lower density regions leads to the "transition" to a vacuum value $\theta_{V}$: if 
this value is small, then $\nu_e$ is near its lowest energy state and even the flavour mixing 
is small.\\ 
The probability of converting a $\nu_e$ to a different flavour in the Sun is therefore strictly  
related to the electron density profile, in fact $\nu_\odot$ can resonantly oscillate only when 
peculiar values of electron density occur.\\
The computations become much more complex in three flavour oscillation approach: two 
oscillations can occur with different flight times (or frequencies). The probability to observe a 
a different neutrino flavour has a structure similar to eq. \ref{eq_probappear2vac} but the 
amplitude depends on two mixing angles. In this case, the interval [0,$\pi$/2] has to be considered 
and the description in terms of $\tan^2 \theta$ is needed. The unitary
mixing matrix is written as 
a product of three rotation: $\theta_{12}$ is connected to the solar data, $\theta_{23}$ is due
to the atmospheric sector, $\theta_{13}$ is presently constrained by measurements at reactors.
A further term is related to CP sector. In this description the matrix is:
\beqar
\left ( \matrix{
c_{12}c_{13} ~~~~~~~~~~~~~~~~~~~~~~~~~~~ s_{12}c_{13} ~~~~~~~~~~~~~~~~~~~~~~~~ s_{13}e^{-i\delta} \cr 
-s_{12}c_{23} - c_{12}s_{23}s_{13}e^{i\delta} ~~~~~ c_{12}c_{23} - s_{12}s_{23}s_{13}e^{i\delta}  
~~~~~~~ s_{23}c_{13}\cr
s_{12}s_{23} - c_{12}c_{23}s_{13}e^{i\delta} ~~~~~~~-c_{12}s_{23} - s_{12}c_{23}s_{13}e^{i\delta}
~~~~ c_{23}c_{13} \cr 
} \right) 
\label{eq_matrix3}
\feqar
where c$_{ij}$ or s$_{ij}$ are the periodic functions of angles $\theta_{ij}$ ($i$,$j$=1,2,3)
while $\delta$ is a phase related to CP.\\
In this case the survival probability is:
\beq
P^{(3)}(\nu_e \rightarrow \nu_e) = cos^4 \theta_{13} P^{(2)}(\nu_e \rightarrow \nu_e) + 
sin^4 \theta_{13}
\label{eq_probsurv3}
\feq
where $P^{(2)}(\nu_e \rightarrow \nu_e)$ is the survival probability within a two flavour
mixing description.\\
A realistic flavour oscillation picture foresees that $\nu_e$'s are produced in the solar core, 
then they can oscillate into active $\nu$ and/or sterile $\nu$'s. If this process happens 
before $\nu_e$'s reach the Earth, these $\nu$'s are undetectable in experiments based on 
captures on nuclei (Cl and Ga). Only SNO can presently identify $\nu_\odot$ 
component other than electronic one.\\ 
In 1990, the use of the luminosity constraint as a test independent of SMs of the  
hypothesis of $\nu_\odot$ flavour oscillations was proposed, \cite{SPI90},(many papers described 
the $\nu_\odot$  spectrum as sum of 3 terms, low-intermediate-high energy $\nu_\odot$'s, 
\cite{DAR91,HAT94a,CAS94,BER94,FOG95,PAR95,BAH96b,HEE96}). \\
Different patterns were considered in $\nu$ oscillation analysis. 
\bit 
\item  
{\bf{ Vacuum oscillations (VAC)}}.  As suggested by Pontecorvo, \cite{PON67}, the  $\nu$ 
oscillation length is the Sun-Earth distance and the $\nu$ survival probability is energy 
dependent. The $\nu_\odot$ flux is differently suppressed due to the possibility of a 
fine-tuning. A distortion of the energy spectrum is expected. A strong seasonal modulation 
should be present  because of the variation of the distance between the Sun and Earth, 
\cite{BER95}. Concerning the D-N asymmetry, VAC solution at E$\geq$ 2 MeV has rapid 
oscillations but the detectability of fluctuations both at high and low energy is very 
difficult due to the rough resolution in energy the detectors presently have.\\
In the case of the \(^7\)Be flux a variation of  10-30 $\%$ is foreseen while a reduction 
lesser than 10 \% should occur in Ga and Cl experiments. The suggested mass difference 
values are in the region $\Delta m^2 \approx 10^{-10}$(eV)$^2$.  
\item 
{\bf{ MSW solution}}. Flavour oscillations can be resonantly enhanced by "matter effect",  or 
Mikheyev-Smirnov-Wolfenstein (MSW) effect, \cite{WOL78,WOL79,MIK85,MIK86}. The (even complete) 
flavour conversion can occur through coherent $\nu - e^-$  scattering at the very high electron 
density prevailing in the solar interior and the resonance condition is:   
\beq 
\frac {n_{e,res}}{N_{Av}} \approx 66 \cos 2\theta_V  \left(\frac {\Delta m^2} {10^{-4}\,eV} 
\right) \left(\frac {10\,MeV}{E} \right) 
\label{eq_resonancecondit}
\feq 
where ${n_{e,res}}$ is the electron density at resonance in cm$^{-3}$, $N_{Av}$ is the  
Avogadro number, ${\Delta\,m^2}$ is absolute value of the difference between the masses of 
two different flavours, $E$ is the neutrino energy. In fig. \ref{fig:COUeldens} the electron 
density profiles as computed in \cite{COU02} is shown. 
\\
If right handed $\nu$'s are taken into account, another density has to be considered: 
\beq 
n_{ster} \,=\,n_e - 0.5 n_n \,=\,n_e \left( \frac {1+3X} {2(1+X)} \right) 
\label{eq_nusterile}
\feq 
where $n_n$ is the neutron density and X is the H fraction. In fig. \ref{fig:COUprofneutron} the ratio between 
electron and neutron density as a function of the distance from the solar center is shown.
\\
For a detailed analysis of survival probability computation in different solutions see 
{\it{e.g.}} \cite{BAH01d}.\\ 
Experimental results suggested different sets of MSW parameters which can account for the  
$\nu_\odot$ fluxes, energy spectra and D-N asymmetry: {\it{large mixing angle}} (LMA),  
{\it{small mixing angle}} (SMA), {\it{low mass}} (LOW). The area above 
$\Delta m^2 \approx 10^{-5}$eV$^2$ near the maximal mixing is called LMA solution; the SMA 
solution is located between $\Delta m^2 \approx 10^{-5} - 10^{-6}$eV$^2$ and $\tan^2\theta \approx 10^{-3}$. 
The LOW solution ranges in the mass region $\Delta m^2 \approx 10^{-7} - 10^{-9}$eV$^2$ while the lower part is called 
quasi-Vacuum (Q-VAC) solution.\\    
Depending on $\Delta m^2$, $\nu_\odot$ oscillations can be driven by the oscillation
phase or by matter effects. In the former strong spectral distortions are expected while in 
the latter case spectral distortions can still occur due to a resonance, which is energy 
dependent, caused by the density of the solar matter. The Earth's matter density can also
affect the conversion probability: in fact a "regeneration" of "disappeared" $\nu_e$'s during 
the night is possible: the flux during N-time (Earth crossing) and D-time (no Earth crossing)
should be different.\\ 
LMA solution predicts rapid oscillations in Earth and a rather flat distribution in zenith 
angle. 
LOW solution corresponds to the matter dominated regime of oscillations and predicts a structure of peaks in the 
zenith distribution, \cite{BAL88,GON01d}. The D-N asymmetry is detectable for the LMA
solution only at E$\geq$ 5 MeV; moreover, its value increases with E.  LOW solution shows the opposite behaviour.\\
LMA predicts a small reduction of the p-p $\nu_\odot$ flux, a strong depletion of the $^7$Be $\nu_\odot$ contribution 
and an energy independent reduction of the $^8$B $\nu_\odot$ component.\\ 
SMA solution requires an almost complete 
suppression of $^7$Be component,  a reduction of the $^8$B term with deformation on the  energy spectrum and no 
practically effect on p-p $\nu_\odot$'s, \cite{HAT94b}.\\
LMA and LOW solutions predict an increase of the survival probability at low energies (below the
SK and SNO thresholds). The LMA solution does not foresees the matter resonance inside the Sun
at low energy so that the probability increases up to the averaged vacuum oscillation value; on
the contrary in the LOW solution the survival probability raises its value because of the
Earth matter effect. These values are different so that experiments sensitive to low energy
$\nu_\odot$'s can check the right solution.  
The shape of the spectrum for LOW solution shows a weak positive slope. 
\item
{\bf{Just So$^2$.}} An early solution, called Just So, was proposed in 1987 to describe vacuum 
oscillations in vacuum with a mass value, $\sim 10^{-11}$ (eV)$^2$, which  reduces the $^8$B contribution to Cl 
experiment; this is an $ad\,hoc$ solution, \cite{GLA87}. Recently, this approach has been renewed but new mass value of 
$\sim 10^{-12}$ (eV)$^2$  and a $^8$B $\nu_\odot$ flux at a level as great as half of the SMs predictions are required,
\cite{RAG95,KRA96,LIU97,BAH00b,FOG00}. 
\item 
{\bf{Resonance Spin Flavour Precession}} (RSFP). 
If lepton flavour is not conserved, $\nu$'s must have flavour-off-diagonal (transition) 
magnetic moments, which applies either to Dirac and Majorana $\nu$'s and trasversal magnetic 
field will produce a simultaneous rotation of $\nu$ spin and flavour, a spin-flavour 
precession which can be resonantly enhanced in matter in a very similar way to the MSW effect,
\cite{CIS71,SCH81b,SCH82,VOL86a,VOL86b,AKH88,AKH88a,LIM88,AKH95,LIM95,AKH01}.
For Majorana $\nu$'s the resulting $\overline\nu$ can still be detected in electron 
scattering experiments, while in the Dirac case the final state remains undetectable. 
The conversion mechanism is dependent on $\nu$ energy. RSFP requires either 
$B_\odot \sim 100$ kG and $\mu_\nu \sim 10^{-11}\mu_B$, a value not experimentally 
excluded but hard to achieve in a simplest extension of the standard particle model. 
The available information on B$_\odot$ is presently limited; a large field in the 
convective zone may not be possible, since it would show up as an 11 year cycle in the 
SK data which is known not to be the case. Instead a large B$_\odot$ in the lower 
radiative zone and the core where most $\nu_\odot$'s are produced has been proposed. 
It remains unclear however whether the sunspots cycle effect extends down to the bottom 
of the convective zone. Altogether, radiative zone and core field profiles on one hand 
and convective zone ones on the other are equally favoured by the data: a strong central 
B$_\odot$ with a rapid decrease thereafter is preferred. The shape of these profiles 
follows a dipole structure centered at the solar centre and closely resembles the solar 
density profile. Furthermore RSFP provides a close relationship between the energy shape 
of the survival probability and B$_\odot$ profile: the most suppressed $\nu_\odot$'s 
have their resonance located in the region where B$_\odot$ is the strongest.   
Specific time signatures of the RSFP mechanism may be related with the 
possible non-axially symmetric character of B$_\odot$ or the inclination of the
Earth's orbit. In the first case a time dependence would appear as a variation 
of the event rate with a period of 28 days, while in the second the possible
polar angle dependence of B$_\odot$ would cause a seasonal variation of the rate. 
Since only the product of $\mu_\nu$ and B$_\odot$ enters in the $\nu$ evolution equation, 
the analysis can be applied to 
any other value of ${\mu_{\nu}}$ provided that B$_\odot$ is rescaled accordingly. \\
It is assumed that $\nu$'s have Majorana-like transition magnetic moments $\mu_\nu$ which cause the 
transitions $\nu_{eL}\to \bar{\nu}_{\mu R}$ or $\nu_{eL}\to \bar{\nu}_{\tau R}$ in the solar magnetic 
field, (the Majorana $\nu$ gives a better fit of $\nu_\odot$ data than the Dirac $\nu$ does). 
The transition probability depends crucially on the shape and on the B$_\odot$ strength  which 
are essentially unknown. Many B$_\odot$ profiles were proposed but the most part of them 
in general allows marginal fits of experimental data: at present only few profiles enables 
fits in agreement with experiments; we mention profiles 1 and 6 from \cite{PUL00}, 
and profile 4 from \cite{PUL01}. 
The RSFP solution of the SNP is hard to establish experimentally. Except for predicting reduced $\nu_\odot$ flux, it 
has negative signatures: no time variations beyond the usual d$^{-2}$ variation due to the eccentricity of the 
Earth's orbit (B$_\odot$ does not vary with time); no D-N effect; no significant distortions of 
the energy spectrum.
In \cite{MIR01} it has been shown that even a non-resonant SFP  should be able to reproduce the experimental data.
\item 
Non-standard interactions inducing $\nu$ oscillations were suggested, 
\cite{VAL87,GUZ91,ROU91,BAR91,BER00,GUZ01}. The Hamiltonian in evolutionary equations shows the usual structure: 
the diagonal term depends either on fermion and electron density radial profiles and on a phenomenological parameter 
characterizing the 
strength of the $\nu$ interaction. The off-diagonal term, which has the same action of the  mixing term in standard 
MSW solution, is strictly related to the fermion density and to a second  parameter responsible for flavour changing. 
In this approach the conversion probability  is energy independent but it is possible to reproduce the 
detected  spectra because the production distributions for $\nu_\odot$'s are different so  that a resonant conversion 
does occur in solar interior. 
\item 
The violation of the weak equivalence principle (VEP) of the general relativity 
has  been proposed as mechanism inducing flavour oscillation even if $\nu$'s are massless, 
\cite{GAS88,HAL91,PAN93,GAG00a,GAG00b,GAG01,NUN01b}. In this case $\nu$ mixing and flavour 
oscillation  are due to different gravitational interactions so that a violation to the general  
relativity is needed. Weak and gravitational interacting eigenstates differ so that a mixing 
angle $\theta_G$ is introduced as in the $\nu$ mixing in vacuum due to the mass. 
The evolution equations, describing degenerate mass propagating through gravitational 
potential, show similar components both in diagonal and off-diagonal terms: they are depending on 
the energy, on the difference of the gravitational interaction and on the  gravitational 
potential. The mixing term, which depends on $\sin 2\theta_G$, is linearly related to the energy 
so that the oscillation length is inversely proportional to the $\nu$ energy. 
Two different mechanisms are possible: VEP resonant conversion (a MSW-like process), which badly 
reproduces experimental data, \cite{PAN00}, and  VEP vacuum conversion which better agrees 
with measurements, \cite{GAG00b,GAG01,RAY02}. 
\item 
Other suggested mechanisms foresee neutrino decay, \cite{CHO01b,CHO01c,CHO02}, and CP and T  
violation, \cite{KUO87,DER99,DIC99,ARA97,BIL98}. In the latter only appearance experiments  
should be able to observe this effect which is beyond the presently available technical  
capabilities. Even neutrino decay in vacuum followed by an oscillatory $scenario$ was 
proposed, \cite{LIN01}, but present experimental results do not support this suggestion. In 
\cite{BAR00} correlations between $\nu$ properties and extra dimensions have been analysed.   
\fit 

\subsection{The pre-SNO situation.} 
\label{sect:presnosit}
Complete analyses of SNP in terms of $\nu$ flavour mixing were done by many authors, 
\cite{FOG00,GIU00,GON01a,GON01b,GON01d,GON01f,SCH01} and quoted references.
Let us underline that slightly different assumptions can be adopted and consequently results 
are similar but not equal. 
Ga experimental results can be separately inserted instead of their average value; 
KAMIOKANDE results are usually excluded; energy spectra from SK are 
sometimes added to SK flux but these values are not independent from the SK rate; 
D-N or zenithal energy spectra are included, sometimes the constraints given by CHOOZ 
experiment are added. \\
Let us remember that the total event rate gives information on average oscillation probability
while the energy spectrum specifies the dependence of the probability from energy, oscillation
length and time.
If only fluxes and interaction rates from different experiments are used the "RATES" analysis is given while
"GLOBAL SOLUTION" occurs when all parameters concerning $\nu_\odot$ flux are included. 
SK experiment gives a dominant contribution due to its large, statistics 
and zenith angle distribution combined with energy spectra.

\subsubsection{Two flavour analysis.} 
\label{sect:twoflavours}
\begin{figure}[ht]
\begin{center} 
\mbox{\epsfig{file=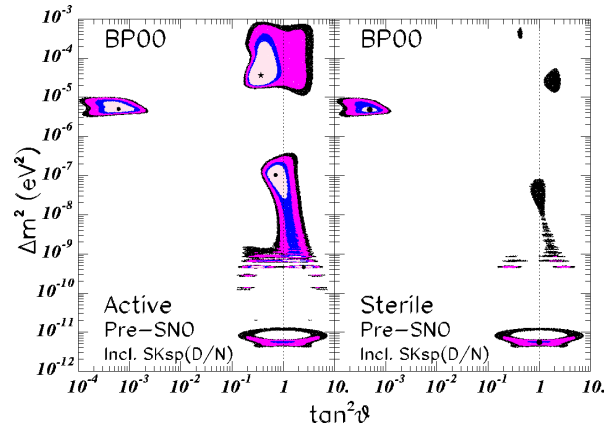,width=0.5\textwidth}} 
\end{center}
\caption{ GLOBAL analysis solutions at the beginning of 2001: input data include the total rates from the
Cl and Ga averaged experiments, the recoil electron D and N energy spectrum measured by SK. 
The C.L. contours are $90$\%, $95$\%, $99$\%, and $99.73$\% (3$\sigma$). The allowed regions are limited by
the CHOOZ reactor measurements, $\sim$ 7-8 $\cdot 10^{-4}$ (eV)$^2$. The local best-fit points
are marked by dark circles. The theoretical errors for $\nu_\odot$ fluxes from \cite{BAH01a} are included, 
from \cite{BAH01c}. } 
\label{fig:BAH258presno}
\end{figure}

\begin{figure}[ht]
\begin{center} 
\mbox{\epsfig{file=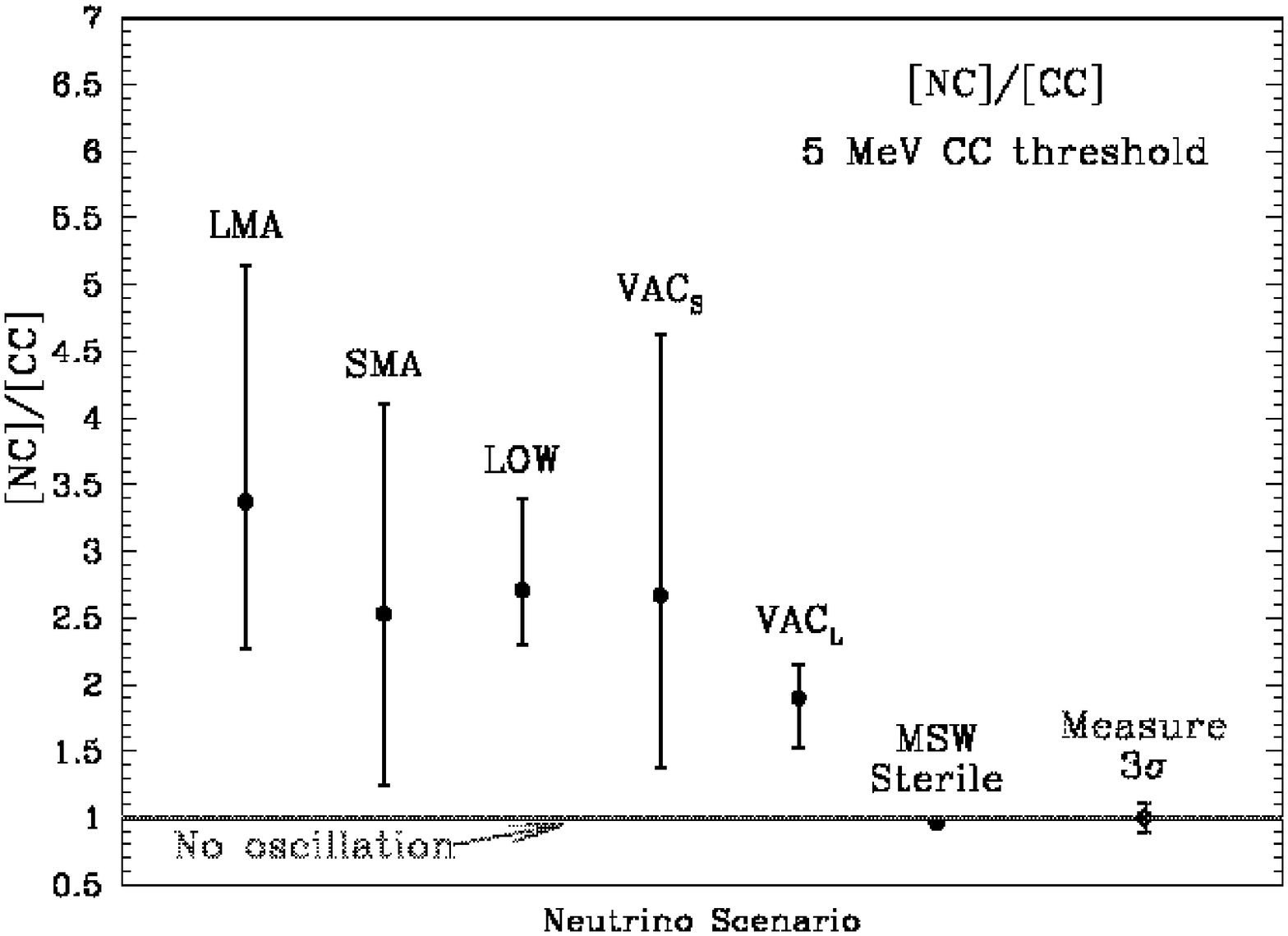,angle=-90.,width=0.5\textwidth}} 
\end{center}
\caption{ The NC/CC double ratio predictions in SNO at E$\geq$ 5 MeV for different
oscillation solutions (the standard model foresees 1.0). 
The solid error bars represent the $99$\% C.L. for the allowed regions. The dashed error bar labeled 
"Measure 3$\sigma$" represents the uncertainty in interpreting the measurements, from \cite{BAH00a}. }
\label{fig:BAH293nccc}
\end{figure}

\begin{figure}[ht]
\begin{center} 
\mbox{\epsfig{file=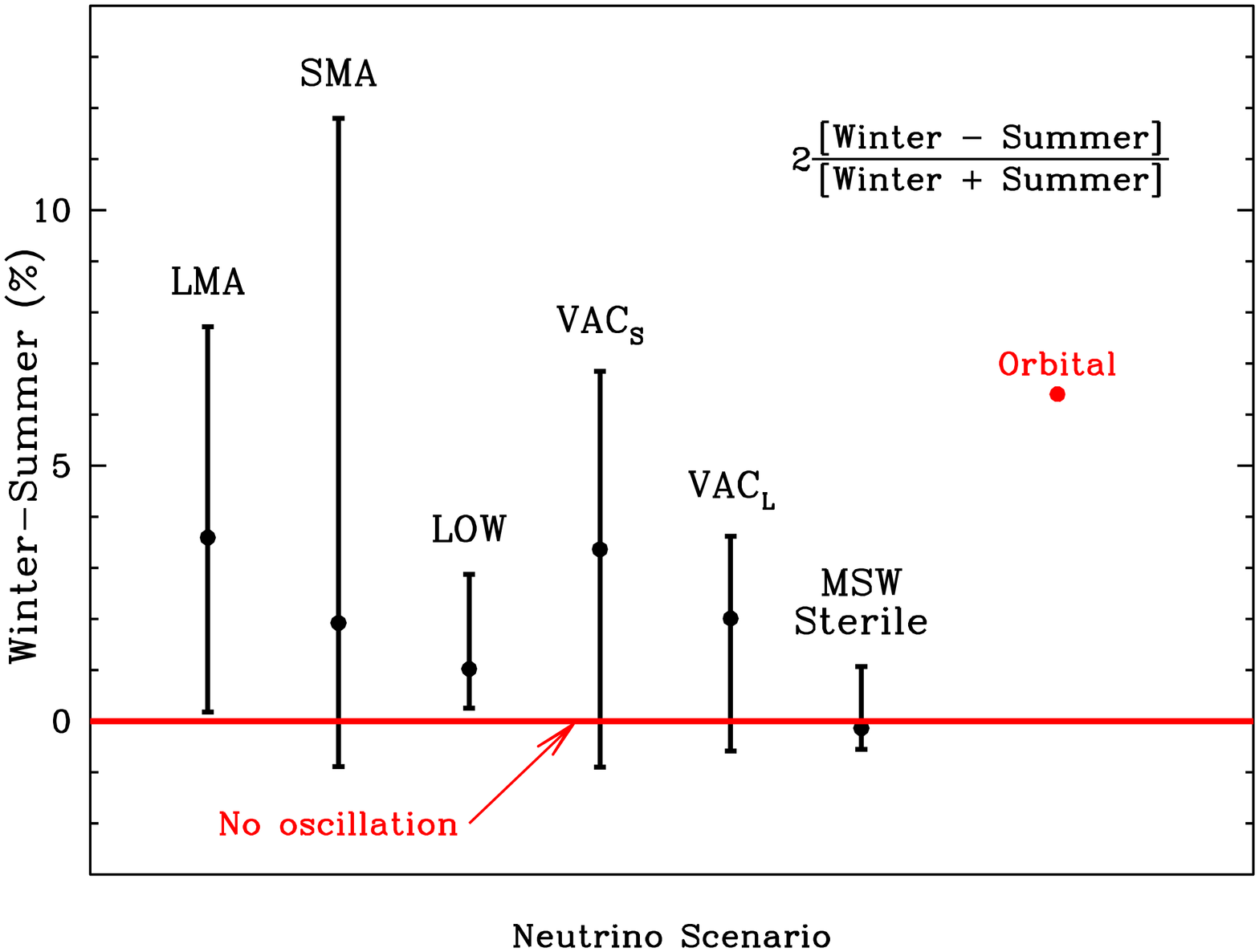,angle=-90.,width=0.5\textwidth}} 
\end{center}
\caption{ The figure shows for different oscillation solutions the difference (\%) between 
the predicted CC rate in SNO, at E$\geq$ 5 MeV, for a 45 day interval during winter and summer. 
"Orbital" represents the 45 day winter-summer difference due to the Earth's motion around the 
sun; the amplitude of this orbital motion has been removed from the neutrino oscillation 
points.  The error bars represent the $99$\% C.L. for the allowed regions, from \cite{BAH00a}.  }
\label{fig:BAH293winsum}
\end{figure}

In terms of 2 flavour mixing the standard procedure uses least-squares analysis so that  the authors calculated the 
allowed regions in ($\Delta m^2,\theta$) plane, see \cite{FOG96,BAH98c}. This is an approximation of rigorous 
frequentist methods, \cite{CRE01,GAR01,GAR02a,GAR02b}, which can offer results even different 
from the standard least-squares approach. We underline that several frequentist 
analyses are available so that it is not easy to select the right procedure.\\ 
In the case of active-active $\nu$ oscillation different regions are 
possible for the GLOBAL fit in ($\Delta m^2,\theta$) plot: SMA, LMA, LOW, VAC and Just So$^2$; 
solutions into sterile $\nu$'s are also allowed (SMA,VAC and Just So$^2$).  D-N asymmetry is small and the survival  
probability at higher energy, with the exception of VAC solution, assumes a practically flat 
dependence. The Earth regeneration  
effects are relevant only within the mass range $10^{-5}$ - $10^{-7}$ (eV)$^2$. 
An analytical approach, following \cite{MIK86,KAN87,TOS87,ITO88a,ITO88b,KRA88a,KRA88b,PET88}, is usually adopted.\\ 
At the beginning of 2001, when SK measurements based on 1258 days of data 
taking were available, the conclusions quoted in \cite{SMY00,SUZ01a} were:
\benu
\item   
The $^8$B spectrum, which does not show a strong suppression at large energies as expected in MSW adiabatic solutions, 
seems to be undisturbed, even if the uncertainties are large.  
The low-energy part of the spectrum plays a crucial role therefore a  lowering of the threshold down to 4.0 
MeV seems to be necessary (but impossible after the incident occurred in late 2001).  
\item 
SK D-N spectrum and the total flux define an enlarged LMA region; the LOW region extends down to the VAC one. 
LMA solution is the favoured one at 95$\%$ C.L. because of the flatness of energy distribution 
and of the smallness of the D-N effect which takes long time to  obtain positive evidence, 
if any. It is hard to distinguish between LMA and LOW  solutions. SMA, Just So solutions and 
Sterile neutrinos are disfavoured at 95$\%$.  
\item 
Measurements of NC interactions are needed. 
\fenu  
LMA, SMA and LOW solutions in the case of active $\nu$'s  and SMA sterile solution were 
estimated in \cite{BAH01b} as relatively robust solutions, while the VAC  solution was 
classified as fragile.  The Just So$^2$ 
solution, which is allowed both for active and sterile $\nu$'s, showed a relative robustness: 
in fact it did not well 
reproduce the Cl rate, otherwise it was in a very good agreement with Ga and SK data. \\ 
The results obtained in GLOBAL analysis before SNO data can be summarized in this way: 
\bit 
\item 
The LMA solution gave the best fit, mainly thanks to the constraints done by  SK data on D-N 
and the flatness of the energy spectrum. 
\item 
The LOW solution, which is connected with Q-VAC region at 99$\%$ CL and extends into the 
second octant, well described the spectrum data, but it was weaker than LMA because of the 
rates. 
\item 
The SMA solution showed worst results due to its difficulty in the spectrum analysis and
the conclusions were similar for sterile solutions.
\fit 
In practice all these solutions were allowed but LMA was slightly  preferred, 
see fig. \ref{fig:BAH258presno}.\\
We remember that the LMA solution was shown to be the best oscillation solution for the first 
time in \cite{GON00}.
\\
In fig. \ref{fig:BAH293nccc} the NC/CC ratio expected in SNO is shown while in fig. 
\ref{fig:BAH293winsum} the predictions concerning seasonal difference are drawn.
\\

\subsubsection{Three flavour analysis.} 
\label{sect:threeflavours}
A more general description is needed when atmospheric, reactor and solar neutrino results are 
analysed: oscillations occur among three flavours and the unitary matrix relating flavours and 
masses contains 2 mass differences, 3 mixing angles and 1 or 3 CP violating  phases, depending on 
Dirac or Majorana nature of $\nu$'s, \cite{FOG00,GON01a,GON01f}.  The CP phases produce effects 
not accessible to the present experiments so that their contributions are usually excluded from 
the results. The transition probability has 2  oscillation lengths and shows an oscillatory 
behaviour. Usually, a hierarchy  $\Delta m^2_{atm} \gg \Delta m^2_\odot$ is present. The analysis 
of $\nu_\odot$ data  constrains 3 parameters ($\Delta m^2$ and two mixing angles) while 
$\nu_{atm}$  results pose boundaries on the remaining mass difference and on 2 mixing angles.  
Only a parameter is common, the  $\theta_{13}$ mixing angle, which is strongly  constrained by 
CHOOZ reactor experiment, \cite{APO99}. Its final result suggests  $\theta_{13} <$ 15$^{\circ}$ 
at 99 $\%$ CL; furthermore, oscillations down to $\Delta\,m^2 \sim$ 7$\cdot 10^{-4}$ (eV)$^2$ for 
full mixing are excluded. As $\tan^2 \theta_{13}$ increases, all the allowed regions in the 
($\Delta m^2$,$\tan^2 \theta_{12}$) plot disappear so that an upper bound to $\tan^2 \theta_{13}$ 
is present. Moreover, no $\overline {\nu_e} \leftrightarrow \overline {\nu_\mu}$ oscillations 
were found by PALO VERDE reactor experiment; the ratio between observed and calculated rate is
$R\,=\,1.01 \pm 0.024(stat.) \pm 0.053 (syst.)$ for $\Delta\,m^2 \geq 1.1\cdot 10^{-3}$ (eV)$^2$ 
at full mixing and $\sin^2\,2\theta \geq$ 0.27 at  large $\Delta\,m^2$, \cite{BOE01}. Other 
parameters do not influence the mixing matrix elements. \\
The results can be summarized in this way:
\bit
\item
LMA is the preferred solution; SMA and LOW solutions are less favoured even if LOW is more 
acceptable and can provide maximal mixing ($\theta_{12}$= $\pi$/4) for non-zero $\theta_{13}$.
Some interesting perspectives survive for VAC solution.
\item
CHOOZ provides an upper limit on $\sin^2 \theta_{13}$ and $\Delta m^2$; solar experimental results
prefer $\sin^2 \theta_{13} \sim$0 in a good agreement with CHOOZ, but there it no reason for it
to be zero.
\item
Unambiguous selection of ONE solution will not be possible in SK data for several years. Much 
higher statistics is needed for instance in the spectrum.
\fit

\subsubsection{Four flavour analysis.}
\label{sect:fourflavours} 
\begin{figure}[ht]
\begin{center} 
\mbox{\epsfig{file=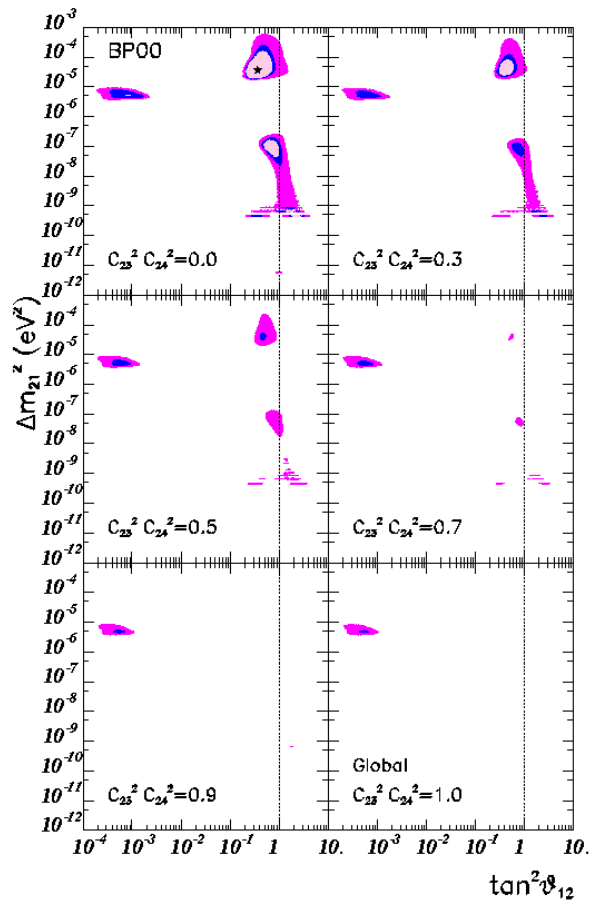,width=0.5\textwidth}} 
\end{center}
\caption{ Results of the GLOBAL analysis of $\nu_\odot$ data for the allowed regions in 
$\Delta{m}^2_{21}$ and $\tan^2 \theta_{12}$ for the 4 $\nu$ description. The different 
panels represent sections at a given value of the active--sterile admixture $|U_{s1}|^2 + 
|U_{s2}|^2 = c_{23}^2 c_{24}^2$ of the three-dimensional allowed regions at 90\%,
95\% and 99\% CL. The best-fit point in the three parameter space is
plotted as a star, from \cite{GON01e}. }
\label{fig:GG4glo}
\end{figure}

More complicated  analyses are needed when LSND results are included: $\nu$ fields are 
connected to 4 mass eigenstates $\it{via}$ a 4x4 unitary mixing matrix  containing 6 mixing 
angles (in any case the CP violating phases are disregarded), \cite{DOO00,GIU00,GON01e}.\\  
The reason usually adopted to introduce a new $\nu$ flavour is the appearance probability in 
LSND experiment: three oscillations are needed  to explain all the available experimental data.  
The presently detected $\nu$ flavours are three with two independent oscillations, therefore 
a new flavour has to be added. It is useful to remember that the invisible width of Z-boson 
as measured at  LEP poses a strong limit to weakly interacting neutrino number (3 flavours) 
so that the new neutrino is supposed to be "sterile", or without interactions with matter. 
In fact, if $\nu_4\simeq \nu_e$ then P($\nu_\odot$) $\simeq$ 1 but the $\nu_\odot$ flux deficit
is incompatible; if $\nu_4\simeq \nu_\mu$ then P($\nu_{atm}$) $\simeq$ 1 but atmospheric
results from SK, MACRO, Soudan rule out this solution; if $\nu_4\simeq \nu_\tau$ then 
for the atmospheric sector $\nu_\mu \rightarrow \nu_e + \nu_s$ with strong matter effect 
never detected. Therefore, $\nu_4 = \nu_s$ is the only viable solution.\\   
There are 6 schemes compatible with solar, atmospheric and LSND results: 
in the so  called "3+1" schemes 3 $\nu$ masses are similar, the remaining 
one is separated by a gap of $\sim$ 1 (eV)$^2$; otherwise "2+2" schemes with the 
same energy gap are foreseen. The analyses are strongly model dependent and the experimental data do not give 
sure indications on the right solution.\\ 
The main conclusions were, \cite{GON01e}, see also fig. \ref{fig:GG4glo}: 
\bit 
\item  
The analysis of $\nu_\odot$ data with 4 flavours was the same as in 2 flavour case but a 3-D parameter 
space was required. Physical space was described if  $\theta_{12}$,  $\theta_{23}$ and $\theta_{24}$ vary within 
the (0,$\pi$/2) interval, the last two  angles entering only by the 
combination $\cos^2 \theta_{23}\cos^2 \theta_{24}$. 
\item 
A simultaneous analysis of active-sterile and active-active solutions was allowed. 
\item 
LMA, LOW and Q-VAC solutions disappeared with increasing values of mixing parameter 
$\cos^2 \theta_{23}\cos^2 \theta_{24}$.
\item 
LMA solution was allowed at 95 $\%$ if $\cos^2 \theta_{23}\cos^2 \theta_{24}=0.5$;  
each solution was estimated possible at 99$\%$ for the maximal mixing case.   
\fit 

\subsection{2001: After SNO (I).} 
\label{sect:aftersno1}
\begin{figure}[ht]
\begin{center} 
\mbox{\epsfig{file=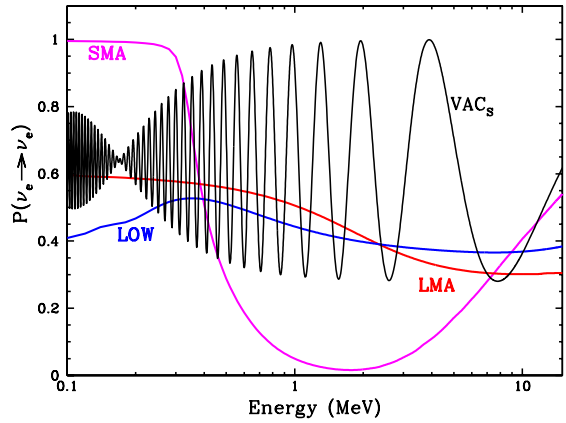,angle=-90.,width=0.5\textwidth}} 
\end{center}
\caption{ Comparison of survival probabilities for flavour oscillation solutions, 
a linear energy scale is used, from \cite{BAH00a}.  }.
\label{fig:BAH293surv}
\end{figure}

\begin{figure}[ht]
\begin{center} 
\mbox{\epsfig{file=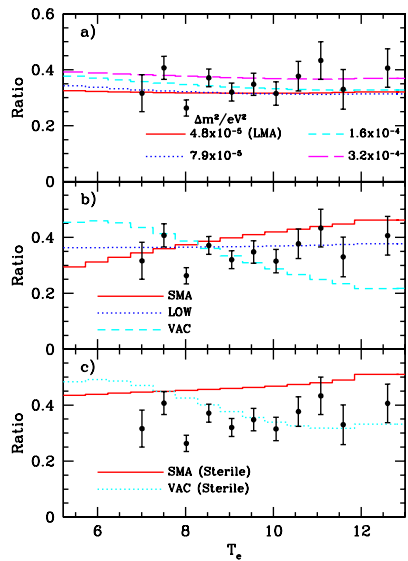,width=0.5\textwidth}} 
\end{center}
\caption{ The recoil electron energy spectra of the CC events in SNO for different GLOBAL solutions
compared with experimental data, from \cite{AHM01}. Shown are the ratio of the number of events 
with and without conversion as a function of the electron kinetic energy.  
In a) LMA solutions for different values of $\Delta m^2$ and $\tan^2 \theta = 0.35$;
in b) SMA, LOW and VAC (active) solutions; in c) SMA(sterile) and 
VAC(sterile) solutions, from \cite{KRA02}. }
\label{fig:KSSnospec}
\end{figure}

\begin{figure}[ht]
\begin{center} 
\mbox{\epsfig{file=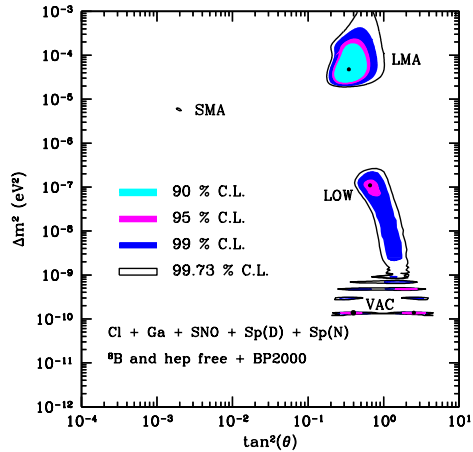,width=0.5\textwidth}} 
\end{center}
\caption{ GLOBAL analysis solutions with $^8$B and Hep $\nu_\odot$ fluxes as free parameters. 
The best fit points are marked by dark circles. The absolute minimum of the $\chi^2$ is 
in the LMA region. The allowed regions are shown at 90\%, 95\%, 99\% and 99.73\% C.L.  
with respect to the minimum in the LMA region, from \cite{KRA02}. }
\label{fig:KSB8hep}
\end{figure}

\begin{figure}[[ht]
\begin{center} 
\mbox{\epsfig{file=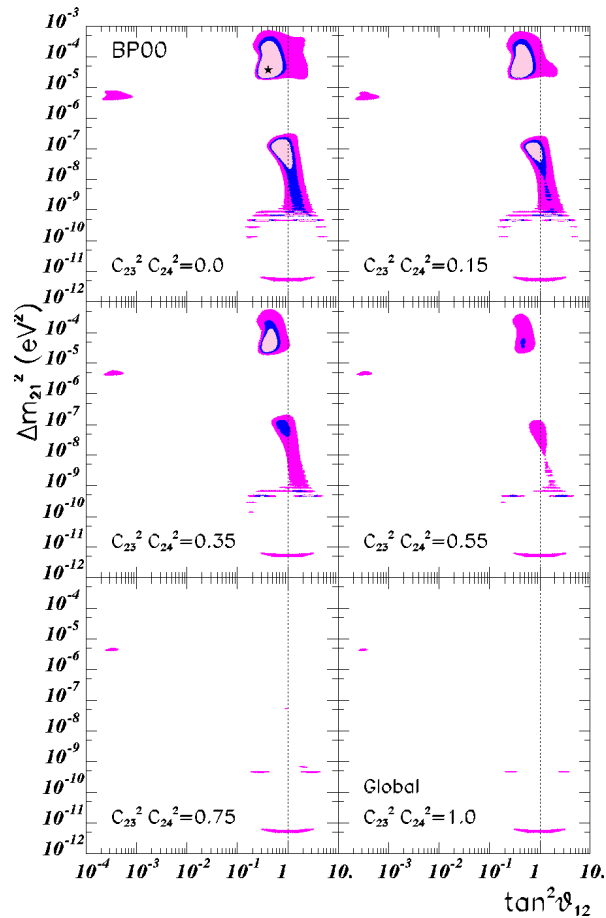,width=0.5\textwidth}} 
\end{center}
\caption{ GLOBAL analysis solutions with 4 $\nu$'s as in fig. \ref{fig:GG4glo} but SNO data, from \cite{AHM01}, 
are included, from \cite{GON01c}. }
\label{fig:GG4glosno}
\end{figure}

Since first results concerning CC and ES interactions in SNO were available, \cite{AHM01},  
and a common energy threshold for SK and SNO experiments was computed, new and more complete 
analyses became possible, \cite{BAN01,FOG01,BAH02c,BAN02a,BAR02a,FOG02a,FOG02b,KRA02}. \\
In summer 2001 the main conclusions were: 
\bit 
\item 
There is a deficit in $\nu_e$ flux with respect to the SMs predictions; moreover,  there is an evidence 
that a $\nu_e$ conversion into $\nu_{\mu,\tau}$ does occur. 
\item 
No astrophysical solution of SNP seemed to be viable. 
\item 
Solutions based on pure active-sterile conversion were strongly depressed. 
\item 
At $E\sim$ 6 MeV the $\nu_e$ survival probability is less than 0.5 (from \cite{AHM01} one finds 
P = 0.334 $\pm$ 0.22 in the case of pure active transition). An even smaller value was found
when sterile components were included.  
\fit 
The combination of SNO and SK results suggested that the $\nu_\odot$ flux depletion is due to the 
particle physics sector or, in other words, not only $\nu_{\odot,e}$'s interact inside terrestrial detectors. 
It seemed therefore (highly) probable that $\nu_e$'s produced in solar core undergo flavour conversion, but with 
an unknown mechanism. \\
The most part of published papers searched for an answer to the SNP within a flavour oscillation solutions. 
Let us underline again that if the same input data and analysis are used, different authors obtain essentially identical
results, but when different strategies are adopted the predictions are not fully coincident, see for instance 
\cite{BAH02c,KRA02} for a comparison. 
The most restricted regions are obtained when all available rates are combined with SK D-N energy 
spectra, while the largest areas are present if $^8$B $\nu_\odot$ flux (or even Hep component) is treated as a free 
parameter, while SK rates are excluded because of D-N energy spectra.\\
We show in table \ref{tabpresnoks} the best-fit GLOBAL solution parameters as done in 
\cite{BAH02c,KRA02}; details concerning the strategy adopted in the analysis are shown 
in captions. \\
\begin{table}[ht]
\caption{\it{ Best-fit GLOBAL analysis oscillation parameters with all solar neutrino data.   
Best-fit values of the parameters $\Delta m^2$, $\tan^2\theta$;  $f_B$ and $f_{Hep}$ 
are the factors multiplying the $^8$B and Hep $\nu_\odot$ fluxes as given in \cite{BAH01a}.  
(s) in the first column means sterile solution. 
The first 5 columns show results quoted in \cite{KRA02}. 
The number of degrees of freedom is 38:  4 rates (Homestake, SAGE, GALLEX+GNO, SNO)  
+ 38 SK spectra points  - 4 parameters.
Last 3 columns are concerning results from \cite{BAH02c}. The number of degrees of freedom 
is 39 [38 SK spectra + 3 rates - 2 parameters ($\Delta\,m^2$,$\theta$)]. 
The  best-fit fluxes and their estimated errors from \cite{BAH01a,BAH02c} are  
included. Interaction rates from the GALLEX+GNO and SAGE experiments provide a unique data 
point. } }
\begin{center} 
\begin{tabular}{|l|c|c|c|c|c||c|c|c|}
\hline  
 & $\Delta m^2$(eV)$^2$ &    $\tan^2\theta$ & $f_B$ & $f_{Hep}$ & $\chi^2_{min}$  
 &$\Delta m^2$(eV)$^2$&$\tan^2 \theta$& $\chi^2_{min}$  \\ \hline 
LMA      & $4.8\cdot 10^{-5}$ &  0.35   & 1.12 & 4 &   29.2  
&$3.7\cdot10^{-5}$  &$0.37$ & 34.5 \\\hline 
VAC      & $1.4\cdot 10^{-10}$ & 0.40 (2.5)  & 0.53 & 6 & 32.0 
& $4.6\cdot10^{-10}$ &$2.5$ & 42.3  \\\hline 
LOW      & 1.1$\cdot 10^{-7}$ &   0.66  & 0.88  & 2   & 34.3 
& $1.0\cdot10^{-7}$  &$0.67$ & 40.8 \\\hline 
SMA      & $6.0\cdot 10^{-6}$ & 0.0019 & 1.12  & 4 & 40.9  
& $5.2\cdot10^{-6}$  &$0.0018$ & 49.9  \\\hline 
J.So$^2$ & $5.5\cdot 10^{-12}$ & 1.0 & 0.44 & 0 &   45.8  
& $5.5\cdot10^{-12}$ &\hskip -6pt$0.61  $ & 52.1  \\\hline\hline\hline 
VAC(s) & $1.4\cdot 10^{-10}$ & 0.38 (2.6) & 0.54 & 9 & 35.1 
& $4.7\cdot10^{-10}$ & $3.0$ & 49.1 \\\hline 
J.So$^2$(s) & $5.5\cdot 10^{-12}$ & 1.0 & 0.44 & 0 & 46.2  
& $5.5\cdot10^{-12}$  &\hskip -6pt$0.61  $ & 52.1 \\\hline 
SMA(s) & $3.8\cdot 10^{-6}$ &  0.00042 & 0.52 & 0.2 &48.2  
& $4.6\cdot10^{-6}$ & 0.00034 & 52.3  \\\hline 
LMA(s) & $1.0\cdot 10^{-4}$ & 0.33 & 1.14 & 0 & 49.0  
&-- &-- &-- \\  \hline 
LOW(s) & $2.0\cdot 10^{-8}$ & 1.05  & 0.83 & 0 & 49.2  
&-- &-- &-- \\\hline 
\end{tabular} 
\end{center}  
\label{tabpresnoks}
\end{table}
\\
The main conclusions the GLOBAL analyses put in evidence were, see \cite{KRA02}: 
\bit 
\item 
The LMA solution foresaw as ratio between CC interactions rate and the value predicted in 
\cite{BAH01a} R$_{CC}$ = 0.20-0.41; the best theoretical values was slightly lower than the SNO mean value, 1 $\sigma$.  
This solution reproduced the physical observables at a level of (or better than) 1 $\sigma$, 
the largest deviation occurring in Cl flux calculations, 1.4 $\sigma$  higher. SNO measurements 
shifted the LMA solution 
region to higher mixing angle values where  the survival probability is greater. A slightly higher $^8$B 
$\nu_\odot$ flux was allowed (by a factor of 1.1). 
\item 
The LOW solution predicted R$_{CC}$ = 0.36-0.42, 2 ${\sigma}$  above the SNO result. 
There was a shift of allowed region toward smaller mixing angles corresponding to smaller survival probability. 
\item 
R$_{CC}$ = 0.33-0.42 was the estimated value for the VAC solution, a 1 $\sigma$ higher mean 
value while VAC oscillations to $\nu_s$ indicated R$_{CC}$= 0.36-0.41.
\item 
Remaining solutions, in particular SMA, were less probable because of their high R$_{CC}$ 
value (0.37-0.50). Moreover, the suppression showed an energy dependence of  opposite sign with respect 
to the SNO data. The Just-So$^2$ (active), SMA (sterile) and Just-So$^2$ (sterile) were strongly depressed 
because they foresee the same rate in SK as in SNO. 
\fit 
The SK and SNO results were well within the region where an oscillation from $\nu_e$ to 
$\nu_{\mu,\tau}$ does occur, independently from any sterile component, \cite{FOG01,FOG02a}. 
The $\nu_e$ survival probability took the lowest value allowed by pre-SNO results. 
Assuming a purely active oscillation, all the main predictions 
given by SMs were confirmed and the $\nu_e$ survival probability was found to be $\sim$ 1/3 
in the SK-SNO energy range, see fig. \ref{fig:BAH293surv}. 
\\
The allowed solutions before SNO data, see fig. \ref{fig:BAH258presno},
were still present at level greater than 3 $\sigma$; the best-fit values are in LMA region 
but remaining solutions (LOW, VAC, Q-VAC) were not excluded. In fig. \ref{fig:KSSnospec} 
experimental energy spectrum is fitted by using different oscillation solutions.
\\
The LMA region was enlarged: SNO results suggested relatively small values of $\nu_e$ survival 
probability consequently LMA solution was favoured in the GLOBAL fit. The SMA solution 
vanished at level greater  than 3 $\sigma$ when SK D-N spectra were included in the analyses. 
The LOW solution was less favoured than the LMA due to Ga experiments which supported 
an increase of the $\nu_e$  survival probability at low energy. \\
SNO results did not modify qualitatively the solution for $\nu_\odot$'s but strengthened the 
case for active oscillations with LMA, see fig. \ref{fig:KSB8hep}.
\\
Many authors discussed the possibility of bi-maximal neutrino oscillations ($\tan^2 \theta$ = 1), 
\cite{MIN97,BAR98,XIN00,CHO01a,GHO01}: this option was not favoured by  LMA solution at a level ranging 
from 98.95\% C.L. to 99.95\% C.L., depending on the adopted strategy.\\ 
When the temperature scaling of the nuclear reactions giving rise to
$^7$Be and $^8$B  $\nu_\odot$'s was imposed as an external condition on the fitting procedure, the assumption  
of no-oscillation is excluded at 7.4 $\sigma$, \cite{BAH01c} see also \cite{BAH02a,BAH02b}. \\
We underline that in 4 flavour oscillation analysis, \cite{GON01c}, the minimum was found to be 
in the LMA region, with $\cos^2 \theta_{23}\cos^2 \theta_{24}$= 0., $\tan^2 \theta_{21}$= 0.41,  
$\Delta m^2_{21}$=$4.1\cdot 10^{-5}(eV)^2$, see also fig. \ref{fig:GG4glosno}.  

\subsubsection{Waiting for Neutral Current results.} 
\label{sect:waitingnc}
\begin{figure}[ht]
\begin{center} 
\mbox{\epsfig{file=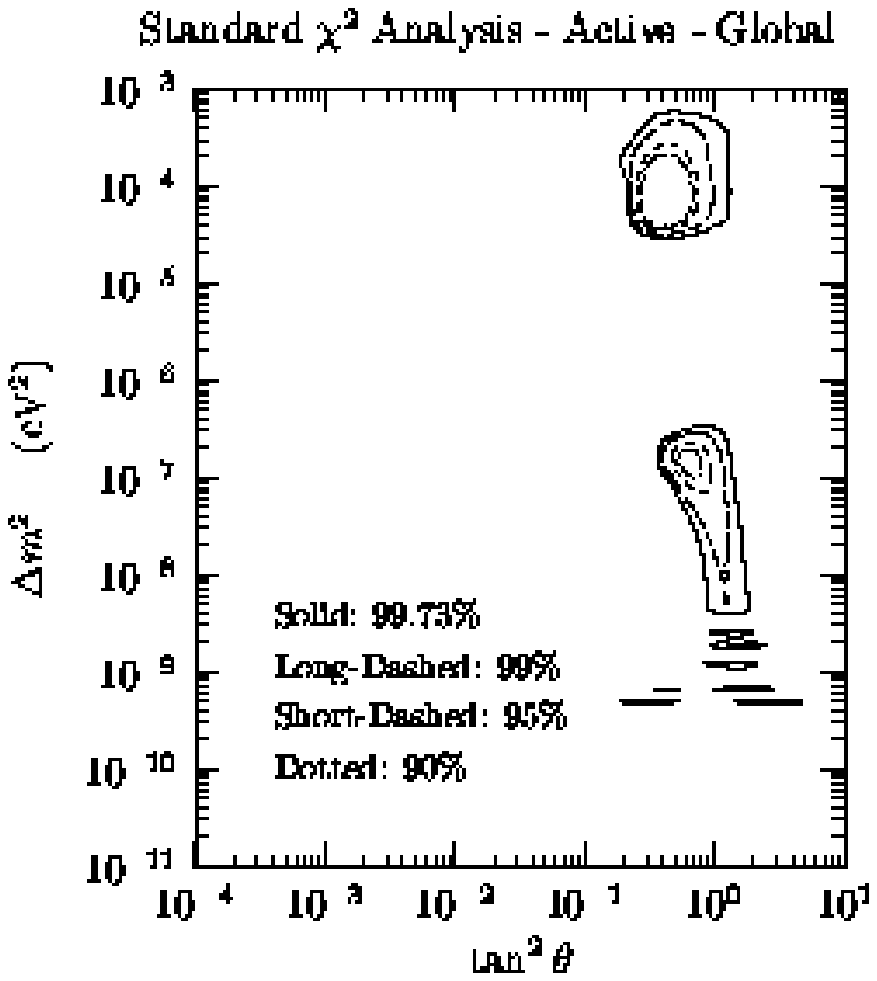,width=0.7\textwidth}} 
\end{center}
\caption{ Allowed regions for active transitions in standard analyses including Cl, Ga averaged, 
SNO rates, SK D and N electron energy spectra and CHOOZ result, from \cite{GAR01}.    }
\label{fig:GIUglochi}
\end{figure}

\begin{figure}[ht]
\begin{center} 
\mbox{\epsfig{file=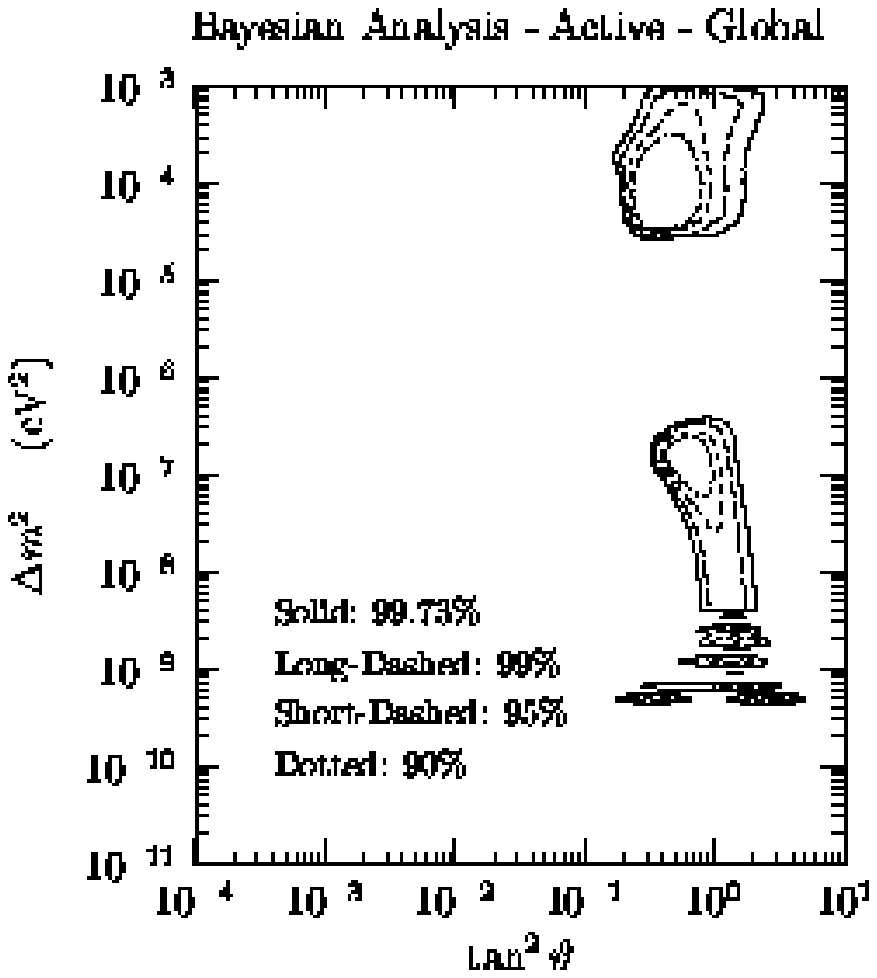,width=0.7\textwidth}} 
\end{center}
\caption{ Credible regions obtained with a Bayesian GLOBAL analysis using the same input data as in
\ref{fig:GIUglochi}. The drawn lines represent the posterior probability to contain the true values
of $\tan^2 \theta$ and $\Delta m^2$, from \cite{GAR01}. }
\label{fig:GIUglobay}
\end{figure}

\begin{figure}[ht]
\begin{center} 
\mbox{\epsfig{file=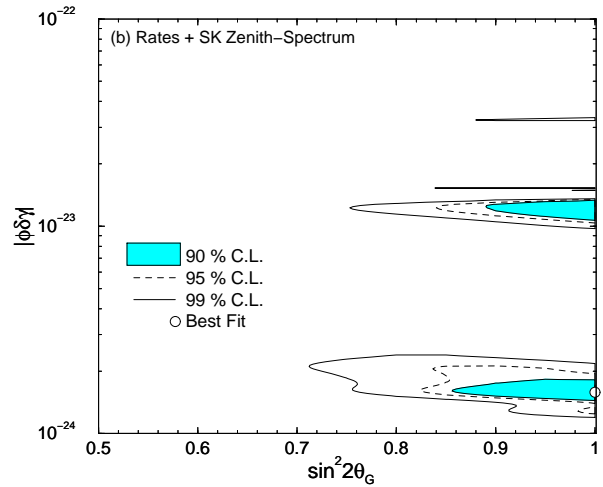,width=0.5\textwidth}} 
\end{center}
\caption{ GLOBAL analysis best-fit solution in violation of equivalence principle description 
as a function of parameters entering the mixing matrix, from \cite{GAG02}. }
\label{fig:GAGvep}
\end{figure}

\begin{figure}[ht]
\begin{center} 
\mbox{\epsfig{file=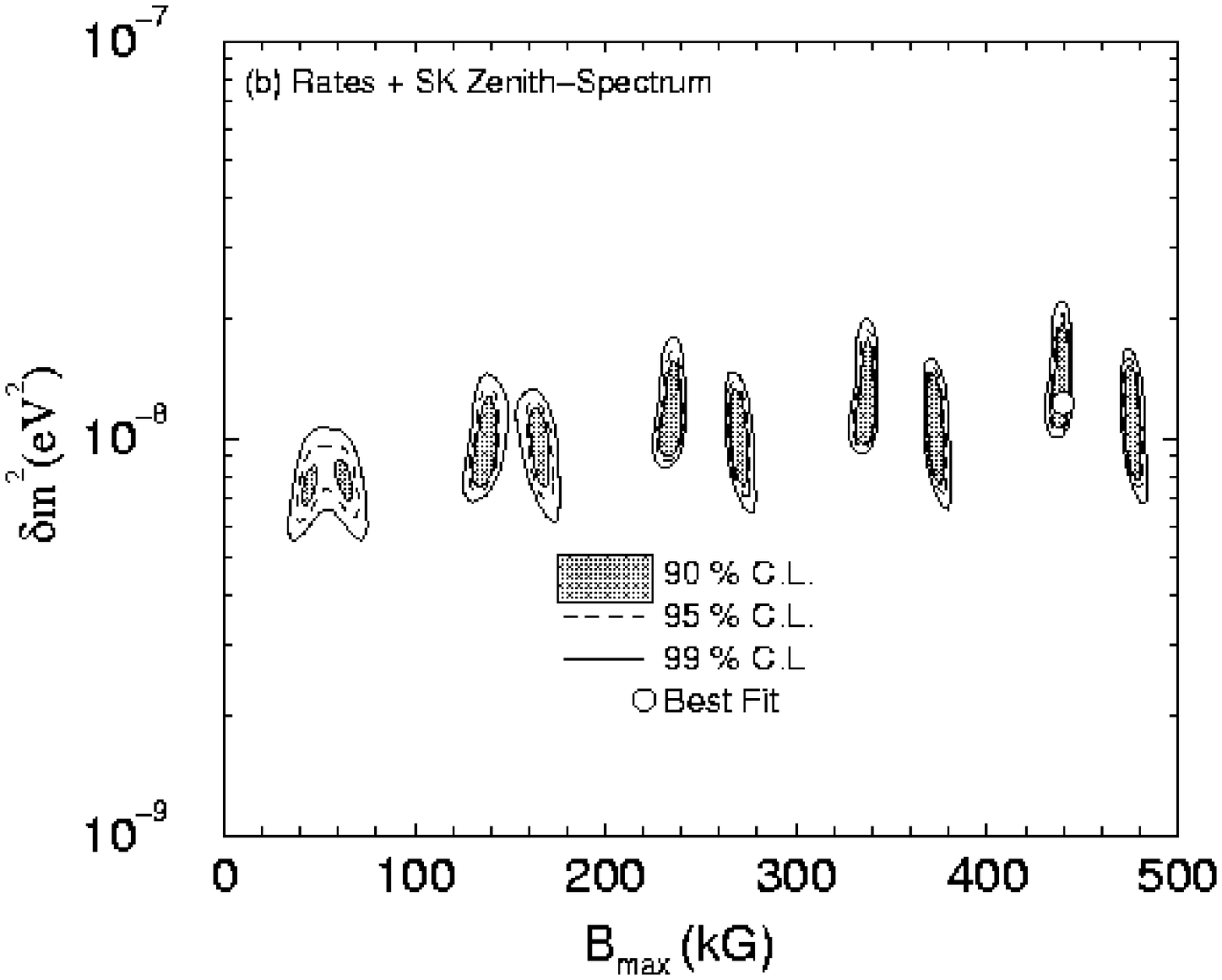,width=0.5\textwidth}} 
\end{center}
\caption{ GLOBAL analysis best-fit solution within a RSFP description. On the X-axis the 
magnetic solar field intensity is shown, from \cite{GAG02}.}
\label{fig:GAGrsfp}
\end{figure}

\begin{figure}[ht]
\begin{center} 
\mbox{\epsfig{file=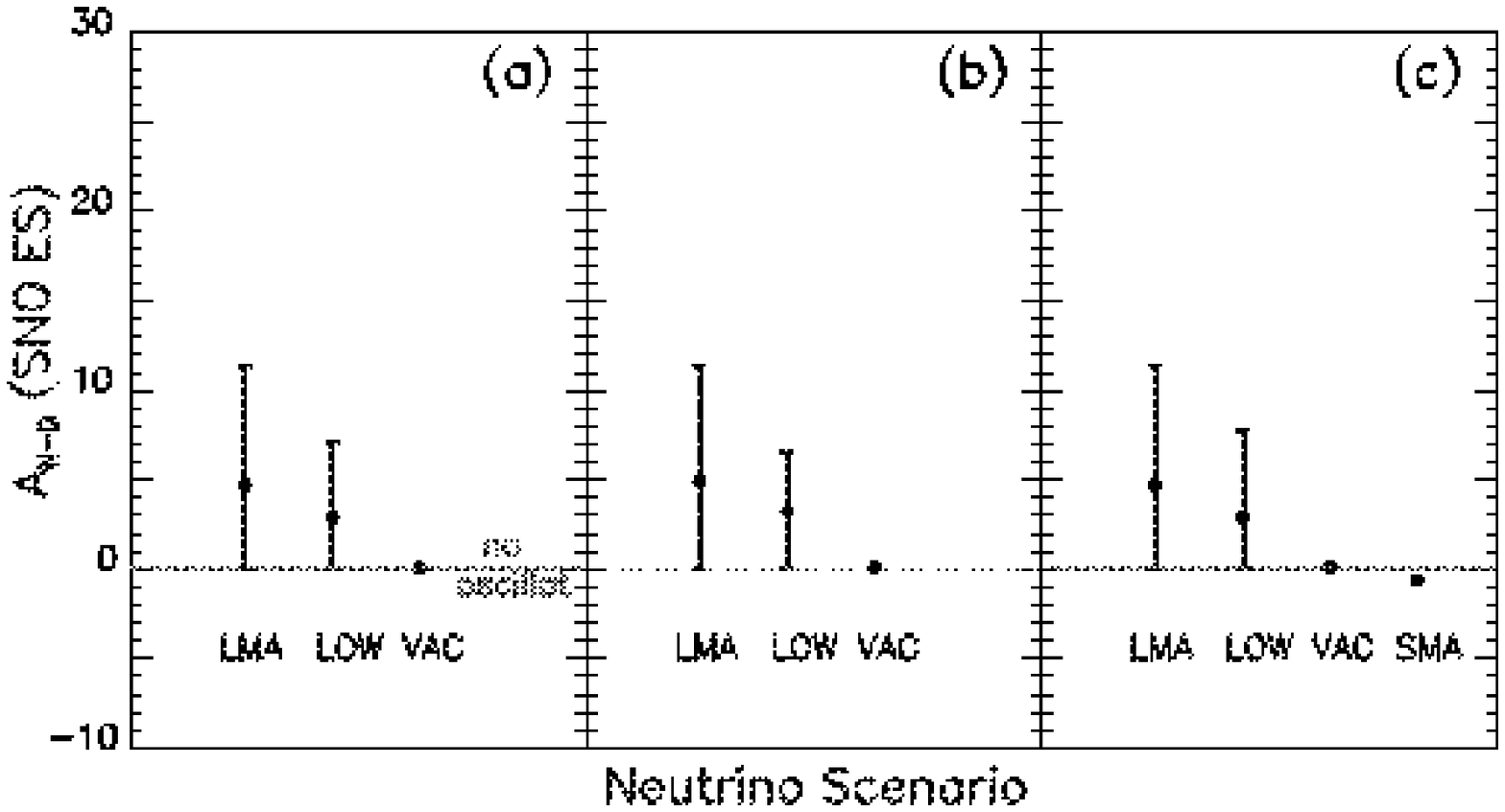,width=0.5\textwidth}} 
\end{center}
\caption{ Predictions for SNO experiment: the percentage differences between D and N ES 
interactions (for flavour oscillation solutions allowed at $3\sigma$ C.L.). The recoil 
electron kinetic energy threshold is 6.75 MeV. The three panels refer to results for different 
analysis strategies: in a) modified $^8$B flux is used, \cite{BAH02c}, in b) SK event rate is 
also included, in c) the $^8$B $\nu$ flux is a free parameter, from \cite{BAH02c}. }
\label{fig:BAH150snodnes}
\end{figure}

\begin{figure}[ht]
\begin{center} 
\mbox{\epsfig{file=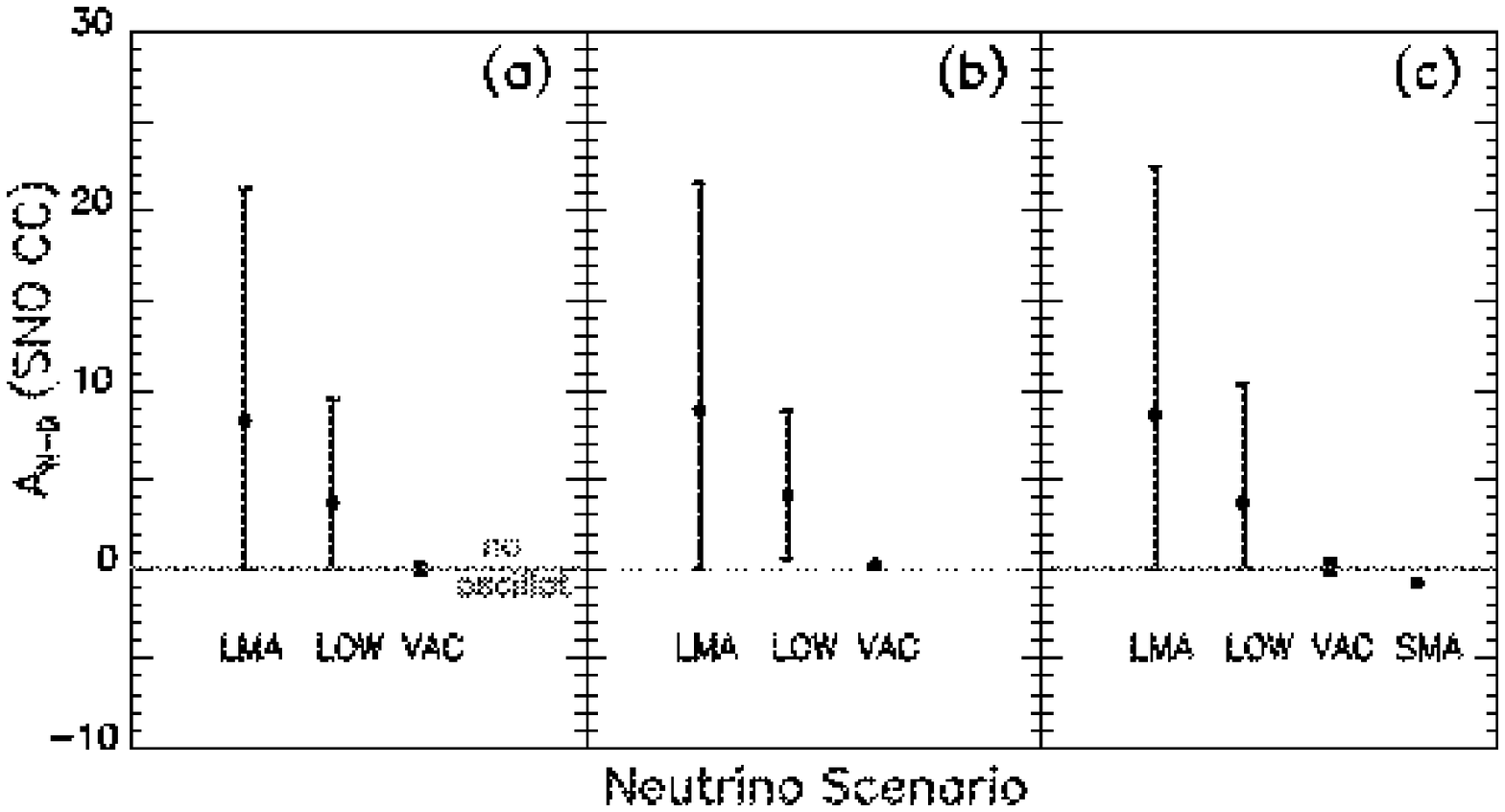,width=0.5\textwidth}} 
\end{center}
\caption{ Predictions for SNO experiment: as in fig. \ref{fig:BAH150snodnes} but for 
CC interactions in SNO (for flavour oscillation solutions allowed at $3\sigma$ C.L.).
The recoil electron kinetic energy threshold is $6.75$ MeV. 
The three panels refer to results for different analysis strategies: in a) modified $^8$B flux is used, 
\cite{BAH02c}, in b) SK event rate is also included, in c) the $^8$B $\nu$ flux is a free parameter,     
from \cite{BAH02c}. }
\label{fig:BAH150snodncc}
\end{figure}

\begin{figure}[ht]
\begin{center} 
\mbox{\epsfig{file=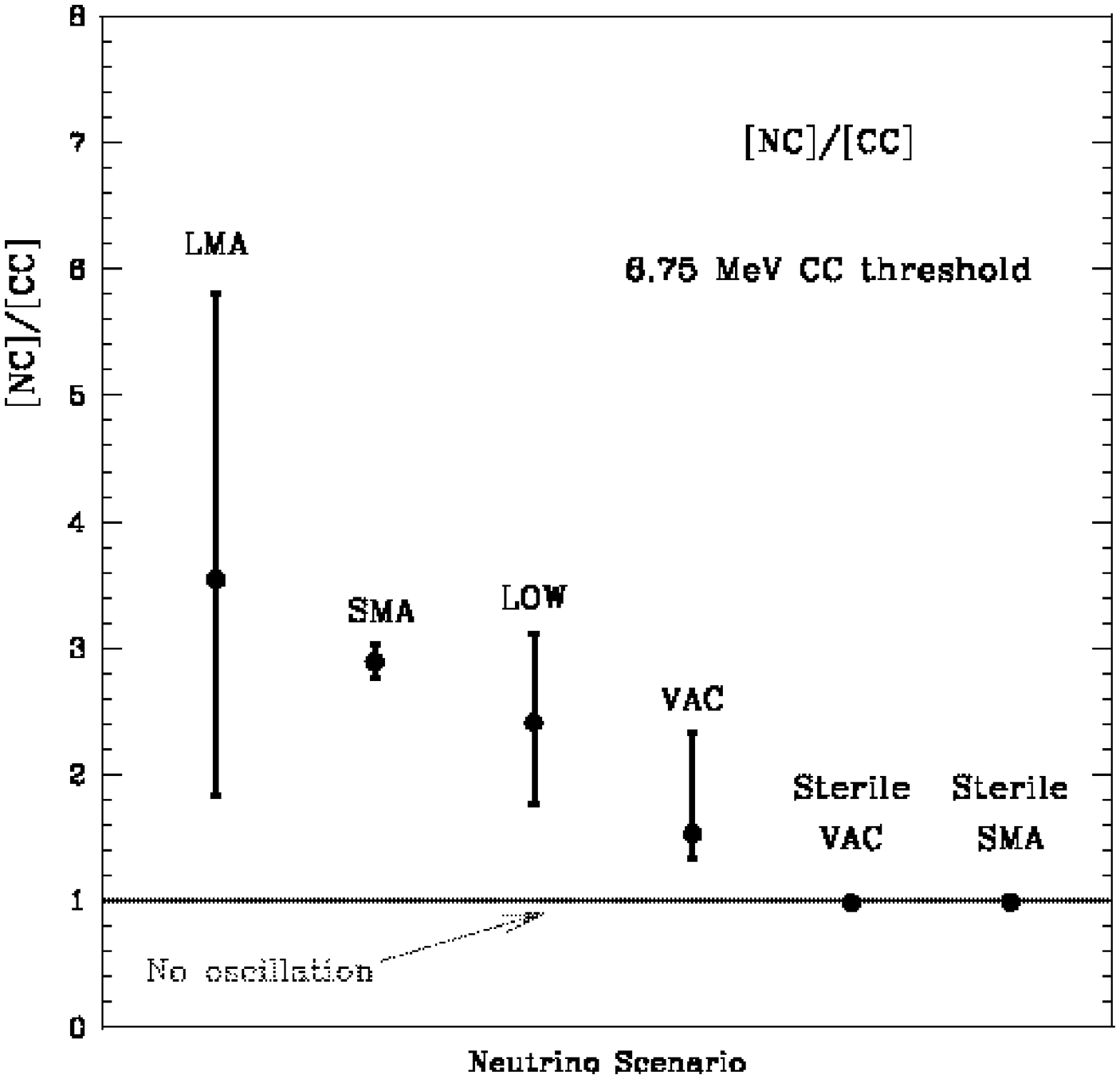,width=0.5\textwidth}} 
\end{center}
\caption{ The NC/CC ratio in the best-fit points of the GLOBAL analysis solutions with free fluxes analysis: 
the error bars show the intervals of NC/CC which correspond to the 3$\sigma$ allowed regions, 
from \cite{KRA02}.  }
\label{fig:KSnccc}
\end{figure}

We report in this section a short chronicle through the predictions before the longed for
SNO results on NC interactions. 
In table \ref{tabbah02crates} and table \ref{tabksrates} the expected $\nu_\odot$ fluxes  
provided by different GLOBAL analysis solutions are shown, from \cite{BAH02c,KRA02}.\\
\begin{table}[ht]
\caption{\it{ Interaction rates (in SNU for Cl and Ga experiments, in $10^6\,\nu\,cm^{-2}s^{-1}$
for SK and SNO detectors) as expected for three different flavour oscillation solutions. 
The first 3 columns are concerning Cl values, in the remaining ones Ga rates are shown. 
The rates are computed for the best-fit values of the allowed solutions by using the
astrophysical factor as quoted in \cite{JUN02} , adapted from \cite{BAH02c}. 
See table \ref{tabsmflux}, table \ref{tabrateclga}  and table \ref{tabexpflux}  
for a comparison with SMs predictions and experimental results.}}
\begin{center}
\begin{tabular}{|l|c|c|c||c|c|c|} 
\hline Source&LMA &LOW &VAC &LMA&LOW&VAC\\ \hline
p-p&0.00&0.00&0.00&41.80&38.70&44.30\\\hline 
p-e-p&0.12&0.11&0.16&1.49&1.35&1.95\\\hline 
Hep&0.01&0.02&0.03&0.02&0.03&0.04\\\hline 
$^7$Be&0.62&0.58&0.54&18.70&17.50&16.40\\\hline 
$^8$B&2.05&2.94&3.95&4.27&6.13&8.33\\\hline 
$^{13}$N&0.05&0.04&0.05&1.80&1.69&2.01\\\hline 
$^{15}$O&0.17&0.16&0.20&2.77&2.64&3.29\\\hline 
$^{17}$F&0.00&0.00&0.00&0.03&0.03&0.04\\\hline\hline
Total&3.02$\pm$0.3&3.85$\pm$0.5&4.93$\pm$0.6&70.9$\pm$2.6&68.1$\pm$2.8&76.4 $\pm$3.2\\\hline \hline
SK&2.39$^{+0.33}_{-0.36}$&3.02$^{+0.42}_{-0.45}$&3.81$^{+0.53}_{-0.57}$ &-- &-- &-- \\\hline 
SNO(CC)&1.72$^{+0.24}_{-0.26}$&2.50$^{+0.35}_{-0.38}$&3.26$^{+0.46}_{-0.49}$ &-- &-- &-- \\\hline 
\end{tabular} 
\end{center} 
\label{tabbah02crates}
\end{table}
\\
\begin{table}[ht]
\caption{\it{ Predictions for Cl, Ga, SK and SNO experiments in the best fit points of 
GLOBAL solutions found in the free flux analysis. For SK and SNO detectors 
the fraction (in percent) with 
respect to the value quoted in \cite{BAH01a} is given, adapted from \cite{KRA02}.}}
\begin{center}
\begin{tabular}{|l|c|c||c|c|}
\hline
 &  Cl(SNU) & Ga(SNU) & SK (\%) & SNO (\%) \\ \hline
LMA   & 2.89 & 71.3 & 45.2 & 32.3 \\ \hline
SMA   & 2.26 & 74.4 & 46.3 & 39.6 \\ \hline
LOW   & 3.12 & 68.5 & 44.6 & 36.8 \\ \hline
VAC   & 3.13 & 70.2 & 42.3 & 36.4 \\ \hline
Just So$^2$ & 3.00 & 70.8 & 43.4 & 43.4 \\\hline\hline\hline
SMA(s) & 2.93 & 75.5 & 43.5 & 44.5 \\ \hline
VAC(s) & 3.24 & 69.9 & 41.4 & 38.1 \\ \hline
Just So$^2$ (s)& 3.01 & 70.9 & 43.4 & 43.5 \\\hline
\end{tabular}
\end{center}
\label{tabksrates}
\end{table}
\\
In \cite{GAR01}, Bayesian and standard least-squares techniques were applied in a GLOBAL 
analysis context.
The $\chi^2$ minimum value was in LMA region if only transition into an active $\nu$ is considered, see fig. 
\ref{fig:GIUglochi}, in VAC region when transition into sterile $\nu$'s were allowed. 
SMA solution was strongly 
disfavoured and only a small region was allowed for VAC solution. As a marginal conclusion, the 
authors deduced that the likelihood ratio is a more powerful test in the analysis with 
respect to the usually adopted "goodness of the fit".
\\
When Bayesian method was applied, either in RATES only or in GLOBAL case, the allowed regions 
were larger than in the standard analysis: in any case, LMA solution gave the best result and was 
strongly suggested to be the right solution, the LOW solution was marginally acceptable, see 
fig.  \ref{fig:GIUglobay}. VAC and SMA solutions were in practice excluded, see also 
\cite{CRE01} for further analyses based on Bayesian approach.   
\\
In table \ref{tabrsfprates} predictions concerning the expected interaction rates in different detectors based on RSFP 
solution to the SNP as computed in \cite{AKH02} are shown.   \\
\begin{table}[ht]
\caption{\it{ The ratio between the expected rates corresponding to best-fit GLOBAL analysis, from \cite{AKH02}, and 
the SSM predictions, from \cite{BAH01a,BAH02c}, are shown. In the first column the selected $B_\odot$ profiles: 1 and 6 
from \cite{PUL00}, 4 from \cite{PUL01}. The values of $\chi_{min}^2$ correspond to 39 d.o.f., adapted
from \cite{AKH02}.} }
\begin{center}
\begin{tabular}{|c|c|c|c|c|c|c|c|} 
\hline \hline
Profile & $R_{Ga}$ & $R_{Cl}$ &  $R_{SK}$  & $R_{SNO}$ & $\Delta m^2$ (eV)$^2$ & $B_\odot$ (kG) & $\chi_{min}^2$ \\
\hline
1 & 0.59 & 0.30 & 0.41 & 0.35 & $7.65\cdot 10^{-9}$ & 45  & 37.8 \\
6 & 0.58 & 0.30 & 0.39 & 0.33 & $1.60\cdot 10^{-8}$ & 113 & 36.1 \\
4 & 0.58 & 0.30 & 0.40 & 0.33 & $1.48\cdot 10^{-8}$ & 101 & 35.5 \\
\hline
\end{tabular}
\end{center}
\label{tabrsfprates} 
\end{table}
\\
In \cite{GAG02}, different standard and non-standard solutions were analysed. In a first step 
a RATES analysis with the usual least-squares method was done: for RSFP and non-standard $\nu$ 
interaction solutions a high confidence level was found, even if a very strong $B_\odot$ is 
required ($\sim$ 100 kG); VAC oscillation gave the best 
result among standard solutions. When energy spectrum and zenithal dependence were included 
and a GLOBAL analysis was done, then LMA, RSFP and non-standard $\nu$ interactions provided 
the  same confidence level, see table \ref{tabnonstandard} where results computed in 
\cite{MIR01} concerning resonant and non-resonant SFP solutions are added in the last raws.
Flavour oscillation solutions and RFSP approach provided results explaining to the solar, 
atmospheric and accelerator data, see fig. \ref{fig:GAGvep} and fig. \ref{fig:GAGrsfp}.\\
\begin{table}[ht]   
\caption{\it{ Comparison of existing solutions to the SNP when total rates, energy spectrum 
and zenithal dependence are used. For each one of the indicated mechanisms 
the best fit values of the relevant parameters are shown, the corresponding $\chi^2_{min}$ 
and C.L. level, adapted from \cite{MIR01,GAG02}. }}
\begin{center} 
\begin{tabular}{|l|c|c|c|c|}
\hline 
 & & & $\chi^2_{min}$ & C.L.\\\hline
NO OSC. & & & 100.0 (48 d.o.f) & $1.6\cdot 10^{-5}$ \\ \hline\hline\hline 
        &$\Delta m^2$(eV)$^2$& $\tan^2\theta$&  & \\
LMA     &  $6.15\cdot 10^{-5}$ & 0.349  & 38.7 & 75\%  \\
VAC     &$4.65\cdot 10^{-10}$& 1.89 & 46.1 & 47\%  \\
LOW     &  $1.01\cdot 10^{-7}$ & 0.783  & 45.0 & 38\%  \\
SMA     &  $4.93\cdot 10^{-6}$ & $4.35\cdot 10^{-4}$ & 61.5 & 6.3\%  \\ \hline\hline\hline 
RSFP    & $\Delta m^2$ (eV)$^2$ & $B_{\odot ,max}$ (kG) &  &  \\ 
        & $1.22\cdot 10^{-8}$ & 440 & 38.4 & 78\%  \\ \hline\hline\hline
Non-standard Int. & $\epsilon' $ & $\epsilon$ & & \\
$d$-quarks    & 0.599 & $3.23 \cdot 10^{-3}$ & 37.9 & 80\% \\
$u$-quarks    & 0.428 & $1.40 \cdot 10^{-3}$ & 37.9 & 80\% \\ \hline\hline\hline 
VEP          & $|\phi\Delta\gamma| $ & $\sin^2 2\theta_G$ & &  \\
             & $1.59\cdot 10^{-24}$ & 1.0 & 42.9 & 60\% \\ \hline\hline\hline
             & $\Delta m^2$ (eV)$^2$ &$\tan^2\theta$ &  & $B_{\perp}$ (kG) \\
Non-Resonant SFP	& $4.0\cdot 10^{-9}$& $3.5\cdot 10^3$   & 3.83(4 d.o.f.)& 84 \\
RSFP            & $8.9\cdot 10^{-9}$& $1.1\cdot 10^{-3}$& 2.98(4 d.o.f.)& 84 \\
\hline
\end{tabular} 
\end{center} 
\label{tabnonstandard}   
\end{table} 
\\
\\
The main conclusions based on the latest SK data (1496 days of running time) were, \cite{SMY02a}:
\bit 
\item 
SMA and VAC solutions were excluded: in fact $^8$B $\nu_\odot$ spectrum seemed to be without distortions. 
A smaller D-N asymmetry and a slightly flatter recoil energy spectrum produced a small reduction in the lower-mass  
region of the LMA solution and in the upper-mass region of the LOW solution, see also\cite{BAH02c}.  
\item 
SK rate required a $^8$B $\nu_\odot$ flux larger than Cl rate allows so almost the Q-VAC 
solutions disappeared: only the upper part of LMA and two very small Q-VAC solutions, which 
better describe SK zenith angle spectrum, survived. The LMA solutions $\Delta\,m^2 \geq$ 
3$\cdot 10^{-5}$ (eV)$^2$ were found to be the only realistic solution 
at 95$\%$. In fact Q-VAC solutions were rejected because of the rates of all experiments.
\item 
The excluded areas did not depend on $\nu_\odot$ flux computed with SMs.
\item
The mixing angle was large but not maximal; moreover, $\Delta m^2 < 10^{-4}$ (eV)$^2$. In three 
flavour oscillation solutions the computed mixing angle values were: $\theta_{12}\approx 
\frac {\pi}{6}$, $\theta_{23}\approx \frac {\pi}{4}$, $\theta_{13}\approx 0$.        
\item 
When SK and SNO results were combined the appearance of $\nu_\odot$ flavours other than 
electronic one was strengthened. 
\fit 
In \cite{FRI02} $\overline \nu_\mu$ and $\overline \nu_\tau$ components were proposed to
be present in $\nu_\odot$ flux, as predicted under spin flavour flip assumption. This process
could be induced by a strong relic B$_\odot$ acting in the radiative zone: its 
strenght should be lower than 0.55 MGauss, mainly due to helioseismological constraints. \\  
Values of quantities measurable by different experiments were newly computed:  
we report in table \ref{tabsnopredbah02c}  [NC]/[CC] ratio and D-N  asymmetries, 
both for ES and CC interactions, \cite{BAH02c}, see also fig. \ref{fig:BAH150snodnes},
fig. \ref{fig:BAH150snodncc} and fig. \ref{fig:KSnccc}.
We remark that the predictions concerning the D-N asymmetry for ES interactions  
in SK and SNO experiments are very similar. \\
\begin{table}[ht]
\caption{\it{ Predictions for NC/CC double ratio, D-N asymmetry for CC and ES events in SNO.  
Two different thresholds of the recoil electron kinetic energy are used: 4.5 MeV and 6.75 MeV.  
The 3$\sigma$ regions are obtained within a GLOBAL analysis using modified
$^8$B $\nu_\odot$ flux, adapted from \cite{BAH02c}.}}
\begin{center}
\begin{tabular}{|l|c|c|}
\hline
 & 4.5 MeV & 6.75  MeV\\\hline          
&[NC]/[CC] &[NC]/[CC]\\\hline  
LMA     & 3.44 (1.79,5.28)   &3.45  (1.82,5.28) \\  
LOW     & 2.39 (1.71,3.22)   &2.37  (1.71,3.20) \\ 
VAC     & 1.76 (1.43,2.06)   &1.82  (1.46,2.17) \\ \hline          
& A$_{D-N}^{CC}$ &A$_{D-N}^{CC}$ \\ \hline  
LMA   (\%)  & 7.4  (0.0,+19.5)&  8.3 (0.0,+21.4)   \\  
LOW   (\%)  & 4.3  (0.0,+10.4)&  3.7 (0.0,+9.5)    \\  
VAC   (\%)  & 0.1  (-0.2,+0.3)& 0.2 (-0.3,+0.5)  \\ \hline          
& A$_{D-N}^{ES}$ &A$_{D-N}^{ES}$ \\ \hline  
LMA   (\%)  & 4.1  (0.0,+10.1)&  4.7  (0.0,+11.4)   \\  
LOW   (\%)  & 3.3  (0.0,+7.8)&   2.9  (0.0,+7.1)    \\  
VAC   (\%)  & 0.0  (-0.1,+0.1)&  0.1  (-0.2,+0.3)   \\ \hline 
\end{tabular} 
\end{center} 
\label{tabsnopredbah02c}
\end{table}
\\
\\
\\

\subsection{2002: After SNO (II).} 
\label{aftersno2}
\begin{figure}[ht]
\begin{center} 
\mbox{\epsfig{file=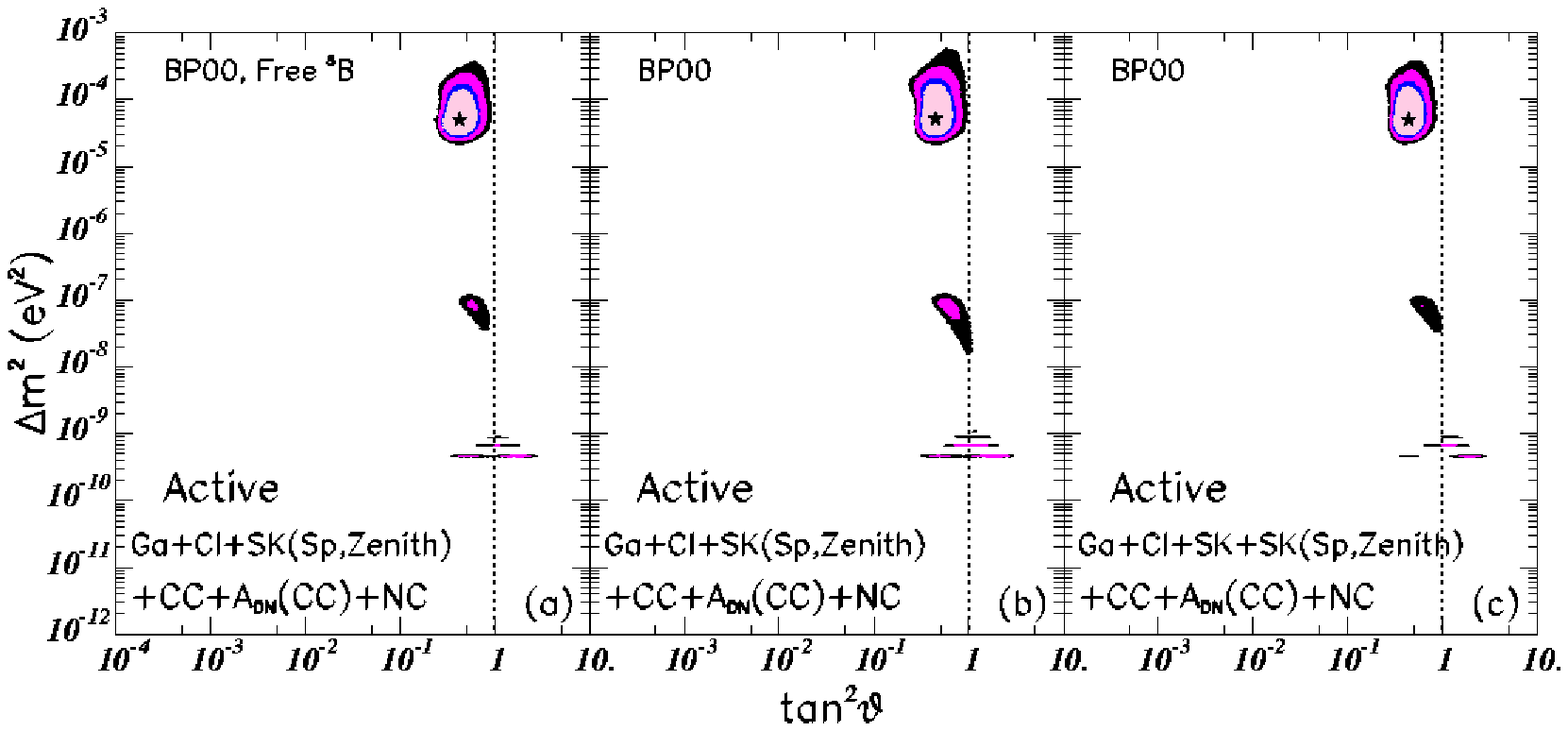,width=0.6\textwidth}} 
\end{center}
\caption{ GLOBAL analysis solutions for three different strategies.  
The input data include the $\nu_\odot$ fluxes and uncertainties predicted by \cite{BAH01a}, 
the total measured  CC and NC event rates  and the D-N asymmetry from SNO,
the Cl and Ga averaged rates, the zenith angle-recoil energy spectrum from SK
The C.L. contours shown in the figure are $90$\%, $95$\%, $99$\%, and $99.73$\% ($3\sigma$). 
The GLOBAL analysis best-fit points are marked by a star.\\   
In a) $^8$B $\nu_\odot$ flux is a free parameter to be determined by the experimental data together with 
$\Delta m^2$ and $\tan^2\theta$. In b) and c) the theoretical uncertainty in the $^8$B
$\nu_\odot$ flux is included in analysis; in b) the total SK rate is included explicitly together with 
a free normalization factor for the zenith angle-recoil energy spectrum of the electrons, from \cite{BAH02f}. }
\label{fig:BAH314_3str}
\end{figure}

\begin{figure}[ht]
\begin{center} 
\mbox{\epsfig{file=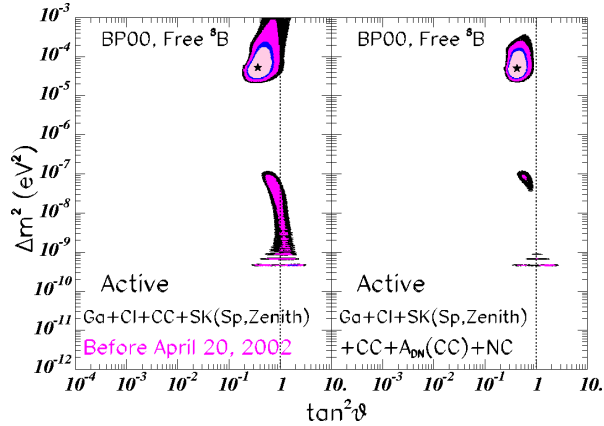,width=0.5\textwidth}} 
\end{center}
\caption{ GLOBAL analysis solutions results within a free $^8$B $\nu_\odot$ flux strategy before and after 
SNO data concerning NC events and D-N asymmetry. In the left panel, the LOW solution is allowed at $97.4$\%; 
on the contrary in the right one LOW is allowed at the $98.8$\%, from \cite{BAH02f}. }
\label{fig:BAH314bef_aft}
\end{figure}

\begin{figure}[ht]
\begin{center} 
\mbox{\epsfig{file=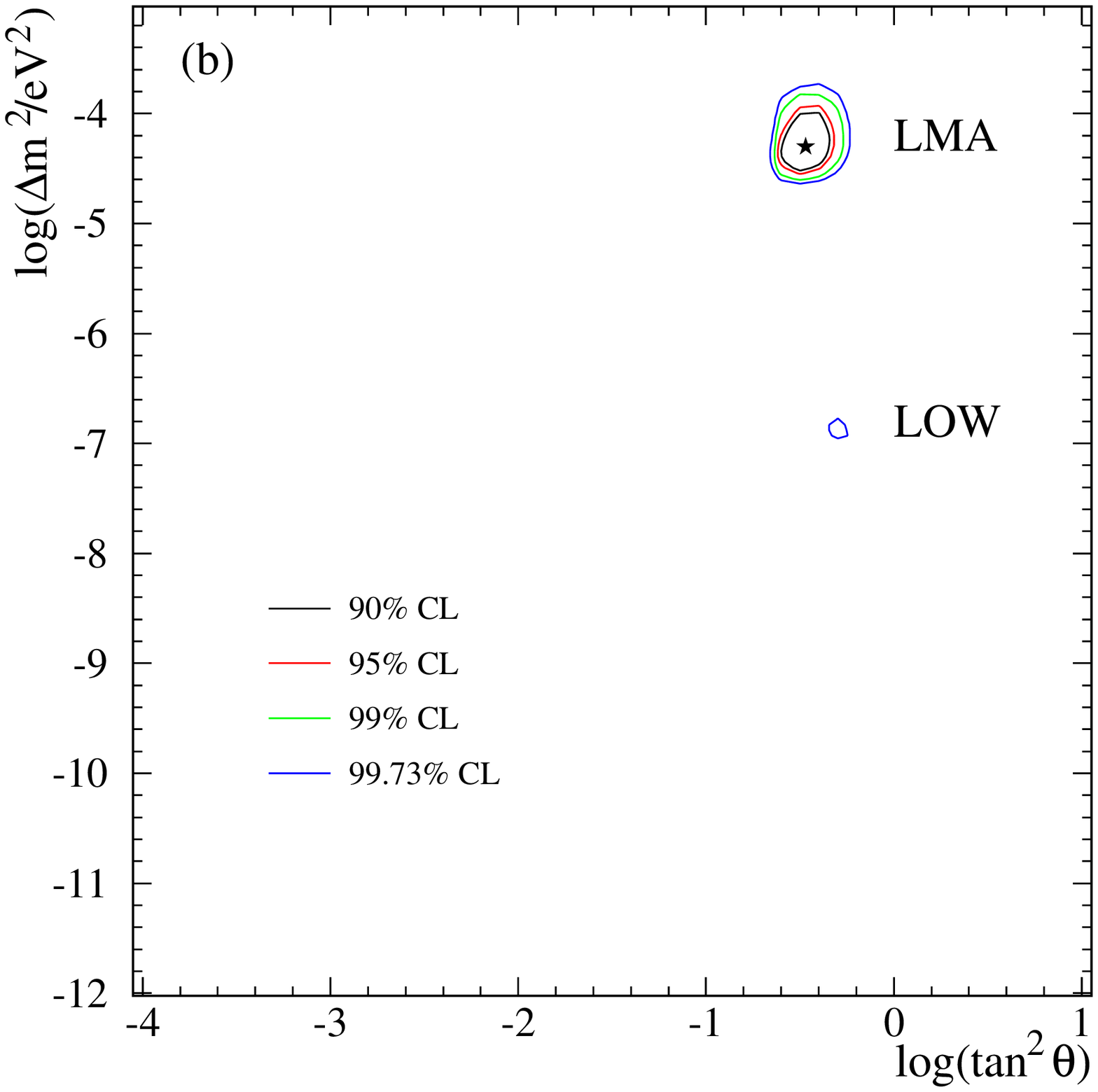,width=0.5\textwidth}} 
\end{center}
\caption{ GLOBAL analysis solutions as computed by SNO coll.; the star indicates the best fit value, 
from \cite{AHM02b}. } 
\label{fig:SNOglobal}
\end{figure}

\begin{figure}[ht]
\begin{center} 
\mbox{\epsfig{file=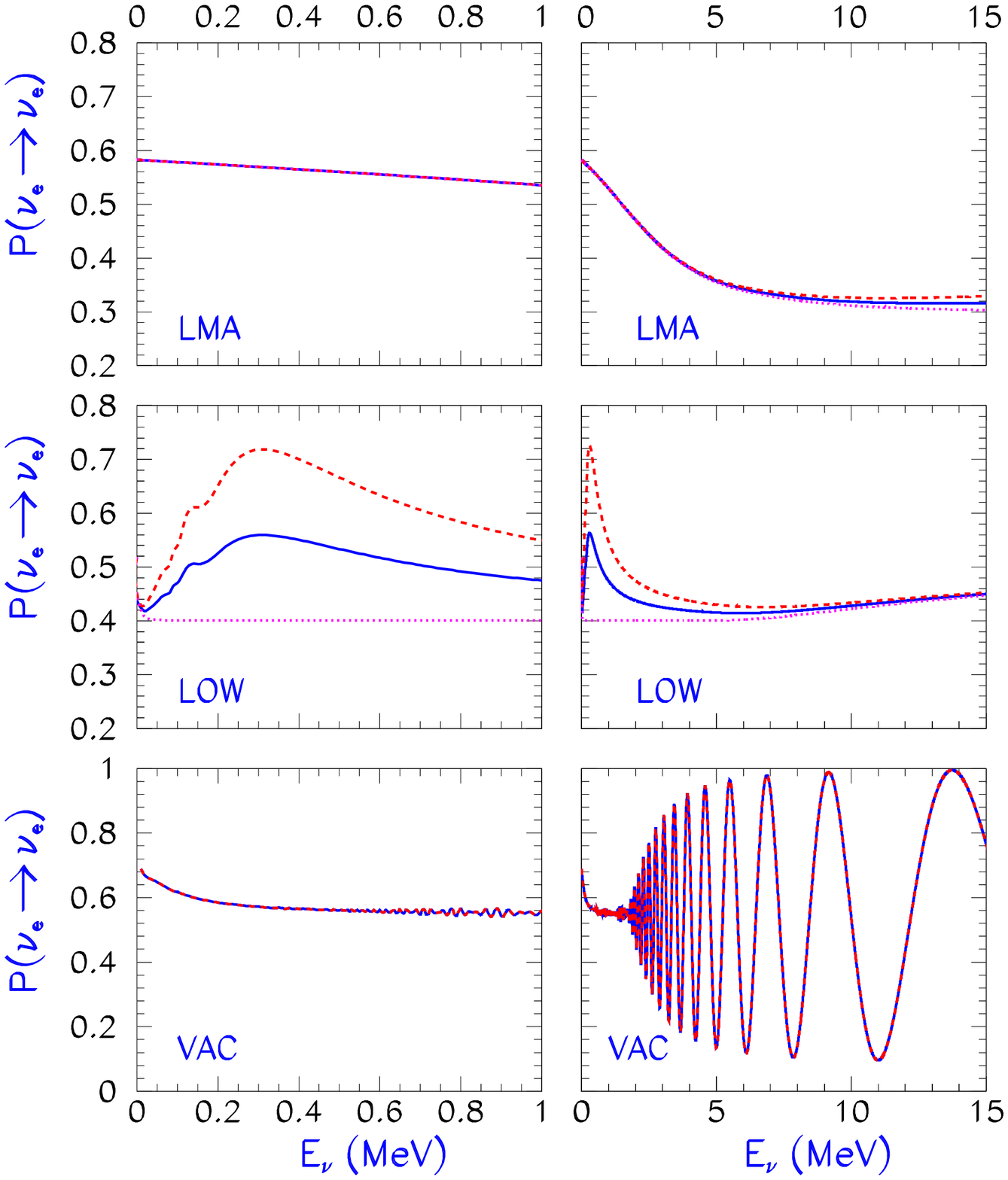,width=0.5\textwidth}} 
\end{center}
\caption{ Survival probabilities (yearly-averaged) computed within a free flux analysis 
for a $\nu_e$ produced in the solar centre: average survival probabilities 
with regeneration in the Earth (full line); D-time probabilities, without regeneration effect (dotted line);
N-time probabilities (dashed line).
The regeneration effects are computed for SNO (right hand panels) and for LNGS (left side panels)
detectors; in any case the differences among SK, SNO and LNGS are very small.
LOW solutions in the right-hand panel are averaged over a small energy band (0.1 MeV) to suppress 
rapid oscillations because of the sensitive dependence upon the Earth-Sun distance. VAC
solutions are averaged over an energy band of $\pm$ 0.05 MeV, \cite{BAH02f}.   }
\label{fig:BAH314surv}
\end{figure}

\begin{figure}[ht]
\begin{center} 
\mbox{\epsfig{file=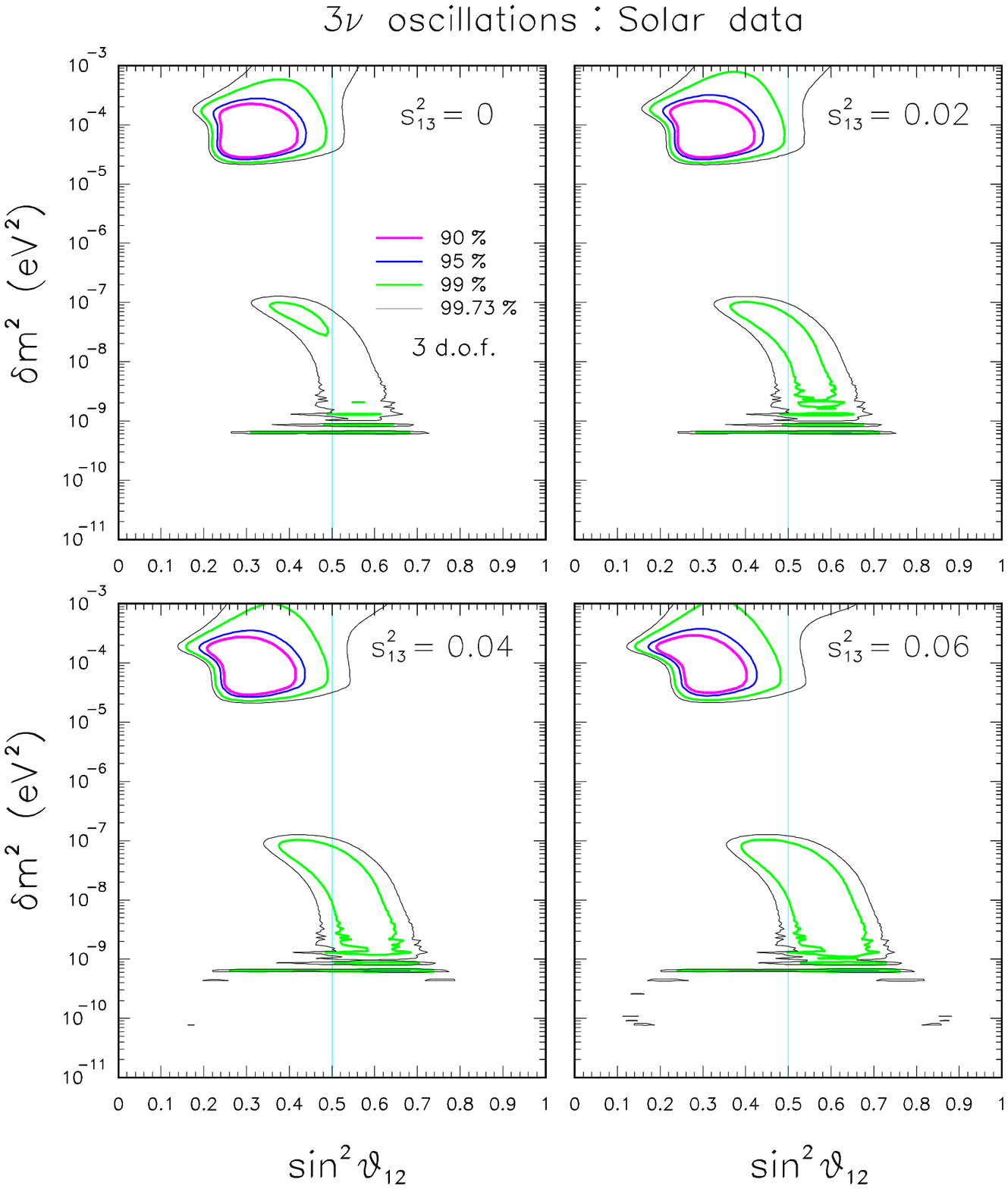,width=0.5\textwidth}} 
\end{center}
\caption{ Three flavour GLOBAL analysis of $\nu_\odot$ data: all experimental rates combined with
winter-summer difference from Ga experiments, energy and zenith spectra and D-N asymmetry.   
Sections are concerning different value of $s^2_{13}= \sin^2\theta_{13}$ parameter. 
C.L. contours are shown: 90$\%$ C.L.(full line), 95$\%$ (thick dot line), 99$\%$ (gray line), 
99.73$\%$ (thin dot line); from \cite{FOG02d}.}
\label{fig:FOG3sol}
\end{figure}

\begin{figure}[ht]
\begin{center} 
\mbox{\epsfig{file=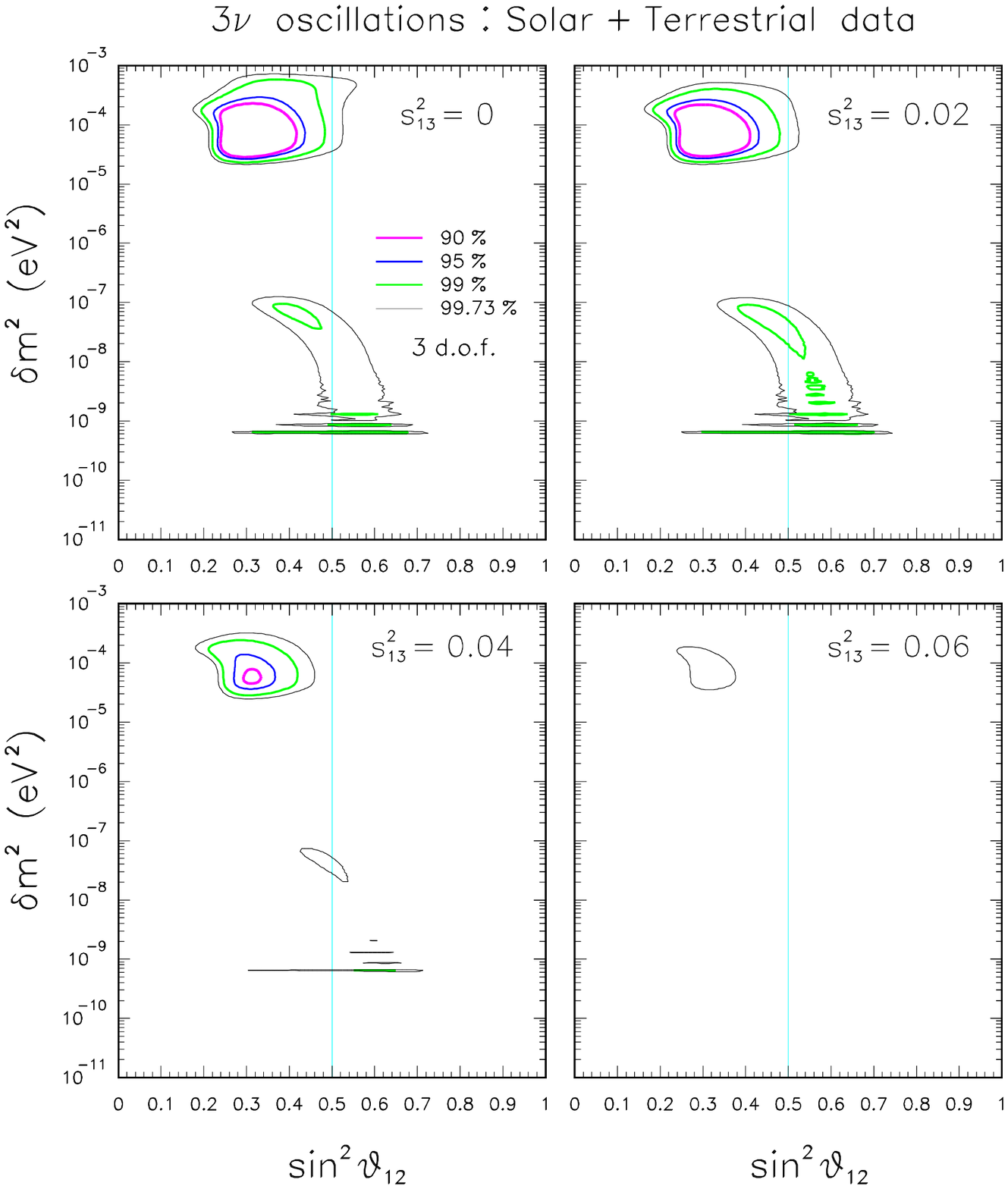,width=0.5\textwidth}} 
\end{center}
\caption{ As in fig. \ref{fig:FOG3sol} but the constraints from CHOOZ, SK atmospheric and
K2K $\nu$ are added, from \cite{FOG02d}.}
\label{fig:FOG3all}
\end{figure}

After the SNO results on NC interactions and D-N asymmetry, a lot of new articles was 
prepared. A great part of published papers involves only active $\nu$ transitions; in 
\cite{BAH02e} the possible contribution of a $\nu$ sterile component is also analysed. 
Usually, the standard least-square technique approximation for the definition of the allowed 
regions with a given confidence level has been applied.\\ 
SNO collaboration has presented results for the CC, NC, and ES fluxes under the assumption that the CC and ES 
recoil energy spectra are undistorted by $\nu$ oscillations or any other new physics. This hypothesis is a good 
approximation for the LMA and LOW solutions but is less accurate for the remaining ones. \\
In \cite{ALI02a,ALI02b,BAH02f,DEH02,FOG02c,KRA02}, results and predictions obtained 
within a GLOBAL analysis approach are exposed (it has to be underlined that used techniques are slightly
different).  \\
In \cite{BAH02f}, the energy dependence and correlations of the errors in the $\nu$ absorption 
cross-sections for the Cl and Ga
experiments as computed in \cite{BAH02e}, the latest data from SAGE, \cite{ABD02}, zenith angle-recoil energy 
spectrum data from SK after $1496$ days, \cite{SMY02a}, the predictions and the uncertainties 
from \cite{BAH01a}, different strategies in analyses are used.\\
The strategy (a) selects  $\nu_\odot$ fluxes from \cite{BAH01a} (with the exception of
$^8$B component which is a free parameter), the experimental rates (with the exception of 
SK total rate), the zenith angle-recoil energy spectrum data for SK, 
fluxes and asymmetries from SNO. On the contrary, strategy (b) includes the predictions concerning $^8$B 
$\nu_\odot$ flux and uncertainty from \cite{BAH01a} while in strategy (c) the total SK 
rate is included together with a free 
normalization factor for the zenith angle-recoil energy spectrum of the recoil electrons.
Strategy (b) allows a slightly larger region for the LOW solution mainly due to the 
uncertainty for the $^8$B $\nu_\odot$ flux, see also fig. \ref{fig:BAH314_3str}.
The comments from \cite{BAH02f} are based on strategy (a).\\
In \cite{DEH02}, the rates measured by Homestake, SAGE, GALLEX+GNO are separately introduced in
computations. The SK values, based on 1496 days of data taking, include 8 energy bins with 7
zenith angles bins in each, except for the first and the last energy bins (44 points). The
experimental errors and systematic uncertainties treatment is respectively from \cite{SMY02a}
and from \cite{GAG02}. In \cite{FOG02c} 81 observables (Cl and Ga average rates, 
winter-summer difference 
from GALLEX+GNO, 44 absolute event rates for the energy-nadir differential spectrum 
of electrons from SK, 34 D-N energy spectrum bins from SNO experiment) and 31
sources of correlated systematics (12 uncertainties related to SSM input, the $^8$B
$\nu$ shape uncertainty, 11 SK error sources and 7 SNO sources) are considered in 
$\chi^2$ modified analysis. 
\benu
\item 
SNO results have reduced the area for the LMA and LOW solutions, see fig. \ref{fig:BAH314bef_aft}, which are 
allowed at 3$\sigma$ C.L. (for VAC solution 2.1 $\sigma$, but SNO coll. does not find this 
solution at the 3$\sigma$ level in its analysis, see fig. \ref{fig:SNOglobal}). 
If $\nu_\odot$ fluxes quoted in \cite{AHM02a,AHM02b} are introduced in computations
with the assumptions of undisturbed recoil energy spectrum and without statistical correlations, 
VAC solution is not allowed at $3.1\sigma$, \cite{BAH02f}. 
\item 
Latest results from SAGE and GNO experiments lower the averaged Ga rate, \cite{ABD02,KIR02}, but
these values are consistent (for instance the latest GNO results are 1.5$\sigma$ below the 
previous GALLEX data). It has to be recalled that troubles in acquisition are reported from
SAGE  coll. so that an "{\it{a posteriori}}" corrective factor is applied to 1996-1999
measurements while GNO experiment has a completely new electronic setup and a different data 
acquisition and a lower background is obtained. Both the experiments have improved their
systematics: consequently the statistical errors are larger while the systematic ones are
reduced. In any case, lower rates in Ga experiments reinforce the LOW solution, \cite{STR02}.
\\
\item 
For MSW solutions the energy dependence of the survival probability is weak at high energy,
see fig. \ref{fig:BAH314surv} : 
SK results do not show statistically significant distortion of the recoil energy spectrum. 
SNO and SK data have provided a survival probability $P (\nu_e \rightarrow \nu_e)\approx 1/3$; at $E\leq$ 1 MeV 
this value should be $\sim$ 1/2, but there are no experimental measurements.
The results from SNO have modified at a small level solutions having a survival probability not 
strongly dependent upon energy (LMA and LOW), while their impact is large on SMA and VAC solutions. A shift of the 
predicted central values for CC and NC interaction rates is also present, \cite{BAH02f}. 
SK and SNO results do not suggest any statistical significant hint for Earth matter effect or
for modifications in $^8$B $\nu_\odot$ spectra; both LMA and LOW solutions can fit the D-N
asymmetry but different energy values and zenith-angle spectra are predicted. At present, SK 
spectral data suggest LMA as favoured solution, \cite{STR02}.
\item 
The allowed regions obtained in VAC and Q-VAC solutions depend upon details of the analysis: if the 
anti-correlation between statistical errors of NC and CC interaction rates is not included, 
VAC solution is disfavoured at 3$\sigma$ level, \cite{BAH02f}. 
\item 
Present 3$\sigma$ upper limits  for $\Delta\,m^2$ in LMA solution are: $\Delta m^2 < 3.7 \cdot 
10^{-4} (eV)^2$ (3.6 in \cite{DEH02}, 2.3 in \cite{BAR02b}). The LMA region does not reach 
the upper bound imposed by the CHOOZ reactor data ($\Delta m^2\leq 8\cdot 10^{-4}$ 
(eV)$^2$), \cite{BAH02f}.  
\item 
If one admits oscillations into either active and sterile $\nu$'s, the total $^8$B $\nu_\odot$ flux has to be 
increased. If a $^8$B factor is defined:
\beq
f_B\,=\,\frac {\Phi(^8B)_{exp}} {\Phi(^8B)_{SSM}}
\label{eq_boronfactor}
\feq
the SNO collaboration experimentally deduces f$_B$ = 1.01 $\pm$ 0.12. 
Its best-fit in LMA region is obtained with f$_B$ = 1.07 $\pm$ 0.08 
(LOW solution gives f$_B$ = 0.91$^{+0.03}_{-0.02}$), \cite{BAH02f} and similar values 
in \cite{DEH02}.\\
According to SMs, the $^8$B $\nu_\odot$ flux is proportional to $S_{17}(0)$. Latest values 
deduced from measurements on cross-sections entering this parameter do not fully agree. 
The choice is relevant in analyses including $\nu_s$'s, \cite{BAH02f,BAR02b,CRE02}.\\
Since $^7$Be and $^8$B $\nu_\odot$'s are generated by different processes involving $^7$Be, 
the agreement between the predictions of SMs and the value deduced from SNO results suggests 
(but does not imply) a $^7$Be $\nu_\odot$ flux also in agreement with SMs, \cite{CRE02}. 
\item 
Bi-maximal mixing solution ($\tan^2 \theta$ = 1) is disfavoured for LMA (3.3$\sigma$ C.L.), LOW (3.2$\sigma$ C.L.) 
and VAC (2.8$\sigma$ C.L.) solutions. On the contrary, approximate bi-maximal mixing solution is favoured: at 3$\sigma$ 
level LMA solution admits 0.24 $< \tan^2 \theta <$ 0.89  (for LOW solution 0.43 $< \tan^2 \theta <$ 0.86), 
\cite{BAH02f}. 
\item 
Among the MSW solutions, the situation has also been clarified: LMA solution is the only viable 
solution at a level of 2.5$\sigma$; LOW solution is excluded at the 98.8\% C.L., SMA solution 
at more than 3.7$\sigma$, pure sterile oscillations at 5.4$\sigma$, \cite{BAH02f}.  
If sterile $\nu$'s are still included, VAC solution has the best-fit but it is excluded at 
5.4$\sigma$ C.L. (for 3 d.o.f.) while SMA sterile solution was acceptable at 3.6$\sigma$ before
results from SNO, \cite{BAH02f}. 
\item 
Assuming no spectral distortion in $^8$B $\nu_\odot$ component, NC and CC results exclude the 
case of no oscillations at a level of $5.1\sigma$, or, in other words, SNO data show a 
5.1$\sigma$ evidence for $\nu_{\mu,\tau}$ appearance even if this value slows down to 
3$\sigma$ when generic spectral distortions are allowed. SK data forbids significant spectral 
distortions while SMA sterile solution predicts this effect: SNO results increase 
the $\nu_{\mu,\tau}$ appearance consequently sterile solutions are excluded in two different ways, \cite{CRE02}. 
\item 
Sterile $\nu$'s can generate solar, atmospheric and LSND oscillations. Atmospheric data give a 6$\sigma$ evidence 
for $\nu_\mu\to \nu_\tau$ versus $\nu_\mu\rightarrow \nu_s$. "2+2" schemes predict that the fraction of $\nu_s$ in 
$\nu_\odot$ and $\nu_{atm}$ oscillations adds to one but both $\nu_\odot$ and $\nu_{atm}$ measurements do not 
require $\nu_s$, \cite{CRE02}. 
\item 
In \cite{DEH02}, oscillations with three flavours are analysed. The number of degrees of freedom 
is the same as in the 2-$\nu$ analysis. The fit parameters are tan$^2\,\theta_{12}$, 
sin $\theta_{13}$, $\Delta m^2_{12}$ and f$_B$. Following the CHOOZ results sin $\theta_{13}
\simeq$ 0.04 therefore the survival probability is:
\beq
P^{(3)}\approx (1 - 2 sin^2 \theta_{13})P^{(2)}
\label{eq_probsurv3a}
\feq
where $P^{(2)}$ is the 2-flavour computed probability. The main effect of this angle is an
overall suppression of the survival probability.
The best fit point is in LMA region (for comparison two flavour analysis result is 
quoted):
\beqar 
3\nu ~ \Delta m^2 = 6.8 \cdot 10^{-5}(eV)^2 ~\tan^2\theta = 0.41 ~f_B = 1.09 ~\chi^2 = 66.2 
\nonumber \\
2\nu ~ \Delta m^2 = 6.15 \cdot 10^{-5}(eV)^2 ~\tan^2\theta = 0.41 ~f_B = 1.05 ~\chi^2 = 65.2  
\nonumber
\label{eq_23results}
\feqar 
The solution requires slightly higher value of the $^8$B $\nu_\odot$ flux.
The changes are small: an increase of $\theta_{13}$ worsens the 2 flavour results.
For LOW and SMA solutions an increase of $\theta_{13}$ leads to a slight improvement of the fit.
\item
In \cite{DOL02,RAF02} implications of LMA and other flavour oscillation solutions on cosmology
and astrophysics and $viceversa$ constraints on oscillation analysis deduced from
astrophysical data are nicely analysed.
\item
In \cite{FOG02c}, a quasi-independent analysis is proposed in the context of active $\nu$'s 
oscillations. The best-fit parameters are in LMA region while Q-VAC and LOW solutions are 
still acceptable at the 99\% and 99.73\% C.L., respectively; SMA solution is practically 
ruled out. LMA bounds appear to be more conservative than in  analyses done by remaining authors.
In particular, (at the 99.73\% C.L.), maximal mixing is marginally allowed in LMA region.
The inclusion of the winter-summer difference from GALLEX+GNO results decreases the likelihood 
of LOW solution because of the smallness of the measured value. 
Furthermore, probable changes in solar T$_c$ values induced by each $\nu_\odot$
component have been estimated in the case of LMA and LOW solutions. \\
In \cite{FOG02d} a GLOBAL analysis in term of three flavour oscillation is presented.
Upgraded solar experimental results, including winter-summer difference from both the Ga 
experiments and the latest SK complete data, are combined with CHOOZ, SK atmospheric and K2K 
measurements. It has to be underlined that terrestrial experiments pose a strong upper limit to
the mass value. The figures \ref{fig:FOG3sol} and \ref{fig:FOG3all} show the main results in usual
plots as a function of increasing value of $sin^2 \theta_{13}$ parameter. The best fit point is in
the same LMA point as for the two flavour analysis but LOW and Q-VAC solutions become less
disfavoured when $sin^2 \theta_{13}$ goes up. Similar conclusions have been obtained in 
\cite{DEH02}. These analyses have been done under the hypothesis of direct $\nu$ mass spectrum 
hierarchy, but negligible variations are present when inverse hierarchy is used.
\\
\item 
Experimental measurements strongly suggest $\theta_{12}\simeq 30^{\circ}$, 
$\theta_{23}\simeq 45^{\circ}$ and $\theta_{13} < 10^{\circ}$.
\item
The four flavour oscillation solution still appears to be highly disfavoured.
In a context including $\nu_e$ disappearance results (from reactor experiments) and $\nu_\mu$
disappearance searches (accelerators), there is a strong conflict (at a level of 3$\sigma$) with
LSND appearance results in "3+1" schemes. If "2+2" schemes are taken into account, including mixed
active+sterile oscillation, atmospheric data strongly support a pure 
$\nu_\mu \rightarrow \nu_\tau$ oscillation; on the contrary solar measurements prefer a pure
$\nu_\mu \rightarrow \nu_s$ solution. These results are clearly incompatible. For a detailed 
analysis see \cite{MAL02}.
\fenu
In table \ref{tabglobah02f} and table \ref{tabglobar02b} the parameters corresponding to 
several GLOBAL solutions as computed by different authors are shown. In table 
\ref{tabratesbah02f} the expected interaction rates for Cl and Ga experiments are also 
reported.\\
\begin{table}[ht]
\caption{\it{ Best-fit GLOBAL analysis oscillation parameters with all solar neutrino data.  
f$_B$ is the "corrective" factor with respect to the $^8$B $\nu_\odot$ flux predicted 
in \cite{BAH01a}. (s) means sterile solution. 
The number of degrees of freedom is  46 [44 (zenith spectrum) + 4 (rates) + 1 ($A_{DN} 
(CC)$) $-$3 (parameters: $\Delta m^2$, $\theta$, and f$_B$)]. Solutions having 
$\chi_{min}^2 \geq$ 45.2 + 11.8 = 57.0 are not allowed at the 3$\sigma$ C.L., adapted 
from \cite{BAH02f}.}}
\begin{center}
\begin{tabular}{|l|c|c|c|c|}
\hline 
 &$\Delta m^2$ (eV)$^2$&$\tan^2(\theta)$& f$_B$& $\chi^2_{min}$ \\ \hline 
LMA& $5.0\cdot 10^{-5}$  &0.42& 1.07 & 45.5 \\ \hline
LOW& $7.9\cdot 10^{-8}$  &0.61& 0.91& 54.3 \\ \hline
VAC& $4.6\cdot 10^{-10}$  &1.8& 0.77& 52.0 \\ \hline
SMA& $5.0\cdot 10^{-6}$  &0.0015& 0.89& 62.7 \\ \hline
Just So$^2$ & $5.8\cdot 10^{-12}$  & 1.0 & 0.46& 86.3  \\ \hline\hline\hline
VAC (s)& $4.6\cdot 10^{-10}$ & 2.3& 0.81& 81.6  \\ \hline
Just So$^2$ (s)& $5.8\cdot 10^{-12}$  & 1.0& 0.46& 87.1 \\ \hline
SMA (s)& $3.7\cdot 10^{-6}$ & 0.0047 & 0.55& 89.3 \\ \hline
\end{tabular}
\end{center}
\label{tabglobah02f}
\end{table}
\\
\begin{table}[ht]
\caption{\it{ The same as in table \ref{tabglobah02f}  but the quoted results 
are: 
upper part, left side from \cite{BAR02b} (72 degrees of freedom), 
right side from \cite{BAN02b} free rate analysis; 
lower part, left side from \cite{DEH02} (78 degrees of freedom),
right side from \cite{FOG02c} (79 degrees of freedom) 
}}
\begin{center}
\begin{tabular}{|l|c|c|c|c||c|c|c|} 
\hline 
 & $\Delta m^2$ (eV)$^2$ & $\tan^2 \theta$& f$_B$ & $\chi^2_{min}$&
$\Delta m^2$ (eV)$^2$ & $\tan^2 \theta$& $\chi^2_{min}$ \\ \hline
LMA&$5.6\cdot 10^{-5}$ & 0.39   & 1.09 &50.7  
   &$6.07\cdot 10^{-5}$ & 0.41   &40.57  
\\ \hline
LOW &$1.1 \cdot 10^{-7}$ & 0.46 & 1.03  &59.9 
    &$1.02 \cdot 10^{-7}$ & 0.60 &50.62 
\\ \hline
VAC &$1.6 \cdot 10^{-10}$ &  0.25 (3.98)  & 2.46 &  76.3  
    &$4.43 \cdot 10^{-10}$ &  1.1  &  56.11 
\\ \hline
SMA &$7.9 \cdot 10^{-6}$ & 0.002 & 4.85 &   108 
    &$5.05 \cdot 10^{-6}$ & 0.0017 &   70.97 
\\ \hline\hline
LMA &$6.15\cdot 10^{-5}$ & 0.41   & 1.05 &65.2   
    &$5.5\cdot 10^{-5}$ & 0.42   &73.4   
\\ \hline
LOW &$9.3\cdot 10^{-8}$  & 0.63 & 0.91  &77.6   
    &$7.3\cdot 10^{-8}$  & 0.67 &83.8  
\\ \hline
VAC &$4.5\cdot 10^{-10}$ &  2.1        & 0.75 &  74.9  
    &$6.5 \cdot 10^{-10}$ &  1.33      &  81.2   
\\ \hline
SMA &$4.6 \cdot 10^{-6}$ & 0.0005 & 0.57 &    99.7  
    &$5.2 \cdot 10^{-6}$ & 0.0011 &     96.9  
\\ \hline
\end{tabular}
\end{center}
\label{tabglobar02b}
\end{table}
\\
\begin{table}[ht]
\caption{\it{ Interaction rates and flux variations as 
expected for different oscillation solutions:  Cl experiment (left) and
Ga experiments (middle), adapted from \cite{BAH02f}; the right part
shows the fractional variation (\%) with respect to the values quoted in 
\cite{BAH01a} is shown, adapted from \cite{FOG02c}.
See table \ref{tabrateclga}, table \ref{tabgarates}, table \ref{tabexpflux}, 
table \ref{tabbah02crates}, table \ref{tabksrates} for a comparison with SMs 
predictions, experimental results and  predictions based on previous 
oscillation analyses.} }
\begin{center}
\begin{tabular}{|l|c|c|c||c|c|c|||c|c|c|c|} 
\hline 
 & LMA&LOW&VAC&LMA&LOW&VAC&LMA&LOW&VAC&SMA \\\hline 
 & SNU&SNU&SNU&SNU&SNU&SNU&\%&\%&\%&\\\\hline
p-p&0.0&0.0&0.0&40.4&38.2&40.3&0.0&+0.5&+1.1&+0.9\\\hline 
p-e-p&0.1&0.10&0.15&1.51&1.25&1.82&0.0&+0.8&+1.6&+1.3\\\hline 
Hep&0.01&0.02&0.02&0.02&0.03&0.04&-0.8&0.0&-1.8&+1.8\\\hline 
$^7$Be&0.62&0.53&0.46&18.6&16.0&14.0&+0.5&-5.5&-11.5&-9.9\\\hline 
$^8$B&2.05&2.26&2.34&4.35&4.72&5.00&+5.2&-12.2&-22.2&-24.2\\\hline 
$^{13}$N&0.04&0.04&0.05&1.79&1.56&1.83&-1.0&-8.3&-15.1&-11.4\\\hline 
$^{15}$O&0.15&0.15&0.18&2.83&2.44&3.01&-1.2&-8.3&-17.0&-12.8\\\hline 
$^{17}$F&0.00&0.00&0.00&0.03&0.03&0.04&--&--&--&--\\\hline\hline\hline
Total&3.03&3.11&3.21&69.6&64.2&66.0&--&--&--&--\\\hline \hline
\end{tabular} 
\end{center} 
\label{tabratesbah02f}
\end{table}

\subsection{Summary.} 
\label{sect:snpfinal}
Previously shown analyses and results introduce assumptions and solutions beyond the 
particle standard model. A question arises: is this model ruled out by astrophysical 
neutrino experimental results? \\
Accelerator experiments give negative answers: they do not show  violation to the particle 
standard model predictions, see for instance KARMEN, \cite{EIT00,ARM02}: 15 events have been  
detected while 15.8$\pm$0.5 events were expected. On the contrary, LSND's latest result, 
\cite{AGU01}, confirms the previous measurements, \cite{ATH98a,ATH98b}, indicating an
excess of 87.9$\pm$ 22.4$\pm$ 6.0 events: this implies an oscillation probability for $\nu$'s 
of 0.264 $\pm$ 0.067 $\pm$ 0.045 \%.\\  
Interesting results have been obtained in Japan by using KEK and SK facilities. K2K is a long 
baseline experiment aiming to establish $\nu$ oscillation in the $\nu_\mu$-disappearance and 
$\nu_e$-appearance mode with a well known flight distance, \cite{HIL01,HAS02}.  
A beam of $\nu_\mu$ is produced at KEK, by 12 GeV/c protons incident on aluminium target
of 3 cm diameter and 66 cm length; its purity is at the level of 98.2 \% of $\nu_\mu$.
At a distance of 300 m, a set of detectors (a 1 kton 
water-\v Cerenkov detector, a scintillating fiber tracker with a water target, lead-glass
counters and a muon range detector) analyses the main characteristics of the $\nu$ beam, 
whose direction is within 0.01 mrad from the direction to SK, which is 250 km far, \cite{AHN01}.  
The beam-line was aligned by GPS position survey.
80.6$^{+7.3}_{-8.0}$ events were expected in the case of no oscillation but 56 fully contained
events have been observed, \cite{ISH02}: more precisely 24 multi-ring events (expected
32.2$\pm$ 5.3) and 32 1-ring events (expected 48.4$\pm$ 6.8) have been detected.
Under the hypothesis of 2 flavour mixing and the conditions $\tan^2 \theta =\,1$ and 
$\Delta m^2 = 3\cdot 10^{-3}$ (eV)$^2$, 52.4 events are predicted.\\ 
When SK atmospheric data and K2K results are combined, $\Delta m^2$ value is "relatively well"
estimated: 2.7 $\pm$ 0.4$\cdot 10^{-3}$ (eV)$^2$, under the hypothesis of maximal mixing.\\
The main challenges for particle standard model come from astrophysical measurements: 
atmospheric, \cite{FUK98b}, and solar $\nu$ data give results in strong disagreement with 
the predictions.  \\ 
Unfortunately, the presently available experimental results do not allow to select with a 
great confidence "THE" right solution or to rule out definitively the remaining ones: new 
experimental data are needed.\\
Coming back to the $\nu_\odot$ physics, the main questions (problems) on fire are: 
\benu 
\item 
How great is the deficit for p-p and $^7$Be $\nu_\odot$ components?  
\item 
Is the ratio between $\nu_e$ and total $\nu_\odot$ flux, as measured by SNO, 
an "ultimate" signal for new physics of weak interactions? Is the flavour 
oscillation, mainly MSW effect with large mixing angle, the correct solution to the SNP? 
\item 
Are the solar modulations measurable? 
\item 
Are the low-energy interaction cross-sections right? How the nuclear transition calculations 
({\it{e.g.}} in excited states) are correctly computed? 
\item 
Could T$_c$, $\rho$ and their profiles inside the Sun be further constrained as suggested by SMs?
Does further upgrading of SMs (or new helioseismology results) avoid the  presently used 
assumptions (spherical symmetry, no effects due to $B_\odot$,  no mass-loss, no rotation...) 
and change the main results on $\nu_\odot$ flux? 
\fenu  

\section{The incoming future.}
\label{sect:incofuture}

\subsection{2002-2003: what news?}
\label{sect:whatnews}  
At present time (autumn 2002), three detectors are still continuing their data taking: 
GNO, SNO and SAGE. SNO will offer other interesting features of $\nu_\odot$ flux (but only at 
high energy) with new results concerning NC interactions. GNO and SAGE will increase their 
statistics and lower their experimental errors on the Ga interaction rate mainly based on 
low-energy p-p $\nu_\odot$'s. At the end of 2002, when new PMT's will be installed, SK will 
run again but with a higher energy threshold. \\
SK and SNO have shown that the survival probability of $^8$B $\nu_\odot$'s is energy-independent. 
The uncertainty of this component in the Ga interaction rate is $\pm$ 1.5 SNU, mainly due to 
the interaction cross-section on Ga. If Hep and F $\nu_\odot$ components are disregarded and 
when the $^7$Be $\nu_\odot$ flux will be measured, it will also be possible to estimate the 
net p-p $\nu_\odot$ flux by difference. This will strongly constrain the solar physics.\\ 
Next years could be fundamental to have such an answer: three new detectors (KAMLAND, BOREXINO 
and ICARUS) will give their experimental results.\\ 
KAMLAND (which has started its data taking in January 2002) and BOREXINO will detect ES and CC 
interactions and they will measure $^7$Be and CNO $\nu_\odot$ components. 
Then, a good estimate of p-p $\nu_\odot$ flux will be allowed. These experiments will check also 
the robustness of different $\nu$ oscillation solutions.\\
The main features of these detectors are reviewed. 

\subsection{KAMLAND.}  
\label{sect:kamland}
This experiment is mainly dedicated to the analysis of $\overline\nu$'s emitted by 
nuclear reactors but the study of terrestrial and solar $\nu$'s is also possible, \cite{ALI98}. 
In the case of reactor emission, the detected reaction is the inverse $\beta$-decay, a process 
having $E_{thr}=$ 1.804 MeV and a well known cross-section (the uncertainty being less 
than 1$\%$). \\ 
The experimental setup is homed at Kamioka mine and measures the 
$\overline\nu_e$ flux produced by Japanese nuclear reactors which are 
far from the detector no more than 350 km.  The $\overline{\nu_e}$ flux 
above $E_{thr}$ is  1.34$\cdot 10^6\,\nu\,cm^{-2}\,s^{-1}$, with an 
uncertainty of $\sim$ 1.4$\%$; the expected rate is $\sim$ 1100 events per year at an
energy ranging from 2 up to 8 MeV. $\overline{\nu_e}$'s are mainly produced from $^{235}$U, 
$^{239}$Pu and $^{241}$Pu ($\sim$ 90$\%$), the remaining contrubution being associated with 
$^{238}$U.\\
The detector consists of a spherical tank with a diameter of 18 m filled 
with a liquid  scintillator (1,2,4 trimethylbenzene (20$\%$), paraffin (80$\%$), with addition 
of PPO as wavelength shifter at a level of 1.5 g/l) having a fast component of 
5.4 ns and a slow  component of 37.5 ns. The quenching factor is 13.8.
The emitted light is collected by 1325 17"-PMT's and 554 20"-PMT's, \cite{PIE01}. 
The attenuation length is 100 m at 400 nm and 20 m at 450 nm while the light transparency is at 
a level of 93 $\%$ at 400 nm. The expected vertex resolution is better than 10 cm at E = 1 MeV;
the energy resolution is $\sim 5\%$/$\sqrt {E}$. The radiopurity is at a level of few 
$10^{-16}$ g/g for U, Th and $^{40}$K. The PMT coverage is $\sim$ 32$\%$, \cite{DEB00,PIE01}.\\
The predicted interaction rate for terrestrial  $\overline\nu_e$'s is $\sim$ 
800 events per year (2 events per day) with an  estimated background of 
$\sim$ 40 events per year (0.1 event per day). \\ 
The background due to neutrons produced by cosmic rays is suppressed by 
detector  location (a shield of 2700 mwe) and by a cosmic ray veto. \\
Data taking started in January 2002. 
If a radiopurity level lower than $10^{-16}$ g/g will be reached (even by 2 
order of magnitude for $^{40}$K background) KAMLAND will detect $^7$Be
$\nu_\odot$'s through elastic scattering interactions, the energy window
for scattered electrons varying from $\sim$ 300 to 800 keV. The expected
interaction rate is $\sim$ 460 events per year.\\ 
The measured positron energy  spectrum should allow a sensitive probe 
of oscillation effects, \cite{BAR01b}, see also \cite{GOU01,GON02,MUR02} for 
analyses concerning the possible impact of KAMLAND results on SNP. 

\subsection{BOREXINO.} 
\label{sect:borexino}
The detector is under installation in Hall C at LNGS and it is aimed to study purely  leptonic $\nu_\odot$ interactions 
at $E\geq$ 250 keV; it will allow to  measure the  $^7$Be $\nu$ flux, \cite{BEL95,ALI02c}.  It consists of a 
nylon transparent and spherical inner vessel (8.5 m), filled with 321 m$^3$ of liquid scintillator, 
pseudocumene as a solvent and PPO at the concentration of 1.5 g/l as solute. The target mass 
is viewed by  $\sim$ 2200 8" PMT's fixed on a supporting structure (13.7 m of diameter) plunged  in a water tank 
(18 m of diameter). The outer detector is a 200 PMT's muon veto system.\\ 
The effective light coverage will be at a level of $\sim$ 30$\%$. In such a detector the total energy, the position 
(with a mean accuracy of  $\sim$ 15 cm in X, Y and Z directions) and the time of each event will be measured. 
The total energy will be estimated with an accuracy ranging from 18 $\%$ at 250  keV to 5 $\%$ at 1 MeV, \cite{CAL00}.\\ 
The main sources of the background are the rock and  concrete of the 
laboratory and the radioactive contamination 
due to the detector materials. The external tank contains water as a 
shield against gamma rays and neutrons from the 
surrounding rock. Moreover, the water allows the detection of \v Cerenkov 
light emitted by cosmic muons crossing 
the detector. The expected background in the fiducial volume is $\sim$ 11 counts per day to be 
compared with the estimated unperturbed $\nu_\odot$ flux of 55 events per day, \cite{ALI02c}. 
BOREXINO will detect all flavour $\nu_\odot$'s $via$ ES interactions. In the electron kinetic 
energy window $T_e$ = 250 - 800 keV, the major contribution to the signal (78\%) is expected 
from a monochromatic line of $^7$Be $\nu_\odot$'s with the energy 863 keV, the remaining 
contributions being from $^{15}$O, $^{13}$N and p-e-p $\nu_\odot$'s ($10\%$, $7.2\%$ and $3.6\%$ 
respectively). \\
Among the big experimental difficulties the BOREXINO coll. had to get over, 
the scintillator  radiopurity required the major effort. A prototype called Counting Test 
Facility (CTF), with an inner nylon vessel (2 m of diameter), was installed 
at LNGS and their encouraging results concerning the level of $^{14}$C, U and Th, showed the 
feasibility of such experiment, \cite{BOR98}.\\  
At the end of 2002, the vessel will be filled (water and scintillator), then, tests will start;
in spring 2003 measurements should begin.\\  
After two years, BOREXINO should give precise measurements concerning $^7$Be $\nu_\odot$ 
interaction  rate. Owing to the large expected statistics, analyses on seasonal and other 
temporal variations  will be possible. 
Moreover, BOREXINO would allow searches for $\overline\nu_e$ coming from 
the Sun  and from other different cosmic sources, \cite{GIA01,RAN01}. 
The radiopurity of the detector, at level of $10^{-16}$ g/g for U and Th, 
the lowest value presently available, has to be pointed out: the so aquired 
experiences and high radiopurity techniques should be applied to 
different sectors, like in $\beta\beta$-decay experiments.  

\subsection{ICARUS.} 
\label{sect:icarus}
Many years ago a technique combining bubble chamber features with an electronic read-out was 
suggested, \cite{RUB77}, then, a 3000 t liquid Ar time projection chamber detector was 
proposed in 1989, \cite{ICA89,ICA94,ICA95}.   
Its main feature is the high quality  (similar to that of a heavy liquid chamber) of "images" that can be obtained 
thanks to its granularity (1 mm). Such a detector is homogeneous and continuously sensitive,  so that it is perfectly 
suitable to estimate the energy of a contained particle.  It allows particle identification: proton-decay, atmospheric 
$\nu$ interactions, high  energy $\nu$'s from accelerator beams and 
naturally $\nu_\odot$ interactions,   
\cite{RUB96,MON99}. \\ 
A 3 t prototype was built at CERN and the experience gained was the technical  basis for further projects; at present 
a 600 tons module was constructed. \\ 
This detector should be able to measure $\nu_\odot$ interactions in two ways: 
\bit 
\item by neutrino scattering off electron: \hspace*{0.5cm} $\nu_e\,+\,e^-\,\rightarrow \nu_x\,+\,e^-$ at $E\geq$ 5 MeV; 
\item by interaction on $^{40}$Ar: \hspace*{0.5cm} 
\(\nu_e\;+^{40}Ar\,\rightarrow e^-\;+^{40}K^* \rightarrow \;^{40}\!K_{g.s.}\,+  \gamma's\). 
\fit 
The expected number of events per years, under standard assumptions, is  $\sim$ 230 by ES interactions (40 events 
from background) and $\sim$ 1440  by absorption (180 events from background).\\ 
This detector will check the $\nu$ oscillation hypothesis measuring the ratio  between ES and 
$\nu_e$ absorption. This reaction is expected to  proceed through two channels: Fermi transition to the 4.38 MeV 
excited isobaric  analogue K$^*$ state or Gamow-Teller transition to the excited levels below  the 4.38 MeV K$^*$ 
state. It will be possible to distinguish between these processes  by the energy and multiplicity of $\gamma$ ray 
emitted in the de-excitation of the  K$^*$ states and by the energy spectrum of the primary electron.\\ 
The module T600 has been tested at Pavia and it should be installed in LNGS at the end of 2002; 
its startup is foreseen in 2003, \cite{ICA00,ICA01a,ICA01b}.  

\subsection{KAMLAND and BOREXINO: is the "final" answer incoming?}
\label{sect:finalanswer}
\begin{figure}[ht]
\begin{center} 
\mbox{\epsfig{file=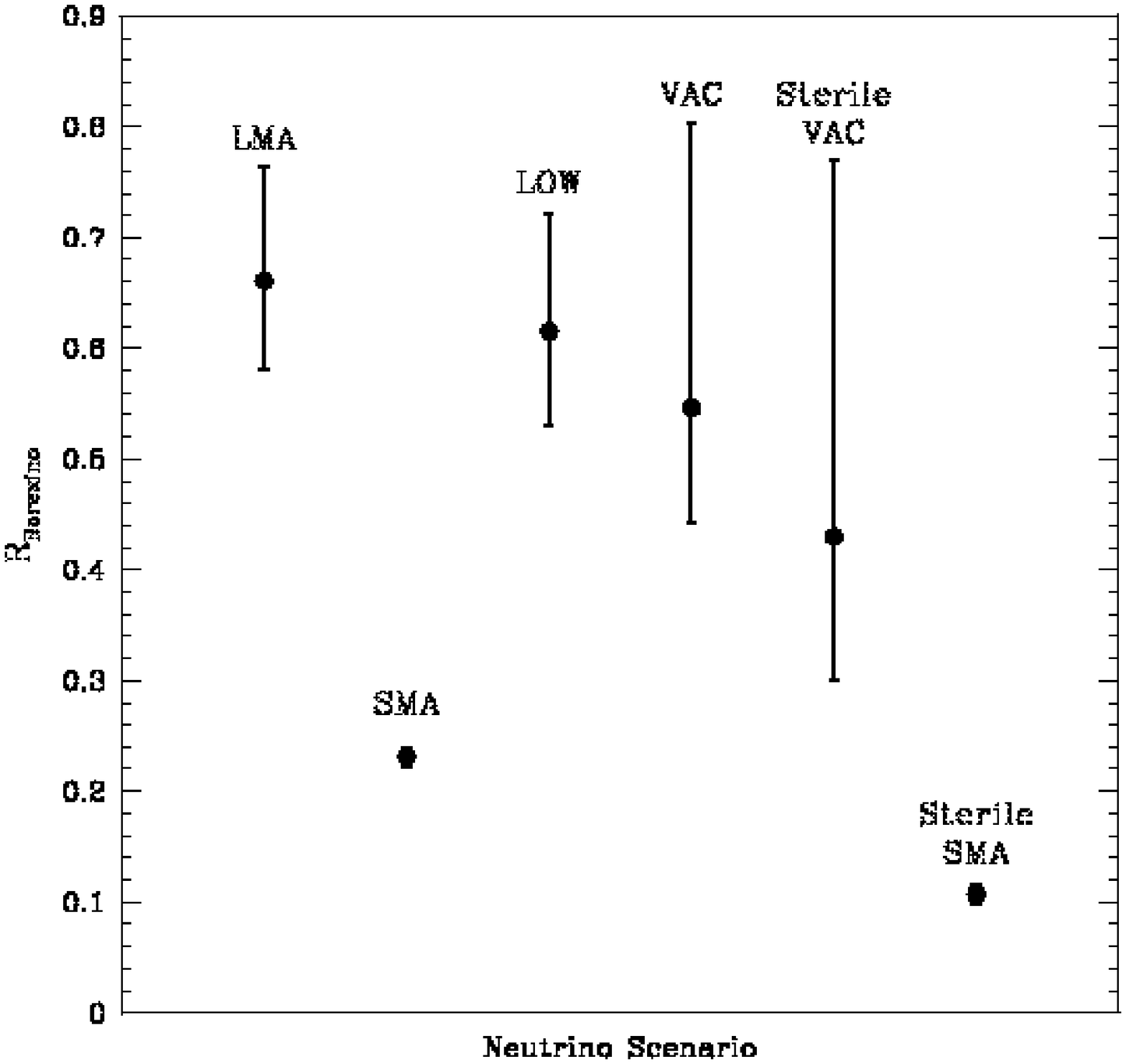,width=0.5\textwidth}} 
\end{center}
\caption{ The event rate predictions for BOREXINO experiment. The circles give values of $R_{BOR}$ 
in the best-fit points of the GLOBAL analysis solutions. The error bars show intervals 
corresponding to the 3$\sigma$ allowed regions, from \cite{KRA02}.  }
\label{fig:KSBe7}
\end{figure}

\begin{figure}[ht]
\begin{center} 
\mbox{\epsfig{file=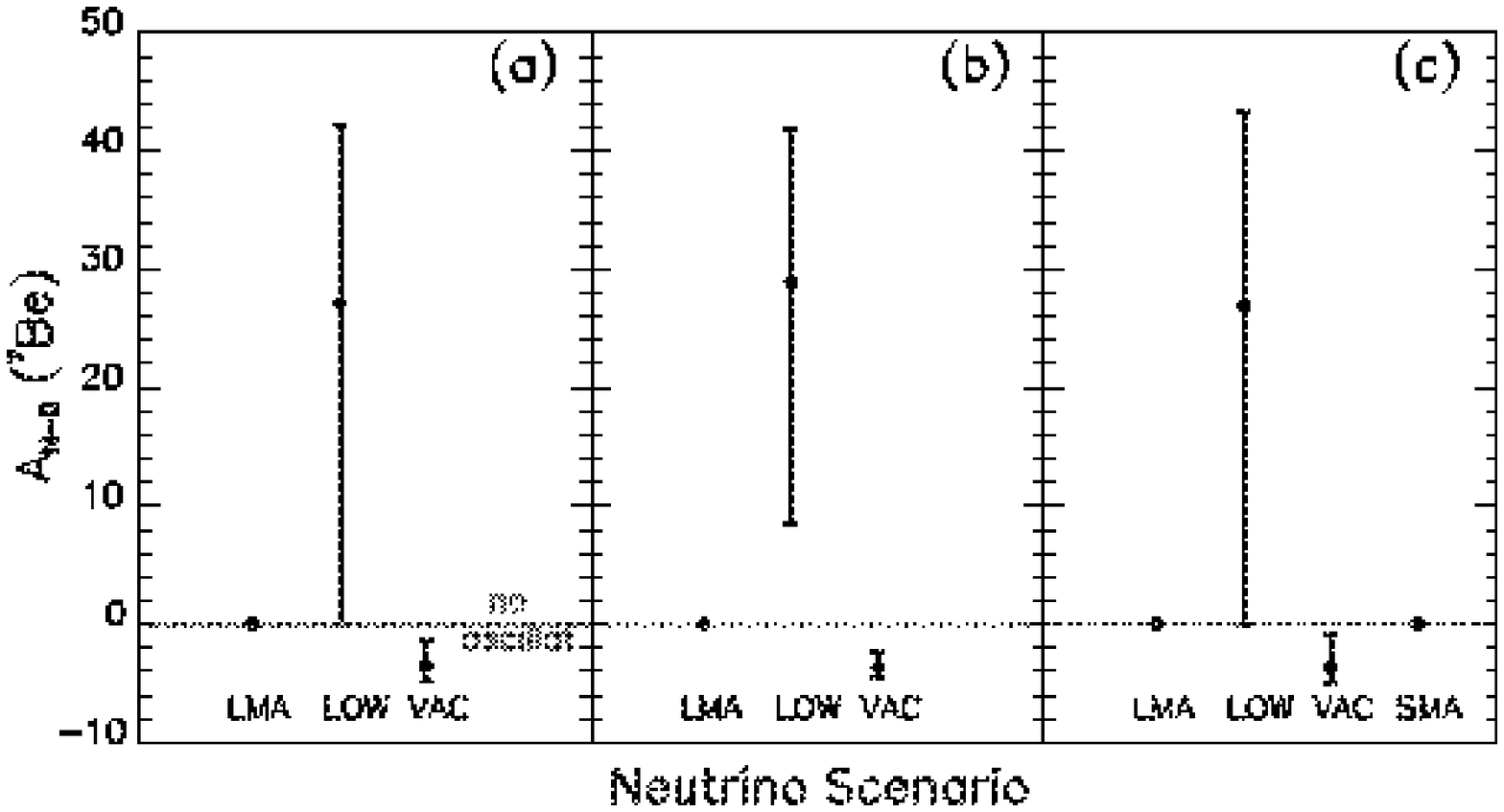,width=0.5\textwidth}} 
\end{center}
\caption{ The predicted percentage differences between D and N rates in BOREXINO for 
recoil electrons with kinetic energies in the range 0.25 - 0.80 MeV.
The three panels refer to results for different analysis strategies: in a) modified $^8$B flux is used, 
\cite{BAH02c}, in b) SK event rate is also included, in c) the $^8$B $\nu$ flux is a free parameter,     
from \cite{BAH02c}. }
\label{fig:BAH150borexdn}
\end{figure}

\begin{figure}[ht]
\begin{center} 
\mbox{\epsfig{file=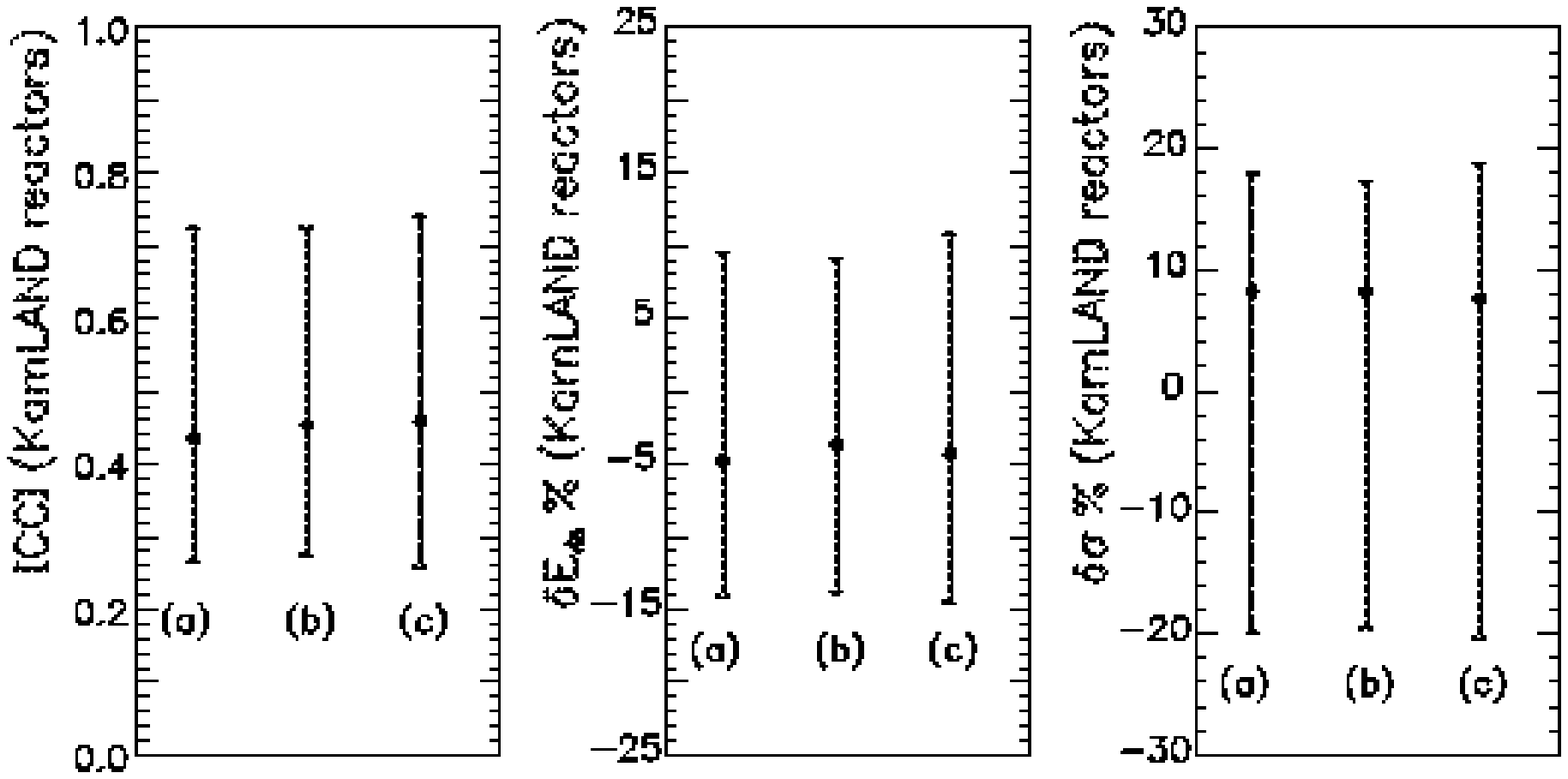,width=0.5\textwidth}} 
\end{center}
\caption{ The $3\sigma$ predictions concerning KAMLAND experiment for the CC and the first
and second moments of the visible energy spectrum with respect to the
expectations based upon \cite{BAH01a,BAH02c}. The visible energy threshold is 1.22 MeV.
Results based on different analysis strategies are shown: in a) modified $^8$B flux is used, 
\cite{BAH02c}, in b) SK event rate is also included, in c) the $^8$B $\nu$ flux is a 
free parameter, from \cite{BAH02c}. }
\label{fig:BAH150kaml}
\end{figure}

KAMLAND and BOREXINO can detect the $^7$Be $\nu_\odot$ flux and strenghten or weaken the presently proposed solution
to the SNP. We shortly summarize in table \ref{tabpredicbah02f} the available predictions 
concerning these experiments, as given in \cite{ALI01,STR01b,BAH02c,BAH02f,BAR02b,KRA02}, and 
the quoted comments.\\
\begin{table}[ht]
\caption{\it{ Best-fit predictions and uncertainties (3$\sigma$ ranges) by using currently available
$\nu_\odot$ data. The threshold of the recoil electron kinetic energy, used in computing the SNO
predictions, is $5$ MeV. For the BOREXINO experiment, electron recoil energies are between 0.25 and 0.8 MeV. 
The results for the KAMLAND reactor observables are computed for E$_{thr}$=1.22 and 2.72 MeV.
Columns nr. 2 and 3 show the predicted values before the latest SNO results,
columns nr. 4 and 5 include NC events in the analysis. 
The p-p $\nu_\odot$ scattering on electron ratio, with respect to the \cite{BAH01a} predictions for future real-time 
detector, is also shown. The analysis allows a $^8$B $\nu_\odot$ free flux,
adapted from \cite{BAH02f}.} }
\begin{center}
\begin{tabular}{|l|c|c||c|c|}
\hline
Observable& LMA $\pm$ 3$\sigma$&LOW $\pm$ 3$\sigma$&LMA $\pm$ 3$\sigma$&LOW $\pm$ 3$\sigma$ \\ \hline
A$_{N-D}$(SNO CC)(\%)&4.4$^{+11.4}_{-4.4}$&1.3$^{+3.9}_{-1.3}$ &4.7$^{+9.1}_{-4.7}$ &2.7$^{+2.7}_{-2.1}$ \\ \hline   
$[$R($^7$Be)$]$& 0.66$^{+0.09}_{-0.07}$ &0.59$^{+0.13}_{-0.06}$ & 0.64$^{+0.09}_{-0.05}$ &0.58$\pm$0.05 \\ \hline
A$_{N-D}$($^7$Be)(\%)& 0.0$^{+0.1}_{-0.0}$ & 15$^{+17}_{-15}$ &0.0$^{+0.1}_{-0.0}$ &23$^{+10}_{-13}$ \\ \hline
$[$CC$]$(KAMLAND)& & & & \\ 
(E$_{thr}$=2.72 MeV)&0.56$^{+0.20}_{-0.34}$ & --&0.49$^{+0.25}_{-0.26}$ &-- \\ 
(E$_{thr}$=1.22 MeV)&0.57$^{+0.16}_{-0.31}$ & --&0.52$^{+0.20}_{-0.25}$ &-- \\ \hline
$\delta E_{vis}$(KAMLAND) (\%) & & & &  \\ 
(E$_{thr}$=2.72 MeV)&-7$^{+14}_{-4}$ & --&-7$^{+14}_{-4}$ &-- \\ 
(E$_{thr}$=1.22 MeV)&-7$^{+15}_{-7}$ & --&-9$^{+17}_{-5}$ &-- \\ \hline
p-p $\nu$'s - e$^-$ scattering & & & & \\
(T$_{thr}$ = 100 keV)& 0.722$^{+0.085}_{-0.067}$ & 0.689$^{+0.058}_{-0.065}$&
0.705$^{+0.073}_{-0.049}$ & 0.683$^{+0.035}_{-0.042}$ \\
(T$_{thr}$ = 50 keV) & 0.718$^{+0.086}_{-0.069}$ & 0.689$^{+0.058}_{-0.068}$&
0.700$^{+0.074}_{-0.050}$ & 0.677$^{+0.038}_{-0.045}$ \\ \hline
\end{tabular} 
\end{center} 
\label{tabpredicbah02f}
\end{table}
\bit
\item 
The predicted ratio for BOREXINO and for CC KAMLAND event rate are not significantly affected by latest 
SNO results, see fig. \ref{fig:KSBe7}.
\item 
Only LOW solution predicts a consistent D-N asymmetry; LMA solution foresees a 
negligible variation while VAC solution has small and negative value due to the Earth-Sun distance and to the  longest 
nights occuring in the Northern hemisphere when the distance diminishes, see fig. \ref{fig:BAH150borexdn}.
\\
In KAMLAND and BOREXINO predicted D-N asymmetry are very similar: for LMA solution 
the values agrees at a level of $0.1$\%; LOW solution has a maximum value for KAMLAND 
which is  $\sim 2$\% less than for BOREXINO. VAC solution 
predicts a minimum value of -3.9\% for KAMLAND which has to be compared with -4.8\% for BOREXINO.\\ 
Fig. \ref{fig:BAH150kaml} shows the computed rate and the distortion of CC interaction spectrum in KAMLAND, which can be 
as large as 20\%: its value, if non-zero, will  strongly constrain the range of the oscillation parameters.
\\
In practice, the measurement of ES interactions is a critical test either for SMs and for 
oscillation solutions: a large value for the D-N effect would imply the correctness of the LOW solution.
\item 
VAC solution would be ruled out if a rate depletion or a spectral distortion in the KAMLAND reactor 
experiment will be observed. Moreover, it predicts a $\Delta m^2$ value too small to lead to an observable effect 
with KAMLAND. A strong signal for VAC solution would be the observation of a large seasonal 
variation in BOREXINO, with a monthly dependence of the observed rate; a D-N effect at a level of $\pm 8$\% 
associated with this seasonal variation is also possible. In BOREXINO the rates predicted by LMA, LOW and VAC solutions
are very similar.
\item 
KAMLAND could indicate LMA as a good solution if a consistent deficit in reactor $\overline \nu$'s will be detected.
\item 
SMA will be the best solution if BOREXINO will detect no $\nu_\odot$'s; 
in any case the SK energy spectrum has to be thought as wrong. 
\item 
Just So$^2$ solution will be the right answer if the ratio CC to NC events as measured at SK and SNO will change, 
indicating no oscillations, and BOREXINO data will show a depression in $^7$Be $\nu_\odot$ flux. 
\item 
NO OSCILLATION will be  the right solution to the SNP if BOREXINO will find no $\nu_\odot$'s 
and if the ratio CC/NC in SNO data is incorrect. 
\item 
BOREXINO should be able to discriminate among the non-standard solutions 
to the SNP, \cite{NUN01a}. Except for the possible direct evidence of the 
LMA solution KAMLAND could provide, BOREXINO is able to distinguish 
between flavour oscillation solutions and RSFP. When latest SNO and Ga 
data are included in analyses, also RSFP show a little shift in the 
best-fit parameter values allowing an even clearer distinction between 
RSFP and oscillation signatures in BOREXINO predictions. In \cite{CHA02} 
such a difference for the LMA solution is possible to more than $5.7\sigma$ whereas for the 
LOW solution all predictions are more than $4.5\sigma$. The only possible model dependence 
of RSFP solution is related to the choice of B$_\odot$ profile, which is constrained by the 
requirement of fitting all $\nu_\odot$ data. In table \ref{tabborexino} the expected interaction
rate in BOREXINO as computed in flavour oscillation and RSFP solutions are compared with the 
prediction in \cite{BAH01a} is shown. \\
In $\overline \nu_{\mu,\tau}$ scenario no D-N asymmetry is foreseen but 
27-day modulation and annual periodicity having the maximum value in September and March 
should constitute a signature of such a mechanism. When the new profiles, proposed in 
\cite{CHA02}, are analysed, a reduction factor of $\sim$ 0.45, for strong 
B$_\odot$ existing in the radiative zone, or of $\sim$ 0.35, if 
B$_\odot$ is acting in convective zone, is predicted. 
\begin{table}[ht]
\caption{\it{ The expected interaction rate (events per day) in BOREXINO 
computed following \cite{BAH01a} and within flavour oscillation and RSFP GLOBAL 
solutions, see also table \ref{tabrsfprates}. The $B_\odot$ profiles (1-6) are from 
\cite{PUL00}, the profile (4) is from \cite{PUL01}}.}
\begin{center}
\begin{tabular}{|c|c|c|c|c|c|} 
\hline 
SSM &LMA &LOW &1   &6   &4   \\\hline
55.2&35.3&32.0&15.5&22.6&19.3\\\hline
\end{tabular}
\end{center}
\label{tabborexino} 
\end{table}
\fit

\section{The next generation of detectors.} 
\label{sect:nextgeneration}
The knowledge on $\nu_\odot$ physics has been greatly enhanced  by SNO's 
results; when BOREXINO and KAMLAND data will be available,  a more 
exhaustive answer will be provided. Perhaps, the main questions could 
have "THE" right way out and only details should remain to be 
clarified. In any case, only next experimental data will give an answer.\\  
Before latest SNO results, many physicists thought that real-time detectors 
measuring p-p $\nu_\odot$ flux and spectra were the "ultimate" weapon to 
solve SNP. Many proposed experiments were based on fascinating techniques, 
but a lot of difficulties has prevented their construction up to now. 
The measurement of energy and direction for each 
event produced by $\nu_\odot$'s should allow 
the on-line estimation of background and a higher signal to noise ratio, 
the spectroscopy of $\nu_\odot$'s, the $\nu_\odot$ astronomy, \cite{BAR01b}. 
We shortly resume, alphabetically ordered, the main features of proposed $\nu_\odot$  
detectors, their performance and present status. 

\subsection{CLEAN.} 
\label{sect:clean}
CLEAN (Cryogenic Low Energy Assay of Neutrinos) experiment, which would detect $\nu_\odot$ 
elastic scattering off electron, proposes to use of liquid He or Ne as a target for 
$\nu_\odot$'s in order to have a good real-time spectroscopy, \cite{MCK00,MCK01}. 
Such a detector should measure the large UV scintillation produced by 
ionizing radiation; a good transparency, with respect to the emitted light, 
and a low background levels are required. UV radiation, $via$ wave length 
shifter, should then be converted to optical photons, allowing the 
detection with PMT's. \\ 
A cryogenic scintillator test facility was proposed: a spherical 
geometry with the external diameter up to 2 m, working temperature of 
$\sim$ 25 K and a central active region of $\sim$ 100 kg are the main 
characteristics. Neon purification, internal background evaluation and 
optical properties of cryogenic scintillator will be attentively analysed.
A radiopurity level as low as in BOREXINO should be needed for U, Th, $^{42}$Ar
and $^{85}$Kr, the most serious one.
\subsection{GENIUS.}
\label{sect:genius} 
GENIUS project (GErmanium in liquid NItrogen Underground Setup)
is based on the powerful capabilities of a set of "naked"  
Ge detectors merged in liquid N$_2$: its main goal is to search for dark 
matter and $\beta \beta$-decay. The minimum external dimension 
of N$_2$ tank will be  13 m diameter and 12 m height. An array of 400 detectors of 
enriched $^{76}$Ge as a  target, 2.5 kg each, is foreseen, \cite{KLA99,BAU00,KLA01b,KLA01a}.
The total target mass of natural Ge corresponds to 3$\cdot 10^{29}$ electrons.\\  
If a background level of 0.001 events per year per kg per keV in the energy  region from 0
to 260 keV will be reached, 
this detector will  be able to measure in real-time low-energy component of $\nu_\odot$'s $via$ 
ES interactions. The maximum electron recoil energy is 261 keV for  p-p $\nu_\odot$'s, with an 
expected rate of 1.8 events per day, and 665 keV for $^7$Be  $\nu_\odot$'s, 0.6 events per day; 
2.4 events per day is the total prediction while the expected background is of $\sim$ 
0.6 events per day. This detector should be installed in LNGS. \\ 
The background is due to both external and internal factors; the former comes 
from to showers, photons, neutrons from rock and direct or secondary 
cosmic ray muon component. After the reduction of the cosmic ray component, 
the main part of the background  comes from N$_2$ and Ge themselves. 
The N$_2$ must reach a radiopurity as low as in BOREXINO's scintillator
for $^{238}$U, $^{232}$Th, $^{40}$K. The Ge presents another challenge, 
due to the possible cosmogenic activation on Earth's surface: the best 
solution should be the complete building phase in underground laboratories. \\ 
If S/N will be greater than 1, the $\nu_\odot$'s detection will be allowed by  spectroscopic 
analysis only, on the contrary the possible solar signature of  each event will be computed.\\ 
A test facility was  approved in 2001 to check the feasibility of the whole detection 
technique and its building step will start as soon as possible, \cite{BAU02}. 

\subsection {HELLAZ.}
\label{sect:hellaz} 
HELLAZ (HELium tpc at Liquid AZote temperature) 
experiment should detect $\nu_\odot$'s in a pressurized He TPC (Time Projection Chamber)
(its volume should be $\sim 2000\,m^3$) having the aim 
of measuring the purely leptonic interactions at E$\geq$ 100 keV. Owing to  the low Z of He, 
it should be possible to deduce the energy of incoming $\nu$,  event by event, and to study 
the flux and the energy spectrum of the p-p  $\nu_\odot$'s with an error 
of $\sim$ 4 $\%$, \cite{BON94}.\\ 
The detector should be filled with 7 t of He + CH$_4$ (95-5 $\%$) added for technical 
reasons. Different combinations of temperature and pressure were presented: 
20 bar/ 300 K, 10 bar/ 140 K, 5 bar/ 77 K \cite{BON01a}. The addition of CH$_4$ is very 
dangerous because of a consistent background due to $^{14}$C $\beta$-decay.\\ 
The calculated rate of ES interaction events from p-p and $^7$Be $\nu_\odot$'s is $\sim$ 10  
events per day taking into account  the electron threshold energy and the detector fiducial 
volume.\\ 
The sensitive volume will be shielded by CO$_2$ blocks, a steel tank and a 
thermal  insulating material. Ionisation electrons, generated by ionizing 
particles which cross the  active volume, are drifted in an uniform axial 
electric field and detected by a MW system. \\ 
The dominant background is $\gamma$-e Compton scattering, its cross-section being  
$10^{20}$ times larger than that of $\nu_e$. HELLAZ would allow a rejection rate of  
$\sim 10^4\,\gamma$ per day mainly thanks to the tracking procedure and to the energy  
measurement. The future of this project is not completely defined. 

\subsection{HERON.}
\label{sect:heron}  
HERON (HElium Roton Observation of Neutrinos) experiment, \cite{LAN01}, is aiming
to realize a real-time detector measuring $\nu_\odot$ elastic scattering off electron 
with a low energy threshold and a high rate. The target is superfluid He 
at T= 0.05 K: a mass of $\sim$ 20 tons, with a fiducial mass of $\sim$ 5 tons, would 
allow an interaction rate of $\sim$ 12 events per day.\\  
When a $\nu$ interacts, a single recoil electron is produced and stopped 
in the liquid.  The $^7$Be $\nu_\odot$'s produce a flat distribution  with 
energy ranging from 0 to 665  keV; on the contrary a p-p $\nu_\odot$'s show 
a slowly falling down spectrum from 0 to  261 keV. The recoil track in He 
is shorter than 2 cm and it seems essentially a point source of radiation.\\ 
The idea is to measure both phonon and UV photon within same calorimetric  
device. In short all the detection process: the initial energy loss of a 
stopping particle is by ionisation; at the ending point, the 
primary energy is distributed among phonons, rotons, photons and some few long-lived dimers,  
depending on the primary ionisation along the track. \\ 
If phonons/rotons reach the liquid free surface and have an energy greater 
than the binding energy of He itself (8$\cdot 10^{-4}$ eV), they should 
begin a quantum evaporation. \\ 
The scintillation UV photons derive from radiative decay of He dimers produced along the track 
and their energy ranges between 14 and 20 eV,  16 eV being the maximum. A silicon and sapphire 
wafer could detect  these photons and produce a prompt heat pulse. Photons and phonons could be  
distinguished by their different arrival times on wafer.\\ 
A 5 l. prototype with Si or Al$_2$O$_3$ readout was built to test the  feasibility of this 
unusual technique and the performance of the detector. \\ 
About the background problem: 
\bit 
\item 
The impurities in superfluid He attach to walls or fall to bottom  (nothing is soluble in He 
and all the elements are frozen; moreover, the  gravitational energy is greater than thermal 
energy so that all the particles fall down). 
\item  
Long-lived isotopes do not exist and the first excited state is at  $\sim$ 20 MeV. 
\item At E$\leq$ 3 MeV 
$\gamma$-conversions are Compton scatters ($>$ 95 $\%$); this is the most dangerous background. 
\item 
Through-going $\mu$'s are rejected by pattern recognition and energy deposit ($\sim$ 450 events 
per day at LNGS and $\sim$ 11000  at Soudan); in any case their interactions in He are negligible. 
\item 
There are $\beta$-decays in Si-wafers ($\sim$ 30 kg) and 1.28 MeV $\gamma$'s  from $^{22}$Na 
($\leq$ 3 events per day).  
\item 
Photons play the main role because the event location and the consistency with signal topology 
are used for energy determination and for background  rejection. 
\fit 
Technological problems are connected with the requirement of low background in  refrigeration 
systems, cryostat and shielding, in the $^3$He purification, in the energy  threshold of 
calorimeter and in the SQUID electronic, which will 
be adapted to large area  Si-wafers.\\ 
MC simulations, in which the distribution of photon incident on the wafers 
from a  source anywhere in the detector is used to determine the most 
probable position of an  event technique, are in progress. The major 
sources of background are Compton  $\gamma$-ray in liquid He from 
cosmogenically produced isotopes in cryostat materials  or in any material 
which contacts He. A MC analysis including many possible background related 
parameters  combined with different shapes and sizes of fiducial-non 
fiducial volume is on run.\\ 
Other topics under analysis are the single 16 eV photon sensitivity on 
wafer sensor, the efficiency of the "coded aperture" for event location 
and background  signature reconstruction and the use of a new magnetic 
calorimeter allowing greater absorbing area. A saphyre readout system with 
a geometry closer to the  final needed version is under construction. 

\subsection{LENS.}
\label{sect:lens}  
LENS (Low Energy Neutrino Scintillator) project was proposed many years ago aiming to 
measure p-p and $^7$Be CC $\nu_\odot$ interaction on isomeric excited states of rare 
earth nuclei. An important step was the discovery of 3 stable target systems 
($^{176}$Yb, $^{160}$Gd, $^{82}$Se) for $\nu$ capture with isomeric tag, 
\cite{RAG97}. The decay of these states produces a distinct, time-delayed 
$\gamma$-ray, located  at, or 
around, the capture side that allow to tag the $\nu_\odot$ interaction in  space and time. 
$\nu_\odot$ interaction produces a transition $0^+\,\rightarrow 1^+$  to an excited state in 
the final nucleus; the emitted $\gamma$ has 
E$\sim$ 100 keV while the mean lifetime of the isomeric states is $\sim$ 100 ns.\\ 
The estimated interaction rate in a 20 tons of natural Yb detector is $\sim$ 370 events per 
year: $\sim$ 180 events from p-p $\nu_\odot$'s, 140 from $^7$Be  $\nu_\odot$'s, 10 from p-e-p 
$\nu_\odot$'s and 40 from NO $\nu_\odot$'s. The reaction is $^{176}$Yb $(\nu_e,e^-)\,^{176}$Lu 
at E$\geq$ 301 keV ($^{176}$Yb has a good isotopic abundance, $\sim$ 13 \%). It is also 
possible a $^7$Be $\nu_\odot$ NC interaction  on $^{173}$Yb at E$\geq$ 482 keV, \cite{CRI00}.\\ 
A 25 tons of Gd should have the same flux, the threshold being E$\geq$ 244 
keV.\\ 
The $\nu$ energy is given by the sum of kinetic energy of electron and the Q value of  the 
reaction. This experiment could have a good energy resolution while the background should 
be rejected thanks to  a  good time resolution. Moreover, electron-$\gamma$ coincidence should 
help to achieve a background reduction at low energy so that it should 
allow a measurement of the low-energy part of the $\nu_\odot$ spectrum. \\ 
Among different aspects, we underline the influence of the use of extractants in rare earth 
technology, \cite{RAG00a,RAG00b}. In 2001 an aromatic scintillation solvent loaded with 
organo-metallic Yb compounds was realized. Background reactions gave serious troubles:
many decays produce two signals correlated in space and time as the $\nu_e$ tag 
($^{231}$Th, $^{169}$Yb, $^{176}$Lu). At present time, the great amount of difficulties 
in electronics and in signal identifications when Yb is used, have stopped the evolution of 
this project. \\ 
Indium liquid scintillator detector was recently proposed, \cite{RAG01}; the first reaction 
is $\nu_e\,+^{115}I \rightarrow ^{115}Sn\,+\,e^-$, then $^{115}$Sn de-excites 
(T=4.76 $\mu$s)  giving rise to a 115.6 keV electron (or $\gamma$ in 4$\%$ of the decays) and to 
a 497.3 keV $\gamma$. The low Q-value (118 keV) allows a practically complete detection of 
$\nu_\odot$ spectrum. Moreover, the delayed coincidence has strong signature, a relatively high 
energy ($\sim$ 613 keV) and a high isotopic abundance ($\sim$ 96 $\%$).\\ 
The electron can be tagged as a product of $\nu_e$ capture by a unique delayed space-time  coincidence of radiations 
de-exciting the isomeric state. In a 4 tons $^{115}$In detector, an interactions rate of 1 p-p $\nu_\odot$ per day is 
foreseen, \cite{RAG76,RAP85}. \\ 
The only but terrible obstacle is the $\beta$ decay, with t= 6.4$\cdot 10^{14}$ y, end-point at 495 keV and activity of 
0.25 Bq/g, which overlaps the p-p and  $^7$Be $\nu_\odot$ signal. The use of In liquid scintillator with specific 
solvents  and "prompt" and "spatial" coincidences should allow a modular detector with a  sufficient efficiency and 
accurate rejection of impurities and background. Among a  lot of technical difficulties, we mention the amount of 
phototubes having small section, 2 cm x 2 cm, ($\sim$ 10000), and the fine resolution needed  to reject background 
events produced by bremsstrahlung.   

\subsection{LESNE.}
\label{sect:lesne}  
LESNE (Lithium Experiment on Solar Neutrinos) is a radiochemical experiment which should use Li 
in metallic form as target nucleus; it was  proposed by a Russian-Italian working group, 
\cite{ZAT97}. A great  advantage of this detector is the compactness; for instance technical 
troubles are due to the required  special handling of Li in the melted form. Thanks to the 
relative low threshold a clear  detection of higher energy line from $^7$Be and from p-e-p 
$\nu_\odot$ should be allowed. The Li mass adequate to perform precision measurements of seasonal 
variations is 100 tons. The $^7$Be $\nu_\odot$ features should be checked by 
means of a cryogenic microcalorimeter, \cite{GAL97a,GAL97b,KOP02}.   

\subsection{MOON.}
\label{sect:moon}
MOON (MOlibdenum Observatory of Neutrinos) experiment is based on a detector using $^{100}$Mo 
as a target, the main aim being the search for $\beta\beta$-decay. 
Among many interesting features, the low energy threshold, 168 keV, the real-time detection 
and a good spatial resolution are the most promising ones. An event rate of $\sim$  5 events 
per day with a 10 tons detector is foreseen, \cite{EJI00a,EJI00b,EJI01b,EJI01a}. \\ 
The physical process is a $\nu$ absorption which induces an inverse $\beta$-decay  followed by 
a $\beta$-decay to $^{100}$Tc with a mean life of 23 s. Two electrons are emitted and the 
identification of the$\nu_\odot$ production reaction is allowed by the measurement of the 
energy of the inverse $\beta^{-}$.
The time window between the  emission of both the electrons allows a rejection  of background 
reactions and accidental $\beta\beta$-decay which is  the main background component and the 
most challenging problem to overcome.\\ 
The detector should consist of 34 tons (3.3 tons of $^{100}$Mo) purified at a level of 
$10^{-3}$ Bq/ton respective of $^{238}$U and $^{232}$Th (similar level was achieved in 
SNO but for different materials). A plastic scintillator set of foils is responsible for the 
electron detection. Wavelength shifter fiber light collectors and PMT's are also present.  
The energy resolution for $^7$Be $\nu_\odot$'s would be of about 15$\%$.\\ 
The originally proposed detector (6x6x5 m) consisted of Mo foils and plains of scintillator  
with an expected time resolution of $\sim$ 1 ns but   
a more compact version (2.5x2.5x1.0 m) enriched in $^{100}$Mo up 
to $85\%$ was recently proposed, \cite{EJI01b}, the lower dimensions allowing a gain in 
energy resolution, in light collection, in cosmogenic background and in signal to noise ratio. 

\subsection{MUNU.}
\label{sect:munu} 
A TPC detector using CF$_4$ were proposed \cite{ARP96} with  essentially the same goals as HELLAZ. Obviously CF$_4$ 
is not as good as He if  one takes into account only angular resolution but a smaller number of 
technical  problems is present, i.e. it is not necessary to use a pressurized vessel.\\  
The MUNU collaboration has realized a TPC looking for the $\nu$ magnetic moment and the 
interactions induced by  reactor $\overline\nu_e$'s: it has published the first measurements of  
energy and direction of the recoil electron at E$\geq$ 0.3 MeV. \\
The main characteristics of the detector are: 
\bit 
\item 
The filling gas has high density and relatively low  atomic number, so that the multiple 
scattering is reduced, being the cosmogenic activation of C and F at a low level. 
\item 
It does not contain free protons.  
\fit 
An unexpected background due to $\alpha$'s and electrons, mainly from Rn,  was reduced by a 
factor of $\sim \,10^3$ after installation of new components  (purifier and cathode). The peak 
of events from the reactor is starting to come out, \cite{BRO02,BRO01b}.\\ 
The present central detector is an acrylic cylinder of 0.9 m inner diameter and 1.62 m long 
filled with CF$_4$ at 3 bar pressure. This vessel is mounted inside a stainless steel tank 
filled with a liquid scintillator, used as a cosmic muons veto and anti-Compton, viewed by 48 
PMT's.\\   
The MUNU detector could be exploited as a low background prototype for the spectroscopy  of 
$\nu_\odot$'s in the energy region below 1 MeV. It has the great advantage to work at atmospheric 
pressure and room temperature. In a detectore having a volume of 200 $m^3$, filled with gas at 
a pressure of 1 bar, the total target mass being of 740 kg, the expected interaction rate per 
day is 0.4 events from p-p $\nu_\odot$'s, 0.4 events from $^7$Be $\nu_\odot$'s and 0.1 events
from p-e-p, $^{13}$N and $^{15}$O $\nu_\odot$'s, \cite{BRO01a}.  The energy threshold is 100 keV.
The background due to Rn and $^{14}$C requires further studies; only a very high radiopurity and a 
sure signal identification procedure will allow the useful feasibility of such a detector 
in $\nu_\odot$ physics. 

\subsection{TPC.}
\label{sect:tpc} 
A proposal based on TPC technique has been done by 
\cite{BON01a,BON01b,BON02b}.  It should be a 
cylinder 20 m long and $\sim$ 20 m in diameter filled with 7-10 tons of gas and separated from the rock by $\sim$ 3 
mwe of high purity shielding.  The gas should be boil-off He in bulk with a small component of natural gas, He(97\%), 
CH$_4$(3\%) at 10 atm pressure: in this way metals  and many radioactive or electronegative gases, from H$_2$O vapor 
to Rn, are  eliminated. The TPC cylinder, having a slight overpressure, is enclosed in an external  pressure vessel. 
Teflon gaskets are used to seal the juncture of the barrel and endcap. The 
philosophy is the one of a device having electronic channels and materials 
coming into contact with the gas as little as possible. The end caps are 
the detector planes and contain a single set of wires, reconstructing the $(x,z)$ profile of a track, and a single 
set of strips, giving the $(y,z)$ profile, $z$ being the drift direction. 
The vertex and energy are reconstructed by combining both track views. More than 
30000 electronic FADC channels are needed for TPC.  100 
keV tracks are considered the TPC ultimate benchmark. Such a track at 10 atm has a total length of 9 cm and generates 
2500 electrons that will drift to the anode while diffusing.\\  
Resolution effects are dominated by angular resolution which depends significantly on multiple scattering: from 
simulations a value of 15$^\circ$ at 100 keV and 10 Atm, and decreasing like $T^{-0.6}$, $T$ being the electron 
kinetic  energy, has been estimated.\\ 
This detector should use MUNU results as starting point for R\&D. 

\subsection{UNO, HYPERKAMIOKANDE, TITANIC.} 
\label{sect:uno}
The philosophy of these proposals is to "enlarge" the SK detector by a large  factor, 
\cite{JUN00}, using the same detecting technique (\v Cerenkov light emission). \\   
Among different proposed geometries for the UNO project (Ultra underground Nucleon decay and 
neutrino Observatory), the most feasible one is a  linear 3-modular detector  60 m 
large, 60 m high and 180 m long. The best motivations  are mainly due to the current PMT 
pressure stress limit (8 bar for 20"-PMT's)  and to the finite light attenuation length in 
pure water, 80 m at a wavelength of $\sim$ 400 nm. This multi-purpose "swimming pool" would 
foresee to use only the medium  tank to $\nu_\odot$ searches 
for, the remaining tanks being dedicated to proton-decay  and galactic $\nu$'s analysis and, 
at the same time, acting as a lateral shield for $\nu_\odot$ central 
detector. The fiducial volume should be of $\sim$ 450000 tons. \\ 
A great advantage of this proposal detector is obvious: new technique is not  needed. The 
problems seem to be (only) the funding and the location but only higher energy $\nu_\odot$'s 
should be detected even with an  energy threshold maybe lower than in SK experiment (4.0 MeV ?). \\ 
HyperKAMIOKANDE is a  second gigantic detector, its proposed mass being 1 Megatons.  
A "linear" modular solution and a "toroidal" version were proposed, \cite{NAK00}.  
The only presently viable location should be the Kamioka mine, with the advantage of the 
existing SK facilities.\\ 
A further and more ambitious project, TITANIC, foresees a fully merged in  sea water, at depth 
ranging from 100 to 200 m, fourfold modular detector, (50x50x100 m)  or 
(70x70x100 m), with a fiducial mass of $\sim$ 2 million tons, \cite{SUZ00b}. 

\subsection{XMASS.}
\label{sect:xmass} 
The experiment XMASS (Xenon MASsive detector for Solar neutrinos) uses
liquid Xe, a sufficiently known scintillator without the background problems related to 
C atoms, \cite{SUZ00a}. \\ 
The interaction process is based on excitation or ionisation of Xe atoms which then go into 
excited Xe$^*_2$ states emitting some time later UV photons at 173 nm.  The recombination 
time is variable, depending on exciting particle, from $\sim$ 40 ns  in the case of electron 
down to 3 ns if $\alpha$'s are present. This feature allows  a pulse shape discrimination. 
The attenuation length is not known but depends upon  the level of purity. The PMT's can be 
placed either in the liquid (at a working  temperature of 165 K) or outside the detector. \\  
A great technical advantage is due to the possibility of using liquid N$_2$ 
to liquefy Xe; furthermore, Xe has both high 
density and high atomic number and it allows a good  self-shielding so that the background problems seem to be not 
too hard to overcome. As a final advantage Xe produces proportional scintillation by multiplication process in  an 
electric field.\\ 
The cosmogenic component of background does not produce long lived Xe-isotopes while the  
$^{85}$Kr and $^{42}$Ar presence is due to their boiling temperatures which are lower  than Xe.
An absorption column removing this component is under analysis. $^3$H can be  removed 
$\it{via}$ chemical process. The U/Th chain should require a contamination level 10 times 
lower than in CTF-Borexino experiment; the feasibility seems to be related to the pulse 
shape discrimination. If the detector will be installed at Kamioka mine, a reduction of $\sim$ 
4 orders of magnitude of neutron flux is needed. A water shield is possible.\\ 
The main difficulties come from the $^{136}$Xe 2$\nu$ $\beta\beta$-decay: the only possible 
solution is an isotopic separation, with enrichment or depletion. A "fair" solution should be 
a $\nu_\odot$'s  module based on odd Xe enriched sample, an even enriched module for 
$\beta \beta$ decay while both the components should be used to search for dark matter.\\ 
A 3 kg detector viewed by new type 2"-PMT's was built to check the feasibility of such a 
technique. A 100 kg facility is presently under construction and installation at Kamioka mine: 
the main goal is to test energy reconstruction, electron/$\gamma$ separation, neutron background, 
self-shielding purification, attenuation length and new PMT's.

\subsection{Further Proposals.}

\label{sect:furtherexp}
Other detectors based on different techniques have been recently suggested; among them we 
mention: 
\bit
\item
COBRA, a $\beta\beta$-decay experiment, which would use CdTe semiconductors and measure in
real-time $\nu_\odot$'s at E$\geq$ 366 keV, \cite{ZUB01,ZUB02}.
\item
SIREN, which should employ Gd as a scintillator: a test module is being designed for 
installation at Boulby mine in U.K., \cite{AKI02,KUD02}.
\item
An hybrid detector, with I and Cl as a target mass, installed at Homestake 
mine, when the future of this mine will be clarified.
\fit
    
\section{Summary and conclusions.} 
\label{sect:summary}
Solar neutrino physics is presently an active and exciting field of 
research: in the last years our knowledge has been greatly enhanced but a 
lot of problems remains unexplained. 
The available experimental results should require a new theoretical effort 
because they imply a particle physics beyond standard model (if the flavour 
oscillation is the right solution to the SNP, then neutrinos do  have a mass).\\ 
We can try to resume the present status in $\nu_\odot$ physics in this way: 
\benu 
\item 
The experiments have detected $\nu_\odot$'s but a deficit is present when  compared to the 
predictions of SMs, which are reinforced by the helioseismological observations.
This deficit is critical mainly for the Ga experiments, which measure low energy component 
of $\nu_\odot$ flux: their interaction rate seems to be lower than the minimum value allowed 
by $L_\odot$ (energy conservation law), which is independent from the adopted SM. 
\item 
The ratio among NC, CC and ES interactions has been (finally) measured with high 
precision by SNO; SK results, which are based on high statistics, show that 
the energy spectrum of the $^8$B component seems to be undisturbed. Being 
true these results, the $\nu_\odot$ flux deduced from NC measurements at 
SNO, well agree with the predictions of SMs; on the contrary either spectral 
distortions or a $\nu$ component without interactions enhances the 
differences. This conclusion is based on data at high energy: no direct 
individual measurements of components at E$\leq$ 1 MeV ($\sim$ 98\% of the 
total $\nu_\odot$ flux) are presently available. 
\item 
It appears very difficult to reproduce the experimental data by changing the
solar physics. The preferred solution to the SNP is in term of $\nu$ flavour oscillations. 
If the detected $\nu$ flavours are not mass eigenstates but a  superposition of 
mass eigenstates, $\nu$'s can change flavour. 
Solar data are well described in terms of 2 flavour oscillations: different regions in 
the parameter space (mixing angle $\theta$ and mass difference $\Delta\,m^2$) were 
obtained at the beginning by experimental measurements. 
Present results seem to be in agreement with predictions given by a solution with 
resonant oscillation in matter and large mixing angle.
\item 
It has to be remarked that first and strong data suggesting $\nu$ flavour oscillations come 
from observation of atmospheric and solar neutrino interactions; 
at present time LSND and K2K experiments confirm these signals but other detectors are still
starting.
\item 
The present detectors (GNO, SAGE, SNO and SK, when operating) will continue their data taking.  
KAMLAND and BOREXINO will measure the $^7$Be $\nu_\odot$ flux; at this point, the  p-p 
$\nu_\odot$ net component will be estimated from GNO-SAGE data and a right solution 
to the SNP will be more evident. Before the end of 2002, KAMLAND measurements concerning
$\overline \nu_e$ from reactors will constraints, if any, oscillation solutions.
\item 
Real-time low-energy detectors, which should allow even a detailed analysis of time dependence 
of $\nu_\odot$ flux, were proposed but many technical problems 
are presently existing (mainly the background estimation and reduction): 
their feasibility and operativity seem to be far. 
\item 
It was proposed, \cite{TUR01b}, to transform GNO experiment in a permanent solar observatory, 
due to its possibility to check the inner solar behaviour by detecting low-energy $\nu_\odot$
component: this should be a return to the beginning of $\nu_\odot$ experimental physics, when
Cl experiment was proposed to study the features of the Sun.
\item 
Present SMs well reproduce solar features despite their assumptions. Numerical techniques and 
self-consistent methods to realize a more refined and complete SM, testable with experimental 
data, are at their beginning. It is needed to have new low-energy nuclear physics experiments 
to better understand and estimate nuclear parameters entering the SMs, even if the laboratory 
conditions are different from the solar plasma. 
\item 
In two flavour $scenario$ among active neutrinos, an important task is to
refine the neutrino squared mass difference and the mixing angle. In order 
to accomplish this task more precise measurements and calculations of the 
$\nu_e$ survival probability $P_{ee}$ and of related observable quantities are needed. 
It is important to stress that improving the statistical analysis and the 
evaluation of the uncertainties is an important task in $\nu_\odot$ physics.
After the observation of the disappearance of atmospheric $\nu_\mu$ and of solar 
$\nu_e$ and the upcoming tests at long-baseline accelerator and reactor 
experiments, an era of searches for smaller effects is starting. 
Moreover, there is no "direct" proof for oscillation patterns in vacuum or in matter.
\fenu 
As a concluding remark it could be useful to remember the final comment to the neutrino 
"birth":\\  
{\it{"...I admit that my expedient may seem rather improbable from the
first, because if neutrons ($\nu$, of course) existed they would have been
discovered long since. Nevertheless, nothing ventured nothing gained..."}} 
(Wolfgang Pauli)\\  
The right philosophy in searching for solutions could be suggested by 
a Sir A. Conan Doyle's sentence: \\ 
{\it{"..It is an old maxim of mine that 
when You have excluded the impossible, whatever remains, however improbable, must be the 
truth}}."\\ 
Therefore, let us underline the proposal quoted in \cite{OHL02}: 
if (or when) neutrino oscillation will be the right answer to the SNP, a neutrino 
oscillation tomography could be in a far future a way to look for the density 
inside the Earth and, maybe, to discover metals and petroleum.\\ 
A worthy conclusion is in our opinion an early prophecy done by L.A.Saeneca in his book 
"Quaestiones naturales":\\ 
{\it{"..Veniet tempus quo ista quae nunc latent in lucem dies extrahat et longioris aevi 
diligentiae.... Veniet tempus quo posteri nostri tam aperta nos nescisse mirentur.}}"  

\section{Acknowledgements.} 
We wish to thank many people: among them we mention E.Bellotti, C.Cattadori, G.L.Fogli and
E.Lisi for useful discussions, S.Couvidat and S.Turck-Chi\`eze  
for information on their solar models, C.Broggini, O.Cremonesi, A.Strumia and 
F.Vissani for their suggestions. \\
Very special thanks to N.Ferrari, B.Ricci and L.Zanotti for a lot of precious comments 
during the manuscript preparation. \\
Thanks again to J.N.Bahcall, E.Bellotti for GNO coll., M.Boulay and S.Oser 
for SNO coll., C.Giunti, M.C.Gonzalez-Garcia, E.Lisi, O.L.G.Peres, 
B.Ricci, A.Y.Smirnov, M.B.Smy for SK coll. and S.Turck-Chi\`eze which have authorized the 
reproduction of plots and figures from their articles.\\
Figure nr. 1 is reproduced by permission of Elsevier Science.\\
Figures nr. 2, 3, 4, 5, 6, 7, 8, 9, 10, 11, 12, 17, 18 are reproduced by 
permission of American Astronomical Society. \\
Figures nr. 15, 16, 20, 21, 22, 23, 24, 25, 29, 30, 33, 36, 40 are reproduced 
by permission of American Physical Society. \\
Figures nr. 19, 27, 28, 31, 32, 34, 35, 37, 38, 39, 41, 42 are reproduced by permission of 
Journal of High Energy Physics.\\
Figures nr. 13, 14, 26, are reproduced by courtesy and permission of the authors.

\newpage
\section{Appendix: WEB pages.}

We also list useful web-pages in $\nu_\odot$ physics:
\benu
\item J.N.BAHCALL = www.sns.ias.edu/~jnb
\item BAKSAN = www.inr.ac.ru/INR/Baksan
\item KAMIOKA = www-sk.icrr.u-tokyo.ac.jp/
\item LNGS = www.lngs.infn.it
\item Neutrino Oscillatory Industry = www.hep.anl.gov/ndk/hypertext/nuindustry
\item Neutrino Unbound = www.to.infn.it/~giunti/NU/
\item BOREXINO = borex.lngs.infn.it
\item EXO = hep.stanford.edu/neutrino/exo
\item GALLEX = www.lngs.infn.it/site/exppro/gallex
\item GENIUS = www.mpi-hd.mpg.de/non\_acc
\item GNO = www.lngs.infn.it/site/exppro/gno
\item HELLAZ = cdfinfo.in2p3.fr/experiences/hellaz
\item HERON = www.physics.brown.edu/research/cme/heron
\item HOMESTAKE = cpt1.dur.ac.uk/scripts/explist
\item ICARUS = www.aquila.infn.it/icarus
\item KAMLAND = www.awa.tohoku.ac.jp/kamland
\item LESNE = al20.inr.troitsk.ru/~beril
\item MOON = ewi.npl.washington.edu/moon
\item MUNU = www.unine.ch/phys/corpus/munu
\item SAGE = ewi.npl/washington/edu/sage
\item SK = www-sk.icrr.u-tokyo.ac.jp
\item SNO = www.sno.phy.queensu.ca
\item UNO = superk.physics.sunysb.edu/nngroup/uno
\fenu

\end{document}